\documentclass[12pt]{report}
\usepackage{epsfig}

\newcommand{\beq}{\begin{equation}}
\newcommand{\eeq}{\end{equation}}
\newcommand{\beqn}{\begin{eqnarray}}
\newcommand{\eeqn}{\end{eqnarray}}

\newcommand{\e}{\mathrm e}

\catcode`\@=11
\def\lsim{\mathrel{\mathpalette\@versim<}}
\def\gsim{\mathrel{\mathpalette\@versim>}}
\def\@versim#1#2{\vcenter{\offinterlineskip
    \ialign{$\m@th#1\hfil##\hfil$\crcr#2\crcr\sim\crcr } }}

\def\be{\begin{equation}}
\def\ee{\end{equation}}
\def\bea{\begin{eqnarray}}
\def\eea{\end{eqnarray}}
\def\ba{\begin{array}}
\def\ea{\end{array}}

\def\0{$\Gamma_0$}

\newcounter{saveeqn}

\begin{document}
\begin{titlepage}
\begin{center}
{\LARGE \rm {COUPLINGS IN SO(10) GRAND UNIFICATION}} \vskip 0.5in
A dissertation presented  by \vskip 0.25in by \vskip 0.25in
{\Large {\it Raza M. Syed}}
 \vskip 0.25in
to

The Department of Physics \vskip 0.25in in partial fulfillment of
the requirements for the degree of

Doctor of Philosophy \vskip 0.25in
 in the field of

\vskip 0.25in
Theoretical Particle Physics \vskip 1.25in
Northeastern University, Boston, Massachusetts

 August, 2005
\end{center}
\end{titlepage}
\newpage
\begin{center}
{\LARGE COUPLINGS IN SO(10) GRAND UNIFICATION} \vskip 0.5in by
 \vskip 0.5in {\Large
{\it Raza M. Syed}} \vskip 3.0in {\Large {ABSTRACT OF
DISSERTATION}} \vskip 0.5in
 Submitted in partial fulfillment of the requirements for the
 degree of Doctor of Philosophy in Physics in the Graduate
 School of Arts and Sciences of Northeastern University, August
 2005.
\end{center}
\vskip3.0in

 \newpage

\begin{center}
{\LARGE ABSTRACT}
\end{center}
In this thesis, we develop techniques for the analysis of $SO(2N)$
invariant couplings which allows a full exhibition of the $SU(N)$
invariant content of the  spinor and tensor representations. The
techniques utilize a so called Basic Theorem which we first
derive. Using this method an evaluation of the trilinear couplings
of the 16 plet of matter and of the 16 and $\overline{16}$ plets
Higgs  is given. In particular, we give a full determination of
couplings in their $SU(5)$ decomposed form, involving $16~16$
 and the 10, 120 and $\overline{126}$ tensor fields together with $\overline{16}~16$
($16~\overline{16}$) and the 1, 45 and 210 tensor fields. We also
compute the vector couplings of $16~16$
($\overline{16}~\overline{16}$) and the 1, 45, 210 gauge fields.
Computation of dimension-5 operators formed from $16$ and
$\overline {16}$ arising from the mediation of 1, 10, 45 and 210
plet of heavy Higgs, are also analyzed. Complete supersymmetric
vector couplings belonging to the singlet and the adjoint
representation of the $SO(10)$ gauge group are computed in the
Wess-Zumino gauge. An $SU(5)\times U(1)$ decomposition of
 the  vector couplings ${16}-16-210$ is completely
 carried out using the Wess-Zumino gauge and its transition to a non
 linear sigma type  model is shown. The utility of these results
 in the analysis of quark-lepton textures and of proton decay is
 briefly discussed. Possible future applications of the techniques
 given here such as in the development of string landscape $SO(10)$
type models is briefly pointed out.

\newpage

\begin{center}
\large  Acknowledgements
\end{center}
I wish to thank my thesis advisor, Professor Pran Nath, for whom I
have the greatest admiration, not only for his immense powers as a
theoretical particle physicist but also for his wisdom. It is only
because of his constant guidance and invaluable knowledge that I
am able to complete my dissertation work.

I would like to dedicate this dissertation to my parents Mr. Syed
Junaid Gauhar and Mrs. Syeda Khalda Gauhar.  They instilled in me
the importance of education and have given me unconditional love
and support throughout my educational career.  I would also like
to thank my sisters Syeda Farah Gauhar and Syeda Deeba Rashid for
their encouragement over the years.


\tableofcontents

\vspace{0.2in}

\noindent {\bf
~~~~~~Bibliography~~~~~~~~~~~~~~~~~~~~~~~~~~~~~~~~~~~~~~~~~~~~~~~~~~~~~~~~~~~~~~~~~~~~~~~~~~~~~~~~~~~~~~~~~~~~214}
\vspace{-0.5in}

\listoftables

 4.1~~~~Particle content of the MSSM

4.2~~~~Anomaly cancellations in MSSM spectrum

\vspace{0.2in}

\chapter{Introduction}\label{First chapter}
\section{The group SO(10) }

The group $SO(10)$ is an interesting possible candidate for
unification of interactions\cite{georgi,fm} and there has been
considerable interest recently in investigating specific grand
unified models based on this group. Thus  models based on the
$SO(10)$ gauge group have many desirable features allowing for all
the quarks and leptons of one generation to reside in its
irreducible 16 plet spinor representation  and allowing for a
natural splitting of Higgs doublets and Higgs triplets. Further,
the complex spinor representation naturally contains the right
handed neutrino which is needed in See-Saw analyses of Majorana
masses. The $SO(10)$  group also offers an enormous freedom in
choosing symmetry breaking patterns.

Progress on the explicit computation of $SO(10)$ couplings has
been less dramatic. Thus while good initial progress occurred in
the early nineteen eighties in the introduction of oscillator
techniques\cite{sakita,wilczek,nandi}, there was little further
progress on this front till recently when a technique was
developed by Pran Nath and the author of this thesis. This method
was devised by using the oscillator method  which allows for the
 explicit computation of $SO(2N)$ invariant couplings\cite{ns1}
 in terms of irreducible $SU(N)$ tensors. It was also
shown in Ref.\cite{ns1} that the new technique is specially useful
in the analysis of couplings involving large tensor
representations. These large tensor representations such as 120,
126 and 210, have already surfaced in the analyses of quark,
charged lepton and neutrino mass textures in several unified
models based on
$SO(10)$\cite{gn,hrr1,hrr2,m,bm,abmrs,cm,bpw,bhrr,ab,brt}.
However, a full analysis of the matter interactions of such large
representations does not exist in the literature in any explicit
form. Therefore one needs to address the question of fully
evaluating these couplings in terms of Standard Model particle
states.

The focus of this thesis is the analytic determination of
supersymmetric $SO(10)$ interactions in terms of irreducible
$SU(5)$ fields.  We develop here a simple technique for the
explicit evaluation of the couplings in terms of the physical
degrees of freedom even for the case when the tensor
representation that couples has a large dimensionality. As
mentioned above, our technique is a natural extension of the work
of Refs.\cite{sakita,wilczek} which introduced the oscillator
expansion in the analysis of $SO(2N)$ interactions. We wish to
point out that there are other techniques \cite
{xi,Fukuyama:2004ps} and in particular purely group theoretic
methods to compute the Clebsch-Gordon coefficients in the
expansion of $SO(10)$ invariant couplings. Such group theoretic
approach was used in Ref.\cite{anderson} to compute the $E_6$
couplings. Our approach is field theoretic and is specially suited
for the computation of $SO(2N)$ couplings. The analysis given here
will greatly facilitate further $SO(10)$ model building and
investigation of the detailed properties of baryon and lepton
number violation, proton decay, neutrino oscillations and other
low energy phenomena in this class of models.

\section{Outline of the thesis }

The  outline of  the thesis is as follows: In order to fully
understand and appreciate the properties associated with $SO(N)$
groups, we give a thorough and rigorous presentation on these
groups in chapter 2. In particular, we discuss vector, tensor and
spinor representations of $SO(N)$ group, their complex and reality
properties, $SO(2N)$ group algebra in $SU(N)$ basis, $SO(2N)$
invariants and specialization to $SO(10)$ case. In chapter 3 we
outline an overview of the need for grand unification and its
attempts.

In chapter 4, we develop techniques for the decomposition of
$SO(2N)$ interactions in terms of $SU(N)$ invariants \cite{ns1}.
This technique (\emph{The Basic Theorem }) uses a unique  set of
reducible $SU(N)$ tensors in terms of which the $SO(2N)$
invariants have a straight forward decomposition. The Basic
Theorem is specially useful for couplings involving large tensor
representations and is central to the computation of
 any $SO(2N)$ invariant couplings.

In chapter 5, we exhibit the technique developed in chapter 4 by
performing a complete determination\cite{ns1} of the trilinear
couplings in the superpotential  and the Lagrangian for the case
of $SO(10)$ involving the 16 plet of matter and $\overline{16}$
plet of Higgs, i.e., we give a full determination of the
$16-16-10$, $16-16-120$ and $16-16-\overline{126}$ couplings. The
possible role of large tensor representations in the generation of
quark lepton textures is discussed\cite{ns1}. It is shown that the
couplings involving $\overline{126}$ dimensional representation
generate extra zeros in the Higgs triplet textures which can lead
to an enhancement of the proton decay lifetime. These results also
have implications for neutrino mass textures.

In chapter 6, we carry the analysis a step further and give a
complete evaluation\cite{ns2} of the $\overline{16}-16$ couplings
which involve the $SO(10)$ tensors 1, 45 and 210. Further,
technically the couplings of $\overline{16}-16$ are not
necessarily the same as of $16-\overline{16}$ (as will be
explained later). Thus, we give a full evaluation of the following
couplings in their $SU(5)$ decomposed form: $\overline{16}-16-1$,
$16-\overline{16}-1$, $\overline{16}-16-45$,
$16-\overline{16}-45$, $\overline{16}-16-210$ and
$16-\overline{16}-210$. An analysis of vector couplings:
${16}^{\dagger}- 16-1$, $\overline{16}^{\dagger}-\overline{16}-
1$, ${16}^{\dagger}- 16-45$,
$\overline{16}^{\dagger}-\overline{16}- 45$, ${16}^{\dagger}-16-
210$ and $\overline{16}^{\dagger}-\overline{16}- 210$, is also
given.

In chapter 7, an analysis is given of the quartic
couplings\cite{ns2} in the superpotential obtained from the
elimination of the singlet, the 45 plet and 210 plet of heavy
Higgs fields from the cubic superpotential. Specifically, we
compute $[\overline {{16}}~{16}]_1[\overline {{16}}~{16}]_1$,
$[\overline {{16}}~{16}]_{45}[\overline {{16}}~{16}]_{45}$ and
$[\overline {{16}}~{16}]_{210}[\overline {{16}}~{16}]_{210}$.
Further, the possible role of large tensor representations in
model building is discussed\cite{ns2}.

In chapter 8, we present the complete supersymmetric couplings in
the Wess-Zumino gauge containing the singlet and 45 of $SO(10)$
and decompose the result into $SU(5)$$\times$U(1) particle
states\cite{ns3}. The question arises how one may construct the
couplings of a 210 vector multiplet which does not belong to the
adjoint representation. Here we need a new technique to  address
this question. The purpose of the next chapter is to do just that.

In chapter 9,  we consider the couplings of the supersymmetric
vector 210 multiplet\cite{ns3} in $SO(10)$. We focus on this
construction both for the theoretical challenge of constructing
such couplings as well as for the possibility that such
interactions may surface in some future effective theories to
describe fully all the degrees of freedom at some relevant energy
scale. We follow the conventional approach and give the coupling
of the $210$ multiplet with $16$ plet of matter, i.e., we compute
the couplings ${16}^{\dagger}-16-210$ and
$\overline{16}^{\dagger}-\overline{16}-210$ in the Wess-Zumino
gauge and carry out a full $SU(5)\times U(1)$ decomposition of it
and elimination of the auxiliary fields. At the very outset we
discard the constraint of gauge invariance since the imposition of
such a constraint is untenable for the 210 multiplet. We consider
the more general couplings of the $210$ multiplet retaining all
the components of the vector multiplet, i.e, we do not impose the
Wess-Zumino gauge constraint\cite{wz}. In this case elimination of
the auxiliary fields leads to a non-linear Lagrangian with
infinite order of nonlinearities in it. The general technique
underlying this procedure is illustrated in Appendix C for the
U(1) case. This analysis has some resemblance to the analysis of
Ref.\cite{fayet1,fayet2} which also used a unconstrained vector
multiplet. However, the analysis of Ref.\cite{fayet1,fayet2} did
not include an explicit mass term for the vector multiplet, nor
the self interactions of the vector fields and it did not
integrate the auxiliarly fields. In fact the motivation of the
work of the above reference was very different in the sense that
it was geared to study spontaneous symmetry breaking and
generation of vector boson masses. Finally, in chapter 10 we
present our conclusions.

\chapter{The SO(N) Group}
In this chapter we present in detail the properties associated
with $SO(N)$ groups\cite{sakita,wilczek,nandi,liealgebra,van} that
are of interest to a theoretical particle physicist.
Specialization to $SO(10)$ gauge group is also discussed.

\section{{SO(N)} algebra. Vector representation}

\noindent\textsc{Definition}\\
 A \emph{group} $G$ is a collection of elements $g_i$ such that \\
(1)~~ There is an identity element, ${\bf{1}}$\\
(2)~~ There is closure under multiplication: $g_1\times g_2=
g_3$\\
(3)~~ Every element has an inverse: $g_i \times g_i^{-1}={\bf{1}}$\\
(4)~~ Multiplication is associative: $(g_i \times g_j)\times
g_k=g_i\times (g_j\times g_k)$\\

Specifically, we have the continuous groups which have an infinite
number of elements. The most important of the continuous groups
are the Lie groups. The Lie groups of particular interest to us
are all given as concrete groups of matrices. Clearly, a set of
arbitrary $N\times N$ invertible matrices satisfies the definition
of group.
\\
\\
\noindent \textsc{Definition}\\
The \emph{orthogonal group, $O(N)$} is a group of $N\times N$ real
orthogonal matrices, $R$ obeying
\begin{equation}
R^TR=RR^T={\bf{1}}
\end{equation}

\noindent It is a group of rotation matrices in an $N$-dimensional
coordinate space. An arbitrary $N$-dimensional real column vector,
$X_{\mu}$ transforms as:
\begin{equation}
X_{\mu}'=R_{\mu\nu}X_{\nu};~~\mu,~\nu=1,...,N
\end{equation}
Note that the group $O(N)$ leaves invariant the scalar product:
$X'^{T}Y'$$ =(RX)^T RY=$$X^T(R^TR)Y=X^TY$. Further, the group
$O(N)$ has two disconnected parts as can be seen by taking the
determinant of both sides of the above equation:
$\det(R^TR)=[\det(R)]^2$ $=\det(\bf{1})$ $\Rightarrow$
$\det(R)=\pm 1$.
\\
\\
\noindent \textsc{Definition}\\
The \emph{special orthogonal group, $SO(N)$} is a group of
$N\times N$ real orthogonal matrices, $R$ obeying

\begin{eqnarray}
R^TR=RR^T={\bf{1}};~~~ \det(R)=+1
\end{eqnarray}

Now consider the group element, $R(a)$ of $SO(N)$ which differ
infinitesimally from the identity: $R(a)\approx {\bf{1}}+a$ for
$a\ll {\bf{1}}$, that is
\begin{equation}
R_{\mu\nu}(a)=\delta_{\mu\nu}+a_{\mu\nu}
\end{equation}
The infinitesimal numbers, $a_{\mu\nu}$ are antisymmetric in their
indices since from Eq.(2.3) we get
$({\bf{1}}+a)^T({\bf{1}}+a)={\bf{1}}+a^T+a+O(a^2)={\bf{1}}$. Thus
$a^T=-a$, that is $a_{\mu\nu}=-a_{\nu\mu}$.

\noindent Also, Eqs.(2.2)and (2.4) implies
\begin{equation}
\delta X'_{\mu}\equiv X'_{\mu}-X_{\mu}=a_{\mu\nu}X_{\nu}
\end{equation}
 Next, make a Taylor expansion of the group element, $R(b)$:
$R(b)= R(0)+b_{\mu\nu}M_{\mu\nu}+O(b^2)$ where
$M_{\mu\nu}=\left(\frac{\partial R(b)}{\partial
b_{\mu\nu}}\right)_{b_{\mu\nu}=0}$. To explore the neighborhood of
the identity, we only need to retain terms linear in $b$: $R(b)=
{\bf{1}}+b_{\mu\nu}M_{\mu\nu}$. This expression is for
infinitesimal transformations. For finite transformations, let
$b_{\mu\nu}=\frac{i}{2}\frac{a_{\mu\nu}}{n}$,
where $n$ is an arbitrarily large number and define\\
$R(a)=\lim_{n\to \infty}(R(b))^n=\lim_{n\to
\infty}\left({\bf{1}}+\frac{i}{2}(\frac{a_{\mu\nu}}{n})M_{\mu\nu}\right)^n$
giving
\begin{eqnarray}
R(a)=e^{\frac{i}{2}a_{\mu\nu}M_{\mu\nu}}\nonumber\\
{\textnormal{with}}~~~~~~a_{\mu\nu}=-a_{\mu\nu}
\end{eqnarray}
The real numbers $a_{\mu\nu}$ are the \emph{parameters} of the
group and specify rotation. $M_{\mu\nu}$'s are linearly
independent matrices and are the \emph{generators}. They are
necessarily imaginary as $R$ is real. The factor of $\frac{1}{2}$
is chosen for convenience and the significance of $i$ will be
clarified below. Further from Eqs.(2.3) and (2.6) one can easily
prove that the generators are antisymmetric: ${\bf{1}}=
({\bf{1}}+a_{\mu\nu}M_{\mu\nu}^T+...)({\bf{1}}+a_{\mu\nu}M_{\mu\nu}+...)={\bf{1}}+a_{\mu\nu}(M_{\mu\nu}^T
+M_{\mu\nu})+O(a^2)$. Thus $M_{\mu\nu}^T=-M_{\mu\nu}$.

The reason for inserting $i$ in Eq.(2.6) is because it is more
convenient in quantum mechanics to use the anti-Hermitian
generators ($M_{\alpha\beta}^{\dagger}=-M_{\alpha\beta}$) rather
than antisymmetric ($M_{\alpha\beta}^{T}=-M_{\alpha\beta}$). Then
$R(a)$ in Eq.(2.6) is unitary ($R^{\dagger}=R^{-1}$). Of course
this does not change the fact that $SO(N)$ is a real Lie algebra.

For infinitesimal transformation, Eqs.(2.2) and (2.6) gives
$X'_{\mu}\approx$ \\ $[\delta_{\mu\nu} +\frac{i}{2}a_{\alpha\beta}
(M_{\alpha\beta})_{\mu\nu}]X_{\nu}$, thus
\begin{equation}
\delta X'_{\mu}\equiv
X'_{\mu}-X_{\mu}=\frac{i}{2}a_{\alpha\beta}(M_{\alpha\beta})_{\mu\nu}X_{\nu}
\end{equation}
Comparing Eqs.(2.5) and (2.7) we get
$a_{\alpha\beta}(M_{\alpha\beta})_{\mu\nu}=-2ia_{\mu\nu}=-i(a_{\mu\nu}-a_{\nu\mu})\\
=-i(\delta_{\mu\alpha}\delta_{\nu\beta}-
\delta_{\nu\alpha}\delta_{\mu\beta})a_{\alpha\beta}$ and therefore

\begin{equation}
(M_{\alpha\beta})_{\mu\nu}=-i(\delta_{\mu\alpha}\delta_{\nu\beta}-
\delta_{\nu\alpha}\delta_{\mu\beta});~~~1\leq \mu < \nu \leq N
\end{equation}
where the indices $\alpha$ and $\beta$ are the \emph{ordinal}
indices of the generators, while $\mu$ and $\nu$ are the row and
column indices of the matrix. It can also be easily shown that

\begin{equation}
tr(M_{\alpha\beta}M_{\mu\nu})=-2(\delta_{\mu\alpha}\delta_{\nu\beta}-\delta_{\mu\beta}\delta_{\nu\alpha})
\end{equation}

 \noindent Note that from Eq.(2.8) for $\mu=\nu$:
$(M_{\alpha\beta})_{\mu\mu}=0$, that is the diagonal elements of
the antisymmetric matrix $M_{\alpha\beta}$ are zero. Thus,
obviously $tr(M_{\alpha\beta})=0$. However,  this trace condition
follows from the condition $\det(R)=+1$ is automatically fulfilled
for antisymmetric matrices: $1=\det(R)=e^{tr(\ln
R)}=e^{tr(\ln[e^{\frac{i}{2}a_{\mu\nu}M_{\mu\nu}}])}=e^{\frac{i}{2}a_{\mu\nu}tr(M_{\mu\nu})}$
and hence $tr(M_{\mu\nu})=0$. The matrix $M_{\alpha\beta}$ is also
antisymmetric with respect to the ordinal indices.

 Eq.(2.8) represents the vector (fundamental or
defining) representation of the generators. It also shows that the
only non-vanishing elements of the matrix $M_{\alpha\beta}$ are
$-i$ and $+i$ given  by the intersection of the $\alpha^{th}$ row,
$\beta^{th}$ column $(\alpha\neq \beta)$ and $\beta^{th}$ row,
$\alpha^{th}$ column, respectively:
\begin{eqnarray}
M_{\alpha\beta}=\left(\matrix{0&.&.&.&.&0\cr .&.&&-i&&.\cr
.&&.&&&. \cr .&+i&&.&&. \cr .&&&&.&.\cr 0&.&.&.&.&0
  \cr}
 \right);~~~M_{\mu\nu}^T=-M_{\mu\nu}
\end{eqnarray}

The commutation relation satisfied by the matrices
$M_{\alpha\beta}$ can be easily calculated using Eq.(2.8):
\begin{equation}
[M_{\alpha\beta},M_{\mu\nu}]=-i(\delta_{\beta\mu}M_{\alpha\nu}+\delta_{\alpha\nu}M_{\beta\mu}
-\delta_{\alpha\mu}M_{\beta\nu}-\delta_{\beta\nu}M_{\alpha\mu})
\end{equation}
For $\alpha=\mu$, we obtain
\begin{equation}
[M_{\alpha\beta},M_{\alpha\nu}]=iM_{\beta\nu};~~{\textnormal
{if}}~\alpha\neq \beta\neq \nu~({\textnormal {no~sum~on}}~ \alpha)
\end{equation}
and if all indices $\alpha$, $\beta$, $\mu$, and $\nu$ are unequal
then Eq.(2.11) gives
\begin{equation}
[M_{\alpha\beta},M_{\mu\nu}]=0;~~{\textnormal {if}}~\alpha\neq
\beta \neq \mu \neq\nu
\end{equation}
All non-zero commutators can be obtained from from Eq.(2.12) and
the antisymmetry property: $M_{\alpha\beta}=-M_{\beta\alpha}$,
while Eq.(2.13) states that the following generators commute with
each other
\begin{eqnarray}
M_{12}, ~M_{34}, ~M_{56},.....,~M_{N~N-1}~~~{\textnormal {for}}
~N~
{\textnormal {even}}\nonumber\\
M_{12},~ M_{34}, ~M_{56},.....,~M_{N-2~N-1}~~~ {\textnormal{for}}
~N~ {\textnormal{odd}}
\end{eqnarray}
The set $\frac{N}{2}$ (or $\frac{N-1}{2}$ for odd $N$) mutually
commuting generators is said to form \emph{Cartan subalgera}. Thus
the \emph{rank} of $SO(N)$ group is $\frac{N}{2}$ or
$\frac{N-1}{2}$ depending on whether $N$ is even or odd.

 Note that we have used $N\times N$ transformation matrices,
$R$ to obtain the defining representation,$M_{\alpha\beta}$, and
commutaion relations of $SO(N)$. Other realizations are also
possible. For example, under the transformation (2.7), an
arbitrary function, $f(X_{\mu})$ goes over to $f'(X_{\mu})\equiv
f(X'_{\mu})=f(X_{\mu}\\
+\frac{i}{2}a_{\alpha\beta}(M_{\alpha\beta})_{\mu\nu}X_{\nu})\approx
f(X_{\mu})+\frac{i}{2}a_{\alpha\beta}(M_{\alpha\beta})_{\mu\nu}X_{\nu}\frac{\partial
f(X_{\mu})}{\partial X_{\mu}}\equiv (1+\frac{i}{2}{\hat
L}_{\alpha\beta})f(X_{\mu})$, where ${\hat
L}_{\alpha\beta}=(M_{\alpha\beta})_{\mu\nu}X_{\nu}\frac{\partial}{\partial
X_{\mu}}$. Using Eq.(2.8), we get
\begin{equation}
{\hat L}_{\alpha\beta}=-i(X_{\beta}\frac{\partial}{\partial
X_{\alpha}}-X_{\alpha}\frac{\partial}{\partial X_{\beta}})
\end{equation}
This is the generalization of the angular momentum operator for
arbitrary space dimension $N$ and satisfies the same commutation
relations as Eq.(2.11). These operators act on wave function,
$\Psi(X_{\mu})$, while the $M_{\mu\nu}$ are the matrix
representations (in the $N$ dimensional Cartesian basis) of the
components of angular momentum.

The \emph{dimension} of this Lie algebra is the number of
independent parameters, $a_{\mu\nu}$. Recall that
$a_{\mu\nu}=-a_{\nu\mu}$ for $\mu<\nu$ and $a_{\mu\mu}=0$, so the
dimension is the number of pairs, that is ${N\choose
2}=\frac{1}{2}N(N-1)$.

\section{Tensors of {SO(N)} group}

Recall that the orthogonal transformations preserve the scalar
product: $X^TY\\
\equiv I$ is an invariant. This invariant can be
written using upper indices (or equally with lower indices) as
\begin{equation}
I=X^{\mu}Y^{\mu}=X^{\mu}\delta_{\mu\nu}Y^{\nu}
\end{equation}

Given a set of basis vectors $\{e_{\mu}\}$,  a \emph{metric
tensor}
 $g_{\mu\nu}$, in an $N$ dimensional space is  defined through the
equation: $g_{\mu\nu}=e_{\mu}e_{\nu}$ satisfying
$g_{\alpha\beta}g^{\gamma\beta}=\delta^{\gamma}_{\alpha}$. An
invariant is specified in the form
\begin{equation}
I=X^{\mu}g_{\mu\nu}X^{\nu}
\end{equation}
Eqs.(2.16) and (2.17) implies that the metric tensor
\begin{equation}
g_{\mu\nu}=\delta_{\mu\nu}=e_{\mu}e_{\nu}
\end{equation}
corresponds to the orthogonal group. Further, the metric tensor
can be used to raise or lower indices of vectors (tensors):
$V_{\mu}=g_{\mu\nu}V^{\nu}$, $V^{\mu}=g^{\mu\nu}V_{\nu}$. Using
Eq.(2.18), we get $V_{\mu}=V^{\mu}$. In other words the covariant
and contravariant vectors coincide for orthonormal basis. From now
on we do not distinguish between superscripts and subscripts.

In general we can define a tensor, $T_{\mu_1\mu_2...\mu_p}$, of
rank $p$ that transforms as the product of $p$ ordinary vectors,
$X_{\mu_i}$:
\begin{equation}
T_{\nu_1\nu_2...\nu_p}'=R_{\nu_1\mu_1}R_{\nu_2\mu_2}...R_{\nu_p\mu_p}T_{\mu_1\mu_2...\mu_p}
\end{equation}
We now define a special kind of a unit totally antisymmetric
tensor, $\epsilon_{\mu_1\mu_2...\mu_N}$, called the
\emph{Levi-Civita }tensor as follows:
\begin{eqnarray}
\epsilon_{\mu_1\mu_2...\mu_N}=\left\{\matrix{+1&&&{\textnormal{even
permutation of indices}}\cr -1&&&{\textnormal{odd permutation of
indices}}\cr 0&&&{\textnormal{if any two indices equal}}
  \cr}\right\}
\end{eqnarray}
As a consequence of its antisymmetry:
\begin{eqnarray}
\epsilon_{\mu_1\mu_2...\mu_N}\epsilon^{\nu_1\nu_2...\nu_N}=\delta^{\nu_1}_{[\mu_1}...\delta^{\nu_N}_{\mu_N]}\\
\epsilon_{\alpha_1..\alpha_m\mu_1\mu_2...\nu_{N-m}}\epsilon^{\alpha_1..\alpha_m\nu_1\nu_2...\nu_{N-m}}
=m!~\delta^{\nu_1}_{[\mu_1}...\delta^{\nu_{N-m}}_{\mu_{N-m}]}
\end{eqnarray}
Using Eq.(2.22) we can define the determinant of a matrix $A$ as
\begin{equation}
\det(A)=\frac{1}{N!}\epsilon_{\mu_1\mu_2...\mu_N}\epsilon^{\nu_1\nu_2...\nu_N}A^{\mu_1}_{\nu_1}...A^{\mu_N}_{\nu_N}
\end{equation}
Finally, multiplying both sides of this equation by
$\epsilon^{\mu_1\mu_2...\mu_N}$ and use Eq.(2.22) to get
\begin{eqnarray}
\epsilon^{\nu_1\nu_2...\nu_N}A^{\mu_1}_{\nu_1}...A^{\mu_N}_{\nu_N}=\epsilon^{\mu_1\mu_2...\mu_N}\det(A)
\end{eqnarray}
\\
\textsc{invariants of the SO(N) group }\\
\\
1.~~ \underline{$2^{nd}$ rank isotropic tensor: Kronecker symbol}\\
\begin{eqnarray}
\delta_{\mu\nu}'=R_{\mu\alpha}R_{\nu\beta}\delta_{\alpha\beta}\nonumber\\
=R_{\mu\alpha}R_{\nu\alpha}\nonumber\\
=(RR^T)_{\mu\nu}\nonumber\\
=\delta_{\mu\nu}
\end{eqnarray}\\
\\
2.~~ \underline{$N^{th}$ rank isotropic tensor: Levi-Civita tensor}\\
\begin{eqnarray}
\epsilon_{\mu_1\mu_2...\mu_N}'=R_{\mu_1\nu_1}R_{\mu_2\nu_2}...R_{\mu_N\nu_N}\epsilon_{\nu_1\nu_2...\nu_N}\nonumber\\
=\det(R)\epsilon_{\mu_1\mu_2...\mu_N}\nonumber\\
=\epsilon_{\mu_1\mu_2...\mu_N}
\end{eqnarray}
 where we have used Eq.(2.24) and the fact $\det(R)=+1$.

 Starting from a tensor of rank $p$ and using one of the above
 invariant tensors, one can construct a new tensor, of smaller
 rank, by the operation of \emph{contraction}. This consists of
 multiplying a rank $p$ tensor by one of the invariant tensors and
 summing over the repeated indices. There are two possibilities:\\

\noindent (a)~~ Contraction with $\delta_{\mu\nu}$ is a trace operation.\\
$~~~~~~~\delta_{\mu_1\mu_2}T_{\mu_1\mu_2\mu_3..\mu_p}=T_{\mu\mu\mu_3..\mu_p}\equiv
{\hat
 T}_{\mu_3\mu_4..\mu_p}$. Here the first two indices have been
 contracted but the operation can be applied to any pair. This
 leads to a tensor of rank $p-2$. A tensor is said to be
 \emph{traceless} if the contraction with a Kronecker symbol of any
 pair of indices vanishes.\\
 \\
(b)~~ Contraction with $\epsilon_{\mu_1\mu_2...\mu_N}$. \\
 $~~~~~~~~\epsilon_{\mu_1\mu_2...\mu_N}T_{\mu_1\mu_2\mu_3}\equiv{\hat
 T}_{\mu_4..\mu_p}$\\

 The contraction operation is related to reducibility. By
 definition, a tensor is \emph{reducible}, if by one of the above
 contractions, one can get a new non-vanishing tensor of smaller
 rank. If that is not possible, the tensor is \emph{irreducible}. As a
 consequence of (a), an irreducible tensor must satisfy
$\delta_{\mu_1\mu_2}T_{\mu_1\mu_2\mu_3..\mu_p}=0$ and from (b) it
follows that the tensor must be symmetric with respect to the
indices $\mu_1,\mu_2,\mu_3$ on which the sum has been
performed.\\
\\
\textsc{global symmetries ($a_{\alpha\beta}$ {\textnormal{independent of space-time coordinate}}, $x$)}\\
\\
1.~~ \underline{Scalar boson in the vector representation}\\
\\
Introduce a set of $N$ scalar bosonic fields by means of an $N$
dimensional column vector, $\phi(x)$:\\
\begin{eqnarray}
\phi(x)=\left(\matrix{\phi_1(x)\cr \phi_2(x)\cr .\cr .\cr .\cr
\phi_N(x)
  \cr}\right)
\end{eqnarray}
The transformation law is the same as before (see Eqs. (2.2)and
(2.6)):
\begin{equation}
\phi(x)\rightarrow \phi'(x)=R\phi(x)
\end{equation}
in terms of its components the above yields
$\phi_{\mu}'(x)=R_{\mu\nu}\phi_{\nu}(x)\approx ({\bf
1}+\frac{i}{2}a_{\alpha\beta}M_{\alpha\beta})_{\mu\nu}\phi_{\nu}$
and on using Eq.(2.8) we get:
\begin{equation}
\phi_{\mu}(x)\rightarrow
\phi_{\mu}'(x)=\phi_{\mu}(x)+a_{\mu\nu}\phi_{\nu}(x)
\end{equation}
Obviously, the dimensionality of the vector representation is
${N\choose 1}=~N$. The generators in the $N$ dimensional vector
representation are given by Eq.(2.8).

 The Lagrangian for $N$ real scalar bosonic
fields given by
\begin{equation}
{\mathsf L}=\frac{1}{2}\sum_{\mu=1}^N
\partial_A\phi_{\mu}(x)\partial^A\phi_{\mu}(x)=\frac{1}{2}
\partial_A\phi^T(x)\partial^A\phi(x)
\end{equation}
is invariant under \emph{global rotations}, since ${\mathsf
L}'=\frac{1}{2}
\partial_A\phi'^T(x)\partial^A\phi'(x)
=\\\frac{1}{2}
\partial_A[\phi^T(x)R^T]
\times\partial^A[R\phi(x)]=\frac{1}{2}
\partial_A\phi^T(x)\partial^A\phi(x)={\mathsf L}$.
Here $A$ is the Dirac index ($A=0-3$) and where we are using the metric $\eta=diag(1,-1,-1,-1)$.\\
\\
2.~~ \underline{Scalar boson in the $2^{nd}$ rank (adjoint) antisymmetric tensor representation}\\
\\
One way to define a second rank antisymmetric tensor,
$\Phi_{\mu\nu}^{({\cal A})}(=-\Phi_{\nu\mu}^{({\cal A})})$, is
\begin{equation}
\Phi_{\mu\nu}^{({\cal A})}=\left(\phi_{\mu}\otimes
\phi_{\nu}\right)_{{{antisymmetric}}}=\frac{1}{2}\left(\phi_{\mu}\otimes\phi_{\nu}-\phi_{\nu}\otimes\phi_{\mu}\right)
\end{equation}
Then Eq.(2.28) gives $\Phi_{\mu\nu}^{({\cal
A})'}=(R_{\mu\rho}\phi_{\rho}\otimes R_{\nu\lambda}\phi_{\lambda}-
R_{\nu\lambda}\phi_{\lambda}\otimes
R_{\mu\rho}\phi_{\rho})$=$R_{\mu\rho}\Phi_{\rho\lambda}^{({\cal
A})}R_{\lambda\nu}^T$. Hence the transformation law takes the form
\begin{equation}
\Phi^{({\cal A})}\rightarrow \Phi'^{({\cal A})}=R\Phi^{({\cal A})}
R^T
\end{equation}
Writing this transformation law explicitly, we get
$\Phi_{\mu\nu}^{({\cal A})'}\approx
[\delta_{\mu\rho}+\frac{i}{2}a_{\alpha\beta}(M_{\alpha\beta})_{\mu\rho}]\\
\times \Phi_{\rho\lambda}^{({\cal A})}
[\delta_{\nu\lambda}+\frac{i}{2}a_{\alpha\beta}(M_{\alpha\beta})_{\nu\lambda}^{({\cal
A})}]$. Expanding this expression using Eq.(2.8) we obtain
\begin{equation}
\Phi_{\mu\nu}^{({\cal A})}\rightarrow \Phi_{\mu\nu}^{({\cal
A})'}=\Phi_{\mu\nu}^{({\cal A})}+a_{\mu\rho}\Phi_{\rho\nu}^{({\cal
A})}+a_{\nu\lambda}\Phi_{\mu\lambda}^{({\cal A})}
\end{equation}

Dimensionality of the second rank antisymmetric tensor
representation is ${N\choose 2}=\frac{1}{2}N(N-1)$. This is also
the number of  generators of the $SO(N)$ group. Hence this
representation is also the \emph{adjoint} representation of the
group. Further, it implies that we have $\frac{1}{2}N(N-1)$
\emph{vector gauge bosons} denoted by $W^A_{\mu\nu}$ having the
global transformation law (2.33):
\begin{equation}
W_{\mu\nu}^A\rightarrow
W_{\mu\nu}^{A'}=W_{\mu\nu}^A+a_{\mu\rho}W_{\rho\nu}^A+a_{\nu\lambda}W_{\mu\lambda}^A
\end{equation}

In analogy with Eq.(2.30) we define the Lagrangian for $N$ scalar
bosonic fields
\begin{equation}
{\mathsf L}=\frac{1}{4}\sum_{\mu,\nu}^N
\partial_A\Phi_{\mu\nu}^{({\cal
A})}\partial^A\Phi_{\mu\nu}^{({\cal A})}=\frac{1}{4}tr\left(
\partial_A\Phi^{({\cal
A})T}\partial^A\Phi^{({\cal A})}\right)
\end{equation}
and it is easily seen to be invariant under the global
transformation (2.32) using the properties of trace.

To find the generators in the adjoint representation it is
convenient to write the transformation (2.32) in the following
fashion
\begin{equation}
\Phi_{\mu_1\mu_2}^{({\cal A})}\rightarrow
\Phi_{\mu_1\mu_2}^{({\cal
A})'}=\frac{1}{2}\left(R_{\mu_1\nu_1}R_{\mu_2\nu_2}-R_{\mu_1\nu_2}R_{\mu_2\nu_1}\right)\Phi_{\nu_1\nu_2}^{({\cal
A})}
\end{equation}

\noindent Inserting Eqs.(2.4) or (2.6) and (2.8) above, we get,
$\Phi_{\mu_1\mu_2}^{({\cal A})'}-\Phi_{\mu_1\mu_2}^{({\cal
A})}=1/2[(\delta_{\mu_1\nu_1} a_{\mu_2\nu_2}\\
+\delta_{\mu_2\nu_2}
a_{\mu_1\nu_1})-(\nu_1\leftrightarrow
\nu_2)]\Phi_{\nu_1\nu_2}^{({\cal A})}\equiv
a_{\alpha\beta}/2(M_{\alpha\beta})_{\mu_1\mu_2,\nu_1\nu_2}\Phi_{\nu_1\nu_2}^{({\cal
A})}$, where
\begin{eqnarray}
(M^{({\cal
A})}_{\alpha\beta})_{\mu_1\mu_2,\nu_1\nu_2}=\frac{1}{2}\left\{\left[\delta_{\mu_1\nu_1}
\left(\delta_{\alpha\mu_2}\delta_{\beta\nu_2}
-\delta_{\alpha\nu_2}\delta_{\beta\mu_2}\right)+\delta_{\mu_2\nu_2}
\left(\delta_{\alpha\mu_1}\delta_{\beta\nu_1}\right.\right.\right.\nonumber\\
\left.\left.\left.-\delta_{\alpha\nu_1}\delta_{\beta\mu_1}\right)\right]
-\left[\nu_1\leftrightarrow \nu_2\right]\right\}
\end{eqnarray}
are the generators of $SO(N)$ group in the adjoint representation.
\newpage
\noindent 3.~~ \underline{Scalar boson in the $2^{nd}$ rank
symmetric and
traceless tensor representation}\\
\\
Second rank symmetric ($\Phi_{\mu\nu}^{({\cal
S})}=\Phi_{\nu\mu}^{({\cal S})}$) and traceless
($\Phi_{\mu\mu}^{({\cal S})}=0$) tensor can be formed from the
product of two $SO(N)$ vectors:
\begin{equation}
\Phi_{\mu\nu}^{({\cal S})}=\left(\phi_{\mu}\otimes
\phi_{\nu}\right)_{{{{symmetric}}}\choose
{traceless}}=\frac{1}{2}\left(\phi_{\mu}\otimes\phi_{\nu}+\phi_{\nu}\otimes\phi_{\mu}\right)
-\frac{\delta_{\mu\nu}}{N}\phi_{\mu}\otimes\phi_{\mu}
\end{equation}
Here $\phi_{\mu}\otimes\phi_{\mu}(\equiv \Phi_{\mu\mu}^{({\cal
S})})$ is the singlet of $SO(N)$ group. The dimensionality of
$\Phi^{({\cal S})}$ is ${N+1\choose 2}-1=\frac{1}{2}N(N+1)-1$.

 The transformation law is given by
\begin{equation}
\Phi^{(S)}\rightarrow \Phi^{({\cal S})'}=R\Phi^{({\cal S})} R^T
\end{equation}
Writing this transformation explicitly in terms of its parameters,
we get
\begin{equation}
\Phi_{\mu\nu}^{({\cal S})}\rightarrow \Phi_{\mu\nu}^{({\cal
S})'}=\Phi_{\mu\nu}^{({\cal S})}+a_{\mu\rho}\Phi_{\rho\nu}^{({\cal
S})}+a_{\nu\lambda}\Phi_{\mu\lambda}^{({\cal S})}
\end{equation}

The globally invariant Lagrangian for the symmetric, traceless
tensor is given by
\begin{equation}
{\mathsf L}=\frac{1}{4}tr\left(
\partial_A\Phi^{({\cal
S})T}\partial^A\Phi^{({\cal S})}\right)
\end{equation}

Another useful form of the transformation law for the evaluation
of the generators in the $\frac{1}{2}N(N+1)-1$ representation is
\begin{equation}
\Phi_{\mu_1\mu_2}^{({\cal S})}\rightarrow
\Phi_{\mu_1\mu_2}^{({\cal
S})'}=\frac{1}{2}\left(R_{\mu_1\nu_1}R_{\mu_2\nu_2}+R_{\mu_1\nu_2}R_{\mu_2\nu_1}\right)\Phi_{\nu_1\nu_2}^{({\cal
S})}
\end{equation}
and as in the case of antisymmetric representation, the generators
are found to be
\begin{eqnarray}
(M^{({\cal
S})}_{\alpha\beta})_{\mu_1\mu_2,\nu_1\nu_2}=\frac{1}{2}\left\{\left[\delta_{\mu_1\nu_1}
\left(\delta_{\alpha\mu_2}\delta_{\beta\nu_2}
-\delta_{\alpha\nu_2}\delta_{\beta\mu_2}\right)+\delta_{\mu_2\nu_2}
\left(\delta_{\alpha\mu_1}\delta_{\beta\nu_1}\right.\right.\right.\nonumber\\
\left.\left.\left.-\delta_{\alpha\nu_1}\delta_{\beta\mu_1}\right)\right]
+\left[\nu_1\leftrightarrow \nu_2\right]\right\}
\end{eqnarray}
\\
4.~~ \underline{Scalar boson in the general $r^{th}$ rank tensor representation}\\
\\
In general, an $r^{th}$ rank antisymmetric  and symmetric tensor
representation of $SO(2N)$ group can be formed from the
antisymmetric and symmetric product of $\phi$'s as follows

\begin{eqnarray}
\Phi^{({\cal A})}_{\mu_1\mu_2...\mu_r}
=\frac{1}{r!}\sum_P(-1)^{\delta_P}
\phi_{{\mu_1}_{P(1)}}\otimes\phi_{{\mu_2}_{P(2)}}\otimes...\otimes
\phi_{{\mu_r}_{P(r)}}\nonumber\\
\Phi^{({\cal S})}_{\mu_1\mu_2...\mu_r}=\frac{1}{r!}\sum_P
\phi_{{\mu_1}_{P(1)}}\otimes\phi_{{\mu_2}_{P(2)}}\otimes...\otimes
\phi_{{\mu_r}_{P(r)}}
\end{eqnarray}
with $\sum_P$ denoting the sum over all permutations and
$\delta_P$ takes on the value $0$ and $1$ for even and odd
permutations, respectively. The dimensionality of the
antisymmetric tensor is $N\choose r$, while the dimensionality of
the symmetric tensor is ${N+r-1}\choose r$.

 For illustration purposes, consider the case of irreducible $3^{rd}$ and $4^{th}$
rank symmetric and traceless tensors. The above equation (with
trace subtracted) at once gives
\begin{eqnarray}
\Phi^{({\cal S})}_{\mu_1\mu_2\mu_3}=\frac{1}{3!}\left(\phi_{\mu_1}
\phi_{\mu_2}\phi_{\mu_3}\right)_{symmetric}-\frac{1}{N+2}\left(
\delta_{\mu_1\mu_2}\phi_{\nu}
\phi_{\nu}\phi_{\mu_3}\nonumber \right.\\
\left.+\delta_{\mu_1\mu_3}\phi_{\nu}
\phi_{\mu_2}\phi_{\nu}+\delta_{\mu_2\mu_3}\phi_{\mu_1}
\phi_{\nu}\phi_{\nu}\right)
\end{eqnarray}
\begin{eqnarray}
\Phi^{({\cal
S})}_{\mu_1\mu_2\mu_3\mu_3}=\frac{1}{4!}\left(\phi_{\mu_1}
\phi_{\mu_2}\phi_{\mu_3}\phi_{\mu_4}\right)_{symmetric}-\frac{1}{N+4}\left(
\delta_{\mu_1\mu_2}\phi_{\nu}\phi_{\nu}
\phi_{\mu_3}\phi_{\mu_4}\nonumber \right.\nonumber\\
\left.+\delta_{\mu_1\mu_3}\phi_{\nu}\phi_{\mu_2}
\phi_{\nu}\phi_{\mu_4}+\delta_{\mu_1\mu_4}\phi_{\nu}\phi_{\mu_2}
\phi_{\mu_3}\phi_{\nu}+\delta_{\mu_2\mu_3}\phi_{\mu_1}\phi_{\nu}
\phi_{\nu}\phi_{\mu_4}\right.\nonumber\\
\left.+\delta_{\mu_2\mu_4}\phi_{\mu_1}\phi_{\nu}
\phi_{\mu_3}\phi_{\nu}+\delta_{\mu_3\mu_4}\phi_{\mu_1}\phi_{\mu_2}
\phi_{\nu}\phi_{\nu}\right)\nonumber\\
+\frac{1}{(N+2)(N+4)}\left(\delta_{\mu_1\mu_2}\delta_{\mu_3\mu_4}\phi_{\mu}\phi_{\mu}
\phi_{\nu}\phi_{\nu}+\delta_{\mu_1\mu_3}\delta_{\mu_2\mu_4}\phi_{\mu}\phi_{\nu}
\phi_{\mu}\phi_{\nu}\right.\nonumber\\
\left.+\delta_{\mu_1\mu_4}\delta_{\mu_2\mu_3}\phi_{\mu}\phi_{\nu}
\phi_{\nu}\phi_{\mu}\right)
\end{eqnarray}
where, for brevity, we have removed the direct product symbols.

In general, a $p^{th}$ rank antisymmetric tensor gives rise to an
irreducible $(N-p)^{th}$ rank antisymmetric tensor through
contraction with the Levi-Civita tensor:
\begin{equation}
\Phi^{({\cal
A})}_{\mu_1\mu_2...\mu_{N-p}}=\frac{1}{p!}\epsilon_{\mu_1\mu_2...\mu_{N-p}\nu_1...\nu_p}\Phi^{({\cal
A})}_{\nu_1...\nu_{p}}
\end{equation}
For $N$ even ($N=2k$), the  tensor $\Phi^{({\cal
A})}_{\mu_1\mu_2...\mu_{k}}$ of dimension ${2k\choose {k}}$ splits
into two irreducible tensors
$\Omega^{({+\alpha})}_{\mu_1\mu_2...\mu_{k}}$ and
$\Omega^{({-\alpha})}_{\mu_1\mu_2...\mu_{k}}$ each of dimension
$\frac{1}{2} {2k\choose {k}}$ according to the $SO(2k)$ invariant
decomposition of a tensor of rank $k$,
\begin{eqnarray}
\Phi^{({\cal
A})}_{\nu_1\nu_2...\nu_{k}}=\Omega^{({+\alpha})}_{\nu_1\nu_2...\nu_{k}}+\Omega^{({-\alpha})}_{\nu_1\nu_2...\nu_{k}}
,~~~~~~~~~~~~~~~~~~~~~~\nonumber\\
\nonumber\\
\textnormal{with}~~~~~~~~\Omega^{({\pm\alpha})}_{\mu_1\mu_2...\mu_{k}}=\frac{1}{2}\left(\delta_{\mu_1\nu_1}
\delta_{\mu_2\nu_2}...\delta_{\mu_{k}\nu_{k}}\pm
\frac{\alpha}{k!}\epsilon_{\mu_1\mu_2...\mu_k\nu_1\nu_2...\nu_k}\right)\Phi^{({\cal
A})}_{\nu_1\nu_2...\nu_{k}},\nonumber\\
\nonumber\\
\textnormal{where}~~~~~~~~\alpha=\left\{\matrix{1;&{\textnormal{k=even}}\cr
i;&{\textnormal{k=odd}}
  \cr}\right\}~~~~~~~~~~~~~~~~~~~~~~
\end{eqnarray}
Further,  $\Omega^{({\pm\alpha})}_{\mu_1\mu_2...\mu_{k}}$ also
satisfies self and anti-self duality conditions:
\begin{equation}
\Omega^{({\pm\alpha})}_{\mu_1\mu_2...\mu_{k}}=\pm
\frac{\alpha}{k!}\epsilon_{\mu_1\mu_2...\mu_k\nu_1\nu_2...\nu_k}\Omega^{({\pm\alpha})}_{\nu_1\nu_2...\nu_{k}}
\end{equation}
Eqs.(2.48) and (2.49) will be proved later in a section related to
spinor representations.

 The transformation laws for the arbitrary rank antisymmetric and symmetric tensors can be easily extended and put in
useful forms such as
\begin{eqnarray}
\Phi^{{({\cal A},{\cal S})}}_{\mu_1\mu_2...\mu_r}\rightarrow
\Phi^{({\cal A},{\cal
S})'}_{\mu_1\mu_2...\mu_r}=R_{\mu_1\nu_1}R_{\mu_2\nu_2}...R_{\mu_r\nu_r}\Phi^{({\cal
A,{\cal S}})}_{\nu_1\nu_2...\nu_r}\nonumber\\
 \Phi^{({\cal
A},{\cal
S})'}_{\mu_1\mu_2...\mu_r}=\frac{1}{r!}\left[\sum_P(-1)^{\delta_P}
R_{{\mu_1}{\nu_1}_{P(1)}}R_{{\mu_2}{\nu_2}_{P(2)}}...
R_{{\mu_r}{\nu_r}_{P(r)}}\right]\Phi^{({\cal
A,{\cal S}})}_{\nu_1\nu_2...\nu_r}\nonumber\\
\Phi^{({\cal A},{\cal S})'}_{\mu_1\mu_2...\mu_r}\approx
\Phi^{({\cal A},{\cal S})}_{\mu_1\mu_2...\mu_r}+\sum_{i=1}^r
a_{\mu_i\nu_i}\Phi^{({\cal A},{\cal
S})}_{\mu_1\mu_2..\nu_i..\mu_{r-1}\mu_r}
\end{eqnarray}

The generators of the $SO(N)$ group in the $r^{th}$ rank
antisymmetric and symmetric representations are given by
\begin{equation}
\left(M^{{{\cal A}\choose {\cal
S}}}_{\alpha\beta}\right)_{\mu_1...\mu_r,\nu_1...\nu_r}=\frac{1}{r!}\left\{\left[\sum_{i=1}^r
~\prod_{\stackrel{j=1}{i\neq j}}^r
\delta_{\mu_j\nu_j}\left(\delta_{\alpha\mu_i}\delta_{\beta\nu_i}-\delta_{\alpha\nu_i}\delta_{\beta\mu_i}\right)\right]
\mp \left[\nu_i \leftrightarrow \nu_j \right]\right\}
\end{equation}

The invariant Lagrangian under global transformations is given by
\begin{equation}
{\mathsf L}=\frac{1}{2r!}
\partial_A\Phi^{({\cal
A},{\cal S})}_{\mu_1\mu_2...\mu_r}\partial^A\Phi^{({\cal A},{\cal
S})}_{\mu_1\mu_2...\mu_r}
\end{equation}
\\
\textsc{local symmetries ($a_{\alpha\beta}$ {\textnormal{dependent on space-time coordinate}}, $x$)}\\
\\
1.~~ \underline{Scalar boson in the vector representation}\\
\\
This time the scalar fields introduced through Eq.(2.27) must
transform as
\begin{eqnarray}
\phi_{\mu}(x)\rightarrow   \phi_{\mu}'(x)=R_{\mu\nu}(x)\phi_{\nu}(x)\nonumber\\
R_{\mu\nu}(x)=\left(e^{\frac{i}{2}a_{\alpha\beta}(x)M_{\alpha\beta}}\right)_{\mu\nu}
\end{eqnarray}
The kinetic energy given by Eq.(2.30): ${\mathsf
L}=\frac{1}{2}\sum_{\mu=1}^N
\partial_A\phi_{\mu}(x)\partial^A\phi_{\mu}(x)$ is no longer
invariant under the \emph{local rotations }(2.53),  because
$\partial_A \phi'(x)=R(x)\partial_A \phi(x)\\+[\partial_a
R(x)]\phi(x)\neq R(x)\partial_A \phi(x)$. Just as in QED it is the
transformation of the derivatives that will cause difficulties. By
analogy with that case we require a \emph{gauge covariant
derivative}, $D_A$, to transform as
\begin{equation}
D_A\phi(x)\rightarrow D_A'\phi'(x)=R(x)D_A\phi(x)
\end{equation}
where $D_A$ is to be understood as a $N \times N$ matrix carrying
a Dirac index, $A$, and operating on the $N$ component scalar
bosonic field, $\phi(x)$.

There are $\frac{1}{2}N(N-1)$ group generators and we introduce
one vector gauge boson, $W_{\mu\nu}^A(x)$ for each and define
\begin{eqnarray}
D^A\phi(x)=\left[\partial^A-\frac{ig}{2}M_{\mu\nu}W_{\mu\nu}^A(x)\right]\phi(x)
=\left[\partial^A-\frac{ig}{2}{\hat{W}}^A(x)\right]\phi(x)\nonumber\\
{\hat{W}}^A(x)\equiv M_{\mu\nu}W_{\mu\nu}^A(x)~~~~~~~~~~~~~~~~~~
\end{eqnarray}
where $g$ is a coupling constant.

 Next, we determine the transformation
law for the vector gauge boson. Substituting Eqs.(2.53) and (2.55)
in Eq.(2.54) and using the fact $R(x)R^T(x)=1$ which implies
$[\partial_A R(x)]R^T(x)=-R(x)\partial_A R^T(x)$, we get
\begin{equation}
{\hat{W}}'_A(x)=R(x)\left[
{\hat{W}}_A(x)+\frac{2i}{g}\partial_A\right]R^T(x)
\end{equation}
Using Eq.(2.8) and the expression, $R(x)\approx ({\bf
1}+\frac{i}{2}a_{\alpha\beta}M_{\alpha\beta})$, we obtain after
some algebra, the local transformation law for the vector bosons
\begin{eqnarray}
W_{\mu\nu}^A(x)\rightarrow
W_{\mu\nu}^{'A}(x)=W_{\mu\nu}^A(x)+a_{\mu\rho}(x)W_{\rho\nu}^A(x)\nonumber\\
+a_{\nu\lambda}(x)W_{\mu\lambda}^A(x)
+\frac{1}{g}\partial^Aa_{\mu\nu}(x),\nonumber\\
{\textnormal{with}}~~~~~W_{\mu\nu}^A(x)=-W_{\nu\mu}^A(x)~~~~~~~~~~~~~~~~~~~~~~~~~~~~
\end{eqnarray}

In analogy with QED, we define the \emph{field strength tensor},
$F_{\mu\nu}^{AB}(x)$ as
\begin{eqnarray}
\left[D_A,D_B\right]=-\frac{ig}{2}{\hat{F}}_{AB}(x),\nonumber\\
{\textnormal{where}}~~~~~{\hat{F}}^{AB}(x)=M_{\mu\nu}F_{\mu\nu}^{AB}(x)~~~~~~~~
\end{eqnarray}
Substituting Eq.(2.55) into (2.58) and together with Eq.(2.8) and
the infinitesimal expression for $R(x)$, we get
\begin{eqnarray}
F_{\mu\nu}^{AB}(x)=\partial^AW_{\mu\nu}^B(x)-\partial^BW_{\mu\nu}^A(x)
-g\left[W_{\mu\sigma}^A(x)W_{\sigma\nu}^B(x)\right.\nonumber\\
\left.-W_{\mu\sigma}^B(x)W_{\sigma\nu}^A(x)\right]
\end{eqnarray}

To find the transformation law for $F^{AB}$ we left and right
multiply Eq.(2.58) by $R(x)$ and $R^T(x)$, respectively, to obtain
\begin{equation}
{\hat{F}}'_{AB}(x)=R(x){\hat{F}}_{AB}(x)R^T{x}
\end{equation}
where we have used the fact that $D_A'(x)=R(x)D_AR^T(x)$ which
follows directly from Eqs.(2.53) and (2.54). Note that Eq.(2.60)
is in the form of Eq.(2.56), hence the corresponding infinitesimal
result (2.57) applies without the derivative term:
\begin{equation}
F_{\mu\nu}^{AB}(x)\rightarrow
F_{\mu\nu}^{'AB}(x)=F_{\mu\nu}^{AB}(x)+a_{\mu\rho}(x)F_{\rho\nu}^{AB}(x)+a_{\nu\lambda}(x)F_{\mu\lambda}^{AB}(x)
\end{equation}

After the introduction of local transformation the Lagrangian
(2.30) must be replaced by
\begin{eqnarray}
{\mathsf L}= \frac{1}{2}D_A\phi^T(x)D^A\phi (x)~~~~~~~~~~~~~~~~~~~~~~~~~\nonumber\\
~~~~=\frac{1}{2}\partial_A\phi_{\mu}\partial^A\phi_{\mu}-g\left(\partial_A\phi_{\mu}\right)W_{\mu\nu}^{A}\phi_{\nu}
-\frac{g^2}{2}\phi_{\mu}W_{A\mu\nu}W_{\nu\sigma}^{A}\phi_{\sigma}
\end{eqnarray}

In order to define the system including the new gauge field,
$W_{\mu\nu}^A(x)$, it is necessary to include a kinetic energy
term for $W_{\mu\nu}^A(x)$:
\begin{equation}
{\mathsf
L}=-\frac{1}{4}F_{AB\mu\nu}(x)F_{\mu\nu}^{AB}(x)+\frac{1}{2}D_A\phi^T(x)D^A\phi
(x)
\end{equation}
\\
\noindent 2.~~ \underline{Scalar boson in the $2^{nd}$ rank antisymmetric tensor representation}\\
\\
Recall from Eq.(2.32) that the second rank antisymmetric tensor,
$\Phi^{({\cal A})}_{\mu\nu}$ transforms as $\Phi'^{({\cal
A})}=R\Phi^{({\cal A})} R^T$. Then just as in the case of the
vector representation  we want the covariant derivative to
transform like Eq.(2.32):
\begin{equation}
\left(D_A\Phi^{({\cal A})}\right)'=R(x)\left(D_A\Phi^{({\cal
A})}\right)R^T(x)
\end{equation}
Then the covariant derivative, $D_A\Phi^{({\cal A})}$ in terms of
\emph{Lie-valued} gauge fields, which has the transformation
property (2.64), is given by
\begin{equation}
D_A\Phi^{({\cal A})}=\partial_A\Phi^{({\cal
A})}-\frac{ig}{2}\left({\hat{W}}_A\Phi^{({\cal
A})}+{\hat{W}}_A^T\Phi^{({\cal A})}\right)
\end{equation}
Inserting the generators, we find the expression for the covariant
derivative to be
\begin{equation}
\left(D^A\Phi^{({\cal A})}\right)_{\mu\nu}=\partial^A\Phi^{({\cal
A})}_{\mu\nu}-g\left({{W}}^A_{\mu\sigma}\Phi^{({\cal
A})}_{\sigma\nu}-{{W}}^A_{\nu\sigma}\Phi^{({\cal
A})}_{\sigma\mu}\right)
\end{equation}

The total Lagrangian is
\begin{equation}
{\mathsf
L}=-\frac{1}{4}F_{AB\mu\nu}F_{\mu\nu}^{AB}+\frac{1}{4}tr\left(D_A\Phi^{({\cal
A})T}D^A\Phi^{({\cal A})}\right)
\end{equation}
\newpage
\noindent 3.~~ \underline{Scalar boson in the $2^{nd}$ rank symmetric tensor representation}\\
\\
Recall from Eq.(2.39) that the $2^{nd}$ rank symmetric tensor,
$\Phi^{({\cal S})}$, transforms as $\Phi'^{({\cal
S})}=R\Phi^{({\cal S})} R^T$. Hence, the results for this case
will be identical to that for the $2^{nd}$ rank antisymmetric
tensor case. Therefore
\begin{equation}
\left(D^A\Phi^{({\cal S})}\right)_{\mu\nu}=\partial^A\Phi^{({\cal
S})}_{\mu\nu}-g\left({{W}}^A_{\mu\sigma}\Phi^{({\cal
S})}_{\sigma\nu}-{{W}}^A_{\nu\sigma}\Phi^{({\cal
S})}_{\sigma\mu}\right)
\end{equation}
\begin{equation}
{\mathsf
L}=-\frac{1}{4}F_{AB\mu\nu}F_{\mu\nu}^{AB}+\frac{1}{4}tr\left(D_A\Phi^{({\cal
S})T}D^A\Phi^{({\cal S})}\right)
\end{equation}
and of course
\begin{equation}
\left(D_A\Phi^{({\cal S})}\right)'=R(x)\left(D_A\Phi^{({\cal
S})}\right)R^T(x)
\end{equation}
\\
4.~~ \underline{Scalar boson in the general $r^{th}$ rank tensor representation}\\
\\
Recall from Eq.(2.50) the transformation law for an arbitrary
antisymmetric and symmetric tensor of rank $r$: $ \Phi^{({\cal
A},{\cal
S})'}_{\mu_1\mu_2...\mu_r}=R_{\mu_1\nu_1}R_{\mu_2\nu_2}...R_{\mu_r\nu_r}\Phi^{({\cal
A,{\cal S}})}_{\nu_1\nu_2...\nu_r}$. Hence, we require that the
corresponding covariant derivative, $D_A\Phi^{({\cal A},{\cal
S})'}_{\mu_1\mu_2...\mu_r}$, transforms as
\begin{equation}
\left(D_A\Phi^{({\cal A},{\cal
S})'}\right)_{\mu_1\mu_2...\mu_r}=R_{\mu_1\nu_1}R_{\mu_2\nu_2}...R_{\mu_r\nu_r}\left(D_A\Phi^{({\cal
A},{\cal S})'}\right)_{\nu_1\nu_2...\nu_r}
\end{equation}
The expression for the covariant derivative is then given by
\begin{equation}
\left(D^A\Phi^{({\cal A},{\cal
S})}\right)_{\mu_1\mu_2...\mu_r}=\partial^A\Phi^{({\cal A},{\cal
S})'}_{\mu_1\mu_2...\mu_r} -g\sum_P(-1)^{\delta_P}
W^A_{{\mu_1}_{P(1)}\nu} \Phi^{({\cal A},{\cal
S})'}_{\nu{\mu_2}_{P(2)}{\mu_3}_{P(3)}...{\mu_r}_{P(r)}}
\end{equation}
For example, in the case of $3^{rd}$ rank tensor, the above result
takes the form
\begin{equation}
\left(D^A\Phi^{({\cal A},{\cal
S})}\right)_{\mu_1\mu_2\mu_3}=\partial^A\Phi^{({\cal A},{\cal
S})'}_{\mu_1\mu_2\mu_r} -g\left( W^A_{{\mu_1}\nu} \Phi^{({\cal
A},{\cal S})'}_{\nu{\mu_2}{\mu_3}}-W^A_{{\mu_2}\nu} \Phi^{({\cal
A},{\cal S})'}_{\nu{\mu_1}{\mu_3}}+W^A_{{\mu_3}\nu} \Phi^{({\cal
A},{\cal S})'}_{\nu{\mu_1}{\mu_2}}\right)
\end{equation}

\section{Digression: general construction of matrix generators of {U(N)}and {SU(N)}}

A general, \emph{unitary group}, $U(N)$ has $N^2$ generators each
of which is an $N\times N$ Hermitian matrix (unitarity of the
group element implies Hermiticity of the generators). On the other
hand, a general \emph{special unitary group}, $SU(N)$ has $N^2-1$,
$N\times N$ Hermitian and traceless (unimodularity of the group
element implies tracelessness  of the generators) matrices as
generators.\\

\noindent\textsc{unitary group, U(N)}\\

\noindent The simplest representation of an $N\times N$  matrix is
one that has all entries zero except one non-vanishing element
with value 1:
\begin{eqnarray}
{ U}_{j}^i=\left(\matrix{0&.&.&.&.&0\cr .&.&&1&&.\cr .&&.&&&. \cr
.&&&.&&. \cr .&&&&.&.\cr 0&.&.&.&.&0
  \cr}
 \right);~~~i,~j=1,2,...,N
\end{eqnarray}
The matrix ${ U}_{i}^j$ has one matrix element with value 1 which
is given by the intersection of $i^{th}$ row and $j^{th}$ column:
\begin{equation}
({ U}_{j}^i)^k_l=\delta_{ik}\delta_{jl}
\end{equation}
Note that the off-diagonal matrices, ${ U}_{i}^j$ ($i\neq j$),
 are non-Hermitian since ${({ U}_{j}^i)^k_l}^{\dagger}={({
U}_{j}^i)^k_l}^{*T}={({ U}_{j}^i)^l_k}=\delta_{il}\delta_{jk}={({
U}_{i}^j)^k_l}$. However, the diagonal matrices, ${ U}_{i}^i$ are
Hermitian:
\begin{eqnarray}
{{ U}_{j}^i}^{\dagger}={ U}_{i}^j;~~~i\neq j\nonumber\\
{{ U}_{i}^i}^{\dagger}={ U}_{i}^i~~~~~~~~~~~
\end{eqnarray}
Collecting other properties of the matrix ${ U}_{i}^j$, we have
\begin{eqnarray}
tr\left({ U}_{j}^i\right)=0;~~~i\neq j\nonumber\\
tr\left({ U}_{i}^i\right)=1~~~~~~~~~~~
\end{eqnarray}
Now we  build generators of $U(N)$ by forming linear combinations
of the matrices ${ U}_{j}^i$ for $i\neq j$ so that they are
necessarily Hermitian. The choice are the combinations
\begin{equation}
{U}_{j}^i+{ U}_{i}^j~~~~{\textnormal{and}}~~~~-i\left({ U}_{j}^i-{
U}_{i}^j\right);~~~\forall i\neq j
\end{equation}
One can easily verify the following properties of these matrix
combinations:
\begin{eqnarray}
\left[{ U}_{j}^i+{ U}_{i}^j\right]^{\dagger}={ U}_{j}^i+{
U}_{i}^j\nonumber\\
 \left[-i\left({ U}_{j}^i-{
U}_{i}^j\right)\right]^{\dagger}=-i\left({ U}_{j}^i-{
U}_{i}^j\right)\nonumber\\
tr\left[{ U}_{j}^i+{ U}_{i}^j\right]=0,~~~~~~tr\left[-i\left({
U}_{j}^i-{ U}_{i}^j\right)\right]=0
\end{eqnarray}
There are now a total of $N^2$, $N\times N$ Hermitian matrix
generators of which there are $N$ diagonal generators, ${
U}_{i}^i$ and ($2{N\choose 2}=$) $N(N-1)$ non-diagonal generators,
${U}_{j}^i+{ U}_{i}^j$ and $-i\left({ U}_{j}^i-{ U}_{i}^j\right)$.
They are all linearly independent. Summarizing,
\begin{eqnarray}
\forall i\neq j=1,2,...,N;~~~a=1,2,...N^2~~~~\nonumber\\
  \nonumber\\
 {
U}_a=\left\{\matrix{{U}_{j}^i+{
U}_{i}^j&&&\frac{N(N-1)}{2}~{\textnormal{generators}}\cr -i\left({
U}_{j}^i-{
U}_{i}^j\right)&&&\frac{N(N-1)}{2}~{\textnormal{generators}}\cr
U^i_i&&&N~{\textnormal{generators}}
  \cr}\right\}
  \end{eqnarray}
An arbitrary group element of the $U(N)$ group is then given by
\begin{eqnarray}
e^{-i\sum_{a=1}^{N^2}\vartheta_a{U}_a}
\end{eqnarray}
where the angles, $\vartheta_a$ parametrize the group
transformation.\\

\noindent\textsc{special unitary group, SU(N)}\\

\noindent We begin here by constructing traceless matrices,
$T^i_j$ from $U^i_j$ as follows:
\begin{equation}
T^i_j=U^i_j-\frac{\delta^i_j}{N}{\bf{1}}
\end{equation}
These are traceless because
$(T^i_j)^k_k=(U^i_j)^k_k-{\delta^i_j}={{1-1;~i=
j}\choose{0-0;~i\neq j}}$, where we have used Eq.(2.77). However,
the matrices $T^i_j$, for $i\neq j$ are not Hermitian because from
Eq.(2.76), $U^i_j$'s are non-Hermitian for $i\neq j$. As before,
we construct linear combinations of $T^i_j$'s which are Hermitian:
\begin{equation}
{ T}_{j}^i+{ T}_{i}^j~~~{\textnormal{and}}~~~
-i\left({T}_{j}^i-{T}_{i}^j\right);~~~i\neq j
\end{equation}
Note that ${ T}_{j}^i+{ T}_{i}^j={U}_{j}^i+{U}_{i}^j$ and
$-i\left({T}_{j}^i-{T}_{i}^j\right)=-i\left({U}_{j}^i-{U}_{i}^j\right)$.

The  N diagonal Hermitian traceless generators are
$T^i_i=U^i_i-\frac{1}{N}{\bf{1}}$ (${T^i_i}^{\dagger}=T^i_i$).
Since the rank of $SU(N)$ group is $N-1$, not all of the $N$
diagonal generators are independent and $N-1$ of them can be
simultaneously diagonalized. It is common to form the following
linear combinations:
\begin{equation}
n_1\left(T^i_i+T^{i+1}_{i+1}\right)~~~{\textnormal{and}}~~~n_2\left(T^i_i-T^{i+1}_{i+1}\right);~~i=1,2,...,N-1
\end{equation}
where $n_1$ and $n_2$ are normalization factors to be determined.
Note that these are a total of $2(N-1)$ matrices and we may choose
any $N-1$ matrices. Next, we determine the normalization factors
from the convention that the traces of squared matrices of $SU(N)$
are equal to 2:
\begin{equation}
tr\left({T^i_j}^2\right)=2
\end{equation}
From Eqs.(2.74) and (2.82),
$n_1\left(T^i_i+T^{i+1}_{i+1}\right)=n_1
\left(U^i_i+U^{i+1}_{i+1}-\frac{2}{N}{\bf{1}}\right)$
$=n_1\times\\diag\left(-\frac{2}{N},...,-\frac{2}{N},
1-\frac{2}{N},1-\frac{2}{N},
-\frac{2}{N},...,-\frac{2}{N}\right)$, where the element
$1-\frac{2}{N}$
 occurs twice at the intersections of $\alpha^{th}$ row ,
 $\alpha^{th}$ column and $(\alpha+1)^{th}$ row ,
 $(\alpha+1)^{th}$ column. Squaring this matrix gives $n_1^2\left(T^i_i+T^{i+1}_{i+1}\right)^2$
$=n_1^2diag\left(\frac{4}{N^2},...,\right.
\\ \left.-\frac{4}{N^2},(1-\frac{2}{N})^2,\\(1-\frac{2}{N})^2,
\frac{4}{N^2},...,\frac{4}{N^2}\right)$. Finally, $tr
\left[n_1^2\left(T^i_i+T^{i+1}_{i+1}\right)^2\right]=n_1^2[2(1-\frac{2}{N})^2+(N-2)\frac{4}{N^2}]$
$=2n_1^2(\frac{N-2}{N})$. On using Eq.(2.85), we get
$n_1=\sqrt{\frac{N}{N-2}}$.

The second normalization is rather easy to determine. From
Eqs.(2.74) and (2.82),
$n_2\left(T^i_i-T^{i+1}_{i+1}\right)=n_1\left(U^i_i-U^{i+1}_{i+1}\right)$
$=n_1diag\left(0,..., 1,-1,0,...,0\right)$. Squaring and taking
the trace of this matrix gives $
tr\left[n_2^2\left(T^i_i-T^{i+1}_{i+1}\right)^2\right]=2$. Thus,
from Eq.(2.85), we have $n_2=1$.
\newpage
Summarizing,
\begin{eqnarray}
a=1,2,...N^2-1~~~~~~~~~~~~~~~~~~~~~~~~~~~~~~~~~~~~~~~~~~~~~\nonumber\\
  \nonumber\\
 {
T}_a=\left\{\matrix{\left[\matrix{{T}_{j}^i+{
T}_{i}^j&&&\frac{N(N-1)}{2}~{\textnormal{generators}}\cr -i\left({
T}_{j}^i-{
T}_{i}^j\right)&&&\frac{N(N-1)}{2}~{\textnormal{generators}}\cr}\right];~\forall i\neq j=1,2,...,N\nonumber\\
\left[\matrix{\sqrt{\frac{N}{N-2}}\left(T^i_i+T^{i+1}_{i+1}\right)&&&N-1~{\textnormal{generators}}\cr
\left(T^i_i-T^{i+1}_{i+1}\right)&&&N-1~{\textnormal{generators}}\cr}\right];~i=1,...N-1\cr}\right.
\end{eqnarray}\\
Any $N-1$ of the total $2(N-1)$ diagonal generators above  are
allowed. It is also instructive to express ${ T}_a$'s in terms of
$U^i_j$'s,
\begin{eqnarray}
a=1,2,...N^2-1~~~~~~~~~~~~~~~~~~~~~~~~~~~~~~~~~~~~~~~~~~~~~\nonumber\\
  \nonumber\\
 {T}_a=\left\{\matrix{\left[\matrix{{U}_{j}^i+{
U}_{i}^j&\frac{N(N-1)}{2}~{\textnormal{generators}}\cr -i\left({
U}_{j}^i-{
U}_{i}^j\right)&\frac{N(N-1)}{2}~{\textnormal{generators}}\cr}\right];~\forall i\neq j=1,2,...,N\nonumber\\
\left[\matrix{\sqrt{\frac{N}{N-2}}\left(U^i_i+U^{i+1}_{i+1}-\frac{2}{N}{\bf{1}}\right)&N-1~{\textnormal{generators}}\cr
\left(U^i_i-U^{i+1}_{i+1}\right)&N-1~{\textnormal{generators}}\cr}\right];~i=1,...N-1\cr}\right.
\end{eqnarray}\\
The generators $T_a$ satisfy
\begin{equation}
[{ T}_a,{T}_b]=if_{abc}{ T}_c
\end{equation}
 An arbitrary group element of the
$SU(N)$ group is then given by
\begin{eqnarray}
e^{-i\sum_{a=1}^{N^2-1}\varphi_a{ T}_a}
\end{eqnarray}
where the angles, $\varphi_a$, parametrize  the group
transformation.\\
\\

\section{SO(2N) algebra in a U(N) basis}

\textsc{complete embedding of U(N) into SO(2N)}\\

\noindent Let $P$ and $Q$ given by
\begin{equation}
P=W+iX ~~\textnormal{and}~~Q=Y+iZ
\end{equation}
be $N$ dimensional complex column vectors of the $U(N)$ group
where $W$, $X$, $Y$ and $Z$ are real vectors. Then the $U(N)$
group transformations,
\begin{eqnarray}
 P'=SP,~~~Q'=SQ ~~~~~~~~~~~~~~~~~~~~\nonumber\\
S=e^{ib.M_{U(N)}};~~S^{\dagger}S=SS^{\dagger}={\bf 1};~~
M_{U(N)}=M_{U(N)}^{\dagger}
\end{eqnarray}
 leaves the
following scalar products invariant:
\begin{eqnarray}
P^{\dagger}P=W^TW+X^TX\nonumber\\
Q^{\dagger}Q=Y^TY+Z^TZ\nonumber\\
Q^{\dagger}P=Y^TW+Z^TX+i\left(Y^TX-Z^TW\right)
\end{eqnarray}
Now, define two $2N$ dimensional vectors as follows:
\begin{equation}
\rho=\left(\matrix{W\cr X\cr}\right),~~~\xi=\left(\matrix{Y\cr
Z\cr}\right)
\end{equation}
The above $U(N)$ invariants can now be expressed in terms of
$\rho$ and $\xi$
\begin{eqnarray}
\rho^T\rho=W^TW+X^TX\nonumber\\
 \xi^T\xi=Y^TY+Z^TZ\nonumber\\
\xi^T\rho=Y^TW+Z^TX\nonumber\\
\xi^T {\bf{J}}\rho=Y^TX-Z^TW,~~\bf{J}=\left(\matrix{0&1\cr-1&0
\cr}\right)
\end{eqnarray}

Next, consider the $SO(2N)$ group acting on the real $2N$
dimensional vectors $\rho$ and $\xi$. Then, the $SO(2N)$ group
transformations
\begin{eqnarray}
 \rho'=R\rho,~~~\xi'=R\xi~~~~~~~~~~~~~~~~~~~~~~~~~~~~~~~~~\nonumber\\
\textnormal{where}~~~R=e^{ia.M_{SO(2N)}}~~~~~~~~~~~~~~~~~~~~~~~~~~~~~~~~~~~~~~~\nonumber\\
~~\textnormal{with}
~~~R^{T}R=RR^T={\bf
1}~~\textnormal{and}~~M_{SO(2N)}=-M_{SO(2N)}^{T}
\end{eqnarray}
 leaves the following scalar products invariant:
\begin{equation}
\rho^T\rho,~~~\xi^T\xi,~~~\xi^T\rho
\end{equation}
Since these quantities in Eq.(2.96), which are $SO(2N)$ invariants
and are also $U(N)$ invariants (see Eq.(2.94)), $U(N)$ is a
"natural" subgroup of $SO(2N)$.

Note that since $R\in SO(2N)$, the antisymmetric generators
$M_{SO(2N)}$ in the basis of Eq.(2.93), can be written as
\begin{eqnarray}
M_{SO (2N)}=i\left(\matrix{\bf {A}&\bf{B}\cr-\bf{B}^T&\bf{C}
\cr}\right)
\end{eqnarray}
where $\bf{A}$ and $\bf{C}$ are real antisymmetric
($\bf{A}=-\bf{A}^T$, $\bf{C}=-\bf{C}^T$) $N\times N$ matrices
while $\bf{B}$ is an arbitrary real $N\times N$ matrix.

Additionally, if we  impose that $R\in U(N)$, then $M_{SO(2N)}$ is
also a generator of $U(N)$: $M_{SO(2N)}\supset M_{U(N)}$. Then the
corresponding transformation must also leave the fourth quantity
in Eq.(2.94) invariant: $\xi'^T {\bf{J}}\rho'=\xi^T
{\bf{J}}\rho~\Rightarrow~\\e^{ia.M_{U(N)}^T}{\bf{J}}
e^{ia.M_{U(N)}} $, which under infinitesimal transformations takes
the form
\begin{equation}
M_{U(N)}^T {\bf{J}}+{\bf{J}} M_{U(N)}=0
\end{equation}
Inserting Eq.(2.97) into Eq.(2.98) gives
\begin{eqnarray}
M_{U(N)}=\left(\matrix{\bf{A}&\bf{B}\cr-\bf{B}&\bf{A} \cr}\right)
\end{eqnarray}
where $\bf{A}$ is a real $N\times N$ antisymmetric matrix and
$\bf{B}$ is a real $N\times N$ symmetric matrix.

The number of independent elements in $\bf{A}$ and $\bf{B}$ are
$\frac{1}{2}N(N-1)$ and $\frac{1}{2}N(N+1)$, respectively, giving
a total of $N^2$ independent elements in $M_{U(N)}$. The traceless
matrices $\bf{A}$ (since $\bf{A}$ is antisymmetric) and the
traceless part of matrices $i\bf{B}$:
$i[{\bf{B}}-\frac{1}{N}tr({\bf{B}})\bf{1}]$ will form the adjoint
$N^2-1$ dimensional representation of the $SU(N)$ group and the
trace of $\bf{B}$: $\frac{i}{N}tr(\bf{B})\bf{1}$ will be an
$SU(N)$ singlet. This term generates the $U(1)$ group of complex
phase transformations. Thus, we have the decomposition
$SO(2N)\rightarrow U(N)\rightarrow SU(N)\otimes U(1)$.

However, the adjoint and  singlet representations of $SU(N)$ is
not the full story. There are other generators of $SO(2N)$ that
are not in $M_{U(N)}$. These remaining ${{2N}\choose 2} -N^2=
N(N-1)$ generators of $SO(2N)$ form two antisymmetric tensor
representations of $U(N)$: $-{\bf{K}}\pm i{\bf{L}}$ each of
dimension $\frac{1}{2}N(N-1)$:
\begin{eqnarray}
\left(\matrix{\bf{K}&{\bf{L}}\cr \bf{L}&-\bf{K} \cr}\right)
\end{eqnarray}
This can be seen from the following argument. From above we have
learned that $2N$ dimensional real vector, $\left(\matrix{W\cr
X\cr}\right)$, of $SO(2N)$, decomposes into $N\oplus {\overline
N}$ dimensional vectors of $SU(N)$ corresponding to $W\pm iX$.
Further, the (antisymmetric) generators of $SO(2N)$ can be
associated with $2^{nd}$ rank antisymmetric tensors. Thus, under
$SO(2N)\supset SU(N)\otimes U(1)$ decomposition:
\begin{eqnarray}
\left [2N\otimes 2N\right]_{as}\supset \left\{(N\oplus {\overline
N})\otimes (N\oplus {\overline N})\right\}_{as}\nonumber\\
\left[{{2N}\choose 2}\right]\supset \left \{N\otimes
N\right\}_{as}\oplus \left \{{\overline N}\otimes {\overline
N}\right\}_{as}\oplus \left
\{N\otimes {\overline N}\right\}\nonumber\\
\left[N(2N-1)\right]\supset\left\{\frac{1}{2}N(N-1)\right\}\oplus\left\{{\overline{\frac{1}{2}N(N-1)}}\right\}\oplus
\left\{N^2-1\right\}\oplus \left\{1\right\}\nonumber\\
\end{eqnarray}

Altogether,
\begin{eqnarray}
M_{SO(2N)}=\left(\matrix{{\bf{A}}+{\bf{K}}&{\bf{B}}+{\bf{L}}\cr-{\bf{B}}+{\bf{L}}&{\bf{A}}-{\bf{K}}
\cr}\right)
\end{eqnarray}

\noindent Summarizing: \\
$\bullet$~~$N^2-1$ dimensional adjoint of $SU(N)$ is formed from
${\bf{A}}+i\left({\bf{B}}-\frac{1}{N}tr({{\bf{B}}}){\bf 1}\right)$\\
$\bullet$~~ Singlet of $SU(N)$ is formed from $\frac{1}{N}tr({{\bf{B}}}){\bf 1}$\\
 $\bullet$~~Antisymmetric representations of $SU(N)$ is formed from
$-{\bf{K}}+ i{\bf{L}}$ and $-{\bf{K}}- i{\bf{L}}$ each of
dimensionality $\frac{1}{2}N(N-1)$\\
$\bullet$~~${\bf{A}}$, ${\bf{K}}$, ${\bf{L}}$ are real $N\times N$
antisymmetric matrices and ${\bf{B}}$ is a real $N\times N$
symmetric matrix.\\
\\
\textsc{generators of SU(N) in terms of SO(2N)}\\
\\
For compactness and clarity we drop the subscript from
$M_{SO(2N)}$ and write them simply as $M$. Looking at the block
structure of $M$ in Eq.(2.102), we can make the following
assignments ($i,j=1,..N$)
\begin{eqnarray}
{\bf{A}}^i_j=-\frac{1}{2}\left(M_{ij}+M_{i+N~j+N}\right),~~~
{\bf{B}}^i_j=\frac{1}{2}\left(M_{i~j+N}+M_{j~i+N}\right)\nonumber\\
{\bf{K}}^{ij}=\frac{1}{2}\left(M_{ij}-M_{i+N~j+N}\right),~~~
{\bf{L}}^{ij}=\frac{1}{2}\left(M_{i~j+N}-M_{j~i+N}\right)
\end{eqnarray}
The generators of $U(N)$ group are $U^i_j$, defined by
\begin{equation}
U^i_j={\bf{A}}^i_j+i{\bf{B}}^i_j
\end{equation}
while the $SU(N)$ generators, $T^i_j$, are given by
\begin{equation}
T^i_j=U^i_j-\frac{1}{N}U^k_k\delta^i_j
\end{equation}
while the $U(1)$ generator, $Y$, is given by
\begin{equation}
Y=U^k_k
\end{equation}
Lastly, the broken generators of $SO(2N)$ group, $V^{ij}$ and
${\overline V}_{ij}$ are
\begin{eqnarray}
V^{ij}=-{\bf{K}}^{ij}-i{\bf{L}}^{ij}\nonumber\\
 {\overline
V}_{ij}=-{\bf{K}}^{ij}+i{\bf{L}}^{ij}
\end{eqnarray}\\
\textsc{decomposition of tensor representations of SO(2N) under
the subgroup SU(N) $\otimes$
U(1)}\\
\\
The irreducible tensor representations of $SO(2N)$ can be
decomposed under $SO(2N)\supset SU(N)\otimes U(1)$ by forming
tensor products and using Young tableau.
\newpage
\noindent 1.~~ \underline{Vector of $SO(2N)$}\\
\\
This case was already considered in the previous subsections:
\begin{equation}
[2N]\supset \{N\}\oplus \{{\overline N}\}
\end{equation}
\\
2.~~ \underline{$2^{nd}$ rank tensors of $SO(2N)$}\\
\\
The antisymmetric tensor representation was also considered in the
previous sections
\begin{equation}
[N(2N-1)]\supset \{1/2N(N-1)\}\oplus
\{{\overline{1/2N(N-1)}}\}\oplus \{N^2-1\}\oplus \{1\}
\end{equation}
In the case of $2^{nd}$ rank symmetric traceless tensor of
dimensionality ${{2n+1}\choose 2}-{2N\choose 0}$, we get
$[2N\otimes 2N]_{s}\supset \{ (N\oplus {\overline N})\otimes
(N\oplus {\overline N})\}_{s}$ which on simplifying gives
\begin{eqnarray}
 \left[N(2N+1)-1\right]\supset \{1/2N(N+1)\}\oplus \{ {\overline{1/2N(N+1)}}\}\oplus \{
N^2-1\}\oplus \{1\}\nonumber\\
\end{eqnarray}
\\
3.~~ \underline{$3^{rd}$ rank tensors of $SO(2N)$}\\
\\
Here we form an anti-symmetrized and a symmetrized product of
three vectors and subtract off the trace in the case of a
symmetric tensor representation.

The result for the decomposition of  antisymmetric tensor
representation with dimensionality ${2N\choose 3}$ is given by
\begin{eqnarray}
[2N/3(2N-1)(N-1)]\supset \{1/6N(N-1)(N-2)\}\nonumber\\
\oplus \{{\overline{1/6N(N-1)(N-2)}}\}
 \oplus
\{1/2N(N+1)(N-2)\}\nonumber\\
 \oplus\{{\overline{1/2N(N+1)(N-2)}}\}\oplus
\{N\}\oplus \{{\overline N}\}
\end{eqnarray}
While the result for the symmetric traceless tensor representation
 of dimensionality ${2N+2\choose 3}-{2N\choose 1}$ is
\begin{eqnarray}
[2N/3(2N+1)(N+1)-2N]\supset \{1/6N(N+1)(N+2)\}\nonumber\\
\oplus \{{\overline{1/6N(N+1)(N+2)}}\} \oplus
\{1/2N(N-1)(N+2)\}\nonumber\\
\oplus \{{\overline{1/2N(N-1)(N+2)}}\}
\end{eqnarray}
\\
4.~~ \underline{$4^{th}$ rank tensors of $SO(2N)$}\\
\\
Using the technique as before, we have the following
decomposition of the ${2N\choose 4}$ component $4^{th}$ rank
antisymmetric tensor representation
\begin{eqnarray}
[1/6N(N-1)(2N-1)(2N-3)]\supset
\{1/24N(N-1)(N-2)(N-3)\}\nonumber\\
\oplus \{{\overline{1/24N(N-1)(N-2)(N-3)}}\}
\oplus\{1/6N(N+1)(N-1)(N-3)\}\nonumber\\
\oplus \{{\overline{1/6N(N+1)(N-1)(N-3)}}\} \oplus
\{1/2N(N-1)\}\{{\overline{1/2N(N-1)}}\}\nonumber\\
\oplus\{1/4N^2(N+1)(N-3)\} \oplus \{N^2-1\}\oplus \{1\}\nonumber\\
\end{eqnarray}

For the case of symmetric traceless tensor representation of
dimensionality ${2N+3\choose 4}-{2N+1\choose 2}$, the
decomposition is
\begin{eqnarray}
[1/6N(N+1)(2N+1)(2N+3)-N(2N+1)]\supset \nonumber\\
\{1/24N(N+1)(N+2)(N+3)\} \oplus
\{{\overline{1/24N(N+1)(N+2)(N+3)}}\}\nonumber\\
\oplus\{1/6N(N+1)(N-1)(N+3)\} \oplus
\{{\overline{1/6N(N+1)(N-1)(N+3)}}\} \nonumber\\
\oplus \{1/4N^2(N-1)(N+3)\}{\overline{}}\nonumber\\
\end{eqnarray}

\noindent 5.~~ \underline{$5^{th}$ rank tensors of $SO(2N)$}\\
\\
 The  $5^{th}$ rank
antisymmetric tensor representation has dimensionality is
${2N\choose 5}$ and can be decomposed as follows
\begin{eqnarray}
[1/15N(2N-1)(N-1)(2N-3)(N-2)]\supset\nonumber\\
\{1/120N(N-1)(N-2)(N-3)(N-4)\}\nonumber\\
 \oplus
\{{\overline{1/120N(N-1)(N-2)(N-3)(N-4)}}\}\nonumber\\
\oplus\{1/24N(N+1)(N-1)(N-2)(N-4)\}\nonumber\\ \oplus
\{{\overline{1/24N(N+1)(N-1)(N-2)(N-4)}}\} \nonumber\\
\oplus
\{1/6N(N-1)(N-2)\}\oplus\{{\overline{1/6N(N-1)(N-2)}}\}\nonumber\\
\oplus\{1/12N^2(N+1)(N-1)(N-4)\} \oplus
\{{\overline{1/12N^2(N+1)(N-1)(N-4)}}\}\nonumber\\
\oplus\{1/2N(N+1)(N-2)\} \oplus \{{\overline
{1/2N(N+1)(N-2)}}\}\oplus \{N\}\oplus \{{\overline N}\}\nonumber\\
\end{eqnarray}

For the case of symmetric traceless tensor representation of
dimensionality ${2N+4\choose 5}-{2N+2\choose 3}$, the
decomposition is
\begin{eqnarray}
[1/15N(N+1)(2N+1)(2N+3)(N+2)-2N/3(2N+1)(N+1)]\supset \nonumber\\
\{1/120N(N+1)(N+2)(N+3)(N+4)\} \nonumber\\
\oplus
\{{\overline{1/120N(N+1)(N+2)(N+3)(N+4)}}\}\nonumber\\
\oplus\{1/24N(N+1)(N-1)(N+2)(N+4)\} \nonumber\\
\oplus
\{{\overline{1/24N(N+1)(N-1)(N+2)(N+4)}}\} \nonumber\\
\oplus \{1/4N^2(N-1)(N+3)\} \oplus
\{{\overline{1/4N^2(N-1)(N+3)}}\}\nonumber\\
\end{eqnarray}

\section{Spinor representation of {SO(2N)}}

\textsc{introducing clifford algebra}\\

\noindent The spinor representations of $SO(2N)$ are most easily
studied by the introduction of the generalized Dirac $\Gamma$
matrices. To that end, recall that the special orthogonal group
$SO(2N)$ can be considered as those linear transformation
($X'_{\mu}=R_{\mu\nu}X_{\nu}$) on the coordinates $X_1$,
$X_2$,..., $X_{2N}$ such that the quadratic form:
$X_1^2+X_2^2+...+X_{2N}^2$ is invariant. Now, if we write this
quadratic form as the square of a linear form of $X_{\mu}$'s, then
\begin{equation}
X_1^2+X_2^2+...+X_{2N}^2=\left(\Gamma_1X_1+\Gamma_1X_1+...\Gamma_{2N}X_{2N}\right)^2
\end{equation}
This requires that
\begin{equation}
\left\{\Gamma_{\mu},\Gamma_{\nu}\right\}=2\delta_{\mu\nu};~~\mu,
\nu=1,...,2N
\end{equation}
$\Gamma_{\mu}$ defines a rank $2N$ \emph{Clifford algebra}. It is
obvious from the last equation that
\begin{equation}
\Gamma_{\mu}^2={\bf{1}}~~~~~~{\textnormal{no sum on}}~\mu
\end{equation}
 Multiply both sides of Eq.(2.118) by $\Gamma_{\nu}$
and use Eq.(2.119) to get
$\Gamma_{\nu}\Gamma_{\mu}\Gamma_{\nu}=\Gamma_{\mu}$ (no sum on
$\nu$). If we assume a matrix representation of $\Gamma_{\mu}$'s,
then taking trace of this last expression we get
$tr(\Gamma_{\mu})=tr(\Gamma_{\nu}\Gamma_{\mu}\Gamma_{\nu})=
tr(-\Gamma_{\mu}\Gamma_{\nu}\Gamma_{\nu})=-tr(\Gamma_{\mu})$.
Hence,
\begin{equation}
tr\left(\Gamma_{\mu}\right)=0
\end{equation}
\\
\textsc{existence of spinorial representations}\\
\\
We show the existence of spinorial represenations by explicit
construction of $\Gamma_{\mu}$'s. We necessarily make them
Hermitian: $\Gamma_{\mu}=\Gamma_{\mu}^{\dagger}$. This means we
can choose a representation in which $\Gamma_{\mu}$ is diagonal.
 Thus,  Eq.(2.119) implies that the
(diagonal elements) eigenvalues of $\Gamma_{\mu}$ are either $+1$
or $-1$ and together with Eq.(2.120) we conclude that the number
of $+1$'s and $-1$'s have to be the same. Thus $\Gamma_{\mu}$ are
even dimensional matrices.

 In general there are two ways to
represent the $\Gamma_{\mu}$'s. One is through the  $2^N\times
2^N$ generalized Dirac matrices formed from the direct products of
Pauli matrices. The other is by means of creation and annihilation
operators acting on a Hilbert space. We will discuss both of these
techniques, however, we rely heavily on the latter method.

Recall that the $2\times 2$ Pauli spin matrices ($N=1$)
\begin{eqnarray}
\sigma_1=\left(\matrix{0&1\cr1&0
\cr}\right),~~~\sigma_2=\left(\matrix{0&-i\cr i&0
\cr}\right),~~~\sigma_3=\left(\matrix{1&0\cr0&-1 \cr}\right)
\end{eqnarray}
satisfy the properties
\begin{equation}
\left\{\sigma_{a},\sigma_{b}\right\}=2\delta_{ab};~~a,b=1,2,3
\end{equation}
Thus, one possibility for $\gamma_{\mu}$'s ($\mu=1,2,..,2N$) is to
express them in terms of Pauli matrices in the following specific
way
\begin{eqnarray}
\Gamma_{2i}={\bf 1}\otimes {\bf 1}\otimes...\otimes{\bf
1}\otimes\sigma_1\otimes\sigma_3\otimes\sigma_3\otimes...\otimes\sigma_3;~~i=1,..,N\nonumber\\
\Gamma_{2i-1}={\bf 1}\otimes {\bf 1}\otimes...\otimes{\bf
1}\otimes\sigma_2\otimes\sigma_3\otimes\sigma_3\otimes...\otimes\sigma_3;~~i=1,..,N
\end{eqnarray}
The $2\times 2$ identity matrix, ${\bf 1}$, occurs ($i-1$) times
and $\sigma_3$ occurs ($N-i$) times in each of these expressions
above. Further, each of the $\sigma_1$ and $\sigma_2$ occur once
at the $i^{th}$ position. Each of the Kronecker products contain
$N$, $2\times 2$ matrices. Hence $\Gamma_{\mu}$'s are $2^N\times
2^N$ matrices. It is very easy to verify that the $\Gamma_{\mu}$'s
given by Eq.(2.123) satifies Eq.(2.118).

For the second technique, we define a set of $N$ fermionic
creation and annihilation operators $b_i$ and $b_i^{\dagger}$
($i=1,...,N$), obeying the anti-commutation rules
\begin{equation}
\{b_i,b_j^{\dagger}\}=\delta_{i}^j;~~~\{b_i,b_j\}=0;
~~~\{b_i^{\dagger},b_j^{\dagger}\}=0
\end{equation}
and represent the set of $N$ Hermitian operators $\Gamma_{\mu}$
($\mu=1,2,..,2N$) by
\begin{eqnarray}
\Gamma_{2i}= (b_i+ b_i^{\dagger});~~~ \Gamma_{2i-1}= -i(b_i-
b_i^{\dagger})\nonumber\\
\Gamma_{\mu}=\Gamma_{\mu}^{\dagger}~~~~~~~~~~
\end{eqnarray}
\\
\noindent\textsc{generators of SO(2N) in the spinor representations - construction of spinors }\\
\\
First, recall that the  $2N\times 2N$ matrix, $R$, is an
orthogonal matrix: $R^TR=RR^T={\bf 1}$ and is associated with
rotations through angle $a_{\mu\nu}$ in the $\mu-\nu$ plane in the
$2N$ dimensional coordinate space: $X'_{\mu}=R_{\mu\nu}X_{\nu}$
where $R(a)=e^{\frac{i}{2}a_{\alpha\beta}M_{\alpha\beta}}$ with
$a_{\alpha\beta}=-a_{\beta\alpha}$.

In order to construct the generators and  spinors in $SO(2N$) we
ask the
following two questions\\
\\
\textsl{How does $\Gamma_{\mu}$ transform under the transformation
$X'_{\mu}=R_{\mu\nu}X_{\nu}$?} \\
The answer is simple. The following transformation on
$\Gamma_{\mu}$:
\begin{equation}
\Gamma_{\mu}\rightarrow \Gamma_{\mu}'=R_{\mu\nu}\Gamma_{\nu}
\end{equation}
will generate the Clifford algebra:
$\left\{\Gamma_{\mu}',\Gamma_{\nu}'\right\}=2\delta_{\mu\nu}$.\\
\\
\textsl{How does an $SO(2N)$ spinor, $\Psi(X)$,  transform under
the transformation $X'_{\mu}=R_{\mu\nu}X_{\nu}$?} \\
To answer this
question, we define the following transformation on the spinor
\begin{equation}
\Psi(X)\rightarrow \Psi'(X')=S(R)\Psi(X)
\end{equation}
where we have introduced a spinor by means of an $2^N$ dimensional
column vector
\begin{eqnarray}
\Psi(x)=\left(\matrix{\Psi_1(x)\cr \Psi_2(x)\cr .\cr .\cr .\cr
\Psi_{2^N}(x)
  \cr}\right)
\end{eqnarray}

The unknown transformation matrix, $S(R)$, induces a similarity
transformation on the matrices $\Gamma_{\mu}$:
\begin{equation}
\Gamma_{\mu}'=S(R)\Gamma_{\mu}S^{-1}(R)
\end{equation}
This is true because $\Gamma_{\mu}'$, given by Eq.(2.129), also
satisfies the Clifford algebra. $S(R)$ is called a spinor
representation of the $SO(N)$ group. The notation $\Delta$ is
usually used for the spinor representation.

Eqs.(2.126) and (2.129) implies
\begin{equation}
R_{\mu\nu}\Gamma_{\nu}=S(R)\Gamma_{\mu}S^{-1}(R)
\end{equation}
 Next, we construct the explicit form of the transformation matrix,
$S(R)$. We write
\begin{equation}
S(R)=e^{-\frac{i}{4}a_{\alpha\beta}\Sigma_{\alpha\beta}}
\end{equation}
where $\Sigma_{\alpha\beta}$ are the spinorial generators to be
determined. For infinitesimal rotations Eqs.(2.130) and (2.131)
gives $\left({\bf
1}-\frac{i}{4}a_{\alpha\beta}\Sigma_{\alpha\beta}\right)\Gamma_{\mu}
\left({\bf
1}+\frac{i}{4}a_{\sigma\epsilon}\Sigma_{\sigma\epsilon}\right) =
\Gamma_{\nu}\left(\delta_{\nu\mu}+a_{\nu\mu}\right)$. On
simplification, it gives
\begin{equation}
\Sigma_{\mu\nu}=\frac{i}{2}\left[\Gamma_{\mu},\Gamma_{\nu}\right]
\end{equation}
Note $\Sigma_{\mu\nu}$ is antisymmetric
($\Sigma_{\mu\nu}=-\Sigma_{\nu\mu}$) as expected and Hermitian
($\Sigma_{\mu\nu}^{\dagger}=\Sigma_{\mu\nu}$). One useful form of
$\Sigma_{\mu\nu}$ is in terms of creation and annihilation
operators. Using Eqs.(2.125) and (2.132) the generators can be
easily put in a more useful form
\begin{eqnarray}
\Sigma_{2j-1~2k-1}=\frac{i}{2}\left[b_j,b_k^{\dagger}\right]-\frac{i}{2}\left[b_k,b_j^{\dagger}\right]
-i\left(b_jb_k+b_j^{\dagger}b_k^{\dagger}\right)\nonumber\\
\Sigma_{2j~2k-1}=\frac{1}{2}\left[b_j,b_k^{\dagger}\right]+\frac{1}{2}\left[b_k,b_j^{\dagger}\right]
-\left(b_jb_k-b_j^{\dagger}b_k^{\dagger}\right)\nonumber\\
\Sigma_{2j~2k}=\frac{i}{2}\left[b_j,b_k^{\dagger}\right]-\frac{i}{2}\left[b_k,b_j^{\dagger}\right]
+i\left(b_jb_k+b_j^{\dagger}b_k^{\dagger}\right)
\end{eqnarray}
 Further, these generators satisfy
the usual $SO(2N)$ Lie algebra
\begin{equation}
\left[\Sigma_{\mu\nu},\Sigma_{\rho\sigma}\right]
=-2i\left(\delta_{\mu\rho}\Sigma_{\nu\sigma}-\delta_{\mu\sigma}\Sigma_{\nu\rho}
-\delta_{\nu\rho}\Sigma_{\mu\sigma}+\delta_{\nu\sigma}\Sigma_{\mu\rho}\right)
\end{equation}
Note for $\mu \neq\nu\neq\rho\neq \sigma$,
$\left[\Sigma_{\mu\nu},\Sigma_{\rho\sigma}\right]=0$. This implies
$\Sigma_{\mu\nu}$ and $\Sigma_{\rho\sigma}$ can be simultaneously
diagonalized, that is there exists a basis in which both
$\Sigma_{\mu\nu}$ and $\Sigma_{\rho\sigma}$ are represented by
diagonal matrices. It is clear from Eq.(2.133) that
$\Sigma_{2k~2k-1}=N-2b_i^{\dagger}b_i$ are the $N$ diagonal
generators and form the  Cartan subalgebra of $SO(2N)$.

As a concluding remark, observe that the group element $S(R)$ is
unitary. This is because $\Sigma_{\mu\nu}$'s are Hermitian.
\begin{equation}
\Psi^{\dagger}\rightarrow
\Psi^{'\dagger}=\Psi^{\dagger}S^{-1}(R);~~~S^{-1}(R)=S^{\dagger}(R)
\end{equation}
\newpage
\noindent\textsc{scalar boson in the spinor representation}\\
\\
\underline{global transformation}\\

 \noindent Assuming the parameters, $a_{\alpha\beta}$
to be independent of space-time, the kinetic energy for the
spinors given by
\begin{equation}
{\mathsf L}=\partial_A \Psi^{\dagger}\partial^A \Psi
\end{equation}
is globally invariant because $\partial_A
\Psi^{'\dagger}\partial^A \Psi'=\partial_A
(\Psi^{\dagger}S^{-1})\partial^A (S\Psi) =\partial_A
\Psi^{\dagger}\partial^A \Psi$, and where we have used Eqs.(2.127)
and (2.135).\\
\\
\noindent\underline{local transformation}\\

\noindent Here we introduce the covariant derivative, $D_A$, just
as in the case for transformation of vectors and tensors under
local transformations
\begin{equation}
D^A=\left[\partial^A-\frac{ig}{2}\Sigma_{\mu\nu}W_{\mu\nu}^A\right]
\end{equation}
with $D_A$ transforming
\begin{equation}
D_A\rightarrow D_A'=SD_AS^{-1}
\end{equation}
The  covariant derivative acting on the spinor
\begin{equation}
(D^A\Psi)_{\mu}=\left[\partial^A-\frac{ig}{2}(\Sigma_{\alpha\beta})_{\mu\nu}W_{\alpha\beta}^A\right]\Psi_{\nu}
\end{equation}
The corresponding complete locally invariant kinetic energy for
the spinors is
\begin{equation}
{\mathsf L}=(D_A \Psi)^{\dagger}D^A
\Psi-\frac{1}{4}F_{AB\mu\nu}F_{\mu\nu}^{AB}
\end{equation}
where the second term represents the kinetic energy for the gauge
fields, $W_{\mu\nu}^A$ (see Eq.(2.59)).\\

\noindent\textsc{fermion in the spinor representation}\\
\\
\underline{global transformation}\\

 \noindent In the case of global transformations the kinetic energy is the
Dirac Lagrangian,
\begin{equation}
{\mathsf L}=i{\overline{\Psi}}\gamma^A\partial_A\Psi
\end{equation}
where ${\overline{\Psi}}=\Psi^{\dagger}\gamma^0$ and $\gamma^A$
are the Dirac matrices of the Lorentz group.\\
\\
\underline{local transformation}\\

 \noindent In the case of a locally invariant kinetic energy we must replace
the ordinary space-time derivative in the previous equation by the
covariant derivative and add the kinetic energy for the gauge
vector bosons
\begin{equation}
{\mathsf
L}=i{\overline{\Psi}}\gamma^AD_A\Psi-\frac{1}{4}F_{AB\mu\nu}F_{\mu\nu}^{AB}
\end{equation}
where the interaction gauge-fermionic part of the Lagrangian  is
\begin{equation}
i{\overline{\Psi}}\gamma^AD_A\Psi
=i{\overline{\Psi}}_{\mu}\gamma^A\partial_A\Psi_{\mu}+\frac{g}{4}{\overline{\Psi}}_{\mu}\gamma^A
W_{\alpha\beta}^A(\Sigma_{\alpha\beta})_{\mu\nu}\Psi_{\nu}
\end{equation}
\newpage
\noindent\textsc{abstract SU(N) subgroup }\\
\\
We begin with the identification that the operators $u^i_j$, given
by
\begin{equation}
u^i_j=-\frac{1}{2}\left[b_i^{\dagger},b_j\right]=-b_i^{\dagger}b_j+\delta_{ij}
\end{equation}
satisfy the $U(N)$ algebra
\begin{equation}
\left[u^i_j,u^k_l\right]=\delta_{il}u^k_j-\delta{kj}u^i_l
\end{equation}
That is, they are the generators of $U(N)$ group. Further, the
operators, $u^i_j$ can be easily expressed as linear combinations
of the $SO(2N)$ generators, $\Sigma_{\mu\nu}$'s, as follows
\begin{equation}
u^i_j=\frac{1}{4}\left(i\Sigma_{2i~2j}+i\Sigma_{2i-1~2j-1}+\Sigma_{2i~2j-1}+\Sigma_{2j~2i-1}
\right)
\end{equation}
Hence, the $u^i_j$'s generate a subalgebra, that is, $U(N)$ is a
subgroup of $SO(2N)$. If one recalls from Eq.(2.104), $u^i_j$ is
expressed in terms of an antisymmetric tensor, ${\bf A}^i_j$ and
symmetric tensor, ${\bf B}^i_j$. These are identified here as
\begin{eqnarray}
u^i_j={\bf A}^i_j+i{\bf B}^i_j \nonumber\\
{\bf
A}^i_j=-\frac{1}{4}\left([b_i^{\dagger},b_j]-[b_j^{\dagger},b_i]\right)\nonumber\\
{\bf
B}^i_j=\frac{i}{4}\left([b_i^{\dagger},b_j]+[b_j^{\dagger},b_i]\right)
\end{eqnarray}

Let's now identify all the spinorial generators of the $SO(2N)$
group under the subgroup decomposition:
\begin{eqnarray}
SO(2N)~~~~~~~~\supset ~~~~~~~~SU(N)\times U(1)~~~~~~~~~~~\nonumber\\
\nonumber\\
N(2N-1)~~~\Sigma_{\mu\nu}~{\textnormal{'s}}~~\supset~~\left\{\matrix{N^2-1&&&
t^i_j~{\textnormal{'s}}\cr
\frac{1}{2}N(N-1)&&&V^{ij}~{\textnormal{'s}}\cr
\frac{1}{2}N(N-1)&&&{\overline V}_{ij}~{\textnormal{'s}}\cr 1&&&Y
  \cr}\right\}
\end{eqnarray}\\
\noindent \underline{$t^i_j$: Generators in the adjoint representation}\\
\begin{eqnarray}
t^i_j=u^i_j-\frac{1}{N}\delta^i_j\sum_{k=1}^Nu^k_k ~~~~~~~~~~~~~~~~~~~~~~~~~~~~~\nonumber\\
t^i_j=-b_i^{\dagger}b_j+\frac{1}{N}\delta_{ij}\sum_{k=1}^Nb_k^{\dagger}b_k
~~~~~~~~~~~~~~~~~~~~~~~~~~~\nonumber\\
t^i_j=\frac{1}{4}\left(i\Sigma_{2i~2j}+i\Sigma_{2i-1~2j-1}+\Sigma_{2i~2j-1}+\Sigma_{2j~2i-1}\right)\nonumber\\
-\frac{1}{2N}
\delta_{ij}\sum_{k=1}^N\Sigma_{2k~2k-1}
\end{eqnarray}
One can identify these real ($t^{i\dagger}_j=t^j_i$) operators,
$t^{i}_j$'s of the \textsl{abstract} $SU(N)$ group with  the
generators $T^{i}_j$'s of the \textsl{matrix} $SU(N)$ group (see
section 2.3), and construct Hermitian operators, ${t}_a$, exactly
like Hermitian traceless generators which are given by Eq.(2.86),
\begin{eqnarray}
a=1,2,...N^2-1~~~~~~~~~~~~~~~~~~~~~~~~~~~~~~~~~~~~~~~~~~~~~\nonumber\\
  \nonumber\\
 {t}_a=\left\{\matrix{\left[\matrix{{t}_{j}^i+{
t}_{i}^j&&&\frac{N(N-1)}{2}~{\textnormal{generators}}\cr -i\left({
t}_{j}^i-{
t}_{i}^j\right)&&&\frac{N(N-1)}{2}~{\textnormal{generators}}\cr}\right];~\forall i\neq j=1,2,...,N\nonumber\\
\left[\matrix{\sqrt{\frac{N}{N-2}}\left(t^i_i+t^{i+1}_{i+1}\right)&&&N-1~{\textnormal{generators}}\cr
\left(t^i_i-t^{i+1}_{i+1}\right)&&&N-1~{\textnormal{generators}}\cr}\right];~i=1,...N-1\cr}\right.
\end{eqnarray}

The Hermitian generators, ${t}_a$, can be expressed in terms of
the real generators, $t^{i}_j$, by means of Hermitian traceless
$N\times N$ matrices, $T_{a}$ ($T_{a}=T_{a}^{\dagger},~tr{T_a}=0$)
introduced in section 2.3, through the equation
\begin{equation}
 {t}_a=\frac{1}{2}(T_a)^i_jt^j_i
\end{equation}

Using Eq.(2.149) we get
\begin{equation}
{t}_a=-b^{\dagger}_j(T_a/2)^i_j b_i =-b^{\dagger}_j(T_a^*/2)^j_i
b_i;~~~~a=1,...,N^2-1
\end{equation}
 Clearly, ${ t}_a$'s satisfy
 \begin{equation}
[{ t}_a,{t}_b]=if_{abc}{ t}_c
\end{equation}
 where $f_{abc}$ are
the structure constants of $SU(N)$.\\

\noindent \underline{$Y$: $U(1)_Y$ generator}\\
\begin{equation}
Y=\sum_{k=1}^Nu^k_k=\frac{N}{2}-\sum_{k=1}^N
b_k^{\dagger}b_k=\frac{1}{2}\Sigma_{2k-1~2k}
\end{equation}
$Y$ commutes with all the generators of the $SU(N)$ group and
hence generates a $U(1)$ subalgebra. One can easily verify that
$Y\left(b_{i_1}^{\dagger}b_{i_2}^{\dagger}...b_{i_m}^{\dagger}|0>\right)=\left(\frac{N}{2}-m\right)\times
\left(b_{i_1}^{\dagger}b_{i_2}^{\dagger}...b_{i_m}^{\dagger}|0>\right)$.
This shows that the representation, $\left[m\right]\equiv
b_{i_1}^{\dagger}b_{i_2}^{\dagger}...b_{i_m}^{\dagger}|0>$ in the
spinor representations of $SO(2N)$ has $U(1)_Y$ charges,
$Q_Y=\frac{N}{2}-m$, up to a normalization. Summarizing
\begin{eqnarray}
\left[m\right]\equiv
b_{i_1}^{\dagger}b_{i_2}^{\dagger}...b_{i_m}^{\dagger}|0>\nonumber\\
Y\left[m\right]=(N-2m)\left[m\right]
\end{eqnarray}
That is, the representation, $\left[m\right]$ has $N-2m$, $U(1)$
charges.\\
\\
\underline{ $V^{ij}$ and ${\overline V}_{ij}$: Antisymmetric tensors of $SU(N)$} \\
\begin{eqnarray}
V^{ij}=b^{\dagger}_ib_j^{\dagger}=\frac{1}{4}\left(\Sigma_{2i~2j-1}
-\Sigma_{2j~2i-1}+i\Sigma_{2i-1~2j-1}-i\Sigma_{2i~2j}\right)\nonumber\\
{\overline V}_{ij}=b_ib_j=\frac{1}{4}\left(-\Sigma_{2i~2j-1}
+\Sigma_{2j~2i-1}+i\Sigma_{2i-1~2j-1}-i\Sigma_{2i~2j}\right)
\end{eqnarray}

\section{Decomposition of SO(2N) spinor into antisymmetric tensors
of SU(N)}

Having shown the complete imbedding of $SU(N)$ in $SO(2N)$, we
expect that an arbitrary $2^N$ spinor ${\Psi}$ of $SO(2N)$ can be
represented by $SU(N)$ tensors, $\psi^{i_1i_2...i_r}$ of $r^{th}$
rank:
$\Psi=\left(\psi,\psi^{i_1},\psi^{i_1i_2}...,\psi^{i_1i_2...i_N}\right)$.
These $SU(N)$ tensors have to be necessarily antisymmetric in
their indices as can be seen by a simple dimensionality argument.
Write the $2^N$ dimensionality of $SO(2N)$ spinor as $(1+1)^N$ and
use  binomial theorem to write this further as $\sum_{k=0}^N
{N\choose k}=1+N+\frac{1}{2}N(N-1)+...+1$. These are exactly the
dimensionalities of antisymmetric tensors.

 We want to construct a basis for antisymmetric $SU(N)$ tensors and also
 a basis for the $SO(2N)$ spinor. For this reason we introduce a Fock vacuum, $|0>\equiv|0,0,..,0,0>$
($N$ unoccupied states), annihilated by the destruction operators
\begin{equation}
b_i|0>=0;~~~\forall i=1,...,N
\end{equation}
One $i^{th}$ occupied state for a fermion out of a possible $N$
states could be denoted as a ket state
$\Omega_i\equiv|0,0,..,0,1,0,..0>=b_i^{\dagger}|0>$. One can then,
in principle, generalize it to ket states of the form:
\begin{equation}
\Omega_{m,i_1i_2...i_m}\equiv
b_{i_1}^{\dagger}b_{i_2}^{\dagger}...b_{i_m}^{\dagger}|0>
\end{equation}
denoting a state occupied by $m$ fermions in positions
$i_1,i_2,..,i_m$. These strings of $0$'s and $1$'s span a $2^N$
dimensional Hilbert space. The corresponding bra state is
\begin{equation}
\Omega_{m,i_m...i_2i_1}^{\dagger}=<0|b_{i_m}...b_{i_2}b_{i_1}
\end{equation}
 It is very easy to verify that the ket and bra states satisfy the
orthonormality relations
\begin{eqnarray}
\Omega_{r,i_r...i_2i_1}^{\dagger}\Omega_{s,j_1j_2...i_s}=\left\{\matrix{\delta^{i_1}_{[j_1}\delta^{i_2}_{j_2}
...\delta^{i_r}_{j_s]}&&& {\textnormal{for}}~r=s\cr\cr
0&&&{\textnormal{for}}~r\neq s \cr}\right\}
\end{eqnarray}
Thus, the ket states $\Omega_{m,i_1i_2...i_m}$ form a basis for
the
 $m^{th}$ rank antisymmetric tensors of $SU(N)$ and the
 collection of these states
 $\left\{\Omega_{m}\right\}$ for $0\leq m\leq N$ form
 a basis for the $SO(2N)$ spinor with expansion coefficients
 being the antisymmetric tensors of $SU(N)$: $\Psi=k_0\Omega_{0}\psi+k_1\Omega_{i_1}\psi^{i_1}
+...+k_N\Omega_{i_1i_2...i_N}\psi^{i_1i_2...i_N}$. $k_i$'s are
normalization constants and are determined through the invariant
scalar product (see Eqs.(2.127) and (2.135)):
$\Psi^{'\dagger}\Psi'=\Psi^{\dagger}\Psi$ where
\begin{eqnarray}
\Psi^{\dagger}\Psi=\sum_{m=0}^N\frac{1}{m!}\psi^{\dagger}_{i_m...i_1}\psi^{i_1...i_m}\nonumber\\
{\textnormal
{where}}~~~\psi^{\dagger}_{i_m...i_1}\equiv\psi^{*i_1...i_m}~~~~~~
\end{eqnarray}
Finally,
\begin{equation}
\Psi=\sum_{p=0}^N\frac{1}{p!}\psi^{i_1...i_p}\prod_{q=1}^{p}b_{i_q}^{\dagger}|0>
\end{equation}

\section{SO(2N) chirality  and projection operators}

Recall that $\sigma_{12}$, $\sigma_{34}$,...,$\sigma_{2N-1~2N}$
can be simultaneously diagonalized. Therefore, we define the
\emph{$SO(2N)$ group chirality operator} $\Gamma_{0}$ as
\begin{equation}
\Gamma_0\equiv \prod_{k=1}^N
\Sigma_{2k-1~2k}=(-i)^N\Gamma_1\Gamma_2....\Gamma_{2N}
\end{equation}
Other easily verified forms of $\Gamma_{0}$ which are useful for
later computation are
\begin{equation}
\Gamma_{0}=\prod_{k=1}^{N}[b_k,b^{\dagger}_k]=\prod_{k=1}^N(1-2b_k^{\dagger}b_k)
\end{equation}
 One can further easily verify  the following properties of the
 chirality operator
\begin{eqnarray}
\Gamma_0^{\dagger}=\Gamma_0\nonumber\\
tr(\Gamma_0)=0\nonumber\\
\left\{\Gamma_0,\Gamma_{\mu}\right\}=0 \nonumber\\
\Gamma_0^2={\bf{1}} \nonumber\\
\left[\Gamma_0,\Sigma_{\mu\nu}\right]=0
\end{eqnarray}
Now,  $\Gamma_0$  commutes with all the generators
$\Sigma_{\mu\nu}$'s. This implies that the spinor representation,
$S(R)\left(=e^{i/2a_{\alpha\beta}\Sigma_{\alpha\beta}}\right)$,
also commutes with $\Gamma_0$: $[\Gamma_0,S(R)]=0$. This, together
with the fact that $\Gamma_0\neq$(constant)${\bf{1}}$ (see Eq.
(2.163)), we conclude, using  Schurs lemma, that the $2^N$ space
is reducible. Further, from $\Gamma_0^2={\bf{1}}$ we can decompose
the $2^{N}$ dimensional spinor into two $2^{N-1}$ irreducible
representations: $\Delta_{+}$ and $\Delta_{-}$, corresponding to
the eigenvalues $+1$ and $-1$ of $\Gamma_0$: $\Gamma_0\Psi=\pm
\Psi$. Thus, we write
\begin{eqnarray}
\Delta(2^N)=\Delta_{+}(2^{N-1})\oplus\Delta_{-}(2^{N-1})\nonumber\\
\Psi\equiv\Psi_{(+)}+\Psi_{(-)}\nonumber\\
\Psi_{(\pm)}=\frac{1}{2}\left({\bf{1}}\pm\Gamma_0\right)\Psi\nonumber\\
\Gamma_0\Psi_{(\pm)}=\pm \Psi_{(\pm)}
\end{eqnarray}
Also one has
\begin{equation}
{\textnormal{Generators
of~}}\Delta_{\pm}:~~~~~~~~\frac{1}{2}({\bf{1}}\pm\Gamma_0)\Sigma_{\mu\nu}=\frac{i}{4}({\bf{1}}\pm\Gamma_0)
\left[\Gamma_{\mu},\Gamma_{\nu}\right]
\end{equation}
$\Psi_{(+)}$ and $\Psi_{(-)}$ are called \emph{semi-spinors} and
they transform as two inequivalent representations.

Note that on using Eq.(2.164), one finds
$\Gamma_0b_i^{\dagger}=-b_i^{\dagger}\Gamma_0$. It is then
generalized to
\begin{equation}
\Gamma_0b_{i_1}^{\dagger}b_{i_2}^{\dagger}...b_{i_m}^{\dagger}|0>=
(-1)^mb_{i_1}^{\dagger}b_{i_2}^{\dagger}...b_{i_m}^{\dagger}|0>
\end{equation}
Using Eqs.(2.166) and (2.167) in Eq.(2.162) we get, for $N=$odd,
\begin{eqnarray}
\Psi_+=\sum_{p=0,2,..}^{N-1}\frac{1}{p!}\psi^{i_1...i_p}\prod_{q=1}^{p}b_{i_q}^{\dagger}|0>\nonumber\\
\Psi_-=\sum_{p=1,3,..}^{N}\frac{1}{p!}\psi^{i_1...i_p}\prod_{q=1}^{p}b_{i_q}^{\dagger}|0>
\end{eqnarray}
while for $N$=even, the sums above terminate at $N$ and $N-1$ for
$\Psi_+$ and $\Psi_{(-)}$, respectively. In general, the $p$ index
tensors of $SU(N)$ can be reduced to $N-p$ index tensors using the
Levi Civita tensor of $SU(N)$:
\begin{eqnarray}
\psi_{i_N...i_{p+1}}=\frac{1}{p!}\epsilon_{i_N...i_1}\psi^{i_1...i_p}\nonumber\\
\psi^{i_N...i_{p+1}\dagger}=\frac{1}{p!}\epsilon^{i_N...i_1}\psi_{i_1...i_p}^{\dagger}
\end{eqnarray}
As is evident from Eq.(2.169), positive chirality components of a
spinor contain antisymmetric $SU(N)$ of even rank, while the
negative chirality components have odd rank tensors. Thus, for odd
$N$:
\begin{eqnarray}
SO(2N)~~\supset ~~SU(N)\times U(1)~~~~~~~~~~~~~~~~~~~~~~~\nonumber\\
2^{N-1}_+\supset [{N\choose 0}]\oplus [{N\choose
2}]\oplus...\oplus
[{N\choose N-1}]\nonumber\\
2^{N-1}_-\supset [{N\choose 1}]\oplus [{N\choose
3}]\oplus...\oplus [{N\choose N}]~~~~~~
\end{eqnarray}
While for even $N$, the top and bottom expressions in Eq.(2.171)
terminate at $[{N\choose N}]$ and $[{N\choose N-1}]$ respectively.

The quantities $\frac{1}{2}({\bf{1}}\pm\Gamma_0)$ are
\emph{projection operators} as they possess the following
properties

\begin{eqnarray}
P_{+}=\frac{1}{2}\left({\bf{1}}+\Gamma_0\right),~~P_{-}=\frac{1}{2}\left({\bf{1}}-\Gamma_0\right)\nonumber\\
P_+^2=P_+,~~~~P_-^2=P_-\nonumber\\
P_+P_-=P_-P_+=0\nonumber\\
P_++P_-={\bf{1}}
\end{eqnarray}
Using the infinitesimal form of Eq.(2.131) and the fifth equation
in Eq.(2.165) it is seen that $P_{\pm}$ commute with $S(R)$ and
$S^{-1}(R)$,
\begin{equation}
P_{\pm}S(R)=S(R)P_{\pm},~~~~P_{\pm}S^{-1}(R)=S^{-1}(R)P_{\pm}
\end{equation}

Next, consider the transformation properties of the semi-spinors.
Using Eqs.(2.127) and (2.173) in Eq.(2.166), we can write
$\Psi_{(\pm)}'=P_{\pm}\Psi'=P_{\pm}S(R)\Psi\\
=S(R)(P_{\pm}\Psi)$.
One can consider the same strategy to find the transformation
property for $\Psi_{(\pm)}^{\dagger}$. The results are
\begin{eqnarray}
\Psi_{(\pm)}\rightarrow \Psi_{(\pm)}'=S(R)\Psi_{(\pm)}\nonumber\\
\Psi_{(\pm)}^{\dagger}\rightarrow
\Psi_{(\pm)}^{'\dagger}=\Psi_{(\pm)}^{\dagger}S^{-1}(R)
\end{eqnarray}

We now collect some useful results regarding the projection
operators. Using the definition of $P_{\pm}$ and the fourth
equation in Eq.(2.165) we get
\begin{equation}
P_+=+\Gamma_0P_+,~~~P_-=-\Gamma_0P_-
\end{equation}
Using the third equation in Eq.(2.165) we can show
\begin{equation}
\Gamma_{\mu}P_+=P_-\Gamma_{\mu},~~~
\Gamma_{\mu}P_-=P_+\Gamma_{\mu};~~~\mu=1,2,...,2N
\end{equation}
Using Eq.(2.176) and the fifth equation in Eq.(2.172) it follows
that
\begin{equation}
P_+\Gamma_{\mu_1}\Gamma_{\mu_2}...\Gamma_{\mu_k}P_+=0,~~~P_-\Gamma_{\mu_1}\Gamma_{\mu_2}...\Gamma_{\mu_k}P_-=0;~~~k={\textnormal
{odd}}~\&~k\leq N
\end{equation}
\begin{equation}
P_+\Gamma_{\mu_1}\Gamma_{\mu_2}...\Gamma_{\mu_k}P_-=0,~~~P_-\Gamma_{\mu_1}\Gamma_{\mu_2}...\Gamma_{\mu_k}P_+=0;~~~k={\textnormal
{even}}~\&~k\leq N
\end{equation}

Consider the product $I\equiv P_+\Gamma_1\Gamma_2..\Gamma_NP_+$
($N$ even). Using Eq.(2.175) and the definition of $\Gamma_0$
(Eq.(2.163)), we can write this product as $I=(-i)^NP_+\Gamma_1\\
\times\Gamma_2..\Gamma_N \Gamma_1\Gamma_2..\Gamma_N
\Gamma_{N+1}\Gamma_{N+2}..\Gamma_{2N}P_+$.
 Next, we make use of the Clifford algebra (Eq.(2.118)) to conclude
$I=(-i)^N(-1)^{\sum_{k=1}^{N-1}}P_+\Gamma_{N+1}\Gamma_{N+2}..\Gamma_{2N}P_+$.
In general an ordered product of $N$ distinct $\Gamma$'s,
$\Gamma_{N+1}\Gamma_{N+2}..\Gamma_{2N}$, can be expressed in terms
of the unordered product of the leftover $N$ distinct $\Gamma$'s
using the Levi Civita tensor. This is possible since there are
$2N$ distinct $\Gamma$'s:
\begin{equation}
\Gamma_{N+1}\Gamma_{N+2}..\Gamma_{2N}=\frac{1}{N!}\epsilon_{12...nj_1j_2...j_N}\Gamma_{j_1}\Gamma_{j_2}...\Gamma_{j_N}
\end{equation}
Using Eq.(2.179) and the fact
$(-i)^N(-1)^{\sum_{k=1}^{N-1}}=(-i)^{N^2}$ in the expression for
I, we get
$P_+\Gamma_1\Gamma_2..\Gamma_NP_+=\frac{(-i)^{N^2}}{N!}\epsilon_{12...nj_1j_2...j_N}
P_+\Gamma_{j_1}\Gamma_{j_2}...\Gamma_{j_N}P_+$. One can similarly
consider an identical product with $P_+$'s replaced by $P_-$'s and
also a product with $P$'s of opposite chirality. The results can
now be summarized as
\begin{eqnarray}
N~{\textnormal{even}}~~~~~~~~\nonumber\\
P_{\pm}\Gamma_1\Gamma_2..\Gamma_NP_{\pm}=\pm\frac{
1}{N!}\epsilon_{12...nj_1j_2...j_N}
P_{\pm}\Gamma_{j_1}\Gamma_{j_2}...\Gamma_{j_N}P_{\pm}\nonumber\\
P_{\pm}\Gamma_1\Gamma_2..\Gamma_NP_{\mp}=\mp\frac{
1}{N!}\epsilon_{12...nj_1j_2...j_N}
P_{\pm}\Gamma_{j_1}\Gamma_{j_2}...\Gamma_{j_N}P_{\mp}
\end{eqnarray}
 It is also very easy to see that the result for a product of odd number of $\Gamma_{\mu}$'s sandwiched between
  projection operators of opposite and same chirality will be
\begin{eqnarray}
N~{\textnormal{odd}}~~~~~~~~\nonumber\\
 P_{\pm}\Gamma_1\Gamma_2..\Gamma_NP_{\mp}=\pm\frac{
i}{N!}\epsilon_{12...nj_1j_2...j_N}
P_{\pm}\Gamma_{j_1}\Gamma_{j_2}...\Gamma_{j_N}P_{\mp}\nonumber\\
P_{\pm}\Gamma_1\Gamma_2..\Gamma_NP_{\pm}=\mp\frac{
i}{N!}\epsilon_{12...nj_1j_2...j_N}
P_{\pm}\Gamma_{j_1}\Gamma_{j_2}...\Gamma_{j_N}P_{\pm}
\end{eqnarray}

\section{Anomaly cancellation of the spinor representation}

Using  Eqs.(2.168), (2.172), (2.175) and (2.155) we can write
\cite{van}
\begin{eqnarray}
P_{\pm}Y^k\left(b_{i_1}^{\dagger}b_{i_2}^{\dagger}...b_{i_m}^{\dagger}|0>\right)
=\frac{{1\pm(-1)^m}}{2}(N-2m)^k\left(b_{i_1}^{\dagger}b_{i_2}^{\dagger}...b_{i_m}^{\dagger}|0>\right)
\end{eqnarray}
The $Y^k$-anomaly, $A_{\pm}(Y^k;N)=tr_{\pm}Y^k$ for the $\pm$
chirality spinor representation is given by
\begin{eqnarray}
A_{\pm}(Y^k;N)=\sum_{m=0}^N{N\choose
m}\frac{1\pm(-1)^m}{2}(N-2m)^k
\end{eqnarray}
where we take into consideration that each of the components,
${N\choose m}$, of the $SU(5)$ tensor, $\psi^{i_ii_2...i_m}$, has
charge, $N-2m$.

 One can now easily prove  the following results
\begin{eqnarray}
A_{\pm}(Y;N)=\left\{\matrix{\pm 1&&;N=1\cr 0&&;N\neq 1
  \cr}\right\}
\end{eqnarray}
\begin{eqnarray}
A_{\pm}(Y^3;N)=\left\{\matrix{\pm 24&&;N=3\cr \pm 1 &&;N=1
  \cr 0&&;N\neq1,~N\neq3
  \cr}\right\}
\end{eqnarray}
From this one can conclude that the spinor representations of
$SO(2N)$ are $U(1)$ \emph{anomaly-free} for $N=2$ and $N\geq 4$.
The cases  $N=1$ ($SO(2)\cong U(1)$) and $N=3$ ($SO(6)\cong
SU(4)$) have, indeed, an anomalous spinor representation.

\section{Charge conjugation under SO(2N) group. Complex and real spinor representations  of SO(2N)}

For infinitesimal transformations ($a_{\alpha\beta}\ll 1$),
Eqs.(2.127) and (2.131) give $\delta\Psi'\equiv
\Psi'-\Psi=-\frac{i}{4}a_{\alpha\beta}\Sigma_{\alpha\beta}\Psi$.
This expression, after complex conjugation and the fact that
$\Sigma_{\alpha\beta}$'s are Hermitian
($\Sigma_{\alpha\beta}^*=\Sigma_{\alpha\beta}^T$), takes the form,
$\delta\Psi^{'*}=
\frac{i}{4}a_{\alpha\beta}\Sigma_{\alpha\beta}^T\Psi^*$. Now
consider an invertible matrix, $B$, and left multiply the last
expression for $\delta\Psi^{'*}$ to obtain
\begin{eqnarray}
\delta\Psi^{c'}=
\frac{i}{4}a_{\alpha\beta}\left(B^{-1}\Sigma_{\alpha\beta}^TB\right)\Psi^c\nonumber\\
\Psi^c\equiv B^{-1}\Psi^{*};~~~BB^{-1}=B^{-1}B={\bf{1}}
\end{eqnarray}
If $\Psi^c$ is to remain invariant under $SO(2N)$ transformations,
then $\Psi^c$ must transform in the same way as $\Psi$. Thus
\begin{eqnarray}
\Psi^c \rightarrow \Psi^{c'}=S(R)\Psi^c\nonumber\\
\delta\Psi^{c'}=\frac{i}{4}a_{\alpha\beta}\left(-\Sigma_{\alpha\beta}\right)\Psi^c
\end{eqnarray}
Comparing Eqs.(2.186)
 with (2.187) gives
\begin{equation}
B^{-1}\Sigma_{\alpha\beta}^TB=-\Sigma_{\alpha\beta}
\end{equation}
 The left hand side of Eq.(2.188) using Eq.(2.132) can be written as
$B^{-1}\Sigma_{\alpha\beta}^TB=-\frac{i}{2}[(B^{-1}\Gamma_{\alpha}^TB)(B^{-1}\Gamma_{\beta}^TB)
-(B^{-1}\Gamma_{\beta}^TB)B^{-1}\Gamma_{\alpha}^TB)]$. This would
be equal to the right hand side of Eq. (2.188) if
\begin{equation}
B^{-1}\Gamma_{\alpha}^TB=\pm\Gamma_{\alpha}
\end{equation}
To see how $\Gamma_0$ transforms under charge conjugation we use
Eq.(2.163) to write
$B^{-1}\Gamma_{0}^TB=(-i)^N(B^{-1}\Gamma_{2N}^TB)...(B^{-1}\Gamma_{2}^TB)(B^{-1}\Gamma_{1}^TB)$.
Using Eq.(2.189) we get $B^{-1}\Gamma_{0}^TB=(-i)^N(\pm
1)^{2N}(-1)^{(N-1)+(N-2)+...+2+1}\Gamma_1\Gamma_2...\Gamma_{2N}$.
Thus
\begin{equation}
B^{-1}\Gamma_{0}^TB= B^{-1}\Gamma_{0}^*B=(-1)^N\Gamma_0
\end{equation}
where the second expression in Eq.(2.190) follows from the fact
that $\Gamma_{0}$ is Hermitian (see Eq.(2.165)).

 Next, we take on the task of determining the explicit
expression for the charge conjugation matrix. To determine the
specific representation of $B$ we first have to define the basis
for $\Gamma_{\mu}$'s and  specify the behavior of $\Gamma_{\mu}$'s
under charge conjugation (see Eq.(2.189)). For illustrative
purposes, let's assume the basis to be
$\Gamma_{2i}^T=\Gamma_{2i},~\Gamma_{2i-1}^T=-\Gamma_{2i}$ ($i=N$)
 and lets assume that $\Gamma_{\mu}$ transforms under $B$ as
 $B^{-1}\Gamma_{\mu}^TB=-\Gamma_{\mu}$. Then the correct
 form of $B$ is  $\prod_{i=1}^N\Gamma_{2i-1}$ for odd $N$. This is
 true because $B^{-1}\Gamma_{2i}^TB=(-1)^N
 \Gamma_{2i}=-\Gamma_{2i}$
 and similarly for $B^{-1}\Gamma_{2i-1}^TB=-(-1)^{N-1}
 \Gamma_{2i-1}=-\Gamma_{2i-1}$. Under transposition this specific representation of $B$
 can be evaluated as follows:
 $B^T=\Gamma_{2N-1}^T...\Gamma_3^T\Gamma_1^T=(-1)^{N}\Gamma_{2N-1}...\Gamma_3\Gamma_1=
 (-1)^{N}(-1)^{(N-2)+(N-1)+...+2+1}
\Gamma_{1}\Gamma_3...\Gamma_{2N-1}\\
=(-1)^{\frac{N(N+1)}{2}}B$.

 \noindent There are two choices for the basis of $\Gamma_{\mu}$'s. These are\\
\\
\noindent \underline{(A.)~~~Basis:
$\Gamma_{2i}^T=-\Gamma_{2\i},~~\Gamma_{2i-1}^T=\Gamma_{2i-1}$}\\
\\
\noindent Within this set there are two possible choices for the
representations of $B$ corresponding to $\pm$ sign in Eq.(2.189)

\begin{eqnarray}
{\textnormal{Set
A}}_1\equiv\left\{\matrix{B^{-1}\Gamma_{\mu}^TB=(-1)^{N+1}\Gamma_{\mu}\cr\cr
B=\prod_{i=1}^N\Gamma_{2i-1},~~~B^T=(-1)^{\frac{N(N-1)}{2}}B\cr}\right\}
\end{eqnarray}
\begin{eqnarray}
{\textnormal{Set
A}}_2\equiv\left\{\matrix{B^{-1}\Gamma_{\mu}^TB=(-1)^{N}\Gamma_{\mu}\cr\cr
B=\prod_{i=1}^N\Gamma_{2i},~~~B^T=(-1)^{\frac{N(N+1)}{2}}B\cr}\right\}
\end{eqnarray}
\\
\underline{(B.)~~~Basis:
$\Gamma_{2i}^T=\Gamma_{2i},~~\Gamma_{2i-1}^T=-\Gamma_{2i-1}$}\\

\begin{eqnarray}
{\textnormal{Set
B}}_1\equiv\left\{\matrix{B^{-1}\Gamma_{\mu}^TB=(-1)^{N+1}\Gamma_{\mu}\cr\cr
B=\prod_{\i=1}^N\Gamma_{2i},~~~B^T=(-1)^{\frac{N(N-1)}{2}}B\cr}\right\}
\end{eqnarray}
\begin{eqnarray}
{\textnormal{Set
B}}_2\equiv\left\{\matrix{B^{-1}\Gamma_{\mu}^TB=(-1)^{N}\Gamma_{\mu}\cr\cr
B=\prod_{i=1}^N\Gamma_{2i-1},~~~B^T=(-1)^{\frac{N(N+1)}{2}}B\cr}\right\}
\end{eqnarray}\\

\noindent One can choose any of the four representations of $B$.
The Set B$_2$ is the most frequent choice. Further, representation
free properties of $B$ are
\begin{equation}
B^2=(-1)^{\frac{N(N-1)}{2}}{\bf{1}},~~~~B^{-1}=B^{\dagger}=(-1)^{\frac{N(N-1)}{2}}B
\end{equation}
that is, $B$ is unitary.

To investigate  whether the representations of a particular
$SO(2N)$ group are real or complex, we need a deciding quantity,
$Q\equiv
B^{-1}[\frac{1}{2}({\bf{1}}\pm\Gamma_0)\Sigma_{\mu\nu}]^*B$. More
will be said about the significance of this quantity in the next
section. Here we will simply evaluate the quantity. Noting that
$\Gamma_0^*=\Gamma_0^T$ and $\Sigma_{\mu\nu}^*=\Sigma_{\mu\nu}^T$,
the expression for $Q$ can be simplified to
$Q=\frac{1}{2}[B^{-1}\Sigma_{\mu\nu}^TB\pm(B^{-1}\Gamma_{0}^TB)(B^{-1}\Sigma_{\mu\nu}^TB)]$.
Use of Eq.(2.190) at once gives
\begin{equation}
B^{-1}\left\{-\left[\frac{1}{2}\left({\bf{1}}\pm\Gamma_0\right)\Sigma_{\mu\nu}\right]^*\right\}B=
\left[\frac{1}{2}\left({\bf{1}}\pm(-1\right)^N\Gamma_0)\Sigma_{\mu\nu}\right]
\end{equation}

To find a connection between between the spinor representation,
$\Delta$ (group element: $S$), and its \emph{contragredient}
representation, $\breve{\Delta}$ (group element: \\
$(S^T)^{-1}$),
consider the product, $Z\equiv B^{-1}S^TB$. Use of the Eq.(2.131)
and (2.188) gives $Z\approx
{\bf{1}}+\frac{i}{4}a_{\alpha\beta}\Sigma_{\alpha\beta}\approx
S^{-1}$. Finally, left and right multiplying by $B$ and $B^{-1}$,
respectively, and taking the inverse on both sides of the
equation, we get
\begin{equation}
\left(S^T\right)^{-1}=BSB^{-1}
\end{equation}
The representations $\Delta$ and $\breve{\Delta}$ are unitarily
equivalent as seen by the Eq.(2.197).

 From Eq.(2.187), we can write $B^{-1}\Psi^{*'}=SB^{-1}\Psi$.
Left multiplying by $B$ and using Eq.(2.197) we get the
transformation of a spinor under the contrgredient representation:
\begin{equation}
\Psi^{*}\rightarrow \Psi^{*'}=\left(S^T\right)^{-1}\Psi^{*}
\end{equation}
Transposition of this equation gives:
$\Psi^{\dagger'}=\Psi^{\dagger}S^{-1}$, this is Eq.(2.135), which
was obtained independently.

One last observation about the transformation of spinors is that
there is one more quantity that transforms like $\Psi^{\dagger}$.
From Eq.(2.127), we can write $B\Psi'=BS\Psi$ and  using
Eq.(2.197) we can we obtain
\begin{equation}
\left(B\Psi\right)\rightarrow \left(B\Psi'\right)=(S^{T})^{-1}
\left(B\Psi\right)
\end{equation}
Taking the transpose of this equation we can write
\begin{equation}
\left(B\Psi\right)^T\rightarrow \left(B\Psi'\right)^T=
\left(B\Psi\right)^TS^{-1}
\end{equation}
which transforms like $\Psi^{\dagger}$.

It is worth summarizing the transformation properties of the above
spinors as they will be very useful later on in looking for
$SO(2N)$
invariants.\\
\\
\underline{Spinor (Cogredient) representation, $\Delta$: S(R)}
\begin{eqnarray}
\Psi\rightarrow \Psi^{'}=S\Psi \nonumber\\
\left(B^{-1}\Psi^{*}\right)\rightarrow
\left(B^{-1}\Psi^{*'}\right)=S\left(B^{-1}\Psi^{*}\right)
\end{eqnarray}
\\
\underline{Contragredient representation, $\breve{\Delta}$:
$(S^T(R))^{-1}$}
\begin{eqnarray}
\Psi^{*}\rightarrow \Psi^{*'}=(S^T)^{-1}\Psi^{*}~\Rightarrow~
\Psi^{\dagger}\rightarrow
\Psi^{\dagger'}=\Psi^{\dagger}S^{-1}~~~~~~~~~~~~~~~~~~~~~~~\nonumber\\
\left(B\Psi\right)\rightarrow \left(B\Psi'\right)=(S^{T})^{-1}
\left(B\Psi\right)~\Rightarrow ~\left(B\Psi\right)^T\rightarrow
\left(B\Psi'\right)^T=\left(B\Psi\right)^TS^{-1}~~~~~~
\end{eqnarray}

A useful result for the construction of invariants later is a
property relating charge conjugation operator and the projection
operators: $P_{\pm}B=BP_{\pm}$ or $BP_{\mp}$ according to whether
$N$ is even or odd, respectively. This result does not depend on
the representation of $B$. Using either representation
$(\prod_{\alpha=1}^N\Gamma_{2\alpha-1},~\prod_{\alpha=1}^N\Gamma_{2\alpha})$
and togther with Eq.(2.190), one can very easily prove the above
result. Collecting
\begin{eqnarray}
P_{\pm}B=BP_{\pm};~~~~ N~{\textnormal{even}}\nonumber\\
P_{\pm}B=BP_{\mp};~~~~ N~{\textnormal{odd}}
\end{eqnarray}
Similarly, one can, with relative ease, prove
\begin{eqnarray}
P_{\pm}^TB=BP_{\pm},~~BP_{\pm}^*=P_{\pm}B,~~B^{-1}P_{\pm}^*=P_{\pm}B^{-1};~~~ N~{\textnormal{even}}\nonumber\\
P_{\pm}^TB=BP_{\mp},~~BP_{\pm}^*=P_{\mp}B,~~B^{-1}P_{\pm}^*=P_{\mp}B^{-1};~~~
N~{\textnormal{even}}
\end{eqnarray}
These properties of projection and charge conjuagation operators
are going to be very useful
when looking for non-vanishing $SO(2N)$ invariants.\\
\\
\noindent\textsc{Ccomplex and real spinor representations of SO(2N)}\\
\\
To motivate the discussion under this topic, assume $\Delta_{+}^*$
and $\Delta_{-}^*$ (contragredient representations) to be the
complex conjugate of the irreducible representations, $\Delta_{+}$
and $\Delta_{-}$, respectively. Further, recall from Eq.(2.167)
that
$L_{\mu\nu}^{(\pm)}\equiv[\frac{1}{2}({\bf{1}}\pm\Gamma_0)\Sigma_{\mu\nu}]$
represents the generators of $\Delta_{\pm}$, therefore the
generators of $\Delta_{\pm}^{*}$ are
$-L^{(\pm)*}_{\mu\nu}\equiv-[\frac{1}{2}({\bf{1}}\pm\Gamma_0)\Sigma_{\mu\nu}]^*$.
This is because if under $\Delta_{\pm}$: $\Psi_{(\pm)}' \approx
-\frac{i}{2}a_{\mu\nu}L^{(\pm)}_{\mu\nu}\Psi_{(\pm)}$, then under
${\Delta_{\pm}^{*}}$: $\Psi_{(\pm)}^{*'} \approx
-\frac{i}{2}a_{\mu\nu}[-L^{(\pm)*}_{\mu\nu}]\Psi_{(\pm)}^*$.

If $\Delta_{\pm}$ is \emph{equivalent} to $\Delta_{\pm}^{*}$
(${\Delta_{\pm}^{*}}\sim  \Delta_{\pm}$), then there exists a
charge conjugation (unitary) matrix $B$, such that
$B^{-1}[-L^{(\pm)*}_{\mu\nu}]B=L^{(\pm)}_{\mu\nu}$ and the
representation is said to be \emph{real}.

If, on the other hand, the representation is \emph{complex}:
${\Delta_{\pm}^{*}}\sim  \Delta_{\mp}$  then in this case the
generators in the $\Delta_{\pm}$ are related to the one in
${\Delta_{\pm}^{*}}$ by
$B^{-1}[-L^{(\pm)*}_{\mu\nu}]B=L^{(\mp)}_{\mu\nu}$.

In the case of real representations, the consequences due to the
existence of matrix $B$ are now going to be explored. Start with
 $B^{-1}[-L^{(\pm)*}_{\mu\nu}]B=L^{(\pm)}_{\mu\nu}$ and take complex
conjugate of this equation to get,
 $B^TL^{(\pm)}_{\mu\nu}(B^{-1})^T=-L^{(\pm)*}_{\mu\nu}$. Eliminating
$L^{(\pm)*}_{\mu\nu}$ between these two equations and left
multiplying the resulting equation by $(B^{-1})^TB$, we get
 $[L^{(\pm)}_{\mu\nu},(B^{-1})^TB]=0$. Then Schurs lemma implies
$(B^{-1})^TB=\lambda{\bf{1}}$, with $\lambda$ being a constant.
Left multiplying this last equation by $B^T$ gives $B=\lambda
B^T$. Transposing this equation we obtain $B^T=\lambda B$. It is
now easy to see from the last two equations that $\lambda^2=1$.
Thus we have two cases: $B=+B^T$ with $B^*B=+{\bf{1}}$ and
$B=-B^T$ with $B^*B=-{\bf{1}}$. We now go a step further and
investigate the possibility of finding real representations which
satisfy these two conditions by $B$.

We will now prove that in the first case of unitary symmetric, $B$
($B=+B^T$, $B^{-1}=B^{\dagger}$), one can find a basis such that
representations are real.  To make progress, we begin by writing
$B$ as the product of a unitary matrix, $U$ ($U^{-1}=U^{\dagger}$)
and its transpose: $B=UU^T$ (certainly $B$ is unitary symmetric).
Inserting this expression for $B$ in
$B^{-1}[-L^{(\pm)*}_{\mu\nu}]B=L^{(\pm)}_{\mu\nu}$ and left and
right multiplying the resulting equation by $U$ and $U^{-1}$,
respectively, we obtain
$-(UL^{(\pm)}_{\mu\nu}U^{-1})^T=UL^{(\pm)}_{\mu\nu}U^{-1}$, where
we have used the fact that $L^{(\pm)}_{\mu\nu}$ are Hermitian as
is evidenced by the explicit expression for $L_{\mu\nu}$ above. If
we define a new set of generators in this basis as
$l^{(\pm)}_{\mu\nu}\equiv UL^{(\pm)}_{\mu\nu}U^{-1}$ and take
Hermitian conjugate on both sides of this expression we conclude
that $l^{(\pm)\dagger}_{\mu\nu}=l^{(\pm)}_{\mu\nu}$. Further, it
also follows that $l^{(\pm)T}_{\mu\nu}=-l^{(\pm)}_{\mu\nu}$. Thus,
the generators, $l^{(\pm)}_{\mu\nu}$, are purely imaginary and
antisymmetric. Therefore, the representations
$e^{-\frac{i}{2}a_{\mu\nu}l^{(\pm)}_{\mu\nu}}$ will be real. They
are also referred to as \emph{strictly-real} or
\emph{real-positive} or \emph{orthogonal-like} representations.

 In the case when $B$ is unitary antisymmetric: $B=-B^T$,
 $B^{-1}=B^{\dagger}$, one cannot find any basis in which the
 representations are real. In this case the representation is said
 to be \emph{pseudo-real} or \emph{real-negative} or
 \emph{simplectic-like} representations.\\

 \noindent Let us now summarize all
 our findings:
\begin{equation}
{\textnormal {Generators of}}~\Delta_{\pm}:\nonumber\\
L_{\mu\nu}^{(\pm)}\equiv
\left[\frac{1}{2}({\bf{1}}\pm\Gamma_0)\Sigma_{\mu\nu}\right]
\end{equation}
\begin{equation}
{\textnormal {Generators of}}~\Delta_{\pm}^{*}:\nonumber\\
-L^{(\pm)*}_{\mu\nu}\equiv-\left[\frac{1}{2}({\bf{1}}\pm\Gamma_0)\Sigma_{\mu\nu}\right]^*
\end{equation}\\

 \noindent \underline{Real spinor representations of SO(2N)}
\begin{itemize} \item $\Delta_{+}$ is inequivalent to
$\Delta_{-}$ and $\Delta_{+}^*=\Delta_{+}$,
$\Delta_{-}^*=\Delta_{-}$ \item Relation between $\Delta_{\pm}^*$
and $\Delta_{\pm}$
\begin{equation}
B^{-1}[-L^{(\pm)*}_{\mu\nu}]B=L^{(\pm)}_{\mu\nu};~~~B^{-1}=B^{\dagger}
\end{equation}
\item \emph{Strictly-real} representations
\begin{equation}
B=+B^T;~~~BB^*=+{\bf{1}}
\end{equation}
A basis could be found in which the generators are purely
imaginary and antisymmetric, making the representations completely
real
      \item \emph{Pseudo-real} representations
      \begin{equation}
B=-B^T;~~~BB^*=-{\bf{1}}
\end{equation}
No basis could be found in which the representations are real\\
\end{itemize}

\noindent \underline{Complex spinor representations of SO(2N)}
\begin{itemize}
\item $\Delta_{+}$ is inequivalent to $\Delta_{-}$ and
$\Delta_{+}^*=\Delta_{-}$, $\Delta_{-}^*=\Delta_{+}$ \item
Relation between $\Delta_{\pm}^*$ and $\Delta_{\mp}$
\begin{equation}
B^{-1}[-L^{(\pm)*}_{\mu\nu}]B=L^{(\mp)}_{\mu\nu};~~~B^{-1}=B^{\dagger}
\end{equation}\\
\end{itemize}
From the definitions of $\Psi_{(\pm)}$ (Eq.(2.166)) and $\Psi^c$
(Eq.(2.186))one could write
$(\Psi_{(\pm)})^c=B^{-1}[\frac{1}{2}({\bf{1}}+\Gamma_0)\Psi]^*$.
On using Eq.(2.190) and the fact $BB^{-1}={\bf{1}}$ we get, at
once,
\begin{equation}
\left(\Psi_{(\pm)}\right)^c=\frac{1}{2}\left[{\bf{1}}\pm
(-1)^N\Gamma_0\right]\Psi^c
\end{equation}
Further, recall from Eq.(2.196)
\begin{equation}
B^{-1}\left[-L^{(\pm)*}_{\mu\nu}\right]B=
\left[\frac{1}{2}\left({\bf{1}}\pm(-1\right)^N\Gamma_0)\Sigma_{\mu\nu}\right]
\end{equation}

Equipped with Eqs.(2.207) through (2.212), we are now ready to
classify $SO(2N)$ according to real and complex spinor
representations.
\newpage
  \noindent \underline{Case 1. $N=$odd. Complex
representations. Theory is
\emph{chiral}}\\
\\
For $N$ odd $(N=2k+1)$, Eq.(2.212) gives
$B^{-1}[-L^{(\pm)*}_{\mu\nu}]B=L^{(\mp)}_{\mu\nu}$ which,
according to Eq.(2.210) tells us that the spinor representations
of $SO(4k+2)$ are complex.

Furthermore, for odd $N$, Eq.(2.213) implies
$\left(\Psi_{(\pm)}\right)^c=\frac{1}{2}\left[{\bf{1}}\mp
\Gamma_0\right]\Psi^c$, then from the definition of $\Psi_{(\pm)}$
it implies
\begin{eqnarray}
\left(\Psi_{(+)}\right)^c=\left(\Psi^c\right)_{-}\nonumber\\
\left(\Psi_{(-)}\right)^c=\left(\Psi^c\right)_{+}
\end{eqnarray}

If we, for a moment, assign the semi-spinors $\Psi_{(+)}$ and
$\Psi_{(-)}$ to left ($L$) and right ($R$)-handed fermions,
respectively, then the interpretation of Eq.(2.213) would be that
the charge conjugated state of a left-handed fermion is a
right-handed anti-fermion and vice versa. For example, for fermion
$\alpha$, $\alpha_L^+\equiv \alpha_L^{-c}$ is the anti-fermion
with respect to $\alpha_R^-$.\\
\\
\noindent\underline{Case 2. $N=$even. Real representations. Theory
is
\emph{vector-like}}\\
\\
For $N$ even $(N=2m)$, Eq.(2.212) gives
$B^{-1}[-L^{(\pm)*}_{\mu\nu}]B=L^{(\pm)}_{\mu\nu}$ which according
to Eq.(2.207), tells us that the spinor representations of
$SO(4m)$ are real.

Additionally, for  $N$ even, Eq.(2.211) implies
$\left(\Psi_{(\pm)}\right)^c=\frac{1}{2}\left[{\bf{1}}\pm
\Gamma_0\right]\Psi^c$, then from the definition of $\Psi_{(\pm)}$
it implies
\begin{eqnarray}
\left(\Psi_{(+)}\right)^c=\left(\Psi^c\right)_{+}\nonumber\\
\left(\Psi_{(-)}\right)^c=\left(\Psi^c\right)_{-}
\end{eqnarray}
In this case, the interpretation would be  that the charge
conjugated state of a left-handed particle is a left-handed
anti-particle and similar argument can be made for a right-handed
particle.

The deciding factor for a particular $SO(4m)$ group to have
strictly real or pseudo-real representations is whether $B=+B^T$
or $B=-B^T$. Here, the specific representation for $B$ is not
important, so, for example, we could choose either
$B^T=(-1)^{\frac{N(N+1)}{2}}B$ or $B^T=(-1)^{\frac{N(N-1)}{2}}B$
(see Eqs.(2.191)-(2.195)). Either expression gives, for $m$ even
(m=2k, N=8k), $B^T=+B$ and for $m$ odd (m=2k+1, N=8k+4), $B^T=-B$.

Thus, $SO(8k)$ possesses  strictly real spinor representations,
while $SO(8k+4)$ have pseudo-real spinor representations.

\section{General SO(2N) scalar and vector type couplings }

In this section we form $SO(2N)$ invariants appearing in both the
Lagrangian and the superpotential.  We will construct four
distinct $SO(2N)$ invariants. They will be built from
$\Psi^{\dagger}$ and $\Psi$, $\Psi^{T}$ and $\Psi$,
$\Psi^{\dagger}$ and $\Psi^{*}$, and $\Psi^{T}$ and $\Psi^{*}$.
Further, classification of these $SO(2N)$ invariants as Lorentz
vectors or scalars will require an appropriate insertion of Dirac
charge conjugation matrix, $C$ and Dirac matrices, $\gamma^A$
($A=0-3$). For completeness, we list all the possible Lorentz
scalars and vectors constructed out of Dirac spinors, $\chi$.
\begin{eqnarray}
{{\textnormal{Lorentz Scalars }}}\nonumber\\
{\overline{\chi}}\chi={\overline{\chi^c}}\chi^c={\overline{\chi}}_L\chi_R+{\overline{\chi}}_R\chi_L\nonumber\\
{\overline{\chi^c}}\chi=\chi_L^TC\chi_L+\chi_R^TC\chi_R\nonumber\\
{\overline{\chi}}\chi^c={\overline{\chi_L}}C{\overline{\chi_L}^T}+{\overline{\chi_R}}C{\overline{\chi_R}^T}
\end{eqnarray}
\begin{eqnarray}
{{\textnormal{Lorentz Vectors}}}\nonumber\\
{\overline{\chi}}\gamma^A\chi={\overline{\chi}}_L\gamma^A\chi_L+{\overline{\chi}}_R\gamma^A\chi_R\nonumber\\
{\overline{\chi^c}}\gamma^A\chi^c=\chi_L^TC\gamma^AC{\overline{\chi}}_L^T+\chi_R^TC\gamma^AC{\overline{\chi}}_R^T\nonumber\\
{\overline{\chi^c}}\gamma^A\chi=\chi_L^TC\gamma^A\chi_R+\chi_R^TC\gamma^A\chi_L\nonumber\\
{\overline{\chi}}\gamma^A\chi^c={\overline{\chi}_L}\gamma^AC{\overline{\chi}_R^T}+{\overline{\chi}_R}\gamma^AC{\overline{\chi}_L^T}
\end{eqnarray}
where
\begin{eqnarray}
{\overline{\chi}}=\chi^{\dagger}\gamma^0\nonumber\\
\chi^c=C{\overline{\chi}}^T\nonumber\\
\chi=\chi_L+\chi_R\nonumber\\
\chi_{L\choose R}=\frac{1}{2}(1\pm\gamma_5)\chi
\end{eqnarray}

\begin{eqnarray}
\gamma_A\gamma_B+\gamma_B\gamma_A=2\eta_{AB}\nonumber\\
\eta=diag(1,-1,-1,-1)\nonumber\\
\gamma^{0\dagger}=\gamma^0;~~~\gamma^{A\dagger}=\gamma^0\gamma^A\gamma^0\nonumber\\
\gamma_5=i\gamma^0\gamma^1\gamma^2\gamma^3\nonumber\\
C=-C^{-1}=-C^{\dagger}=-C^T=i\gamma^2\gamma^0\nonumber\\
C(\gamma^{A})^TC^{-1}=-\gamma^A
\end{eqnarray}\\
\noindent \textsc{invariants formed from $\Psi^{\dagger}$ and $\Psi$: Lagrangian case}\\
 \\
We commence  by constructing bilinear spinors formed from the
Kronecker product of $\Psi^{\dagger}$, $\Psi$ and a symmetrized or
an antisymmetrized product of appropriate number of
$\Gamma_{\mu}$'s. These bilinears then transform as vectors and
tensors of $SO(2N)$ as constructed earlier. For example, the
quantity defined by
$\Upsilon_{\{\mu\nu\rho\}}=\Psi^{\dagger}\Gamma_{\{\mu}\Gamma_{\nu}\Gamma_{\rho\}}\Psi$
transforms as a third rank symmetric tensor. This is because
$\Upsilon_{\{\mu\nu\rho\}}'=\Psi^{'\dagger}\Gamma_{\{\mu}\Gamma_{\nu}\Gamma_{\rho\}}\Psi'=
\Psi^{\dagger}S^{-1}\Gamma_{\{\mu}\Gamma_{\nu}\Gamma_{\rho\}}S\Psi
=\Psi^{\dagger}(S^{-1}\Gamma_{\{\mu}S)(S^{-1}\\
\times\Gamma_{\nu}S)(S^{-1}\Gamma_{\rho\}}S)\Psi
=R_{\mu\alpha}R_{\nu\beta}R_{\rho\gamma}\Upsilon_{\{\alpha\beta\gamma\}}$
and where we have used Eqs.(110), (111), (113) and (118). We now
list some bilinears along with their transformation properties
however, we will not specify the symmetry associated with their
indices:
\begin{eqnarray}
{\textnormal{Scalar}}:~\Upsilon=\Psi^{\dagger}\Psi;~~\Upsilon'=\Upsilon\nonumber\\
{\textnormal{Vector}}:~\Upsilon_{\mu_1}=\Psi^{\dagger}\Gamma_{\mu_1}\Psi;~~\Upsilon_{\mu_1}'
=R_{\mu_1\nu_1}\Upsilon_{\nu_1}\nonumber\\
{\textnormal{$2^{nd}$Rank
Tensor}}:~\Upsilon_{\mu_1\mu_2}=\Psi^{\dagger}\Gamma_{\mu_1}\Gamma_{\mu_2}\Psi;~~
\Upsilon_{\mu_1\mu_2}'=R_{\mu_1\nu_1}R_{\mu_2\nu_2}\Upsilon_{\nu_1\nu_2}\nonumber\\
{\textnormal{$r^{th}$Rank
Tensor}}:~\Upsilon_{\mu_1...\mu_r}=\Psi^{\dagger}\Gamma_{\mu_1}...\Gamma_{\mu_r}\Psi;~~
\Upsilon_{\mu_1...\mu_r}'=R_{\mu_1\nu_1}...R_{\mu_r\nu_r}\Upsilon_{\nu_1...\nu_r}\nonumber\\
\end{eqnarray}

An arbitrary $SO(2N)$ tensor of $r^{th}$ rank , can be written as
the sum of four bilinear semi-spinors
($\Psi=\Psi_{(+)}+\Psi_{(-)}$):
$\Psi^{\dagger}\Gamma_{\mu_1}\Gamma_{\mu_2}...\Gamma_{\mu_r}\Psi
=\Psi^{\dagger}_{(\pm)}\Gamma_{\mu_1}\Gamma_{\mu_2}...\\
\times...\Gamma_{\mu_r}\Psi_{(\pm)}+
\Psi^{\dagger}_{(\pm)}\Gamma_{\mu_1}\Gamma_{\mu_2}...\Gamma_{\mu_r}\Psi_{(\mp)}$.
Using  $\Psi_{(\pm)}=P_{\pm}\Psi$, one can write
$\Psi^{\dagger}_{(\pm)}\Gamma_{\mu_1}\Gamma_{\mu_2}...\Gamma_{\mu_r}\Psi_{(\pm)}=
\Psi^{\dagger}[P_{\pm}\Gamma_{\mu_1}\Gamma_{\mu_2}...\Gamma_{\mu_r}P_{\pm}]\Psi$.
From Eq.(2.177), this quantity vanishes if $r$ is odd. Similarly,
one can show using Eq.(2.178) that the quantity
$\Psi^{\dagger}_{(\pm)}\Gamma_{\mu_1}\Gamma_{\mu_2}...\Gamma_{\mu_r}\Psi_{(\mp)}$
vanishes if $r$ is even. Hence, we have the following results
\begin{eqnarray}
\forall N,~r~{\textnormal{odd}}~~~~~~~~~~~~~~~~~~~~~\nonumber\\
\Psi^{\dagger}_{(\pm)}\Gamma_{\mu_1}\Gamma_{\mu_2}...\Gamma_{\mu_r}\Psi_{(\pm)}=0~~~~~~~~~~~~~~~~~
\end{eqnarray}
\begin{eqnarray}
\forall N,~r~{\textnormal{even}}~~~~~~~~~~~~~~~~~~~~~\nonumber\\
\Psi^{\dagger}_{(\pm)}\Gamma_{\mu_1}\Gamma_{\mu_2}...\Gamma_{\mu_r}\Psi_{(\mp)}=0~~~~~~~~~~~~~~~~~
\end{eqnarray}
Recall that the  representation, $\Delta$,  decomposes into two
irreducible representations $\Delta_{\pm}$, while the
contragredient representations, $\breve{\Delta}$ splits into two
irreducible  $\Delta_{\pm}^*$:
\begin{eqnarray}
\Delta=\Delta_{+}\oplus\Delta_{-}\nonumber\\
 \breve{\Delta}=\Delta_{+}^*\oplus\Delta_{-}^*\nonumber\\
\Delta_{+}^*=\Delta_{-},~~\Delta_{-}^*=\Delta_{+}; ~~~{\textnormal
{for}}~N~{\textnormal{odd}}\nonumber\\
\Delta_{+}^*=\Delta_{+},~~\Delta_{-}^*=\Delta_{-};~~~{\textnormal
{for}}~N~{\textnormal{even}}\nonumber\\
\Delta_{+} ~{\textnormal {and}}~ \Delta_{-} ~{\textnormal{are
inequivalent}}
\end{eqnarray}

The surviving even and odd rank tensors of $SO(2N)$ are\\
\\
 \textsl{Even Rank Antisymmetric Tensors}: $\Delta_{\pm}^*\otimes \Delta_{\pm}$
\begin{eqnarray}
\Psi^{\dagger}_{(\pm)}\Psi_{(\pm)}\nonumber\\
\Psi^{\dagger}_{(\pm)}\Gamma_{[\mu_1}\Gamma_{\mu_2]}\Psi_{(\pm)}\nonumber\\
\Psi^{\dagger}_{(\pm)}\Gamma_{[\mu_1}\Gamma_{\mu_2}\Gamma_{\mu_3}\Gamma_{\mu_4]}\Psi_{(\pm)}\nonumber\\
......~~~~~~~......~~~~~~~......\nonumber\\
\Psi^{\dagger}_{(\pm)}\Gamma_{[\mu_1}\Gamma_{\mu_2}\Gamma_{\mu_3}...\Gamma_{\mu_m]}\Psi_{(\pm)}
,~~~\left\{\matrix{m=N-1;&{\textnormal{N odd}}\cr
m=N;&{\textnormal{N even}}
  \cr}\right\}
\end{eqnarray}\\
\\
\textsl{Odd Rank Antisymmetric Tensors}: $\Delta_{\pm}^*\otimes
\Delta_{\mp}$
\begin{eqnarray}
\Psi^{\dagger}_{(\pm)}\Gamma_{\mu_1}\Psi_{(\mp)}\nonumber\\
\Psi^{\dagger}_{(\pm)}\Gamma_{[\mu_1}\Gamma_{\mu_2}\Gamma_{\mu_3]}\Psi_{(\mp)}\nonumber\\
\Psi^{\dagger}_{(\pm)}\Gamma_{[\mu_1}\Gamma_{\mu_2}\Gamma_{\mu_3}\Gamma_{\mu_4}\Gamma_{\mu_5]}\Psi_{(\mp)}\nonumber\\
......~~~~~~~......~~~~~~~......\nonumber\\
\Psi^{\dagger}_{(\pm)}\Gamma_{[\mu_1}\Gamma_{\mu_2}\Gamma_{\mu_3}...\Gamma_{\mu_m]}\Psi_{(\mp)}
,~~~\left\{\matrix{m=N;&{\textnormal{N odd}}\cr
m=N-1;&{\textnormal{N even}}
  \cr}\right\}
\end{eqnarray}

Each of the these irreducible antisymmetric tenors have
dimensionalities ${2N\choose r}$, where $r(<N)$ represents the
rank of the tensor.

We do not get any new tensors for $m>n$. To illustrate this point,
consider a tensor of rank $N+k$ ($k\leq N$), this tensor can be
reduced to a tensor of rank ($2N-[N+k]=$) $N-k$  by contraction
with an $SO(2N)$ Levi-Civita tensor carrying $2N$ distinct
indices.

There is an additional subtlety for tensors of rank $N$. For $N$
even, the real tensor $\Upsilon_{\mu_1...\mu_N}^{(N=even)}$
 of dimensionality ${2N\choose N}$ splits into  two real tensors $\Psi^{\dagger}_{(+)}\Gamma_{[\mu_1}\Gamma_{\mu_2}...\Gamma_{\mu_N]}\Psi_{(+)}$ and
$\Psi^{\dagger}_{(-)}\Gamma_{[\mu_1}\Gamma_{\mu_2}...\Gamma_{\mu_N]}\Psi_{(-)}$
each of dimension $\frac{1}{2}{2N\choose N}$. These decomposed
tensors are \emph{real self-dual and  anti-self-dual},
respectively,

\begin{eqnarray}
\Upsilon_{\mu_1...\mu_N}^{(N=even)}=\Psi^{\dagger}_{(+)}\Gamma_{[\mu_1}\Gamma_{\mu_2}...\Gamma_{\mu_N]}\Psi_{(+)}
+\Psi^{\dagger}_{(-)}\Gamma_{[\mu_1}\Gamma_{\mu_2}...\Gamma_{\mu_N]}\Psi_{(-)}\nonumber\\
\nonumber\\
{{\Psi^{\dagger}_{(+)}\Gamma_{[\mu_1}\Gamma_{\mu_2}...\Gamma_{\mu_N]}\Psi_{(+)}}\choose
{\Psi^{\dagger}_{(-)}\Gamma_{[\mu_1}\Gamma_{\mu_2}...\Gamma_{\mu_N]}\Psi_{(-)}}}~~~~~~~~~~~~~~~~~~~~~~~~~~~~~~~~~~~~~~~~~~~~~~~~~~~~~~~\nonumber\\
=\frac{1}{2}\left(\delta_{\mu_1\nu_1}
\delta_{\mu_2\nu_2}...\delta_{\mu_{N}\nu_{N}}\pm
\frac{1}{N!}\epsilon_{\mu_1\mu_2...\mu_N\nu_1\nu_2...\nu_N}\right)\Upsilon_{\nu_1...\nu_N}^{(N=even)}\nonumber\\
\nonumber\\
{{\Psi^{\dagger}_{(+)}\Gamma_{[\mu_1}\Gamma_{\mu_2}...\Gamma_{\mu_N]}\Psi_{(+)}}\choose
{\Psi^{\dagger}_{(-)}\Gamma_{[\mu_1}\Gamma_{\mu_2}...\Gamma_{\mu_N]}\Psi_{(-)}}}~~~~~~~~~~~~~~~~~~~~~~~~~~~~~~~~~~~~~~~~~~~~~~~~~~~~~~~\nonumber\\
= \pm\frac{
1}{N!}\epsilon_{\mu_1...\mu_N\nu_1...\nu_N}{{\Psi^{\dagger}_{(+)}\Gamma_{[\mu_1}\Gamma_{\mu_2}...\Gamma_{\mu_N]}\Psi_{(+)}}\choose
{\Psi^{\dagger}_{(-)}\Gamma_{[\mu_1}\Gamma_{\mu_2}...\Gamma_{\mu_N]}\Psi_{(-)}}}\nonumber\\
\end{eqnarray}

For $N$ odd, the real tensor $\Upsilon_{\mu_1...\mu_N}^{(N=odd)}$
 of dimensionality ${2N\choose N}$ splits into  two complex tensors $\Psi^{\dagger}_{(+)}\Gamma_{[\mu_1}\Gamma_{\mu_2}...\Gamma_{\mu_N]}\Psi_{(-)}$ and
$\Psi^{\dagger}_{(-)}\Gamma_{[\mu_1}\Gamma_{\mu_2}...\Gamma_{\mu_N]}\Psi_{(+)}$
each of dimension $\frac{1}{2}{2N\choose N}$. These decomposed
tensors are \emph{complex self-dual and anti-self-dual},
respectively,

\begin{eqnarray}
\Upsilon_{\mu_1...\mu_N}^{(N=odd)}=\Psi^{\dagger}_{(+)}\Gamma_{[\mu_1}\Gamma_{\mu_2}...\Gamma_{\mu_N]}\Psi_{(-)}
+\Psi^{\dagger}_{(-)}\Gamma_{[\mu_1}\Gamma_{\mu_2}...\Gamma_{\mu_N]}\Psi_{(+)}~~~~~~~\nonumber\\
\nonumber\\
{{\Psi^{\dagger}_{(+)}\Gamma_{[\mu_1}\Gamma_{\mu_2}...\Gamma_{\mu_N]}\Psi_{(-)}}\choose
{\Psi^{\dagger}_{(-)}\Gamma_{[\mu_1}\Gamma_{\mu_2}...\Gamma_{\mu_N]}\Psi_{(-)}}}~~~~~~~~~~~~~~~~~~~~~~~~~~~~~~~~~~~~~~~~~~~~~~~~~~~~~~~\nonumber\\
=\frac{1}{2}\left(\delta_{\mu_1\nu_1}
\delta_{\mu_2\nu_2}...\delta_{\mu_{N}\nu_{N}}\pm
\frac{i}{N!}\epsilon_{\mu_1\mu_2...\mu_N\nu_1\nu_2...\nu_N}\right)\Upsilon_{\nu_1...\nu_N}^{(N=odd)}\nonumber\\
\nonumber\\
{{\Psi^{\dagger}_{(+)}\Gamma_{[\mu_1}\Gamma_{\mu_2}...\Gamma_{\mu_N]}\Psi_{(-)}}\choose
{\Psi^{\dagger}_{(-)}\Gamma_{[\mu_1}\Gamma_{\mu_2}...\Gamma_{\mu_N]}\Psi_{(+)}}}~~~~~~~~~~~~~~~~~~~~~~~~~~~~~~~~~~~~~~~~~~~~~~~~~~~~~~~\nonumber\\\nonumber\\
= \pm\frac{
i}{N!}\epsilon_{\mu_1...\mu_N\nu_1...\nu_N}{{\Psi^{\dagger}_{(+)}\Gamma_{[\mu_1}\Gamma_{\mu_2}...\Gamma_{\mu_N]}\Psi_{(-)}}\choose
{\Psi^{\dagger}_{(-)}\Gamma_{[\mu_1}\Gamma_{\mu_2}...\Gamma_{\mu_N]}\Psi_{(+)}}}\nonumber\\
\end{eqnarray}
Eqs. (2.225) and (2.226) can be verified at once upon using
$\Psi_{(\pm)}=P_{\pm}\Psi$ and Eqs.(2.180) and (2.181).

If we follow the conventional assignment of the \emph{left-handed
 fermion states} to $\Psi_{(+)}$ then
$\Psi_{(-)}{{=\Psi_{(+)}^*~N~odd}\choose {=\Psi_{(-)}^*~N~even}}$
contains the \emph{right-handed CP conjugate fermion states}. The
product $2^{N-1}_{(\pm)}\otimes 2^{N-1}_{(\mp)}$ ($N$ odd) and
$2^{N-1}_{(\pm)}\otimes 2^{N-1}_{(\pm)}$ ($N$ even) then
transforms as a \emph{Lorentz vector}. Thus, to make these tensors
a Lorentz vector, we must insert the product $\gamma^0\gamma^A$.
Similarly, the product $2^{N-1}_{(\pm)}\otimes 2^{N-1}_{(\pm)}$
($N$ odd) and $2^{N-1}_{(\pm)}\otimes 2^{N-1}_{(\mp)}$ ($N$ even)
transform as a {\emph{Lorentz scalar}} and we must insert a
$\gamma^0$ in these tensors.

In order to form SO(2N) invariants we must contract antisymmetric
bosonic tensors (Eq.(2.50)) of appropriate rank with each of the
above antisymmetric tensors formed from the Kronecker product of
spinors so that the whole object transforms like a scalar. For the
sake of illustration, consider the quantity
$I=\Psi^{\dagger}_{(+)}\Gamma_{[\mu_1}\Gamma_{\mu_2}\Gamma_{\mu_3]}\Psi_{(-)}\Phi^{({\cal
A})}_{\mu_1\mu_2\mu_3}$, which we now show to transform as an
$SO(2N)$ scalar:
$I'=\Psi^{'\dagger}_{+}\Gamma_{[\mu_1}\Gamma_{\mu_2}\Gamma_{\mu_3]}\Psi_{(-)}'\Phi^{({\cal
A})'}_{\mu_1\mu_2\mu_3}=\Psi^{\dagger}_{(+)}(S^{-1}\Gamma_{[\mu_1}S)\\
\times(S^{-1}\Gamma_{\mu_2}S)(S^{-1}\Gamma_{\mu_3]}S)\Psi_{(-)}
 R_{\mu_1\nu_1} R_{\mu_2\nu_2}R_{\mu_3\nu_3}\Phi^{({\cal
A})}_{\nu_1\nu_2\nu_3}=
\Psi^{\dagger}_{(+)}(R_{\mu_1\sigma_1}\Gamma_{[\sigma_1})\\
\times(R_{\mu_2\sigma_2}\Gamma_{\sigma_2})
(R_{\mu_3\sigma_3}\Gamma_{\sigma_3]})\Psi_{(-)}
 R_{\nu_1\mu_1}^T R_{\nu_2\mu_2}^T R_{\nu_3\mu_3}^T\Phi^{({\cal
A})}_{\nu_1\nu_2\nu_3}=\Psi^{\dagger}_{(+)}\Gamma_{[\sigma_1}\Gamma_{\sigma_2}\Gamma_{\sigma_3]}\Psi_{(-)}\\
\times\Phi^{({\cal
A})}_{\nu_1\nu_2\nu_3}\delta_{\nu_1\sigma_1}\delta_{\nu_2\sigma_2}\delta_{\nu_3\sigma_3}=I$.\\
\\
Let us now summarize our results
\begin{eqnarray}
N~{{\textnormal{Even (Vector Couplings) }}}~~~~~~~~~~~~~~~~~~~~~~\nonumber\\
g_{\acute{a}\acute{b}}{\overline{\Psi}}_{(\pm){\acute{a}}}\gamma^A\Psi_{(\pm){\acute{b}}}\Phi^{({\cal
A})}_A\nonumber\\
g_{\acute{a}\acute{b}}{\overline{\Psi}}_{(\pm){\acute{a}}}\gamma^A\Gamma_{[\mu_1}\Gamma_{\mu_2]}\Psi_{(\pm){\acute{b}}}\Phi^{({\cal
A})}_{A\mu_1\mu_2}\nonumber\\
g_{\acute{a}\acute{b}}{\overline{\Psi}}_{(\pm){\acute{a}}}\gamma^A\Gamma_{[\mu_1}\Gamma_{\mu_2}\Gamma_{\mu_3}\Gamma_{\mu_4]}\Psi_{(\pm){\acute{b}}}\Phi^{({\cal
A})}_{A\mu_1\mu_2\mu_3\mu_4}\nonumber\\
......~~~~~~~......~~~~~~~......\nonumber\\
g_{\acute{a}\acute{b}}{\Upsilon}_{\acute{a}\acute{b}~\mu_1\mu_2...\mu_N}^{(\pm
1)A}\Omega_{A\mu_1\mu_2...\mu_N}^{(\pm 1)}\nonumber\\
\nonumber\\
 2^{N-1}_{(+)}\otimes 2^{N-1}_{(+)}={{2N}\choose
{0}}\oplus{{2N}\choose {2}}\oplus...\oplus{{2N}\choose
{N-2}}\oplus \frac{1}{2}{{2N}\choose {N}}^{(+1)}\nonumber\\
2^{N-1}_{(-)}\otimes 2^{N-1}_{(-)}={{2N}\choose
{0}}\oplus{{2N}\choose {2}}\oplus...\oplus{{2N}\choose
{N-2}}\oplus \frac{1}{2}{{2N}\choose {N}}^{(-1)}
\end{eqnarray}

\begin{eqnarray}
N~{{\textnormal{Even (Scalar Couplings)}}}~~~~~~~~~~~~~~~~~~~~~~\nonumber\\
 h_{\acute{a}\acute{b}}{\overline{\Psi}}_{(\pm){\acute{a}}}\Gamma_{\mu_1}\Psi_{(\mp){\acute{b}}}\Phi^{({\cal
A})}_{\mu_1}\nonumber\\
h_{\acute{a}\acute{b}}{\overline{\Psi}}_{(\pm){\acute{a}}}\Gamma_{[\mu_1}\Gamma_{\mu_2}\Gamma_{\mu_3]}\Psi_{(\mp){\acute{b}}}\Phi^{({\cal
A})}_{\mu_1\mu_2\mu_3}\nonumber\\
h_{\acute{a}\acute{b}}{\overline{\Psi}}_{(\pm){\acute{a}}}\Gamma_{[\mu_1}\Gamma_{\mu_2}\Gamma_{\mu_3}\Gamma_{\mu_4}\Gamma_{\mu_5]}\Psi_{(\mp){\acute{b}}}\Phi^{({\cal
A})}_{\mu_1\mu_2\mu_3\mu_4\mu_5}\nonumber\\
......~~~~~~~......~~~~~~~......\nonumber\\
h_{\acute{a}\acute{b}}{\overline{\Psi}}_{(\pm){\acute{a}}}\Gamma_{[\mu_1}\Gamma_{\mu_2}...\Gamma_{\mu_{N-1}]}
\Psi_{(\mp){\acute{b}}}\Phi_{\mu_1\mu_2...\mu_{N-1}}^{{(\cal{A})}}\nonumber\\
\nonumber\\
2^{N-1}_{(\pm)}\otimes 2^{N-1}_{(\mp)}={{2N}\choose
{1}}\oplus{{2N}\choose {3}}\oplus...\oplus{{2N}\choose
{N-3}}\oplus{{2N}\choose {N-1}}
\end{eqnarray}

\begin{eqnarray}
N~{{\textnormal{odd (Vector Couplings) }}}~~~~~~~~~~~~~~~~~~~~~~\nonumber\\
 g_{\acute{a}\acute{b}}{\overline{\Psi}}_{(\pm){\acute{a}}}\gamma^A\Psi_{(\pm){\acute{b}}}\Phi^{({\cal
A})}_A\nonumber\\
g_{\acute{a}\acute{b}}{\overline{\Psi}}_{(\pm){\acute{a}}}\gamma^A\Gamma_{[\mu_1}\Gamma_{\mu_2]}\Psi_{(\pm){\acute{b}}}\Phi^{({\cal
A})}_{A\mu_1\mu_2}\nonumber\\
g_{\acute{a}\acute{b}}{\overline{\Psi}}_{(\pm){\acute{a}}}\gamma^A\Gamma_{[\mu_1}\Gamma_{\mu_2}\Gamma_{\mu_3}\Gamma_{\mu_4]}\Psi_{(\pm){\acute{b}}}\Phi^{({\cal
A})}_{A\mu_1\mu_2\mu_3\mu_4}\nonumber\\
......~~~~~~~......~~~~~~~......\nonumber\\
g_{\acute{a}\acute{b}}{\overline{\Psi}}_{(\pm){\acute{a}}}\gamma^A\Gamma_{[\mu_1}\Gamma_{\mu_2}...\Gamma_{\mu_{N-1}]}\Psi_{(\pm){\acute{b}}}\Phi^{({\cal
A})}_{A\mu_1\mu_2...\mu_{N-1}}\nonumber\\
\nonumber\\
2^{N-1}_{(\pm)}\otimes 2^{N-1}_{(\mp)}={{2N}\choose
{0}}\oplus{{2N}\choose {2}}\oplus...\oplus{{2N}\choose
{N-3}}\oplus{{2N}\choose {N-1}}
\end{eqnarray}

\begin{eqnarray}
N~{{\textnormal{Odd (Scalar Couplings)}}}~~~~~~~~~~~~~~~~~~~~~~\nonumber\\
h_{\acute{a}\acute{b}}
{\overline{\Psi}}_{(\pm){\acute{a}}}\Gamma_{\mu_1}\Psi_{(\mp){\acute{b}}}\Phi^{({\cal
A})}_{\mu_1}\nonumber\\
h_{\acute{a}\acute{b}}{\overline{\Psi}}_{(\pm){\acute{a}}}\Gamma_{[\mu_1}\Gamma_{\mu_2}\Gamma_{\mu_3]}\Psi_{(\mp){\acute{b}}}\Phi^{({\cal
A})}_{\mu_1\mu_2\mu_3}\nonumber\\
h_{\acute{a}\acute{b}}{\overline{\Psi}}_{(\pm){\acute{a}}}\Gamma_{[\mu_1}\Gamma_{\mu_2}\Gamma_{\mu_3}\Gamma_{\mu_4}\Gamma_{\mu_5]}\Psi_{(\mp){\acute{b}}}\Phi^{({\cal
A})}_{\mu_1\mu_2\mu_3\mu_4\mu_5}\nonumber\\
......~~~~~~~......~~~~~~~......\nonumber\\
h_{\acute{a}\acute{b}}{{\Upsilon}}_{\acute{a}\acute{b}~\mu_1\mu_2...\mu_N}^{(\pm
i)}\Omega_{\mu_1\mu_2...\mu_N}^{(\mp i)}\nonumber\\
\nonumber\\
2^{N-1}_{(-)}\otimes 2^{N-1}_{(-)}={{2N}\choose
{1}}\oplus{{2N}\choose {3}}\oplus...\oplus{{2N}\choose
{N-2}}\oplus \frac{1}{2}{{2N}\choose {N}}^{(+i)}\nonumber\\
2^{N-1}_{(+)}\otimes 2^{N-1}_{(+)}={{2N}\choose
{1}}\oplus{{2N}\choose {3}}\oplus...\oplus{{2N}\choose
{N-2}}\oplus \frac{1}{2}{{2N}\choose {N}}^{(-i)}
\end{eqnarray}
The last lines in each of the Eqs.(2.227)-(2.230) represent the
decomposition law for the Kronecker product of the spinor
representations of $SO(2N)$ and each of the numbers appearing on
the right hand side of the equations represent the
dimensionalities of the tensors formed. For example, in Eq.(227),
the rank 2 tensor,
${\overline{\Psi}}_{(\pm){\acute{a}}}\gamma^A\Gamma_{[\mu_1}\Gamma_{\mu_2]}\Psi_{(\pm){\acute{b}}}$
has dimensionality ${2N\choose 2}$ and so on.

A word about our notation is in order. $g_{\acute{a}\acute{b}}$'s
and $h_{\acute{a}\acute{b}}$'s are coupling constants. The indices
$\acute{a}$ and $\acute{b}$ run over different number of
generations carried by the semi-spinors. Here, $\Omega^{\pm}$ are
ordinary (bosonic) rank $N$ tensors and satisfy duality conditions
given by Eqs.(2.48) and (2.49). The other  notation used in the
previous four equations that need clarification are
\begin{equation}
{\overline{\Psi}}_{(\pm)\acute{a}}=\Psi^{\dagger}_{(\pm)\acute{a}}\gamma^0
\end{equation}
\begin{eqnarray}
N~{\textnormal{odd}}~~~~~~~~~~~~~~~~~~\nonumber\\
 \Upsilon_{\acute{a}\acute{b}~\mu_1...\mu_N}=
\Upsilon_{\acute{a}\acute{b}~\mu_1\mu_2...\mu_N}^{(+ i)}+
\Upsilon_{\acute{a}\acute{b}~\mu_1\mu_2...\mu_N}^{(-i)}\nonumber\\
 \Upsilon_{\acute{a}\acute{b}~\mu_1\mu_2...\mu_N}^{(\pm
 i)}=\frac{1}{2}\left(\delta_{\mu_1\nu_1}
\delta_{\mu_2\nu_2}...\delta_{\mu_{N}\nu_{N}}\pm\frac{
i}{N!}\epsilon_{\mu_1...\mu_N\nu_1...\nu_N}\right)\Upsilon_{\acute{a}\acute{b}~\mu_1\mu_2...\mu_N}^{(\pm
 i)}\nonumber\\
 \Upsilon_{\acute{a}\acute{b}~\mu_1\mu_2...\mu_N}^{(\pm i)}=
{\overline{\Psi}}_{(\pm)\acute{a}}\Gamma_{[\mu_1}\Gamma_{\mu_2}...\Gamma_{\mu_N]}\Psi_{(\mp)\acute{b}}\nonumber\\
{{\Upsilon}}_{\acute{a}\acute{b}~\mu_1\mu_2...\mu_N}^{(\pm i)}=
\pm\frac{
i}{N!}\epsilon_{\mu_1...\mu_N\nu_1...\nu_N}{{\Upsilon}}_{\acute{a}\acute{b}~\mu_1\mu_2...\mu_N}^{(\pm
i)}
\end{eqnarray}
\newpage
\begin{eqnarray}
N~{\textnormal{even}}~~~~~~~~~~~~~~~~~~\nonumber\\
{\Upsilon}_{\acute{a}\acute{b}~\mu_1\mu_2...\mu_N}^{A}=\Upsilon_{\acute{a}\acute{b}~\mu_1\mu_2...\mu_N}^{(+
1)}+\Upsilon_{\acute{a}\acute{b}~\mu_1\mu_2...\mu_N}^{(-1)}\nonumber\\
\Upsilon_{\acute{a}\acute{b}~\mu_1\mu_2...\mu_N}^{(\pm
 1)}=\frac{1}{2}\left(\delta_{\mu_1\nu_1}
\delta_{\mu_2\nu_2}...\delta_{\mu_{N}\nu_{N}}\pm\frac{
1}{N!}\epsilon_{\mu_1...\mu_N\nu_1...\nu_N}\right)\Upsilon_{\acute{a}\acute{b}~\mu_1\mu_2...\mu_N}^{(\pm
 1)}\nonumber\\
\Upsilon_{\acute{a}\acute{b}~\mu_1\mu_2...\mu_N}^{(\pm 1)}=
{\overline{\Psi}}_{(\pm)\acute{a}}\gamma^A\Gamma_{[\mu_1}\Gamma_{\mu_2}...\Gamma_{\mu_N]}\Psi_{(\pm)\acute{b}}\nonumber\\
{{\Upsilon}}_{\acute{a}\acute{b}~\mu_1\mu_2...\mu_N}^{(\pm 1)}=
\pm\frac{
1}{N!}\epsilon_{\mu_1...\mu_N\nu_1...\nu_N}{{\Upsilon}}_{\acute{a}\acute{b}~\mu_1\mu_2...\mu_N}^{(\pm
1)}
\end{eqnarray}

The reader must have noticed in Eqs.(2.227) and (2.230) that the
ordinary (bosonic) rank $N$ tensors and the rank $N$ tensors
formed from the Kronecker product of semi-spinors satisfy  same
(opposite) in sign duality conditions for $N$ even (odd). This is
absolutely necessary in order for the invariant to be
non-vanishing. This can be seen by means of a simple example.
Consider two 2-index antisymmetric tensors $A_{\mu_1\mu_2}$ and
$B_{\mu_1\mu_2}$ of $SO(4)$ (N=2) satisfying the duality
conditions:
$A_{\mu_1\mu_2}=+\frac{1}{2!}\epsilon_{\mu_1\mu_2\nu_1\nu_2}A_{\nu_1\nu_2}$
and
$B_{\mu_1\mu_2}=-\frac{1}{2!}\epsilon_{\mu_1\mu_2\nu_1\nu_2}B_{\nu_1\nu_2}$.
The invariant $A_{\mu_1\mu_2}B_{\mu_1\mu_2}$ is zero because
$A_{\mu_1\mu_2}B_{\mu_1\mu_2}=-\frac{1}{(2!)^2}\epsilon_{\mu_1\mu_2\nu_1\nu_2}
\epsilon_{\mu_1\mu_2\sigma_1\sigma_2}A_{\nu_1\nu_2}B_{\sigma_1\sigma_2}$
$=-\frac{1}{(2!)^2}(2!)(\delta_{\nu_1\sigma_1}\delta_{\nu_2\sigma_2}-\delta_{\nu_1\sigma_2}\delta_{\nu_2\sigma_1})$\\
$=-A_{\mu_1\mu_2}B_{\mu_1\mu_2}$.

The next task is to determine the symmetry of the couplings in the
exchange of identical semi-spinors. Looking at Eqs.(2.227) and
(2.229), define a $k$-index ($k=$ odd) Kronecker product of
spinors,
$K_{\acute{a}\acute{b}}={{\Psi}}_{(\pm){\acute{a}}}\gamma^0\gamma^A\Gamma_{\mu_1}...\Gamma_{\mu_k}
\Psi_{(\pm){\acute{b}}}$ and recall that the spinors are
anticommutating fermion fields, $\gamma^{0\dagger}=\gamma^{0}$ and
that $\Gamma_{\mu}^{\dagger}=\Gamma_{\mu}$. From the fact
 $K_{\acute{a}\acute{b}}=K_{\acute{a}\acute{b}}^{\dagger}$ conclude that
  $K_{\acute{a}\acute{b}}=-{{\Psi}}_{(\pm){\acute{b}}}
 \Gamma_{\mu_k}...\Gamma_{\mu_2}\Gamma_{\mu_1}\\\times \gamma^{A\dagger}\gamma^0
\Psi_{(\pm){\acute{a}}}$. Further, from the properties satisfied
by the Dirac matrices, $(\gamma^0)^2={\bf{1}}$ and
$\gamma^{A}=\gamma^0\gamma^{A\dagger}\gamma^0$ and  the
anticommutating property of  $\Gamma_{\mu}$'s, one can easily show
that
$K_{\acute{a}\acute{b}}=-(-1)^{\frac{k(k-1)}{2}}K_{\acute{b}\acute{a}}$.
Thus
\begin{eqnarray}
~~~\forall N~{\textnormal{and}}~k~{\textnormal{even}}~~~~~~~~~~~~~~~~~~~~~~~~~~~~\nonumber\\
 {\mathsf L}\equiv
g_{\acute{a}\acute{b}}{\overline{\Psi}}_{(\pm){\acute{a}}}\gamma^A\Gamma_{\mu_1}\Gamma_{\mu_2}...\Gamma_{\mu_{k}}
\Psi_{(\pm){\acute{b}}}\Phi^{({\cal A})}_{A\mu_1\mu_2...\mu_{k}}\nonumber\\
{\mathsf
L}=\frac{1}{2}\left[g_{\acute{a}\acute{b}}+(-1)^{\frac{k(k-1)+2}{2}}g_{\acute{b}\acute{a}}\right]
{\overline{\Psi}}_{(\pm){\acute{a}}}\gamma^A\Gamma_{\mu_1}\Gamma_{\mu_2}...\Gamma_{\mu_{k}}
\Psi_{(\pm){\acute{b}}}\Phi^{({\cal A})}_{A\mu_1\mu_2...\mu_{k}}
\end{eqnarray}

If the symmetry factor $(-1)^{\frac{k(k-1)+2}{2}}$ is odd, and if
we restrict to one generation or do not allow for mixing between
generations ($\acute{a}=\acute{b}$), then ${\mathsf L}$ would
vanish.

 One can also build symmetric traceless tensors. Here one starts with
a  symmetrized product of $\Gamma_{\mu}$'s and then subtracts off
the trace. For example a $2^{nd}$ rank symmetric traceless tensor
of dimension ${{2N+1}\choose 2}-{{2N}\choose {0}}$ takes the form
\begin{equation}
\Psi^{\dagger}_{(\pm)}\Gamma_{\{\mu_1}\Gamma_{\mu_2\}}\Psi_{(\pm)}-\delta_{\mu_1\mu_2}\Psi^{\dagger}_{(\pm)}\Psi_{(\pm)}
\end{equation}
while a $3^{rd}$ rank symmetric traceless tensor of dimension
${{2N+2}\choose 3}-{{2N}\choose {1}}$ can be written as
\begin{eqnarray}
\Psi^{\dagger}_{(\pm)}\Gamma_{\{\mu_1}\Gamma_{\mu_2}\Gamma_{\mu_3\}}\Psi_{(\mp){\acute{b}}}
-\frac{N}{N+2}\left[\delta_{\mu_2\mu_3}\Psi^{\dagger}_{(\pm)}\Gamma_{\mu_1}\Psi_{(\mp)}
+\delta_{\mu_1\mu_3}\Psi^{\dagger}_{(\pm)}\Gamma_{\mu_2}\Psi_{(\mp)}\right.\nonumber\\
\left.+\delta_{\mu_1\mu_2}\Psi^{\dagger}_{(\pm)}\Gamma_{\mu_3}\Psi_{(\mp)}\right]\nonumber\\
\end{eqnarray}
where we have used the fact that
$tr(\Gamma_{\mu_i}\Gamma_{\mu_j})=N$.\\
\\
\\
\noindent \textsc{invariants formed from $\Psi^{T}$ and $\Psi$: Lagrangian and superpotential cases}\\
 \\
We, again, build bilinear spinors from $\Psi^T$ and $\Psi$ and an
antisymmetric or a symmetric product of $\Gamma_{\mu}$'s. As
before these quantities transform as scalars, vectors and tensors.
For illustration purposes, consider a quantity defined through
$\Xi_{\mu_1\mu_2}\equiv(B\Psi)^T\Gamma_{\mu_1}\Gamma_{\mu_2}\Psi$.
Then from Eqs.(2.127) and (2.200),
$\Xi_{\mu_1\mu_2}'={(B\Psi')}^T\Gamma_{\mu_1}\Gamma_{\mu_2}\Psi'$
$={(B\Psi)}^TS^{-1}\Gamma_{\mu_1}(SS^{-1}){\mu_2}(S\Psi)$$=R_{\mu_1\nu_1}R_{\mu_2\nu_2}\Xi_{\nu_1\nu_2}$.
Hence $\Xi_{\mu_1\mu_2}$ transforms like a second rank tensor. The
following are the bilinear spinors that can be formed in this case
\begin{eqnarray}
{\textnormal{Scalar}}:~\Xi=\Psi^{T}B^T\Psi;~\Xi'=\Xi\nonumber\\
\nonumber\\
{\textnormal{Vector}}:~\Xi_{\mu_1}=\Psi^{T}B^T\Gamma_{\mu_1}\Psi;\nonumber\\
\Xi_{\mu_1}'
=R_{\mu_1\nu_1}\Xi_{\nu_1}\nonumber\\
\nonumber\\
 {\textnormal{$2^{nd}$Rank
Tensor}}:~\Xi_{\mu_1\mu_2}=\Psi^{T}B^T\Gamma_{\mu_1}\Gamma_{\mu_2}\Psi;\nonumber\\
\Xi_{\mu_1\mu_2}'=R_{\mu_1\nu_1}R_{\mu_2\nu_2}\Xi_{\nu_1\nu_2}\nonumber\\
\nonumber\\
 {\textnormal{$r^{th}$Rank
Tensor}}:~\Xi_{\mu_1..\mu_r}=\Psi^{T}B^T\Gamma_{\mu_1}..\Gamma_{\mu_r}\Psi;\nonumber\\
\Xi_{\mu_1...\mu_r}'=R_{\mu_1\nu_1}..R_{\mu_r\nu_r}\Xi_{\nu_1..\nu_r}
\end{eqnarray}

An $r^{th}$ rank tensor can be decomposed in to a sum of four
$r^{th}$ rank tensors involving semi-spinors
$(\Psi=\Psi_{(+)}+\Psi_{(-)})$:
$\Psi^{T}B^T\Gamma_{\mu_1}..\Gamma_{\mu_r}\Psi
=\Psi^{T}_{(\pm)}B^T\Gamma_{\mu_1}\\
\times\Gamma_{\mu_2}...\Gamma_{\mu_r}\Psi_{(\pm)}+
\Psi^{T}_{(\pm)}B^T\Gamma_{\mu_1}\Gamma_{\mu_2}...\Gamma_{\mu_r}\Psi_{(\mp)}$.
Using the fact $\Psi_{(\pm)}=P_{\pm}\Psi$ and Eqs. (2.172),
(2.177), (2.178), (2.193), (2.194) and (2.204) one can easily
verify the following results:
\begin{eqnarray}
N~{\textnormal{odd}},~r~{\textnormal{even}}~~{\textnormal{or}}~~N~{\textnormal{even}},~r~{\textnormal{odd}}\nonumber\\
\Psi^{T}_{(\pm)}B^T\Gamma_{\mu_1}\Gamma_{\mu_2}...\Gamma_{\mu_r}\Psi_{(\pm)}=0
\end{eqnarray}
\begin{eqnarray}
N~{\textnormal{odd}},~r~{\textnormal{odd}}~~{\textnormal{or}}~~N~{\textnormal{even}},~r~{\textnormal{even}}\nonumber\\
\Psi^{T}_{(\pm)}B^T\Gamma_{\mu_1}\Gamma_{\mu_2}...\Gamma_{\mu_r}\Psi_{(\mp)}=0
\end{eqnarray}

If we make the same assignment of the particles as in the previous
case: left-handed states fermion states to $\Psi_{(+)}$ and
right-handed CP conjugate fermion states to $\Psi_{(-)}$, then the
product $2^{N-1}_{(\pm)}\otimes 2^{N-1}_{(\mp)}$ ($N$ odd and
even) transforms as a Lorentz vector. Therefore to make these
tensors a Lorentz vector, we must insert the product $C\gamma^A$,
where $C$ is the Dirac charge conjugation matrix. The product
$2^{N-1}_{(\pm)}\otimes 2^{N-1}_{(\pm)}$ ($N$ odd and even)
transforms as a Lorentz scalar and we must insert  $C$ in these
tensors.

Following the same lines as in the previous case where invariants
were formed from $\Psi^{\dagger}$ and $\Psi$, we get
\begin{eqnarray}
N~{{\textnormal{Even (Vector Couplings) }}}~~~~~~~~~~~~~~~~~~~~~~\nonumber\\
g_{\acute{a}\acute{b}}{\Psi}^T_{(\pm){\acute{a}}}C\gamma^AB^T\Gamma_{\mu_1}\Psi_{(\mp){\acute{b}}}\Phi^{({\cal
A})}_{A\mu_1}\nonumber\\
g_{\acute{a}\acute{b}}{\Psi}^T_{(\pm){\acute{a}}}C\gamma^AB^T\Gamma_{[\mu_1}\Gamma_{\mu_2}\Gamma_{\mu_3]}\Psi_{(\mp){\acute{b}}}\Phi^{({\cal
A})}_{A\mu_1\mu_2\mu_3}\nonumber\\
g_{\acute{a}\acute{b}}{\Psi}^T_{(\pm){\acute{a}}}C\gamma^AB^T\Gamma_{[\mu_1}\Gamma_{\mu_2}\Gamma_{\mu_3}\Gamma_{\mu_4}\Gamma_{\mu_5]}\Psi_{(\mp){\acute{b}}}\Phi^{({\cal
A})}_{A\mu_1\mu_2\mu_3\mu_4\mu_5}\nonumber\\
......~~~~~~~......~~~~~~~......\nonumber\\
g_{\acute{a}\acute{b}}{\Psi}^T_{(\pm){\acute{a}}}C\gamma^AB^T\Gamma_{[\mu_1}\Gamma_{\mu_2}...\Gamma_{\mu_{N-1}]}\Psi_{(\mp){\acute{b}}}\Phi^{({\cal
A})}_{A\mu_1\mu_2...\mu_{N-1}}\nonumber\\
\nonumber\\
 2^{N-1}_{(\pm)}\otimes 2^{N-1}_{(\mp)}={{2N}\choose
{1}}\oplus{{2N}\choose {3}}\oplus...\oplus{{2N}\choose
{N-3}}\oplus{{2N}\choose {N-1}}
\end{eqnarray}

\begin{eqnarray}
N~{{\textnormal{Even (Scalar Couplings)}}}~~~~~~~~~~~~~~~~~~~~~~\nonumber\\
 h_{\acute{a}\acute{b}}{\Psi}^T_{(\pm){\acute{a}}}CB^T\Psi_{(\pm){\acute{b}}}\Phi^{({\cal
A})}_{\mu_1}\nonumber\\
h_{\acute{a}\acute{b}}{\Psi}^T_{(\pm){\acute{a}}}CB^T\Gamma_{[\mu_1}\Gamma_{\mu_2]}\Psi_{(\pm){\acute{b}}}\Phi^{({\cal
A})}_{\mu_1\mu_2}\nonumber\\
h_{\acute{a}\acute{b}}{\Psi}^T_{(\pm){\acute{a}}}CB^T\Gamma_{[\mu_1}\Gamma_{\mu_2}\Gamma_{\mu_3}\Gamma_{\mu_4]}\Psi_{(\pm){\acute{b}}}\Phi^{({\cal
A})}_{\mu_1\mu_2\mu_3\mu_4}\nonumber\\
......~~~~~~~......~~~~~~~......\nonumber\\
h_{\acute{a}\acute{b}}\Xi_{\acute{a}\acute{b}~\mu_1\mu_2...\mu_N}^{(\pm
1)}\Omega_{\mu_1\mu_2...\mu_N}^{(\pm 1)}\nonumber\\
\nonumber\\
2^{N-1}_{(+)}\otimes 2^{N-1}_{(+)}={{2N}\choose
{0}}\oplus{{2N}\choose {2}}\oplus...\oplus{{2N}\choose
{N-2}}\oplus\frac{1}{2}{{2N}\choose {N}}^{+ 1}\nonumber\\
2^{N-1}_{(-)}\otimes 2^{N-1}_{(-)}={{2N}\choose
{0}}\oplus{{2N}\choose {2}}\oplus...\oplus{{2N}\choose
{N-2}}\oplus\frac{1}{2}{{2N}\choose {N}}^{- 1}
\end{eqnarray}

\begin{eqnarray}
N~{{\textnormal{odd (Vector Couplings) }}}~~~~~~~~~~~~~~~~~~~~~~\nonumber\\
 g_{\acute{a}\acute{b}}{\Psi}^T_{(\pm){\acute{a}}}C\gamma^AB^T\Psi_{(\mp){\acute{b}}}\Phi^{({\cal
A})}_A\nonumber\\
g_{\acute{a}\acute{b}}{\Psi}^T_{(\pm){\acute{a}}}C\gamma^AB^T\Gamma_{[\mu_1}\Gamma_{\mu_2]}\Psi_{(\mp){\acute{b}}}\Phi^{({\cal
A})}_{A\mu_1\mu_2}\nonumber\\
g_{\acute{a}\acute{b}}{\Psi}^T_{(\pm){\acute{a}}}C\gamma^AB^T\Gamma_{[\mu_1}\Gamma_{\mu_2}\Gamma_{\mu_3}\Gamma_{\mu_4]}\Psi_{(\mp){\acute{b}}}\Phi^{({\cal
A})}_{A\mu_1\mu_2\mu_3\mu_4}\nonumber\\
......~~~~~~~......~~~~~~~......\nonumber\\
g_{\acute{a}\acute{b}}{\Psi}^T_{(\pm){\acute{a}}}C\gamma^AB^T\Gamma_{[\mu_1}\Gamma_{\mu_2}...\Gamma_{\mu_{N-1}]}\Psi_{(\mp){\acute{b}}}\Phi^{({\cal
A})}_{A\mu_1\mu_2...\mu_{N-1}}\nonumber\\
\nonumber\\
2^{N-1}_{(\pm)}\otimes 2^{N-1}_{(\mp)}={{2N}\choose
{0}}\oplus{{2N}\choose {2}}\oplus...\oplus{{2N}\choose
{N-3}}\oplus{{2N}\choose {N-1}}
\end{eqnarray}

\begin{eqnarray}
N~{{\textnormal{Odd (Scalar Couplings)}}}~~~~~~~~~~~~~~~~~~~~~~\nonumber\\
h_{\acute{a}\acute{b}}
{\Psi}^T_{(\pm){\acute{a}}}CB^T\Gamma_{\mu_1}\Psi_{(\pm){\acute{b}}}\Phi^{({\cal
A})}_{\mu_1}\nonumber\\
h_{\acute{a}\acute{b}}{\Psi}^T_{(\pm){\acute{a}}}CB^T\Gamma_{[\mu_1}\Gamma_{\mu_2}\Gamma_{\mu_3]}\Psi_{(\pm){\acute{b}}}\Phi^{({\cal
A})}_{\mu_1\mu_2\mu_3}\nonumber\\
h_{\acute{a}\acute{b}}{\Psi}^T_{(\pm){\acute{a}}}CB^T\Gamma_{[\mu_1}\Gamma_{\mu_2}\Gamma_{\mu_3}\Gamma_{\mu_4}\Gamma_{\mu_5]}\Psi_{(\pm){\acute{b}}}\Phi^{({\cal
A})}_{\mu_1\mu_2\mu_3\mu_4\mu_5}\nonumber\\
......~~~~~~~......~~~~~~~......\nonumber\\
h_{\acute{a}\acute{b}}\Xi_{\acute{a}\acute{b}~\mu_1\mu_2...\mu_N}^{(\mp
i)}\Omega_{\mu_1\mu_2...\mu_N}^{(\pm i)}\nonumber\\
\nonumber\\
2^{N-1}_{(-)}\otimes 2^{N-1}_{(-)}={{2N}\choose
{1}}\oplus{{2N}\choose {3}}\oplus...\oplus{{2N}\choose
{N-2}}\oplus \frac{1}{2}{{2N}\choose {N}}^{(+i)}\nonumber\\
2^{N-1}_{(+)}\otimes 2^{N-1}_{(+)}={{2N}\choose
{1}}\oplus{{2N}\choose {3}}\oplus...\oplus{{2N}\choose
{N-2}}\oplus \frac{1}{2}{{2N}\choose {N}}^{(-i)}
\end{eqnarray}
where we have defined

\begin{eqnarray}
N~{\textnormal{even}}~~~~~~~~~~~~\nonumber\\
\Xi_{\mu_1...\mu_N}=\Xi_{\mu_1...\mu_N}^{(+
1)}+\Xi_{\mu_1...\mu_N}^{(- 1)}\nonumber\\
\Xi_{\mu_1...\mu_N}^{(\pm 1)}=\frac{1}{2}\left(\delta_{\mu_1\nu_1}
\delta_{\mu_2\nu_2}...\delta_{\mu_{N}\nu_{N}}\pm
\frac{1}{N!}\epsilon_{\mu_1\mu_2...\mu_N\nu_1\nu_2...\nu_N}\right)\Xi_{\nu_1...\nu_N}\nonumber\\
\Xi_{\mu_1...\mu_N}^{(\pm 1)}\equiv
\Psi^{T}_{(\pm)}CB^T\Gamma_{[\mu_1}\Gamma_{\mu_2}...\Gamma_{\mu_N]}\Psi_{(\pm)}\nonumber\\
\Xi_{\mu_1...\mu_N}^{(\pm 1)}= \pm\frac{
1}{N!}\epsilon_{\mu_1...\mu_N\nu_1...\nu_N}\Xi_{\mu_1...\mu_N}^{(\pm
1)}
\end{eqnarray}
\newpage
\begin{eqnarray}
N~{\textnormal{odd}}~~~~~~~~~~~~\nonumber\\
\Xi_{\mu_1...\mu_N}=\Xi_{\mu_1...\mu_N}^{(+
i)}+\Xi_{\mu_1...\mu_N}^{(- i)}\nonumber\\
\Xi_{\mu_1...\mu_N}^{(\pm i)}=\frac{1}{2}\left(\delta_{\mu_1\nu_1}
\delta_{\mu_2\nu_2}...\delta_{\mu_{N}\nu_{N}}\pm
\frac{i}{N!}\epsilon_{\mu_1\mu_2...\mu_N\nu_1\nu_2...\nu_N}\right)\Xi_{\nu_1...\nu_N}\nonumber\\
\Lambda_{\mu_1...\mu_N}^{(\pm i)}\equiv
\Psi^{T}_{(\mp)}CB^T\Gamma_{[\mu_1}\Gamma_{\mu_2}...\Gamma_{\mu_N]}\Psi_{(\mp)}\nonumber\\
\Xi_{\mu_1...\mu_N}^{(\pm i)}= \pm\frac{
i}{N!}\epsilon_{\mu_1...\mu_N\nu_1...\nu_N}\Xi_{\mu_1...\mu_N}^{(\mp
i)}
\end{eqnarray}
These duality conditions follow immediately upon using
$\Psi_{(\pm)}=P_{\pm}\Psi$  and Eqs.(2.180) and (2.181). The
couplings above appear in the Lagrangian. These couplings can also
appear in the superpotential, provided the Dirac charge
conjugation matrix, $C$ is removed.

As a final result, we find the symmetry factor in the exchange of
identical semi-spinors. Looking at Eqs.(2.241) and (2.243). Define
a $k$-index tensor:
$h_{\acute{a}\acute{b}}{\Psi}^T_{(\pm){\acute{a}}}CB^T
\Gamma_{\mu_1}\Gamma_{\mu_2}...\Gamma_{\mu_k]}\Psi_{(\pm){\acute{b}}}$.
Here $k$ is odd for $N$ odd and $k$ is even for $N$ even. As
before, using the fact that the transpose of this defined quantity
is itself and that Dirac charge conjugation matrix, $C$, is
antisymmetric, we conclude that
\begin{eqnarray}
k~{\textnormal{odd}},~
N~{\textnormal{odd}}~{\textnormal{or}}~k~{\textnormal{even}},~
N~{\textnormal{even}}~~~~~~~~~~~~~~~~~\nonumber\\
 {\cal L}\equiv
h_{\acute{a}\acute{b}}{{\Psi}}^T_{(\pm){\acute{a}}}CB^T\Gamma_{\mu_1}\Gamma_{\mu_2}...\Gamma_{\mu_{k}}
\Psi_{(\pm){\acute{b}}}\Phi^{({\cal
A})}_{\mu_1...\mu_{k}}\nonumber\\
{\cal
L}=\frac{1}{2}\left[h_{\acute{a}\acute{b}}+(-1)^{\frac{(N-k)(N-k+1)}{2}}h_{\acute{b}\acute{a}}\right]
{{\Psi}}^T_{(\pm){\acute{a}}}CB^T\Gamma_{\mu_1}\Gamma_{\mu_2}...\Gamma_{\mu_{k}}
\Psi_{(\pm){\acute{b}}}\Phi^{({\cal
A})}_{\mu_1...\mu_{k}}\nonumber\\
\end{eqnarray}

Note this result is independent of the representation of $B$.
Looking at Eqs.(2.191)-(2.194), we have two distinct options:
$B^{-1}\Gamma_{\mu}^TB=(-1)^N\Gamma_{\mu}$,
$B^T=(-1)^{\frac{N(N+1)}{2}}B$ and
$B^{-1}\Gamma_{\mu}^TB=(-1)^{N+1}\Gamma_{\mu}$,
$B^T=(-1)^{\frac{N(N-1)}{2}}B$. Both options give the same
result of Eq.(2.246).\\
\\
\noindent \textsc{Invariants formed from $\Psi^{\dagger}$ and $\Psi^*$: Lagrangian case}\\
 \\
Here we construct couplings out of $\Psi^{\dagger}$ and $\Psi^*$.
This is possible if we write a quantity with two indices (say) as
$\Lambda_{\mu_1\mu_2}=\Psi^{\dagger}\Gamma_{\mu_1}\Gamma_{\mu_2}B\Psi^*$.
From Eq.(2.198),
$\Lambda_{\mu_1\mu_2}'=\Psi^{\dagger'}\Gamma_{\mu_1}\Gamma_{\mu_2}B\Psi^{*'}$=$\Psi^{\dagger}
S^{-1}\Gamma_{\mu_1}\Gamma_{\mu_2}B(S^{-1})^T\Psi^*$$=\Psi^{\dagger}(S^{-1}\Gamma_{\mu_1}S)(S^{-1}\Gamma_{\mu_2}S)\\
\times S^{-1}B(S^{-1})^T\Psi^*$. Using Eqs.(2.130), (2.195) and
(2.197) we get
$\Lambda_{\mu_1\mu_2}'=R_{\mu_1\nu_1}R_{\mu_2\nu_2}(\Psi^{\dagger}\Gamma_{\nu_1}\Gamma_{\nu_2}B\Psi^*)
=R_{\mu_1\nu_1}R_{\mu_2\nu_2}\Lambda_{\nu_1\nu_2}$. Hence, we can
form the following tensors
\begin{eqnarray}
{\textnormal{Scalar}}:~\Lambda=\Psi^{\dagger}B\Psi^*;\nonumber\\
\Lambda'=\Lambda\nonumber\\
\nonumber\\
{\textnormal{Vector}}:~\Lambda_{\mu_1}=\Psi^{\dagger}\Gamma_{\mu_1}B\Psi^*;\nonumber\\
\Lambda_{\mu_1}'
=R_{\mu_1\nu_1}\Lambda_{\nu_1}\nonumber\\
\nonumber\\
 {\textnormal{$2^{nd}$Rank
Tensor}}:~\Lambda_{\mu_1\mu_2}=\Psi^{\dagger}\Gamma_{\mu_1}\Gamma_{\mu_2}B\Psi^*;\nonumber\\
\Lambda_{\mu_1\mu_2}'=R_{\mu_1\nu_1}R_{\mu_2\nu_2}\Lambda_{\nu_1\nu_2}\nonumber\\
\nonumber\\
 {\textnormal{$r^{th}$Rank
Tensor}}:~\Lambda_{\mu_1..\mu_r}=\Psi^{\dagger}\Gamma_{\mu_1}..\Gamma_{\mu_r}B\Psi^*;\nonumber\\
\Lambda_{\mu_1...\mu_r}'=R_{\mu_1\nu_1}..R_{\mu_r\nu_r}\Lambda_{\nu_1..\nu_r}
\end{eqnarray}

One can  very easily prove that in the case of couplings formed
from $\Psi^{\dagger}$ and $\Psi^*$, the following general results
hold true
\begin{eqnarray}
{\textnormal{For}}~r~{\textnormal{odd}},~
N~{\textnormal{even}}~{\textnormal{or}}~r~{\textnormal{even}},~
N~{\textnormal{odd}}~~~~~~~~~~~~~~~~~\nonumber\\
\Psi^{\dagger}_{\pm}\Gamma_{\mu_1}\Gamma_{\mu_2}...\Gamma_{\mu_r}B\Psi^*_{\pm}=0
~~~~~~~~~~~~~~~~~~~~~~~~\end{eqnarray}
\begin{eqnarray}
{\textnormal{For}}~r~{\textnormal{even}},~
N~{\textnormal{even}}~{\textnormal{or}}~r~{\textnormal{odd}},~
N~{\textnormal{odd}}~~~~~~~~~~~~~~~~~\nonumber\\
\Psi^{\dagger}_{\pm}\Gamma_{\mu_1}\Gamma{\mu_2}...\Gamma_{\mu_r}B\Psi^*_{\mp}=0
~~~~~~~~~~~~~~~~~~~~~~~~\end{eqnarray}

We now present a complete sets of invariants. In order for the
couplings to transform correctly as Lorentz vectors and Lorentz
scalars we must insert, by hand, quantities like
$\gamma^0\gamma^AC\gamma^{0T}$ and $\gamma^0\gamma^A\gamma^{0T}$,
respectively. Thus the non-vanishing couplings are

\begin{eqnarray}
N~{{\textnormal{Even (Vector Couplings) }}}~~~~~~~~~~~~~~~~~~~~~~\nonumber\\
g_{\acute{a}\acute{b}}{\overline{\Psi}}_{(\pm){\acute{a}}}\gamma^A\Gamma_{\mu_1}BC{\overline{\Psi}}^T_{(\mp){\acute{b}}}\Phi^{({\cal
A})}_{A\mu_1}\nonumber\\
g_{\acute{a}\acute{b}}{\overline{\Psi}}_{(\pm){\acute{a}}}\gamma^A\Gamma_{[\mu_1}\Gamma_{\mu_2}\Gamma_{\mu_3]}BC{\overline{\Psi}}^T_{(\mp){\acute{b}}}\Phi^{({\cal
A})}_{A\mu_1\mu_2\mu_3}\nonumber\\
g_{\acute{a}\acute{b}}{\overline{\Psi}}_{(\pm){\acute{a}}}\gamma^A\Gamma_{[\mu_1}\Gamma_{\mu_2}\Gamma_{\mu_3}\Gamma_{\mu_4}\Gamma_{\mu_5]}BC{\overline{\Psi}}^T_{(\mp){\acute{b}}}\Phi^{({\cal
A})}_{A\mu_1\mu_2\mu_3\mu_4\mu_5}\nonumber\\
......~~~~~~~......~~~~~~~......\nonumber\\
g_{\acute{a}\acute{b}}{\overline{\Psi}}_{(\pm){\acute{a}}}\gamma^A\Gamma_{[\mu_1}\Gamma_{\mu_2}...\Gamma_{\mu_{N-1}]}BC{\overline{\Psi}}^T_{(\mp){\acute{b}}}\Phi^{({\cal
A})}_{A\mu_1\mu_2...\mu_{N-1}}\nonumber\\
\nonumber\\
 2^{N-1}_{(\pm)}\otimes 2^{N-1}_{(\mp)}={{2N}\choose
{1}}\oplus{{2N}\choose {3}}\oplus...\oplus{{2N}\choose
{N-3}}\oplus{{2N}\choose {N-1}}
\end{eqnarray}

\begin{eqnarray}
N~{{\textnormal{Even (Scalar Couplings)}}}~~~~~~~~~~~~~~~~~~~~~~\nonumber\\
 h_{\acute{a}\acute{b}}{\overline{\Psi}}_{(\pm){\acute{a}}}BC{\overline{\Psi}}^T_{(\pm){\acute{b}}}\Phi^{({\cal
A})}\nonumber\\
h_{\acute{a}\acute{b}}{\overline{\Psi}}_{(\pm){\acute{a}}}\Gamma_{[\mu_1}\Gamma_{\mu_2]}BC{\overline{\Psi}}^T_{(\pm){\acute{b}}}\Phi^{({\cal
A})}_{\mu_1\mu_2}\nonumber\\
h_{\acute{a}\acute{b}}{\overline{\Psi}}
_{(\pm){\acute{a}}}\Gamma_{[\mu_1}\Gamma_{\mu_2}\Gamma_{\mu_3}\Gamma_{\mu_4]}BC{\overline{\Psi}}^T_{(\pm){\acute{b}}}\Phi^{({\cal
A})}_{\mu_1\mu_2\mu_3\mu_4}\nonumber\\
......~~~~~~~......~~~~~~~......\nonumber\\
h_{\acute{a}\acute{b}}\Lambda_{\acute{a}\acute{b}~\mu_1\mu_2...\mu_N}^{(\pm
1)}\Omega_{\mu_1\mu_2...\mu_N}^{(\pm 1)}\nonumber\\
\nonumber\\
2^{N-1}_{(+)}\otimes 2^{N-1}_{(+)}={{2N}\choose
{0}}\oplus{{2N}\choose {2}}\oplus...\oplus{{2N}\choose
{N-2}}\oplus\frac{1}{2}{{2N}\choose {N}}^{+ 1}\nonumber\\
2^{N-1}_{(-)}\otimes 2^{N-1}_{(-)}={{2N}\choose
{0}}\oplus{{2N}\choose {2}}\oplus...\oplus{{2N}\choose
{N-2}}\oplus\frac{1}{2}{{2N}\choose {N}}^{- 1}
\end{eqnarray}

\begin{eqnarray}
N~{{\textnormal{odd (Vector Couplings) }}}~~~~~~~~~~~~~~~~~~~~~~\nonumber\\
 g_{\acute{a}\acute{b}}{\overline{\Psi}}_{(\pm){\acute{a}}}\gamma^ABC{\overline{\Psi}}^T_{(\mp){\acute{b}}}\Phi^{({\cal
A})}_A\nonumber\\
g_{\acute{a}\acute{b}}{\overline{\Psi}}_{(\pm){\acute{a}}}\gamma^A\Gamma_{[\mu_1}\Gamma_{\mu_2]}BC{\overline{\Psi}}^T_{(\mp){\acute{b}}}\Phi^{({\cal
A})}_{A\mu_1\mu_2}\nonumber\\
g_{\acute{a}\acute{b}}{\overline{\Psi}}_{(\pm)}{\acute{a}}\gamma^A\Gamma_{[\mu_1}\Gamma_{\mu_2}\Gamma_{\mu_3}\Gamma_{\mu_4]}BC{\overline{\Psi}}^T_{(\mp){\acute{b}}}\Phi^{({\cal
A})}_{A\mu_1\mu_2\mu_3\mu_4}\nonumber\\
......~~~~~~~......~~~~~~~......\nonumber\\
g_{\acute{a}\acute{b}}{\overline{\Psi}}_{(\pm){\acute{a}}}\gamma^A\Gamma_{[\mu_1}\Gamma_{\mu_2}...\Gamma_{\mu_{N-1}]}BC{\overline{\Psi}}^T_{(\mp){\acute{b}}}\Phi^{({\cal
A})}_{A\mu_1\mu_2...\mu_{N-1}}\nonumber\\
\nonumber\\
2^{N-1}_{(\pm)}\otimes 2^{N-1}_{(\mp)}={{2N}\choose
{0}}\oplus{{2N}\choose {2}}\oplus...\oplus{{2N}\choose
{N-3}}\oplus{{2N}\choose {N-1}}
\end{eqnarray}
\newpage
\begin{eqnarray}
N~{{\textnormal{Odd (Scalar Couplings)}}}~~~~~~~~~~~~~~~~~~~~~~\nonumber\\
h_{\acute{a}\acute{b}}
{\overline{\Psi}}_{(\pm){\acute{a}}}\Gamma_{\mu_1}BC{\overline{\Psi}}^T_{(\pm){\acute{b}}}\Phi^{({\cal
A})}_{\mu_1}\nonumber\\
h_{\acute{a}\acute{b}}{\overline{\Psi}}_{(\pm){\acute{a}}}\Gamma_{[\mu_1}\Gamma_{\mu_2}\Gamma_{\mu_3]}BC{\overline{\Psi}}^T_{(\pm){\acute{b}}}\Phi^{({\cal
A})}_{\mu_1\mu_2\mu_3}\nonumber\\
h_{\acute{a}\acute{b}}{\overline{\Psi}}_{(\pm){\acute{a}}}\Gamma_{[\mu_1}\Gamma_{\mu_2}\Gamma_{\mu_3}\Gamma_{\mu_4}\Gamma_{\mu_5]}BC{\overline{\Psi}}^T_{(\pm){\acute{b}}}\Phi^{({\cal
A})}_{\mu_1\mu_2\mu_3\mu_4\mu_5}\nonumber\\
......~~~~~~~......~~~~~~~......\nonumber\\
h_{\acute{a}\acute{b}}\Lambda_{\acute{a}\acute{b}~\mu_1\mu_2...\mu_N}^{(\mp
i)}\Omega_{\mu_1\mu_2...\mu_N}^{(\pm i)}\nonumber\\
\nonumber\\
2^{N-1}_{(-)}\otimes 2^{N-1}_{(-)}={{2N}\choose
{1}}\oplus{{2N}\choose {3}}\oplus...\oplus{{2N}\choose
{N-2}}\oplus \frac{1}{2}{{2N}\choose {N}}^{(-i)}\nonumber\\
2^{N-1}_{(+)}\otimes 2^{N-1}_{(+)}={{2N}\choose
{1}}\oplus{{2N}\choose {3}}\oplus...\oplus{{2N}\choose
{N-2}}\oplus \frac{1}{2}{{2N}\choose {N}}^{(+i)}
\end{eqnarray}

where we have defined
\begin{eqnarray}
{\overline{\Psi}}_{(\pm)}=\Psi^{\dagger}_{(\pm)}\gamma^0\nonumber\\
{\overline{\Psi}}_{(\pm)}^T=\gamma^{0T}\Psi^{*}_{(\pm)}
\end{eqnarray}

\begin{eqnarray}
N~{\textnormal{even}}~~~~~~~~~~~~\nonumber\\
\Lambda_{\mu_1...\mu_N}=\Lambda_{\mu_1...\mu_N}^{(+
1)}+\Lambda_{\mu_1...\mu_N}^{(- 1)}\nonumber\\
\Lambda_{\mu_1...\mu_N}^{(\pm
1)}=\frac{1}{2}\left(\delta_{\mu_1\nu_1}
\delta_{\mu_2\nu_2}...\delta_{\mu_{N}\nu_{N}}\pm
\frac{1}{N!}\epsilon_{\mu_1\mu_2...\mu_N\nu_1\nu_2...\nu_N}\right)\Lambda_{\nu_1...\nu_N}\nonumber\\
\Lambda_{\mu_1...\mu_N}^{(\pm 1)}\equiv
{\overline{\Psi}}_{(\pm)}\Gamma_{[\mu_1}\Gamma_{\mu_2}...\Gamma_{\mu_N]}BC{\overline{\Psi}}^T_{(\pm)}\nonumber\\
\Lambda_{\mu_1...\mu_N}^{(\pm 1)}= \pm\frac{
1}{N!}\epsilon_{\mu_1...\mu_N\nu_1...\nu_N}\Lambda_{\mu_1...\mu_N}^{(\pm
1)}
\end{eqnarray}
\begin{eqnarray}
N~{\textnormal{odd}}~~~~~~~~~~~~\nonumber\\
\Lambda_{\mu_1...\mu_N}=\Lambda_{\mu_1...\mu_N}^{(+
i)}+\Lambda_{\mu_1...\mu_N}^{(- i)}\nonumber\\
\Lambda_{\mu_1...\mu_N}^{(\pm
i)}=\frac{1}{2}\left(\delta_{\mu_1\nu_1}
\delta_{\mu_2\nu_2}...\delta_{\mu_{N}\nu_{N}}\pm
\frac{i}{N!}\epsilon_{\mu_1\mu_2...\mu_N\nu_1\nu_2...\nu_N}\right)\Lambda_{\nu_1...\nu_N}\nonumber\\
\Lambda_{\mu_1...\mu_N}^{(\pm i)}\equiv
{\overline{\Psi}}_{(\mp)}\Gamma_{[\mu_1}\Gamma_{\mu_2}...\Gamma_{\mu_N]}BC{\overline{\Psi}}^T_{(\mp)}\nonumber\\
\Lambda_{\mu_1...\mu_N}^{(\pm i)}= \pm\frac{
i}{N!}\epsilon_{\mu_1...\mu_N\nu_1...\nu_N}\Lambda_{\mu_1...\mu_N}^{(\mp
i)}
\end{eqnarray}

The symmetry factor is
\begin{eqnarray}
k~{\textnormal{odd}},~
N~{\textnormal{odd}}~{\textnormal{or}}~k~{\textnormal{even}},~
N~{\textnormal{even}}~~~~~~~~~~~~~~~~~\nonumber\\
 {\cal L}\equiv
h_{\acute{a}\acute{b}}{\overline{\Psi}}_{(\pm){\acute{a}}}\Gamma_{\mu_1}\Gamma_{\mu_2}...\Gamma_{\mu_{k}}
BC{\overline{\Psi}}^T_{(\pm){\acute{b}}}\Phi^{({\cal
A})}_{\mu_1...\mu_{k}}\nonumber\\
{\cal
L}=\frac{1}{2}\left[h_{\acute{a}\acute{b}}+(-1)^{\frac{(N-k)(N-k+1)}{2}}h_{\acute{b}\acute{a}}\right]
{\overline{\Psi}}_{(\pm){\acute{a}}}\Gamma_{\mu_1}\Gamma_{\mu_2}...\Gamma_{\mu_{k}}
BC{\overline{\Psi}}^T_{(\pm){\acute{b}}}\Phi^{({\cal
A})}_{\mu_1...\mu_{k}}\nonumber\\
\end{eqnarray}\\
\noindent \textsc{invariants formed from $\Psi^{T}$ and $\Psi^{*}$: Lagrangian case}\\
 \\
In this case, a second rank tensor, for example, is defined by
$\Theta_{\mu\nu}\equiv \Psi^T
B\Gamma_{\mu}\Gamma_{\nu}\nonumber\\
\times B^{-1}\Psi^*$, because $\Theta_{\mu\nu}'=\Psi^{'T}
B\Gamma_{\mu}\Gamma_{\nu}B^{-1}\Psi^{*'}=\Psi^TS^TB\Gamma_{\mu}\Gamma_{\nu}B^{-1}(S^T)^{-1}\Psi^*$,
and where we have used Eq.(2.198). On using Eq.(2.197),
$\Theta_{\mu\nu}'=
\Psi^T(BS^{-1})\\
\times\Gamma_{\mu}(SS^{-1})\Gamma_{\nu}B^{-1}(BSB^{-1})\Psi^*
=\Psi^TB
 (S^{-1}\Gamma_{\mu}S)(S^{-1}\Gamma_{\nu}S)
B^{-1}\Psi^*$
$=R_{\mu\sigma}R_{\nu\rho}\Psi^T\\
\times B\Gamma_{\sigma}\Gamma_{\rho}B^{-1}\Psi^*$
$=R_{\mu\sigma}R_{\nu\rho}\Theta_{\sigma\rho}$. Thus,
\begin{eqnarray}
{\textnormal{Scalar}}:~\Theta=\Psi^{T}BB^{-1}\Psi^*;\nonumber\\
\Theta'=\Theta\nonumber\\
\nonumber\\
{\textnormal{Vector}}:~\Theta_{\mu_1}=\Psi^{T}B\Gamma_{\mu_1}B^{-1}\Psi^*;\nonumber\\
\Lambda_{\mu_1}'
=R_{\mu_1\nu_1}\Theta_{\nu_1}\nonumber\\
\nonumber\\
 {\textnormal{$2^{nd}$Rank
Tensor}}:~\Theta_{\mu_1\mu_2}=\Psi^{T}B\Gamma_{\mu_1}\Gamma_{\mu_2}B^{-1}\Psi^*;\nonumber\\
\Theta_{\mu_1\mu_2}'=R_{\mu_1\nu_1}R_{\mu_2\nu_2}\Theta_{\nu_1\nu_2}\nonumber\\
\nonumber\\
 {\textnormal{$r^{th}$Rank
Tensor}}:~\Theta_{\mu_1..\mu_r}=\Psi^{T}B\Gamma_{\mu_1}..\Gamma_{\mu_r}B^{-1}\Psi^*;\nonumber\\
\Theta_{\mu_1..\mu_r}'=R_{\mu_1\nu_1}..R_{\mu_r\nu_r}\Theta_{\nu_1..\nu_r}
\end{eqnarray}
Using the properties (2.204) and Eqs.(2.177)-(2.178), the reader
can make the following inferences
\begin{eqnarray}
\forall N,~r~{\textnormal{odd}}~~~~~~~~~~~~~\nonumber\\
\Psi^{T}_{\pm}B\Gamma_{\mu_1}\Gamma_{\mu_2}...\Gamma_{\mu_r}B^{-1}\Psi^*_{\pm}=0
\end{eqnarray}
\begin{eqnarray}
\forall N,~r~{\textnormal{even}}~~~~~~~~~~~~~\nonumber\\
\Psi^{T}_{\pm}B\Gamma_{\mu_1}\Gamma{\mu_2}...\Gamma_{\mu_r}B^{-1}\Psi^*_{\mp}=0
\end{eqnarray}
In order for the couplings to correctly exhibit the Lorentz vector
and scalar behavior, we must insert, by hand,
 $C\gamma_AC\gamma^{0T}$ and $CC\gamma^{0T}$, respectively.

We have the following couplings in the Lagrangian,

\begin{eqnarray}
N~{{\textnormal{Even (Vector Couplings) }}}~~~~~~~~~~~~~~~~~~~~~~\nonumber\\
g_{\acute{a}\acute{b}}{{\Psi}}^T_{(\pm){\acute{a}}}C\gamma^ABB^{-1}C{\overline{\Psi}}^T_{(\pm){\acute{b}}}\Phi^{({\cal
A})}_{A}\nonumber\\
g_{\acute{a}\acute{b}}{{\Psi}}^T_{(\pm){\acute{a}}}C\gamma^AB\Gamma_{[\mu_1}\Gamma_{\mu_2]}B^{-1}C{\overline{\Psi}}^T_{(\pm){\acute{b}}}\Phi^{({\cal
A})}_{A\mu_1\mu_2}\nonumber\\
g_{\acute{a}\acute{b}}{{\Psi}}^T_{(\pm){\acute{a}}}C\gamma^AB\Gamma_{[\mu_1}\Gamma_{\mu_2}\Gamma_{\mu_3}\Gamma_{\mu_4]}B^{-1}C{\overline{\Psi}}^T_{(\pm){\acute{b}}}\Phi^{({\cal
A})}_{A\mu_1\mu_2\mu_3\mu_4}\nonumber\\
......~~~~~~~......~~~~~~~......\nonumber\\
g_{\acute{a}\acute{b}}\Theta_{\acute{a}\acute{b}~\mu_1\mu_2...\mu_N}^{(\pm
1)}\Omega_{\mu_1\mu_2...\mu_N}^{(\pm 1)}\nonumber\\
\nonumber\\
 2^{N-1}_{(+)}\otimes 2^{N-1}_{(+)}={{2N}\choose
{0}}\oplus{{2N}\choose {2}}\oplus...\oplus{{2N}\choose
{N-2}}\oplus \frac{1}{2}{{2N}\choose {N}}^{(+1)}\nonumber\\
 2^{N-1}_{(-)}\otimes 2^{N-1}_{(-)}={{2N}\choose
{0}}\oplus{{2N}\choose {2}}\oplus...\oplus{{2N}\choose
{N-2}}\oplus \frac{1}{2}{{2N}\choose {N}}^{(-1)}
\end{eqnarray}

\begin{eqnarray}
N~{{\textnormal{Even (Scalar Couplings)}}}~~~~~~~~~~~~~~~~~~~~~~\nonumber\\
 h_{\acute{a}\acute{b}}{{\Psi}}^T_{(\pm){\acute{a}}}CB\Gamma_{\mu_1}B^{-1}C{\overline{\Psi}}^T_{(\mp){\acute{b}}}\Phi^{({\cal
A})}_{\mu_1}\nonumber\\
h_{\acute{a}\acute{b}}{{\Psi}}^T_{(\pm){\acute{a}}}CB\Gamma_{[\mu_1}\Gamma_{\mu_2}\Gamma_{\mu_3]}B^{-1}C{\overline{\Psi}}^T_{(\mp){\acute{b}}}\Phi^{({\cal
A})}_{\mu_1\mu_2\mu_3}\nonumber\\
h_{\acute{a}\acute{b}}{{\Psi}}^T
_{(\pm){\acute{a}}}CB\Gamma_{[\mu_1}\Gamma_{\mu_2}\Gamma_{\mu_3}\Gamma_{\mu_4}\Gamma_{\mu_5]}B^{-1}C{\overline{\Psi}}^T_{(\mp){\acute{b}}}\Phi^{({\cal
A})}_{\mu_1\mu_2\mu_3\mu_4\mu_5}\nonumber\\
......~~~~~~~......~~~~~~~......\nonumber\\
h_{\acute{a}\acute{b}}{{\Psi}}_{(\pm){\acute{a}}}CB\Gamma_{[\mu_1}\Gamma_{\mu_2}...\Gamma_{\mu_{N-1}]}B^{-1}C{\overline{\Psi}}^T_{(\mp){\acute{b}}}\Phi^{({\cal
A})}_{\mu_1\mu_2...\mu_{N-1}}\nonumber\\
2^{N-1}_{(\pm)}\otimes 2^{N-1}_{(\mp)}={{2N}\choose
{1}}\oplus{{2N}\choose {3}}\oplus...\oplus{{2N}\choose
{N-3}}\oplus{{2N}\choose {N-1}}
\end{eqnarray}

\begin{eqnarray}
N~{{\textnormal{odd (Vector Couplings) }}}~~~~~~~~~~~~~~~~~~~~~~\nonumber\\
 g_{\acute{a}\acute{b}}{{\Psi}}^T_{(\pm){\acute{a}}}C\gamma^ABB^{-1}C{\overline{\Psi}}^T_{(\pm){\acute{b}}}\Phi^{({\cal
A})}_{A}\nonumber\\
g_{\acute{a}\acute{b}}{{\Psi}}^T_{(\pm){\acute{a}}}C\gamma^AB\Gamma_{[\mu_1}\Gamma_{\mu_2]}B^{-1}C{\overline{\Psi}}^T_{(\pm){\acute{b}}}\Phi^{({\cal
A})}_{A\mu_1\mu_2}\nonumber\\
g_{\acute{a}\acute{b}}{{\Psi}}^T_{(\pm){\acute{a}}}C\gamma^AB\Gamma_{[\mu_1}\Gamma_{\mu_2}\Gamma_{\mu_3}\Gamma_{\mu_4]}B^{-1}C{\overline{\Psi}}^T_{(\pm){\acute{b}}}\Phi^{({\cal
A})}_{A\mu_1\mu_2\mu_3\mu_4}\nonumber\\
......~~~~~~~......~~~~~~~......\nonumber\\
g_{\acute{a}\acute{b}}{{\Psi}}^T_{(\pm){\acute{a}}}C\gamma^AB\Gamma_{[\mu_1}\Gamma_{\mu_2}...\Gamma_{\mu_{N-1}]}
B^{-1}C{\overline{\Psi}}^T_{(\pm){\acute{b}}}\Phi^{({\cal
A})}_{A\mu_1\mu_2...\mu_{N-1}}\nonumber\\
\nonumber\\
 2^{N-1}_{(\pm)}\otimes 2^{N-1}_{(\mp)}={{2N}\choose
{0}}\oplus{{2N}\choose {2}}\oplus...\oplus{{2N}\choose
{N-3}}\oplus {{2N}\choose {N-1}}
\end{eqnarray}

\begin{eqnarray}
N~{{\textnormal{Odd (Scalar Couplings)}}}~~~~~~~~~~~~~~~~~~~~~~\nonumber\\
 h_{\acute{a}\acute{b}}{{\Psi}}^T_{(\pm){\acute{a}}}CB\Gamma_{\mu_1}B^{-1}C{\overline{\Psi}}^T_{(\mp){\acute{b}}}\Phi^{({\cal
A})}_{\mu_1}\nonumber\\
h_{\acute{a}\acute{b}}{{\Psi}}^T_{(\pm){\acute{a}}}CB\Gamma_{[\mu_1}\Gamma_{\mu_2}\Gamma_{\mu_3]}B^{-1}C{\overline{\Psi}}^T_{(\mp){\acute{b}}}\Phi^{({\cal
A})}_{\mu_1\mu_2\mu_3}\nonumber\\
h_{\acute{a}\acute{b}}{{\Psi}}^T
_{(\pm){\acute{a}}}CB\Gamma_{[\mu_1}\Gamma_{\mu_2}\Gamma_{\mu_3}\Gamma_{\mu_4}\Gamma_{\mu_5]}B^{-1}C{\overline{\Psi}}^T_{(\mp){\acute{b}}}\Phi^{({\cal
A})}_{\mu_1\mu_2\mu_3\mu_4\mu_5}\nonumber\\
......~~~~~~~......~~~~~~~......\nonumber\\
h_{\acute{a}\acute{b}}\Theta_{\acute{a}\acute{b}~\mu_1\mu_2...\mu_N}^{(\mp
i)}\Omega_{\mu_1\mu_2...\mu_N}^{(\pm i)}\nonumber\\
\nonumber\\
2^{N-1}_{(+)}\otimes 2^{N-1}_{(+)}={{2N}\choose
{1}}\oplus{{2N}\choose {3}}\oplus...\oplus{{2N}\choose
{N-2}}\oplus \frac{1}{2}{{2N}\choose {N}}^{(+i)}\nonumber\\
2^{N-1}_{(-)}\otimes 2^{N-1}_{(-)}={{2N}\choose
{1}}\oplus{{2N}\choose {3}}\oplus...\oplus{{2N}\choose
{N-2}}\oplus \frac{1}{2}{{2N}\choose {N}}^{(-i)}
\end{eqnarray}
where we have defined
\begin{eqnarray}
N~{\textnormal{even}}~~~~~~~~~~~~\nonumber\\
\Theta_{\mu_1...\mu_N}=\Theta_{\mu_1...\mu_N}^{(+
1)}+\Theta_{\mu_1...\mu_N}^{(- 1)}\nonumber\\
\Theta_{\mu_1...\mu_N}^{(\pm
1)}=\frac{1}{2}\left(\delta_{\mu_1\nu_1}
\delta_{\mu_2\nu_2}...\delta_{\mu_{N}\nu_{N}}\pm
\frac{1}{N!}\epsilon_{\mu_1\mu_2...\mu_N\nu_1\nu_2...\nu_N}\right)\Theta_{\nu_1...\nu_N}\nonumber\\
\Theta_{\mu_1...\mu_N}^{(\pm 1)}\equiv
{{\Psi}}^T_{(\pm)}C\gamma^AB\Gamma_{[\mu_1}\Gamma_{\mu_2}...\Gamma_{\mu_N]}B^{-1}C{\overline{\Psi}}^T_{(\pm)}\nonumber\\
\Lambda_{\mu_1...\mu_N}^{(\pm 1)}= \pm\frac{
1}{N!}\epsilon_{\mu_1...\mu_N\nu_1...\nu_N}\Lambda_{\mu_1...\mu_N}^{(\pm
1)}
\end{eqnarray}
\begin{eqnarray}
N~{\textnormal{odd}}~~~~~~~~~~~~\nonumber\\
\Theta_{\mu_1...\mu_N}=\Theta_{\mu_1...\mu_N}^{(+
i)}+\Theta_{\mu_1...\mu_N}^{(- i)}\nonumber\\
\Theta_{\mu_1...\mu_N}^{(\pm
i)}=\frac{1}{2}\left(\delta_{\mu_1\nu_1}
\delta_{\mu_2\nu_2}...\delta_{\mu_{N}\nu_{N}}\pm
\frac{i}{N!}\epsilon_{\mu_1\mu_2...\mu_N\nu_1\nu_2...\nu_N}\right)\Lambda_{\nu_1...\nu_N}\nonumber\\
\Lambda_{\mu_1...\mu_N}^{(\pm i)}\equiv
{{\Psi}}^T_{(\pm)}CB\Gamma_{[\mu_1}\Gamma_{\mu_2}...\Gamma_{\mu_N]}B^{-1}C{\overline{\Psi}}^T_{(\mp)}\nonumber\\
\Lambda_{\mu_1...\mu_N}^{(\pm i)}= \pm\frac{
i}{N!}\epsilon_{\mu_1...\mu_N\nu_1...\nu_N}\Lambda_{\mu_1...\mu_N}^{(\mp
i)}
\end{eqnarray}
The correct symmetry factor can be easily computed. It is found to
be
\begin{eqnarray}
\forall N,~k~{\textnormal{even}}~~~~~~~~~~~~~~~~~~~~~~~\nonumber\\
 {\cal L}\equiv
g_{\acute{a}\acute{b}}{{\Psi}}^T_{(\pm){\acute{a}}}C\gamma^AB\Gamma_{\mu_1}\Gamma_{\mu_2}...\Gamma_{\mu_{k}}
B^{-1}C{\overline{\Psi}}^T_{(\pm){\acute{b}}}\Phi^{({\cal
A})}_{A\mu_1...\mu_{k}}\nonumber\\
{\cal
L}=\frac{1}{2}\left[g_{\acute{a}\acute{b}}+(-1)^{\frac{k(k-1)+2}{2}}g_{\acute{b}\acute{a}}\right]
{{\Psi}}^T_{(\pm){\acute{a}}}C\gamma^AB\Gamma_{\mu_1}\Gamma_{\mu_2}...\Gamma_{\mu_{k}}
B^{-1}C{\overline{\Psi}}^T_{(\pm){\acute{b}}}\Phi^{({\cal
A})}_{A\mu_1...\mu_{k}}\nonumber\\
\end{eqnarray}

\section{Specialization to {SO(10)} gauge group}

\noindent\textsc{decomposition of SO(10) spinor under SU(5) }\\
\\
In the case of $SO(10)$ ($N=5$) group the 32 ($=2^5$) dimensional
spinor splits into  two inequivalent irreducible 16 ($=2^5$)
dimensional representations: $\Psi_{(+)}\sim 16$,
$\Psi_{(-)}\sim{\overline {16}}$. Then Eq.(2.171) gives
\begin{eqnarray}
{\textnormal{under}}~~~SO(10)\supset SU(5)\otimes U(1)~~~~~~~~~~~~~~~~~~~\nonumber\\
16\supset [1](M)\oplus [10](M^{ij})\oplus [{\overline
{5}}](M_i)\nonumber\\
{\overline {16}}\supset [5](N^i)\oplus
[{\overline{10}}](N_{ij})\oplus [1](N)
\end{eqnarray}
where
\begin{eqnarray}
{\overline {5}}\sim \left(\matrix{d_1^c\cr d_2^c\cr d_3^c\cr
e^-\cr  \nu_e
  \cr}\right)_{L}\equiv \left(\matrix{d_{ L\alpha}^c\cr
e^-_L\cr  \nu_{eL}
  \cr}\right)
\end{eqnarray}
\begin{eqnarray}
10 \sim\left(\matrix{0 & u_3^c &-u_2^c &| &-u^1 &-d^1\cr -u_3^c &0
&u_1^c &| &-u^2 &-d^2\cr u_2^c &-u_1^c &0 &| &-u^3 &-d^3 \cr
-&-&-&-&-&-\cr u^1 &u^2 &u^3 &| &0 &e^+ \cr d^1 &d^2 &d^3 &| &-e^+
&0
  \cr}
 \right)_L\nonumber\\
 \\
 \equiv \left(\matrix{\epsilon^{\alpha\beta\xi}u^c_{L\xi}&| &-u_{L}^{\alpha}~-d_{L}^{\alpha}\cr --&-&--& \cr
 ~~u_{L}^{\alpha}~&| &\cr
&|&\epsilon_{ab}e_{L}^{+}\cr ~d_{L}^{\alpha}~&|&
  \cr}
 \right)~~~~~~~~~~~~~
\end{eqnarray}

\begin{eqnarray}
1 \sim \nu_{eL}^c
\end{eqnarray}
Here, $\alpha,\beta,\xi=1,2,3$ are $SU(3)$ color indices, while
$a,b=4,5$ are $SU(2)$ indices and we adopt that all particles are
left-handed ($L$). $\epsilon_{\alpha\beta\xi}$ and $\epsilon_{ab}$
are Levi-Civita tensors of $SU(3)$ and $SU(2)$ subgroups,
respectively.

From Eq.(2.169),
$\Psi_{(+)}=\psi|0>+\frac{1}{2}\psi^{ij}b_i^{\dagger}b_j^{\dagger}|0>
+\frac{1}{24}\psi^{ijkl}b_i^{\dagger}b_j^{\dagger}b_k^{\dagger}b_l^{\dagger}|0>$\\
($i,j,k,l=1,...,5$). Note that $(5\sim)~ \psi^{ijkl}$ of $SU(5)$
is reducible, thus the irreducible $({\overline{5}}\sim)~
\psi^{ijkl}=\frac{1}{1!}\epsilon^{ijklm}\psi_m$.
 Since $\Psi_{(+)}$ is a 16 component vector, we make a slight change of notation:
  $\Psi_{(+)}\rightarrow |\Psi_{(+)}>$. To identify the $SU(5)$ multiplets with matter
particles, we also replace the symbols $\psi$ by ${\bf{M}}$.
Taking, into consideration generation (family) indices (${\acute
{a}}, {\acute {b}}$), we write for the $SO(10)$ spinor,
\begin{equation}
\Psi_{(+){\acute {a}}}\rightarrow|\Psi_{(+){\acute {a}}}>=|0>{\bf
M}_{\acute {a}}+\frac{1}{2}b_i^{\dagger}b_j^{\dagger}|0>{\bf
M}_{\acute {a}}^{ij} +\frac{1}{24}\epsilon^{ijklm}b_j^{\dagger}
b_k^{\dagger}b_l^{\dagger}b_m^{\dagger}|0>{\bf M}_{{\acute
{a}}i}\nonumber\\
\end{equation}
 where
\begin{eqnarray}
{\bf M}_{\acute {a}}=\nu_{L{\acute {a}}}^c;~~~~{\bf M}_{{\acute
{a}}\alpha}=D_{L{\acute {a}}\alpha}^c; ~~~~{\bf M}_{\acute
{a}}^{\alpha\beta}=\epsilon^{\alpha\beta\gamma}U_{L{\acute
{a}}\gamma}^c; ~~~~{\bf M}_{{\acute {a}}4}=E_{L{\acute
{a}}}^-\nonumber\\
{\bf M}_{\acute {a}}^{4\alpha}=U_{L{\acute {a}}}^{\alpha};
~~~~{\bf M}_{{\acute {a}}5}=\nu_{L{\acute {a}}};~~~~~~~{\bf
M}_{\acute {a}}^{5\alpha}=D_{L{\acute
{a}}}^{\alpha};~~~~~~~~~~{\bf M}_{\acute {a}}^{45}=E_{L{\acute
{a}}}^+\nonumber\\
\end{eqnarray}

Similarly, from Eq.(2.169), we have for the negative chirality
states:
$\Psi_{(-)}=\psi^ib_i^{\dagger}|0>+\frac{1}{6}\psi^{ijk}b_i^{\dagger}b_j^{\dagger}b_{k}^{\dagger}|0>
+\frac{1}{120}\psi^{ijklm}b_i^{\dagger}b_j^{\dagger}b_k^{\dagger}b_l^{\dagger}b_m^{\dagger}|0>$.
Again $(10\sim)~ \psi^{ijk}$ is reducible, thus the irreducible
$({\overline{10}}\sim)~
\psi^{ijk}=\frac{1}{2!}\epsilon^{ijklm}\psi_{lm}$. Also
$({\overline{1}}\sim)~
\psi^{ijklm}=\frac{1}{0!}\epsilon^{ijklm}\psi'$ and thus
$\psi^{ijklm}b_i^{\dagger}b_j^{\dagger}b_k^{\dagger}b_l^{\dagger}b_m^{\dagger}$
$=\epsilon^{ijklm}b_i^{\dagger}b_j^{\dagger}b_k^{\dagger}b_l^{\dagger}b_m^{\dagger}\psi'$
$=5!~b_1^{\dagger}b_2^{\dagger}b_3^{\dagger}b_4^{\dagger}b_5^{\dagger}\psi'$.
To distinguish between the two multiplets, we replace the $\psi$'s
by ${\bf{N}}$. Putting everything together, we have for the
${\overline{16}}$ multiplet
\begin{equation}
\Psi_{(-){\acute {b}}}\rightarrow|\Psi_{(-){\acute
{b}}}>=b_1^{\dagger}b_2^{\dagger}
b_3^{\dagger}b_4^{\dagger}b_5^{\dagger}|0>{\bf N}_{\acute {b}}
+\frac{1}{12}\epsilon^{ijklm}b_k^{\dagger}b_l^{\dagger}
b_m^{\dagger}|0>{\bf N}_{{\acute {b}}ij}+b_i^{\dagger}|0>{\bf
N}_{\acute {b}}^i
\end{equation}

The group theoretic difference between the chirality $(+)$ and
$(-)$ states is that chirality $(+)$ states are generated by the
action of an even number of creation operators on the vacuum state
while the chirality $(-)$ states are generated by the action of an
odd number of creation operators. The chirality $(+)$ fields,
$|\Psi_{(+)\acute{a}}>$, can be identified with the three
generations of quarks and leptons, constituted by three sets of
$\bar 5+10$ plets of SU(5) and three $SU(5)$ singlet (right handed
neutrino) fields. The chirality $(-)$ fields,
$|\Psi_{(-)\acute{b}}>$ could be Higgs multiplets that arise in
$SO(10)$ model building. The role of the negative chirality fields
in model building will be discussed elsewhere.

For the sake of completeness, we also give the Hermitian
conjugate, transpose and complex conjugates of the $\Psi_{(\pm)}$
mutiplets:
\begin{eqnarray}
\Psi_{(+){\acute {a}}}^{\dagger}\rightarrow <\Psi_{(+){\acute
{a}}}|={\bf M}_{\acute {a}}^{\dagger}<0|+\frac{1}{2}{\bf
M}_{\acute {a}ij}^{\dagger}<0|b_jb_i +\frac{1}{24}{\bf M}_{{\acute
{a}}}^{i\dagger}\epsilon_{ijklm}<0|b_m b_lb_kb_j\nonumber\\
\Psi_{(+){\acute {a}}}^{T}\rightarrow <\Psi_{(+){\acute
{a}}}^{*}|={\bf M}_{\acute {a}}^{\bf{T}}<0|+\frac{1}{2}{\bf
M}_{\acute {a}}^{ij\bf{T}}<0|b_jb_i +\frac{1}{24}{\bf M}_{{\acute
{a}}i}^{{\bf T}}\epsilon^{ijklm}<0|b_m b_lb_kb_j\nonumber\\
\Psi^*_{(+){\acute {a}}}\rightarrow|\Psi^*_{(+){\acute
{a}}}>=|0>{\bf M}^*_{\acute
{a}}+\frac{1}{2}b_i^{\dagger}b_j^{\dagger}|0>{\bf M}_{\acute
{a}ij}^{*} +\frac{1}{24}\epsilon_{ijklm}b_j^{\dagger}
b_k^{\dagger}b_l^{\dagger}b_m^{\dagger}|0>{\bf M}_{{\acute
{a}}}^{i*}\nonumber\\
\end{eqnarray}
\begin{eqnarray}
\Psi_{(-){\acute {a}}}^{\dagger}\rightarrow <\Psi_{(-){\acute
{a}}}|={\bf N}_{\acute
{a}}^{\dagger}<0|b_5b_4b_3b_2b_1+\frac{1}{12}{\bf N}_{\acute
{a}}^{ij\dagger}\epsilon_{ijklm}<0|b_mb_lb_k +{\bf N}_{{\acute
{a}}i}^{\dagger}<0|b_i\nonumber\\
\Psi_{(-){\acute {a}}}^{T}\rightarrow <\Psi_{(-){\acute
{a}}}^{*}|={\bf N}_{\acute
{a}}^{\bf{T}}<0|b_5b_4b_3b_2b_1+\frac{1}{12}{\bf N}_{\acute
{a}ij}^{{\bf T}}\epsilon^{ijklm}<0|b_mb_lb_k +{\bf N}_{{\acute
{a}}}^{i{\bf T}}<0|b_i\nonumber\\
\Psi^*_{(-){\acute {b}}}\rightarrow|\Psi_{(-){\acute
{b}}}>=b_1^{\dagger}b_2^{\dagger}
b_3^{\dagger}b_4^{\dagger}b_5^{\dagger}|0>{\bf N}_{\acute {b}}^*
+\frac{1}{12}\epsilon_{ijklm}b_k^{\dagger}b_l^{\dagger}
b_m^{\dagger}|0>{\bf N}_{{\acute {b}}}^{ij*}+b_i^{\dagger}|0>{\bf
N}_{\acute {b}i}^*\nonumber
\end{eqnarray}
where we have the used the convention set in Eqs.(2.161) and
(2.170).
\newpage
\noindent\textsc{SO(10) invariants}\\
\\
From the decomposition formulas of Eqs. (2.229), (2.230), (2.242),
(2.243), (2.252), (2.253), (2.263), (2.264) we obtain
\begin{eqnarray}
16 ^{\dagger}\otimes {\overline{16}}=10\oplus 120\oplus
126^{(+i)}\nonumber\\
{\overline{16}}^{\dagger}\otimes 16=10\oplus 120\oplus
126^{(-i)}\nonumber\\
16^{\dagger}\otimes 16=1_{(as)}\oplus45_{(s)}\oplus 210_{(as)} \nonumber\\
{\overline{16}}^{\dagger}\otimes
{\overline{16}}=1_{(as)}\oplus45_{(s)}\oplus
210_{(as)}\\
\nonumber\\
16\otimes 16= 10_{(s)}\oplus 120_{(as)}\oplus
126_{(s)}^{(-i)}\nonumber\\
 {\overline{16}}\otimes
{\overline{16}}=10_{(s)}\oplus
120_{(as)}\oplus 126_{(s)}^{(+i)} \nonumber\\
16\otimes {\overline{16}}=1\oplus45\oplus 210 \nonumber\\
{\overline{16}}\otimes 16=1\oplus45\oplus 210 \\
\nonumber\\
{\overline{16}}^{\dagger}\otimes {\overline{16}}^{\dagger}=10_{(s)}\oplus 120_{(as)}\oplus 126_{(s)}^{(+i)} \nonumber\\
{{16}}^{\dagger}\otimes {{16}}^{\dagger}=10_{(s)}\oplus
120_{(as)}\oplus 126_{(s)}^{(-i)}\nonumber\\
16^{\dagger}\otimes {\overline{16}}^{\dagger}=1\oplus45\oplus 210 \nonumber\\
{\overline{16}}^{\dagger}\otimes 16^{\dagger}=1\oplus45\oplus 210 \\
\nonumber\\
16 \otimes {\overline{16}}^{\dagger}=10\oplus 120\oplus
126^{(+i)}\nonumber\\
{\overline{16}}\otimes 16^{\dagger}=10\oplus 120\oplus
126^{(-i)}\nonumber\\
16\otimes 16^{\dagger}=1_{(as)}\oplus45_{(s)}\oplus 210_{(as)} \nonumber\\
{\overline{16}}\otimes
{\overline{16}}^{\dagger}=1_{(as)}\oplus45_{(s)}\oplus 210_{(as)}
\end{eqnarray}
where the subscripts $s$ and $as$  indicate whether the particular
tensor formed from the Kronecker product of spinors appears
symmetrically or antisymmetrically in the exchange of $16$ plets
(see Eqs.(2.234), (2.246), (2.257) and (2.267)). The superscript
$\pm i$ indicates the duality condition satisfied by the 126
dimensional, $5^{th}$ rank tensor of $SO(10)$ (see Eqs.(2.232),
(2.245), (2.256) and (2.266)). In the literature $126^{(\mp i)}$
are defined as $126^{(-i)}\equiv 126$ and $126^{(+i)}\equiv
{\overline{126}}$. As mentioned earlier, it is imperative that  in
order for the invariant to be non-vanishing the $5^{th}$ rank
bosnic tensor (not shown here) which couples to the Kronecker
products of spinors must satisfy opposite in sign duality
condition.

Counting, the number of $SO(2N)$ invariants in
Eqs.(2.276)-(2.279), gives us a total of 48 couplings. In this
thesis, besides other computations, we have evaluated 15 of these
couplings in terms of $SU(5)$ fields.

Here it is also appropriate  to identify  the particular
representation of the $SO(10)$ charge conjugation matrix that we
are going to use in the subsequent chapters. To that end, we first
choose the basis to be:
$\Gamma_{2i}^T=\Gamma_{2i},~~\Gamma_{2i-1}^T=-\Gamma_{2i-1}$. This
is reasonable because $\Gamma_{2i}(=b_i+b_i^{\dagger})$ are
represented by symmetric matrices and
$\Gamma_{2i-1}=-i(b_i-b_i^{\dagger}))$ are represented by
antisymmetric matrices in the spinor basis. For the particular
representation of B, we choose Eq.(2.194) and along with
Eq.(2.195), we have:
\begin{eqnarray}
B^{-1}\Gamma_{\mu}^TB=-\Gamma_{\mu}\nonumber\\
B=-i\prod_{k=1}^5(b_k-b_k^{\dagger}),~~~B^T=-B,~~~B^{-1}=B
\end{eqnarray}\\
\noindent\textsc{decomposition of ordinary (bosonic) SO(10) vector
and
tensor representations with respect to $SU(5)$}\\
\\
From Eqs.(2.108)-(2.116), we have the following decompositions:
\begin{eqnarray}
SO(10)~~~~~~~~~~\supset~~~~~~~~~~SU(5)\otimes U(1)\nonumber
\end{eqnarray}
\begin{eqnarray}
\Phi_{\mu}:~[10]\supset\{5\}\oplus \{{\overline{5}}\}
\end{eqnarray}
\begin{eqnarray}
\Phi_{\mu\nu}^{({\cal A})}:~[45]\supset\{24\}\oplus\{10\}\oplus \{{\overline{10}}\}\oplus\{1\}\nonumber\\
\Phi_{\mu\nu}^{({\cal S})}:~[54]\supset\{24\}\oplus\{15\}\oplus
\{{\overline{15}}\}~~~~~~~~
\end{eqnarray}
\begin{eqnarray}
 \Phi_{\mu\nu\rho}^{({\cal
A})}:~[120]\supset\{5\}\oplus\{{\overline{5}}\}\oplus\{10\}
\oplus \{{\overline{10}}\}\oplus\{45\}\oplus\{{\overline{45}}\}\nonumber\\
\Phi_{\mu\nu\rho}^{({\cal
S})}:~[210']\supset\{35\}\oplus\{{\overline{35}}\}\oplus\{70\}
\oplus \{{\overline{70}}\}~~~~~~~~~~~~~~~~
\end{eqnarray}
\begin{eqnarray}
\Phi_{\mu\nu\rho\sigma}^{({\cal A})}:~[210]\supset
\{1\}\oplus\{5\}\oplus\{{\overline{5}}\}\oplus\{10\} \oplus
\{{\overline{10}}\}\oplus\{24\}\oplus\{40\}\oplus\{{\overline{40}}\}\nonumber\\
\oplus\{75\}~~~~~~~~~~~~~~~~~~~~~~~~~~~~~~~~~~~~~~~~~~~~~~~~~~~~~~~~~~~~~\nonumber\\
 \Phi_{\mu\nu\rho\sigma}^{({\cal
S})}:~[660]\supset \{70\}\oplus\{{\overline{70}}\}\oplus\{160\}
\oplus \{{\overline{160}}\}\oplus\{200\}~~~~~~~~~~~~~~~~~~~~~
\end{eqnarray}
\begin{eqnarray}
\Phi_{\mu\nu\rho\sigma\lambda}^{({\cal A})}:~[252]\supset
\{1\}\oplus\{1\}\oplus\{5\}\oplus\{{\overline{5}}\}\oplus\{10\}
\oplus
\{{\overline{10}}\}\oplus\{15\}\oplus\{{\overline{15}}\}\nonumber\\
\oplus\{45\}\oplus\{{\overline{45}}\}
\oplus\{50\}\oplus\{{\overline{50}}\}\nonumber\\
\Phi_{\mu\nu\rho\sigma}^{({\cal S})}:~[1782]\supset
\{126\}\oplus\{{\overline{126}}\}\oplus\{315\} \oplus
\{{\overline{315}}\}\oplus\{450\} \oplus \{{\overline{450}}\}~~~
\end{eqnarray}
Recall that the real 252-dimensional $5^{th}$ rank tensor,
$\Phi_{\mu\nu\rho\sigma\lambda}^{({\cal A})}$ decomposes  into two
complex $5^{th}$ tensors,
$\Omega_{\mu\nu\rho\sigma\lambda}^{({\cal A})}$ and
${\overline{\Omega}}_{\mu\nu\rho\sigma\lambda}^{({\cal A})}$ each
of dimensionality 126:
\begin{eqnarray}
\Omega_{\mu\nu\rho\sigma\lambda}^{({\cal A})}:
~[126]\supset\{1\}\oplus\{{\overline{5}}\}\oplus\{10\}\oplus\{{\overline{15}}\}\oplus\{45\}
\oplus\{{\overline{50}}\}\nonumber\\
{\overline{\Omega}}_{\mu\nu\rho\sigma\lambda}^{({\cal
A})}:~[{\overline{126}}]\supset\{1\}\oplus\{{{5}}\}\oplus\{{\overline{10}}\}\oplus\{{{15}}\}\oplus\{{\overline{45}}\}
\oplus\{{{50}}\}
\end{eqnarray}\\
\noindent\textsc{spinorial generators of SO(10)}\\
\\
The two most interesting subgroup decompositions are \\
 $SO(10)\supset SO(6)\otimes SO(4)\cong SU(4)_C\otimes
SU(2)_L\otimes SU(2)_R$\\
and\\
 $SO(10)\supset SU(5)\otimes
U(1)_X$\\
In the case of a full exhibition of $SU(4)_C\otimes SU(2)_L\otimes
SU(2)_R$ invariant content of $SO(10)$ couplings see \cite{ag}.\\
\\
\underline{$SO(6)\cong SU(4)_C\supset SU(3)_C\otimes U(1)_{B-L}$}\\
\\
1.~~$SU(3)_C$ Generators.\\

The general $SU(N)$ spinorial generators are given by Eqs.(2.149)
and (2.150). Here we specialize them for  $N=3$:
\begin{eqnarray}
t^{\alpha}_{\beta}=-b_{\alpha}^{\dagger}b_{\beta}+\frac{\delta_{\alpha\beta}}{3}\sum_{\xi=1}^3b_{\xi}^{\dagger}
b_{\xi};~~~~\alpha,~\beta,~\xi=1,2,3~~~~~~~~~~~~~~~~~~~~~~~~~~~~~~~~\nonumber\\
t^{\alpha}_{\beta}=\frac{i}{4}\left(\Sigma_{2\alpha~2\beta}+\Sigma_{2\alpha-1~2\beta-1}-i\Sigma_{2\alpha~2\beta-1}-i\Sigma_{2\beta~2\alpha-1}\right)
-\frac{\delta_{\alpha\beta}}{6}
\sum_{\xi=1}^3\Sigma_{2\xi~2\xi-1}~~~~~
\end{eqnarray}
\begin{eqnarray}
a=1,2,...8~~~~~~~~~~~~~~~~~~~~~~~~~~~~~~~~~~~~~~~~~~~~~\nonumber\\
  \nonumber\\
 {t}_{a}=2\times\left\{\matrix{\left[\matrix{{t}_{\alpha}^{\beta}+{
t}_{\beta}^{\alpha}&&&3~{\textnormal{generators}}\cr -i\left({
t}_{\beta}^{\alpha}-{
t}_{\alpha}^{\beta}\right)&&&3~{\textnormal{generators}}\cr}\right];~\forall \alpha \neq \beta=1,2,3\nonumber\\
\left[\matrix{\sqrt{3}\left(t^{\alpha}_{\alpha}+t^{\alpha+1}_{\alpha+1}\right)&&&2~{\textnormal{generators}}\cr
\left(t^{\alpha}_{\alpha}-t^{\alpha+1}_{\alpha+1}\right)&&&2
~{\textnormal{generators}}\cr}\right];~\alpha=1,2\cr}\right.
\end{eqnarray}\\

\noindent Factor of 2 inserted above is chosen for convenience. We
choose the following diagonal generators  ($\alpha=1$):
$\sqrt{3}\left(t^{1}_{1}+t^{2}_{2}\right)$ and
$\left(t^{1}_{1}-t^{2}_{2}\right)$. The complete list of all the
$SU(3)$ color generators are
\begin{eqnarray}
2{\sqrt 3}\left(t^1_1+t^2_2\right)=-\frac{2}{\sqrt
3}(b_1^{\dagger}b_1+b_2^{\dagger}b_2-2b_3^{\dagger}b_3)=\frac{1}{\sqrt
3}(\Sigma_{12}+\Sigma_{34}-2\Sigma_{56})\nonumber\\
2\left(t^1_1-t^2_2\right)=-2(b_1^{\dagger}b_1-b_2^{\dagger}b_2)=\Sigma_{34}-\Sigma_{12}\nonumber\\
2\left(t^1_2+t^2_1\right)=-2(b_1^{\dagger}b_2+b_2^{\dagger}b_1)=\Sigma_{23}-\Sigma_{14}\nonumber\\
-2i\left(t^1_2-t^2_1\right)=2i(b_1^{\dagger}b_2-b_2^{\dagger}b_1)=\Sigma_{24}+\Sigma_{13}\nonumber\\
2\left(t^1_3+t^3_1\right)=-2(b_1^{\dagger}b_3+b_3^{\dagger}b_1)=\Sigma_{25}-\Sigma_{16}\nonumber\\
-2i\left(t^1_3-t^3_1\right)=2i(b_1^{\dagger}b_3-b_3^{\dagger}b_1)=\Sigma_{26}+\Sigma_{15}\nonumber\\
2\left(t^2_3+t^3_2\right)=-2(b_2^{\dagger}b_3+b_3^{\dagger}b_2)=\Sigma_{45}-\Sigma_{36}\nonumber\\
-2i\left(t^2_3-t^3_2\right)=2i(b_2^{\dagger}b_3-b_3^{\dagger}b_2)=\Sigma_{46}+\Sigma_{35}\nonumber\\
\end{eqnarray}\\
\newpage
\noindent 2.~~$U(1)_{B-L}$ Generator.\\

From Eq.(2.154), we get
\begin{equation}
B-L=\frac{3}{2}-\sum_{\alpha=1}^3
b_{\alpha}^{\dagger}b_{\alpha}=\frac{1}{2}\Sigma_{2\alpha-1~2\alpha}
\end{equation}
Convention is to put a normalization factor of $\frac{2}{3}$, that
is $B-L\rightarrow \frac{2}{3}(B-L)$. Therefore the new $B-L$ is
given by
\begin{equation}
B-L=1-\frac{2}{3}(b_1^{\dagger}b_1+b_2^{\dagger}b_2+b_3^{\dagger}b_3)=\frac{1}{3}(\Sigma_{12}+\Sigma_{34}+\Sigma_{56})
\end{equation}\\
 3.~~$SU(4)_{C}$ Generators.\\

Using Eq.(2.150), we have a total of  12 non-diagonal generators
of $SU(4)_C$. In addition to the 6 non-diagonal generators of
$SU(3)_C$ (see Eq.(2.289)), we have
\begin{eqnarray}
2\left(t^1_4+t^4_1\right)=-2(b_1^{\dagger}b_4+b_4^{\dagger}b_1)=\Sigma_{27}-\Sigma_{18}\nonumber\\
-2i\left(t^1_4-t^4_1\right)=2i(b_1^{\dagger}b_4-b_4^{\dagger}b_1)=\Sigma_{28}+\Sigma_{17}\nonumber\\
2\left(t^2_4+t^4_2\right)=-2(b_2^{\dagger}b_4+b_4^{\dagger}b_2)=\Sigma_{47}-\Sigma_{38}\nonumber\\
-2i\left(t^2_4-t^4_2\right)=2i(b_2^{\dagger}b_4-b_4^{\dagger}b_2)=\Sigma_{48}+\Sigma_{37}\nonumber\\
2\left(t^3_4+t^4_3\right)=-2(b_3^{\dagger}b_4+b_4^{\dagger}b_3)=\Sigma_{67}-\Sigma_{58}\nonumber\\
-2i\left(t^3_4-t^4_3\right)=2i(b_3^{\dagger}b_4-b_4^{\dagger}b_3)=\Sigma_{68}+\Sigma_{57}
\end{eqnarray}
The 3 diagonal generators of $SU(4)_C$ are chosen to be
\begin{eqnarray}
2{\sqrt 2}\left(t^1_1+t^2_2\right)=-\sqrt
2(b_1^{\dagger}b_1+b_2^{\dagger}b_2-b_3^{\dagger}b_3-b_4^{\dagger}b_4)\nonumber\\
=\frac{1}{\sqrt
2}(\Sigma_{56}+\Sigma_{78}-7\Sigma_{34}-7\Sigma_{12})\nonumber\\
2{\sqrt 2}\left(t^2_2+t^3_3\right) =-\sqrt
2(b_2^{\dagger}b_2+b_3^{\dagger}b_3-b_1^{\dagger}b_1-b_4^{\dagger}b_4)\nonumber\\
=\frac{1}{\sqrt
2}(\Sigma_{12}+\Sigma_{78}-7\Sigma_{34}-7\Sigma_{56})\nonumber\\
2\left(t^1_1-t^2_2\right)=-2(b_1^{\dagger}b_1-b_2^{\dagger}b_2)=4(\Sigma_{34}-\Sigma_{12})~~~~~~~~~~
\end{eqnarray}\\
 \underline{$SO(4)\cong SU(2)_L\otimes SU(2)_R\supset SU(2)_L\otimes U(1)_{R}$}\\
\\
1.~~$SU(2)_L$ Generators.\\

In this case the generators are given by
\begin{eqnarray}
T_{3L}\equiv-\left(t^4_4-t^5_5\right)=\frac{1}{2}(b_4^{\dagger}b_4-b_5^{\dagger}b_5)=\frac{1}{4}\left(\Sigma_{78}-\Sigma_{9~10}\right)\nonumber\\
\frac{1}{2}\left(t^4_5+t^5_4\right)=-\frac{1}{2}(b_4^{\dagger}b_5+b_5^{\dagger}b_4)=\frac{1}{4}\left(\Sigma_{7~10}-\Sigma_{89}\right)\nonumber\\
\frac{i}{2}\left(t^4_5-t^5_4\right)=-\frac{i}{2}(b_4^{\dagger}b_5-b_5^{\dagger}b_4)=-\frac{1}{4}\left(\Sigma_{8~10}+\Sigma_{79}\right)
\end{eqnarray}
Again the overall normalizations of $-1$ and
 $\pm\frac{1}{2}$are inserted for convenience.\\
\\
2.~~$U(1)_R$ Generator.\\

From Eq.(2.154), we get $T_{3R}\equiv
\frac{1}{2}\sum_{p=4}^{5}\Sigma_{2p-1~2p}=2/2-\sum_{p=4}^{5}b_{p}^{\dagger}b_p$.
We again normalize $T_{3R}$ with a factor of $-\frac{1}{2}$, that
is $T_{3R}\rightarrow -\frac{1}{2}T_{3R}$. Therefore,
\begin{equation}
T_{3R}=-\frac{1}{4}\left(\Sigma_{78}+\Sigma_{9~10}\right)=\frac{1}{2}\left(b_4^{\dagger}b_4+b_5^{\dagger}b_5-1\right)
\end{equation}\\
\\
3.~~$SU(2)_R$ Generators.\\

Here the diagonal generator, $T_{3R}$ is given by Eq.(2.295) while
the other two generators are
\begin{eqnarray}
\frac{1}{4}\left(\Sigma_{7~10}+\Sigma_{89}\right)=-\frac{1}{2}(b_4b_5-b_4^{\dagger}b_5^{\dagger})\nonumber\\
-\frac{1}{4}\left(\Sigma_{8~10}-\Sigma_{79}\right)=-\frac{i}{2}(b_4b_5+b_4^{\dagger}b_5^{\dagger})
\end{eqnarray}
Each of these generators is a linear combination of the broken
generators in $SO(10)$ (see Eq.(2.156)). Since $SO(4)$ is
isomorphic to $SU(2)_{L}\times SU(2)_{R}$, the six generators
given by Eqs.(2.294)-(2.296) are also the generators of $SO(4)$.\\
\\
\noindent\underline{$U(1)_R\otimes U(1)_{B-L}\supset U(1)_Y$}\\
\\
We begin by writing, $Y=\alpha T_{3R}+\beta(B-L)$. From the
hypercharge assignments of the known particles one gets $\alpha=1$
and $\beta=\frac{1}{2}$. Thus
\begin{equation}
Y=T_{3R}+\frac{B-L}{2}
\end{equation}
From Eqs.(2.291) and (2.295) we can immediately write down,
\begin{eqnarray}
Y=-\frac{1}{3}\left(b_1^{\dagger} b_1+b_2^{\dagger}
b_2+b_3^{\dagger} b_3\right)+\frac{1}{2}\left(b_4^{\dagger}
b_4+b_5^{\dagger} b_5\right)\nonumber\\
=\frac{1}{6}\left(\Sigma_{12}+\Sigma_{34}+\Sigma_{56}\right)-\frac{1}{4}\left(\Sigma_{78}+\Sigma_{9~10}\right)
\end{eqnarray}\\

\noindent\underline{$U(1)_{L}\otimes U(1)_{Y}\supset U(1)_{em}$}\\
\\
Assume that, $Q= aT_{3L}+bY$. From the charge assignments of the
known particles, we find $a=1$ and $b=1$. Therefore, we write
\begin{equation}
Q= T_{3L}+Y
\end{equation}

From Eqs.(2.294) and (2.298) we can immediately write down
\begin{eqnarray}
Q=-\frac{1}{3}\left(b_1^{\dagger} b_1+b_2^{\dagger}
b_2+b_3^{\dagger} b_3\right)+b_4^{\dagger}
b_4\nonumber\\
=\frac{1}{6}\left(\Sigma_{12}+\Sigma_{34}\Sigma_{56}\right)-\frac{1}{2}\Sigma_{9~10}
\end{eqnarray}\\
\underline{$SO(10)\supset SU(5)\otimes U(1)_X$}\\
\\
Eq.(2.154) can be used to  find the $U(1)_X$ generator.
\begin{eqnarray}
X=\frac{5}{2}-\left(b_1^{\dagger} b_1+b_2^{\dagger}
b_2+b_3^{\dagger} b_3+b_4^{\dagger} b_4+b_5^{\dagger}
b_5\right)\nonumber\\
=\frac{1}{2}\left(\Sigma_{12}+\Sigma_{34}+\Sigma_{56}+\Sigma_{78}+\Sigma_{9~10}\right)
\end{eqnarray}\\

\noindent\textsc{$T_{3L}$, $T_{3R}$, $B-L$, $Y$, $X$ in
the ordinary 10-dimensional representation
of SO(10)}\\
\\
For symmetry breaking purposes, where the spinor and tensor
representations of $SO(10)$ acquire a vacuum expectation value in
either $T_{3L}$, $T_{3R}$, $B-L$, $Y$ or $X$  direction, it is
essential to have these operators in the ordinary 10-dimensional
representation given by Eq.(2.10).

From Eq.(2.10), the generators, $M_{\mu\nu}$ in the 10-dimensional
representation is a $10\times 10$ matrix with $-i$
 and $+i$ as the only non-vanishing elements and are given by the
 intersection of the $\mu^{th}$ row and $\nu^{th}$ column and
$\nu^{th}$ row and $\mu^{th}$ column with $\mu\neq\nu$. If $-i$
 and $+i$ are to lie immediately above and below the diagonal,
 that is when $\mu=\nu-1$, then $M_{\mu\nu}$ can be expressed as
\begin{eqnarray}
M_{\mu-1~\mu}=\sigma_2\otimes diag(0,..,0,1,0,...,0)\nonumber\\
\sigma_2=\left(\matrix{0&-i\cr i&0 \cr}\right)
\end{eqnarray}
That is, we have expressed $10\times 10$ generator,
$M_{\mu-1~\mu}$ in terms $(5\times 5)\otimes (2\times 2)$. Here, 1
appears at the $(\frac{\mu}{2})^{th}$ position along the diagonal
of the $5\times 5$ matrix.

Replacing, the letter $M$ by $\Sigma$ and using Eqs.(2.291),
(2.294), (2.295), (2.298), (2.300) and (2.301), we get

\begin{eqnarray}
B-L=\frac{1}{3}\sigma_2\otimes diag(1,1,1,0,0)\nonumber\\
T_{3L}=\frac{1}{4}\sigma_2\otimes diag(0,0,0,1,-1)\nonumber\\
T_{3R}=-\frac{1}{4}\sigma_2\otimes diag(0,0,0,1,1)\nonumber\\
Y=\frac{1}{4}\sigma_2\otimes diag(\frac{2}{3},\frac{2}{3},\frac{2}{3},-1,-1)\nonumber\\
Q=\frac{1}{4}\sigma_2\otimes diag(\frac{2}{3},\frac{2}{3},\frac{2}{3},0,-2)\nonumber\\
X=\frac{1}{2}\sigma_2\otimes diag(1,1,1,1,1)
\end{eqnarray}\\
\\
\noindent\textsc{$T_{3L}$, $T_{3R}$, $B-L$, $Y$, $X$ quantum
number assignments of the states in the 16 of SO(10)}\\
\\
Let's begin by writing the 16 plet of matter for one generation
(Eq.(2.272)) in terms of Standard Model particle states
(Eq.(2.273)):
\begin{eqnarray}
|\Psi_{(+)}>=|0>\nu_{eL}^c~~~~~~~~~~~~~~~~~~~~~~~~~~~~~~~~~~~~~~~~~~~~~~~~~~~~~~~~~~~~~~~~\nonumber\\
+\frac{1}{2}\epsilon^{\alpha\beta\xi}b_{\alpha}^{\dagger}b_{\beta}^{\dagger}|0>u_{L\xi}^c
+\frac{1}{2}b_{4}^{\dagger}b_{\alpha}^{\dagger}|0>u_{L}^{\alpha}
+\frac{1}{2}b_{5}^{\dagger}b_{\alpha}^{\dagger}|0>d_{L}^{\alpha}
+\frac{1}{2}b_{4}^{\dagger}b_{5}^{\dagger}|0>e_{L}^{+}\nonumber\\
+\frac{1}{24}\epsilon^{\alpha\beta\xi}b_4^{\dagger}
b_5^{\dagger}b_{\alpha}^{\dagger}b_{\beta}^{\dagger}|0>d_{L\xi}^c
+\frac{1}{24}b_5^{\dagger}
b_1^{\dagger}b_{2}^{\dagger}b_{3}^{\dagger}|0>e_{L}^{-}
+\frac{1}{24}b_1^{\dagger}
b_2^{\dagger}b_{3}^{\dagger}b_{4}^{\dagger}|0>\nu_{eL}
\end{eqnarray}

\noindent For example, in order to calculate $B-L$ quantum number
assignment of the state $u_{L}^{\alpha}$, we use Eqs.(2.291) and
(2.304) to write
$(B-L)(b_4^{\dagger}b_{\alpha}^{\dagger}|0>$)$=[1-\frac{2}{3}(b_1^{\dagger}b_1+b_2^{\dagger}b_2+b_3^{\dagger}b_3)](b_4^{\dagger}b_{\alpha}^{\dagger}
|0>)$ $=\frac{1}{3}(b_4^{\dagger}b_{\alpha}^{\dagger}|0>)$.
Therefore, $B-L$ quantum number  of $u_{L}^{\alpha}$ is
$(B-L)_{u_{L}^{\alpha}}=\frac{1}{3}$. We now give a complete list
of all the quantum numbers of Standard Model particle states
appearing in the 16 plet of $SO(10)$:
\begin{eqnarray}
(B-L)_{\nu_{eL}^c}=1,~~(B-L)_{\nu_{eL}}=-1,~~(B-L)_{e_{L}^{+}}=1,~~(B-L)_{e_{L}^{-}}=-1\nonumber\\
(B-L)_{u_{L\alpha}^c}=-\frac{1}{3},~~(B-L)_{u_{L}^{\alpha}}=\frac{1}{3},
~~(B-L)_{d_{L\alpha}^c}=-\frac{1}{3},~~(B-L)_{d_L^{\alpha}}=\frac{1}{3}\nonumber\\
\end{eqnarray}
\begin{eqnarray}
(T_{3L})_{\nu_{eL}^c}=0,~~(T_{3L})_{\nu_{eL}}=\frac{1}{2},~~(T_{3L})_{e_{L}^{+}}=0,~~(T_{3L})_{e_{L}^{-}}=-\frac{1}{2}\nonumber\\
(T_{3L})_{u_{L\alpha}^c}=0,~~(T_{3L})_{u_{L}^{\alpha}}=\frac{1}{2},
~~(T_{3L})_{d_{L\alpha}^c}=0,~~(T_{3L})_{d_L^{\alpha}}=-\frac{1}{2}
\end{eqnarray}
\begin{eqnarray}
(T_{3R})_{\nu_{eL}^c}=-\frac{1}{2},~~(T_{3R})_{\nu_{eL}}=0,~~(T_{3R})_{e_{L}^{+}}=\frac{1}{2},~~(T_{3R})_{e_{L}^{-}}=0\nonumber\\
(T_{3R})_{u_{L\alpha}^c}=-\frac{1}{2},~~(T_{3R})_{u_{L}^{\alpha}}=0,
~~(T_{3R})_{d_{L\alpha}^c}=\frac{1}{2},~~(T_{3R})_{d_L^{\alpha}}=0
\end{eqnarray}
\begin{eqnarray}
(Y)_{\nu_{eL}^c}=0,~~(Y)_{\nu_{eL}}=-\frac{1}{2},~~(Y)_{e_{L}^{+}}=1,~~(Y)_{e_{L}^{-}}=-\frac{1}{2}\nonumber\\
(Y)_{u_{L\alpha}^c}=-\frac{2}{3},~~(Y)_{u_{L}^{\alpha}}=\frac{1}{6},
~~(Y)_{d_{L\alpha}^c}=\frac{1}{3},~~(Y)_{d_L^{\alpha}}=\frac{1}{6}
\end{eqnarray}
\begin{eqnarray}
(Q)_{\nu_{eL}^c}=0,~~(Q)_{\nu_{eL}}=0,~~(Q)_{e_{L}^{+}}=1,~~(Q)_{e_{L}^{-}}=-1\nonumber\\
(Q)_{u_{L\alpha}^c}=-\frac{2}{3},~~(Q)_{u_{L}^{\alpha}}=\frac{2}{3},
~~(Q)_{d_{L\alpha}^c}=\frac{1}{3},~~(Q)_{d_L^{\alpha}}=-\frac{1}{3}
\end{eqnarray}
\begin{eqnarray}
(X)_{\nu_{eL}^c}=\frac{5}{2},~~(X)_{\nu_{eL}}=-\frac{3}{2},~~(X)_{e_{L}^{+}}=\frac{1}{2},~~(X)_{e_{L}^{-}}=-\frac{3}{2}\nonumber\\
(X)_{u_{L\alpha}^c}=\frac{1}{2},~~(X)_{u_{L}^{\alpha}}=\frac{1}{2},
~~(X)_{d_{L\alpha}^c}=-\frac{3}{2},~~(X)_{d_L^{\alpha}}=\frac{1}{2}
\end{eqnarray}

\chapter{Grand Unification}\label{gu_chapter}
One of the main effort in particle physics is to find ever more
unifying principles in particle physics. An early attempt in this
direction after the formulation of the Standard Model was the work
of Pati and Salam\cite{ps} in the unification of quarks and
leptons within the gauge group $SU(2)_L\times SU(2)_R\times
SU(4)_C$. Quickly thereafter the SU(5) unification was proposed by
Georgi and Glashow\cite{gg} wherein  the gauge coupling constants
of $SU(3)_C$, $SU(2)_L$ and $U(1)_Y$ are unified at the grand
unification scale $M_G$. The non-supersymmetric models have severe
fine tuning problems in keeping the Higgs triplet light and needs
supersymmetry to relax this constraint (For a review see,
Ref.\cite{applied,Brazilian,nilles,tasi}). However, in
supersymmetric grand unification one of the main problem concerns
the issue of breaking of supersymmetry. It is difficult to break
supersymmetry within the global supersymmetry framework and one
needs local supersymmetry to do the
breaking\cite{applied,Brazilian,nilles}. This leads one to
contruct supergravity grand unification models\cite{can82}. The
ways to construct supersymmetric theories is well known and the
reader is referred to well known
reports\cite{wess,applied,Brazilian,nilles}. In the analysis to
follow we will start with the particle spectrum of MSSM exhibited
in the Table below.
\begin{center}
\begin{tabular}{p{5.5cm} c c c}
\multicolumn{3}{c}{Table 4.1:~Particle content of the MSSM} \\
\hline \hline
\\
&Vector Multiplets &\\
$j=1$ & & $j=\frac{1}{2}$\\
\hline\\
$g^a_\mu(x),\ a=1\dots 8$ & & $\lambda^a(x),\ a=1\dots 8$\\
gluons $(g)$ & & gluinos $(\tilde g)$\\
$B^\alpha_\mu(x),\;B^Y_\mu(x),\ \alpha=1,2,3$ & &$\lambda^\alpha(x),\;\lambda^Y(x),\ \alpha=1,2,3$\\
$SU(2)_L\times U(1)_Y$ gauge bosons& &$SU(2)_L\times U(1)_Y$ gauginos\\
\hline
\hline
\\
&Chiral Multiplets&\\
$j=\frac{1}{2}$& &$j=0$\\
\hline\\
$(u_{iL},\,d_{iL});\;u_{iR},\,d_{iR}$& &$(\tilde u_{iL},\,\tilde d_{iL});\;\tilde u_{iR},\,\tilde d_{iR}$\\
quarks ($i$=generation)& &squarks\\
$(\nu_{iL},\,e_{iL});\;e_{iR}$& &$(\tilde \nu_{iL},\,\tilde e_{iL});\;\tilde e_{iR}$\\
leptons& &sleptons\\
$\tilde H_1\!\!=\!\!(\tilde H^0_1,\,\tilde H^-_1);\;\tilde H_2\!=\!(\tilde H^+_2,\,\tilde H^0_2)$& &$H_1\!=\!(H^0_1,\,H^-_1);\;H_2=(H^+_2,\,H^0_2)$\\
Higgsinos& &Higgs bosons\\
\hline
\hline
\end{tabular}
\end{center}
\newpage

\begin{center} \begin{tabular}{|c|c|c|c|c|}
\multicolumn{5}{c}{Table 4.2: ~Anomaly cancellations in MSSM spectrum } \\
\hline
 bose & fermi & Y & $tr(Y^3)$& $tr(T_3^2Y)$ \\
\hline
g &$\tilde g$   & 0 & 0 & 0 \\
$W^a$ & $\tilde W_a$ & 0 & 0&0\\
B & $\tilde B$& 0 &0  &0 \\
  $\tilde L$=($\tilde\nu,\tilde e^-)_L$ & $L=(\nu,e^-)_L$ & -1 & -2
 & -$\frac{2}{4}$\\
$\tilde e^c_L$  & $e^c_L$& 2 & 8 & 0\\
 $\tilde q=(\tilde u,\tilde d)_L$ & $q=(u,d)_L$ & $\frac{1}{3}$ &($\frac{3\times 2}{27}$) &
  $\frac{2}{4}$\\
$\tilde u^c_L$ & $u^c_L$& -$\frac {4}{3}$& $-(\frac{3\times 64}{27}$) & 0\\
 $\tilde d^c_L$ & $d^c_L$& $\frac {2}{3}$& ($\frac{3\times 8}{27}$) & 0\\
 $(H^0_1,H^-_1)$ &  $(\tilde H^0_1,\tilde H^-_1)$ & -1 & -2 & $-\frac{2}{4}$ \\
 $(H^+_2,H^0_2)$ &  $(\tilde H^+_2,\tilde H^0_2)$ & 1 & 2 & $\frac{2}{4}$\\
 \hline
total & & & tr($Y^3$)=0& tr($T_3^2Y$)=0 \\
 \hline
\end{tabular}
\end{center}
As exhibited above the spectrum is anomaly free. The above spectrum can be accomodated in
the $16_{+}$ spinor of SO(10). We will now give details of the techniques for the computation
of couplings  of the 16-plet of spinors  with tensor representations.


\chapter{Technique for the Evaluation of $\bf{SO(2N)}$ Invariants. The Basic Theorem}\label{eval basic}

Here we review the recently developed technique
\cite{ns1,ns2,ns3,ns4}
 for the analysis of $SO(2N)$ invariant
couplings which allows a full exhibition of the $SU(N)$ invariant
 content of the spinor and tensor
representations. The technique utilizes a basis consisting of a
specific set of reducible $SU(N)$ tensors in terms of which the
$SO(2N)$ invariant couplings have a simple expansion.

\section{Specific set of {SU(N)} reducible tensors}

We begin with the observation that the natural basis \cite{ns1}
for the expansion of the $SO(2N)$ vertex is in terms of a specific
set of $SU(N)$ reducible tensors, ${\Phi}_{c_k}$ and
${\Phi}_{\overline c_k}$ which we define as
\begin{eqnarray}
A^k\equiv{\Phi}_{c_k}\equiv{\Phi}_{2k}+i{\Phi}_{2k-1}\nonumber\\
A_k\equiv{\Phi}_{\overline c_k}\equiv{\Phi}_{2k}-i{\Phi}_{2k-1}
\end{eqnarray}
This can be extended immediately
 to define the quantity $\Phi_{c_ic_j\bar c_k..}$
with an arbitrary number of unbarred and barred indices where each
 $c$ index can be expanded out so that
 \begin{eqnarray}
A^iA^jA_k...={\Phi}_{c_ic_j\overline
c_k...}={\Phi}_{2ic_j\overline c_k...}+i{\Phi}_{2i-1c_j\overline
 c_k...}~~{\textnormal{etc}}...
 \end{eqnarray}
 Thus, for example, the quantity  $\Phi_{c_{i_1}c_{i_2}\overline c_{i_3}...c_{i_n}}$ is a sum of
 $2^n$ terms gotten by expanding all the c indices.
$\Phi_{c_ic_j\overline c_k...c_n}$ is completely anti-symmetric in
the interchange of its $c$ indices whether unbarred or barred:
\begin{equation}
 {\Phi}_{c_i\overline c_jc_k...\overline c_n}=-{\Phi}_{c_k\overline c_jc_i...\overline c_n}~~{\textnormal{etc}}..
 \end{equation}
Further,
\begin{equation}
 {\Phi}^*_{c_i\overline c_jc_k...\overline c_n}={\Phi}_{\overline c_ic_j\overline
  c_k...c_n} ~~{\textnormal{etc}}...
\end{equation}\\
Use of quantities like ${\Phi}^*_{c_i\overline c_jc_k...\overline
c_n}$ are also useful in evaluating kinetic energy like terms such
as:
$-\frac{1}{2}\partial_A\Phi_{\mu}\partial^A\Phi_{\mu}^{\dagger}$,
$\frac{1}{4}\Phi_{\mu\nu}^{AB}\Phi_{\mu\nu~AB}$, etc.. We give the
general results here,
\begin{eqnarray}
\Phi_{\mu_1\mu_2\mu_3...\mu_r}\Phi_{\mu_1\mu_2\mu_3...\mu_r}
=\frac{1}{2^r}\left[\Phi_{c_{i_1}c_{i_2}c_{i_3}...c_{i_{r-1}}c_{i_r}}
\Phi_{\overline c_{i_1}\overline c_{i_2}\overline
c_{i_3}...\overline c_{i_{r-1}}\overline c_{i_r}}\right.\nonumber\\
\left.+ \Phi_{\overline
c_{i_1}c_{i_2}c_{i_3}...c_{i_{r-1}}c_{i_r}}\Phi_{c_{i_1}\overline
c_{i_2}\overline c_{i_3}...\overline c_{i_{r-1}}\overline c_{i_r}}
+...+\Phi_{ c_{i_1}c_{i_2} c_{i_3}...c_{i_{r-1}}\overline
c_{i_r}}\Phi_{\overline c_{i_1}\overline c_{i_2}\overline c_{i_3}
...\overline c_{i_{r-1}}c_{i_r}}\right.\nonumber\\
\left.+\Phi_{\overline c_{i_1}\overline c_{i_2}
c_{i_3}...c_{i_{r-1}} c_{i_r}}\Phi_{c_{i_1} c_{i_2}\overline
c_{i_3}... \overline c_{i_{r-1}}\overline c_{i_r}}
+...+\Phi_{
c_{i_1} c_{i_2} c_{i_3}... \overline c_{i_{r-1}}\overline
c_{i_r}}\Phi_{\overline c_{i_1} \overline
c_{i_2} \overline c_{i_3}... c_{i_{r-1}}c_{i_r}}\right.\nonumber\\
\left.+...+\Phi_{\overline c_{i_1}\overline c_{i_2} \overline
c_{i_3}...\overline c_{i_{r-1}} \overline c_{i_r}}\Phi_{c_{i_1}
c_{i_2} c_{i_3}...
 c_{i_{r-1}}c_{i_r}}\right]\nonumber\\
\end{eqnarray}
and
\begin{eqnarray}
\Phi_{\mu_1\mu_2\mu_3...\mu_r}\Phi_{\mu_1\mu_2\mu_3...\mu_r}^{\dagger}
=\frac{1}{2^r}\left[\Phi_{c_{i_1}c_{i_2}c_{i_3}...c_{i_{r-1}}c_{i_r}}
\Phi_{ c_{i_1} c_{i_2}
c_{i_3}... c_{i_{r-1}} c_{i_r}}^{\dagger}\right.\nonumber\\
\left.+ \Phi_{\overline
c_{i_1}c_{i_2}c_{i_3}...c_{i_{r-1}}c_{i_r}}\Phi_{\overline c_{i_1}
c_{i_2} c_{i_3}... c_{i_{r-1}} c_{i_r}}^{\dagger} +...+\Phi_{
c_{i_1}c_{i_2} c_{i_3}...c_{i_{r-1}}\overline c_{i_r}}\Phi_{
c_{i_1} c_{i_2}
c_{i_3}... c_{i_{r-1}}\overline c_{i_r}}^{\dagger}\right.\nonumber\\
\left.+\Phi_{\overline c_{i_1}\overline c_{i_2}
c_{i_3}...c_{i_{r-1}} c_{i_r}}\Phi_{\overline c_{i_1} \overline
c_{i_2} c_{i_3}...  c_{i_{r-1}}c_{i_r}}^{\dagger} +...+\Phi_{
c_{i_1} c_{i_2} c_{i_3}... \overline c_{i_{r-1}}\overline
c_{i_r}}\Phi_{c_{i_1}
c_{i_2} c_{i_3}... \overline c_{i_{r-1}}\overline c_{i_r}}^{\dagger}\right.\nonumber\\
\left.+...+\Phi_{\overline c_{i_1}\overline c_{i_2} \overline
c_{i_3}...\overline c_{i_{r-1}} \overline c_{i_r}}\Phi_{\overline
c_{i_1} \overline c_{i_2} \overline c_{i_3}...
 \overline c_{i_{r-1}}\overline c_{i_r}}^{\dagger}\right]\nonumber\\
\end{eqnarray}

 We now make the observation that the object
$\Phi_{c_ic_j\overline c_k...c_n}$ transforms like a reducible
representation of $SU(N)$. Thus if we are able to compute the
$SO(2N)$ invariant couplings
 in terms of these reducible tensors of $SU(N)$ then
there remains only the further step of decomposing the reducible
 tensors into their irreducible parts.

\section{Basic Theorem to evaluate an {SO(2N)} vertex}

 A result
essential to our analysis is the Basic Theorem\cite{ns1} which
states that an
 $SO(2N)$ vertex ${\Gamma}_{\mu_{1}}{\Gamma}_{\mu_{2}}{\Gamma}_{\mu_3}...{\Gamma}_{\mu_r}{\Phi}_{\mu_1\mu_2\mu_3
...{{\mu_r}}}$ can be expanded in the following fashion
\begin{eqnarray}
{\Gamma}_{\mu_{1}}{\Gamma}_{\mu_{2}}{\Gamma}_{\mu_3}...{\Gamma}_{\mu_r}{\Phi}_{\mu_1\mu_2\mu_3
...{{\mu_r}}}=
b_{i_{1}}^{\dagger}b_{i_2}^{\dagger}b_{i_3}^{\dagger}...b_{i_r}^{\dagger}{\Phi}_{c_{i_1}c_{i_2}c_{i_3}...c_{i_r}}\nonumber\\
+(b_{i_1}
b_{i_2}^{\dagger}b_{i_3}^{\dagger}...b_{i_r}^{\dagger}{\Phi}_{\overline
c_{i_1}c_{i_2}c_{i_3}...c_{i_r}}+ \textnormal{perms})
 +(b_{i_1}
b_{i_2}b_{i_3}^{\dagger}...b_{i_r}^{\dagger}{\Phi}_{\overline
c_{i_1}\overline
c_{i_2}c_{i_3}...c_{i_r}}+~\textnormal{perms})\nonumber\\
+...+(b_{i_1}b_{i_2}b_{i_3}...b_{i_{r-1}}b_{i_r}^{\dagger}{\Phi}_{\overline
c_{i_1}\overline
c_{i_2} \overline c_{i_3}...\overline c_{i_{r-1}}c_{i_r}}+~\textnormal{perms})\nonumber\\
+b_{i_1}b_{i_2}b_{i_3}...b_{i_r}{\Phi}_{\overline c_{i_1}\overline
c_{i_2}\overline c_{i_3}...\overline c_{i_r}}\nonumber\\
\end{eqnarray}
Eq.(4.7) is
 the basic result we need in the analysis of the SO(2N) invariant
couplings. It is found convenient to arrange the right hand side
of Eq.(4.7)  in a normal ordered form by which we mean that all
the $b$'s are either to the right or to the left and all the
$b^{\dagger}$'s are either to the left or to the right  using
strictly the anti-commutation relations on the $b$'s and
$b^{\dagger}$'s of Eq.(2.124). When a pair of $b^{\dagger}$ and b
have a summed index such as $b_n^{\dagger}b_n$ we will move them
together either to the left or to the right.
 After normal ordering one decomposes $\Phi_{c_ic_j\bar c_k..c_n}$
into its $SU(N)$ irreducible components. The final step consists
of carrying out
 the process of removing all the $b$ and $b^{\dagger}$ using the
anti-commutation relation (2.124) and the condition $b_i|0>=0$
(Eq.(2.157)), thus allowing one  to compute the couplings in the
SU(N) invariant decomposition.

\chapter{Complete Cubic Couplings of Bilinear Spinors and 10, 120  and 126 plet in
{SO(10)} Unification}\label{Dark_Matter}

We illustrate  the  technique developed in chapter \ref{eval
basic} by giving a complete  analysis of the trilinear couplings
\cite{ns1} in the superpotential with the 16 plet of matter and
${\overline{16}}$ plet of Higgs. Since $16\times 16= 10_s+
120_{as}+ 126_{s}$, we compute explicitly in terms of $SU(5)$
fields the couplings $16-16-10_s$, $16-16-120_{as}$ and
$16-16-\overline{126}_s$. Further, we discuss that the coupling of
the $\overline{126}$ with matter leads to extra zeros in the Higgs
triplet sector and the existence of such zeros can enhance the
proton decay lifetime.

\section{The {\bf{10}} plet, {\bf{120}} plet and the ${\bf{\overline{126}}}$ plet  couplings in the superpotential}

The procedure  of chapter \ref{eval basic} makes it
straightforward to analyze the
 $SO(2N)$ invariant couplings involving large tensor representations.
 As an example of our technique we give  a complete
determination of the superpotential at the trilinear level
involving two spinor representations, which consists of the
couplings ${ 16\times 16\times 10}$, ${ 16\times 16\times 120}$
and ${ 16\times 16\times \overline {126}}$. We also compute the
same interactions with 16 plet of matter replaced by
${\overline{16}}$ plet of Higgs.\\
\\
\textbf{(a)}~~\textsc{$16\times16\times10$
 coupling}\\
\\
We begin by computing
 the $ {16\times 16\times 10}$ coupling\cite{ns1} which is given by
 \begin{equation}
      {\mathsf W}_{(++)}^{(10)}=
{f}^{^{(10)}}_{\acute{a}\acute{b}}<\widehat{\Psi}^{*}_{(+)\acute{a}}|B\Gamma_{\mu}
|\widehat{\Psi}_{(+)\acute{b}}>\Phi_{\mu}\nonumber\\
\end{equation}
 where ${f}^{^{(10)}}_{\acute{a}\acute{b}}$ are the Yukawa coupling constants and semi-spinors $\Psi_{(\pm)}$ with a
$~\widehat{  }~$ stands for chiral superfields.
 Following the procedure of chapter~\ref{eval basic} we decompose the vertex
 so that
 \begin{equation}
\Gamma_{\mu}\Phi_{\mu}= b_i\Phi_{\bar c_i}+b_i^{\dagger}
\Phi_{c_i}
\end{equation}
In Eq.(5.2) the tensors are already in their irreducible form and
one can identify
\begin{equation}
\Phi_{c_i}={\mathsf h}^i;~~~~\Phi_{\bar c_i}={\mathsf h}_i
\end{equation}
with the 5 plet and $\bar 5$ plet of Higgs respectively. To
normalize the tensors we define
\begin{equation}
{\mathsf h}_{i}=\sqrt 2{\mathsf H}_i;~~~~ {\mathsf h}^i=\sqrt
2{\mathsf H}^i
\end{equation}
 so that the kinetic energy
$-\partial_{A}\Phi_{\mu}\partial^{A}\Phi_{\mu}^{\dagger}$ of the
tensor $\Phi_{\mu}$ takes the form
\begin{equation}
\mathsf{L}_{kin}^{10-Higgs}=-\partial_{A}{\mathsf
H}_{i}\partial^{A} {\mathsf H}_{i}^{\dagger}
 -\partial_{A}{\mathsf H}^i\partial^{A} {\mathsf H}^{i\dagger}
\end{equation}
where $A(=0,1,2,3)$ is the Dirac index.

 Using the above  we compute the $16-16-10$ coupling and find
 \begin{eqnarray}
  {\mathsf W}_{(++)}^{(10)}=i2\sqrt 2 {f}^{^{(10)(+)}}_{\acute{a}\acute{b}}\left[\widehat {\bf M}_{\acute{a}}^{ij\bf{T}}
  \widehat {\bf M}_{\acute{b}i}{\mathsf H}_{j}
  +\left(-\widehat {\bf M}_{\acute{a}}^{\bf{T}}\widehat {\bf M}_{\acute{b}m} + \frac{1}{8}\epsilon_{ijklm}
  \widehat {\bf M}_{\acute{a}}^{ij\bf{T}}\widehat {\bf M}_{\acute{b}}^{kl}\right)
  {\mathsf H}^{m}\right]\nonumber\\
\end{eqnarray}
where ${f}^{^{(.)(\pm)}}_{\acute{a}\acute{b}}$ are defined by
\begin{equation}
{f}^{^{(.)(\pm)}}_{\acute{a}\acute{b}}=\frac{1}{2}(
{f}^{^{(.)}}_{\acute{a}\acute{b}}\pm{f}^{^{(.)}}_{\acute{b}\acute{a}})
\end{equation}
In the equation above, for example  $\widehat {\bf
M}_{\acute{a}}^{\bf{T}}$ represents the transpose of the chiral
superfield, $\widehat {\bf M}_{\acute{a}}$ etc..
 Note that ${f}^{^{(.)(\pm)}}_{\acute{a}\acute{b}}$ are
symmetric (antisymmetric) under the interchange of generation
indices $\acute{a}$ and $\acute{b}$. As expected the 16-16-10
couplings given by Eq.(5.6) are correctly symmetric in the
generation indices. We note that the
  couplings have the  $SU(5)$ invariant structure consisting of
  $1_M-\bar 5_M-5_H$,
  $10_M-\bar 5_M-\bar 5_H$ and $10_M -10_M-5_H$.
  \newpage
 \noindent \textbf{(b)}~~\textsc{$16\times16\times120$
 coupling}\\
\\
 Next, we discuss the 16-16-120 coupling\cite{ns1} defined  by
 \begin{equation}
  {\mathsf W}^{(120)}_{(++)}=
\frac{1}{3!}{f}^{^{(120)}}_{\acute{a}\acute{b}}<\widehat{\Psi}^{*}_{(+)\acute{a}}|B
\Gamma_{[\mu}\Gamma_{\nu}\Gamma_{\rho]}
|\widehat{\Psi}_{(+)\acute{b}}>\Phi_{\mu\nu\rho}\nonumber\\
\end{equation}
where
\begin{equation}
\Gamma_{[\mu}\Gamma_{\nu}\Gamma_{\rho]}=\frac{1}{3!}\sum_P(-1)^{\delta_P}
\Gamma_{\mu_{P(1)}}\Gamma_{\nu_{P(2)}}\Gamma_{\rho_{P(3)}}
\end{equation}
with $\sum_P$ denoting the sum over all permutations and
$\delta_P$ takes on the value $0$ and $1$ for even and odd
permutations respectively.

 We expand the vertex using Eq.(4.7) and find
  \begin{eqnarray}
\Gamma_{\mu}\Gamma_{\nu}\Gamma_{\lambda}\Phi_{\mu\nu\lambda} =
b_ib_jb_k \Phi_{\bar c_i\bar c_j\bar c_k} +
b_i^{\dagger}b_j^{\dagger}b_k^{\dagger} \Phi_{c_i c_j c_k} + 3
(b_i^{\dagger}b_jb_k \Phi_{c_i\bar c_j \bar c_k}+
 b_i^{\dagger}b_j^{\dagger}b_k \Phi_{c_i c_j \bar c_k}) \nonumber\\
+3( b_i \Phi_{\bar c_n c_n \bar c_i}+ b_i^{\dagger} \Phi_{\bar c_n
c_n c_i} )\nonumber\\
\end{eqnarray}
The 120 plet of $SO(10)$ has the $SU(5)$ decomposition $120=5+\bar
5+10+\bar{10}+45+\overline{45}$. The tensors in Eq.(5.10) can be
decomposed in terms of the irreducible $SU(5)$ tensors as follows:
\begin{eqnarray}
\Phi_{c_ic_j\bar c_k}={\mathsf
h}^{ij}_k+\frac{1}{4}\left(\delta^i_k {\mathsf h}^j- \delta^j_k
{\mathsf h}^i\right),~~~ \Phi_{c_i\bar c_j\bar c_k}={\mathsf
h}^{i}_{jk}+\frac{1}{4}\left(\delta^i_j {\mathsf h}_k-
\delta^i_k {\mathsf h}_j\right)\nonumber\\
\Phi_{c_ic_jc_k}=\epsilon^{ijklm}{\mathsf h}_{lm},~~~ \Phi_{\bar
c_i\bar c_j\bar c_k}=\epsilon_{ijklm}{\mathsf h}^{lm},
~~~\Phi_{\bar c_nc_n c_i}={\mathsf h}^i, \Phi_{\bar c_nc_n \bar
c_i}={\mathsf h}_i
\end{eqnarray}
 where ${\mathsf h}_{i}$, ${\mathsf h}^{i}$, ${\mathsf h}_{ij}$, ${\mathsf h}^{ij}$, ${\mathsf h}^{ij}_k$, ${\mathsf h}^i_{jk}$
 are the $\bar{5}$, 5, $\overline{10}$, 10, 45 and $\overline{45}$ plet representations of $SU(5)$.
To normalize them we make the following redefinition of fields
 \begin{eqnarray}
{\mathsf h}^{i}=\frac{4}{\sqrt 3} {\mathsf H}^{i},~~~ {\mathsf
h}^{ij}=\frac{1}{\sqrt 3} {\mathsf H}^{ij},~~~
{\mathsf h}^{ij}_k=\frac{2}{\sqrt 3} {\mathsf H}^{ij}_k\nonumber\\
{\mathsf h}_{i}=\frac{4}{\sqrt 3} {\mathsf H}_{i},~~~ {\mathsf
h}_{ij}=\frac{1}{\sqrt 3} {\mathsf H}_{ij},~~~ {\mathsf
h}^{i}_{jk}=\frac{2}{\sqrt 3} {\mathsf H}^{i}_{jk}
\end{eqnarray}
In terms of the redefined fields the kinetic energy term for the
120 multiplet which is given by
$-\partial_{A}\Phi_{\mu\nu\lambda}$
$\partial^{A}\Phi_{\mu\nu\lambda}^{\dagger}$ takes on the form
\begin{eqnarray}
{\mathsf L}_{kin}^{120-Higgs}=-\frac{1}{2!}\partial_{A}{\mathsf H}
^{ij}
\partial^{A} {\mathsf H}^{ij\dagger}
-\frac{1}{2!}\partial_{A} {\mathsf H}_{ij}
\partial^{A} {\mathsf H}_{ij}^{\dagger}
-\frac{1}{2!}\partial_{A} {\mathsf H}^{ij}_k
\partial^{A}{\mathsf H}^{ij\dagger}_k  \nonumber\\
 -\frac{1}{2!}\partial_{A} {\mathsf H}_{jk}^i
\partial^{A}{\mathsf H}_{jk}^{i\dagger}
-\partial_{A} {\mathsf H}^{i}
\partial^{A} {\mathsf H}^{i\dagger}
-\partial_{A} {\mathsf H}_{i}
\partial^{A} {\mathsf H}_{i}^{\dagger}
\end{eqnarray}\\
 A straightforward computation of Eq.(5.8) using Eqs.(5.10)\footnote{The symmetrical arrangement in the first
brace of Eq.(5.10) is necessary for achieving an automatic
anti-symmetry in the generation indices for the $SU(5)$
$10_M-10_M-45_H$ coupling.}, (5.11) and (5.12) gives
 \begin{eqnarray}
 {\mathsf W}_{(++)}^{(120)}=
=i\frac{2}{\sqrt 3}{f}^{^{(120)(-)}}_{\acute{a}\acute{b}}\left[
-2\widehat{\bf M}_{\acute{a}}^{\bf T}\widehat{\bf
M}_{\acute{b}i}{\mathsf H}^{i}-\widehat{\bf M}_{\acute{a}}^{ij\bf
T}\widehat{\bf M}_{\acute{b}i}{\mathsf
H}_{j}\right.\nonumber\\
\left.+\widehat{\bf M}_{\acute{a}i}^{\bf T}\widehat{\bf
M}_{\acute{b}j}{\mathsf H}^{ij} +\widehat{\bf
M}_{\acute{a}}^{ij\bf T}\widehat{\bf M}_{\acute{b}}{\mathsf
H}_{ij}\right.\nonumber\\
\left.-\frac{1}{4}\epsilon_{ijklm}\widehat{\bf
M}_{\acute{a}}^{ij\bf T}\widehat{\bf
M}_{\acute{b}}^{mn}{\mathsf H}^{kl}_{n}\right.\nonumber\\
\left.+\widehat{\bf M}_{\acute{a}i}^{\bf T}\widehat{\bf
M}_{\acute{b}}^{jk}{\mathsf H}^{i}_{jk}\right]
\end{eqnarray}
The front factor ${f}^{^{(120)(-)}}_{\acute{a}\acute{b}}$ in
Eq.(5.14) exhibits
 correctly the  anti-symmetry in the generation indices.
Further, the couplings have the $1_M-\bar 5_M-5_H$,
$1_M-10_M-\bar{10}_H$, $\bar 5_M-\bar 5_M-10_H$, $10_M-\bar
5_M-\bar 5_H$, $\bar 5_M-10_M-\bar{45}_H$ and $10_M-10_M-45_H$
$SU(5)$ invariant structures.\\
\\
\textbf{(c)}~~\textsc{$16\times16\times\overline{126}$
coupling}\\

  We now turn to the most difficult of the three cases,
  i.e., the ${ 16-16-\overline {126}}$ coupling\cite{ns1} which is given by
\begin{equation}
{\mathsf
W}_{(++)}^{(\overline{126})}=\frac{1}{5!}{f}^{^{(\overline{126})}}_{\acute{a}\acute{b}}<\widehat{\Psi}^{*}_{(+)\acute{a}}|B\Gamma_{[\mu}\Gamma_{\nu}\Gamma_{\rho}
\Gamma_{\lambda}\Gamma_{\sigma]}
|\widehat{\Psi}_{(+)\acute{b}}>\overline{\Phi}_{\mu\nu\rho\lambda\sigma}
\end{equation}
 For the reduction of the $SO(10)$ vertex it is more
 convenient initially to work with the full 252 dimensional tensor,
 $\Xi_{\mu\nu\lambda\rho\sigma}$ which can be be decomposed as
$\Xi_{\mu\nu\lambda\rho\sigma}$=$\overline
{\Phi}_{\mu\nu\lambda\rho\sigma}$+
$\Phi_{\mu\nu\lambda\rho\sigma}$, where
\begin{equation}
\left(\matrix{\overline{\Phi}_{\mu\nu\lambda\rho\sigma}\cr
 \Phi_{\mu\nu\lambda\rho\sigma}}\right)=
   \frac{1}{2}\left(\delta_{\mu\alpha}
\delta_{\nu\beta}\delta_{\rho\gamma}\delta_{\lambda\delta}\delta_{\sigma\theta}
\pm
\frac{i}{5!}\epsilon_{\mu\nu\rho\lambda\sigma\alpha\beta\gamma\delta\theta}\right)
\Xi_{\alpha\beta\gamma\delta\theta}
\end{equation}
and where the $\overline{\Phi}_{\mu\nu\lambda\rho\sigma}$ is the
${\overline{126}}$ plet
 and $\Phi_{\mu\nu\lambda\rho\sigma}$ is the ${126}$ plet representation.
 These representations break into
the $SU(5)$ irreducible parts as ${\overline{126}=1+5+
\overline{10}+ 15+\overline{45}+50}$ and ${ 126=1+\bar 5+10+
\overline{15}+ 45 +\overline{50}}$. We begin by expanding
$\Gamma_{\mu}\Gamma_{\nu}\Gamma_{\lambda}
\Gamma_{\rho}\Gamma_{\sigma}\Xi_{\mu\nu\lambda\rho\sigma}$
 using Eq.(4.7)and  following steps similar to the previous case using
  normal ordering:
  \begin{eqnarray}
\Gamma_{\mu}\Gamma_{\nu}\Gamma_{\lambda}
\Gamma_{\rho}\Gamma_{\sigma}\Xi_{\mu\nu\lambda\rho\sigma}=
b_i^{\dagger}b_j^{\dagger}b_k^{\dagger}b_l^{\dagger}b_m^{\dagger}\Xi_{c_ic_jc_kc_lc_m}
+b_ib_jb_kb_lb_m\Xi_{\bar c_i\bar c_j\bar c_k\bar c_l\bar c_m}\nonumber\\
+5b_i^{\dagger}b_j^{\dagger}b_k^{\dagger}b_l^{\dagger}b_m\Xi_{c_ic_jc_kc_l\bar
c_m} +5b_i^{\dagger}b_jb_kb_lb_m\Xi_{ c_i\bar c_j\bar c_k\bar
c_l\bar c_m}\nonumber\\
+10b_i^{\dagger}b_j^{\dagger}b_k^{\dagger}b_lb_m\Xi_{c_ic_jc_k\bar
c_l\bar c_m}
+10b_i^{\dagger}b_j^{\dagger}b_kb_lb_m\Xi_{c_ic_j\bar c_k\bar c_l\bar c_m}\nonumber\\
+10b_i^{\dagger}b_j^{\dagger}b_k^{\dagger}\Xi_{c_ic_jc_k\bar
c_nc_n} +10b_ib_jb_k\Xi_{\bar c_i\bar c_j\bar c_k\bar c_nc_n}\nonumber\\
+30b_i^{\dagger}b_j^{\dagger}b_k\Xi_{c_ic_j\bar c_k\bar
c_nc_n} +30b_i^{\dagger}b_jb_k\Xi_{c_i\bar c_j\bar c_k\bar c_nc_n}\nonumber\\
+15b_i^{\dagger}\Xi_{c_i\bar c_n c_n\bar c_pc_p} +15b_i\Xi_{\bar
c_i\bar c_n c_n\bar c_pc_p}
 \end{eqnarray}
The $SO(10)$ reducible tensors appearing in the equation above can
be expressed in terms of irreducible $SU(5)$ tensors as follows:
\begin{eqnarray}
 \Xi_{c_ic_jc_k\bar c_nc_n}={\mathsf
 h}^{ijk};~~~~~~\Xi_{\bar c_i\bar c_j\bar c_kc_n\bar c_n}={\mathsf
 h}_{ijk}\nonumber\\
\Xi_{c_i\bar c_nc_n\bar c_pc_p}={\mathsf h}^{i};~~~~~~\Xi_{\bar
c_i\bar c_nc_n\bar c_pc_p}={\mathsf h}_{i}\nonumber\\
\Xi_{c_i\bar c_j\bar c_k\bar c_nc_n}={\mathsf
 h}^{i}_{jk}+\frac{1}{4}\left(\delta^i_k{\mathsf h}_j-\delta^i_j{\mathsf h}_k\right)\nonumber\\
\Xi_{\bar c_i c_j c_k c_n\bar c_n}={\mathsf
 h}_{i}^{jk}+\frac{1}{4}\left(\delta_i^k{\mathsf h}^j-\delta_i^j{\mathsf h}^k\right)\nonumber\\
\Xi_{c_ic_jc_kc_l\bar c_m}={\mathsf h}^{ijkl}_m+\frac{1}{2}
\left(\delta^i_m{\mathsf h}^{jkl}-\delta^j_m{\mathsf h}^{ikl}+
\delta^k_m{\mathsf h}^{ijl}-\delta^l_m{\mathsf h}^{ijk} \right)\nonumber\\
\Xi_{c_ic_jc_k\bar c_l\bar c_m}={\mathsf h}^{ijk}_{lm}+\frac{1}{2}
\left(\delta^i_l{\mathsf h}^{jk}_m -\delta^j_l{\mathsf h}^{ik}_m
+\delta^k_l{\mathsf h}^{ij}_m -\delta^i_m{\mathsf h}^{jk}_l
+\delta^j_m{\mathsf h}^{ik}_l -\delta^k_m{\mathsf h}^{ij}_l\right)\nonumber\\
+\frac{1}{12}\left(\delta^i_l\delta^j_m{\mathsf
h}^k-\delta^j_l\delta^i_m{\mathsf h}^k
-\delta^i_l\delta^k_m{\mathsf h}^j+\delta^k_l\delta^i_m{\mathsf
h}^j
+\delta^j_l\delta^k_m{\mathsf h}^i-\delta^k_l\delta^j_m{\mathsf h}^i\right)\nonumber\\
\Xi_{c_ic_j\bar c_k\bar c_l\bar c_m}={\mathsf
h}^{ij}_{klm}+\frac{1}{2} \left(\delta^i_k{\mathsf h}^{j}_{lm}
-\delta^i_l{\mathsf h}^{j}_{km} +\delta^i_m{\mathsf h}^{j}_{kl}
-\delta^j_k{\mathsf h}^{i}_{lm}
+\delta^j_l{\mathsf h}^{i}_{km} -\delta^j_m{\mathsf h}^{i}_{kl}\right)\nonumber\\
+\frac{1}{12}\left(\delta^i_k\delta^j_l{\mathsf
h}_m-\delta^i_k\delta^j_m{\mathsf h}_l
-\delta^i_l\delta^j_k{\mathsf h}_m+\delta^i_l\delta^j_m{\mathsf
h}_l
+\delta^i_m\delta^j_k{\mathsf h}_l-\delta^i_m\delta^j_l{\mathsf h}_k\right)\nonumber\\
\Xi_{c_i\bar c_j\bar c_k\bar c_l\bar c_m}={\mathsf h}^i_{jklm}+
\frac{1}{2} \left(\delta^i_j{\mathsf h}_{klm}-\delta^i_k{\mathsf
h}_{jlm}+
\delta^i_l{\mathsf h}_{jkm}-\delta^i_m{\mathsf h}_{jkl} \right)\nonumber\\
\Xi_{c_ic_jc_kc_lc_m}=\epsilon^{ijklm} {\mathsf h},~~~~~~
\Xi_{\bar c_i\bar c_j\bar c_k\bar c_l\bar c_m}= \epsilon_{ijklm}
\bar {\mathsf h}\nonumber\\
\end{eqnarray}

 The fields that appear above are not yet properly normalized.
 To normalize the fields we carry out a field redefinition
 so that
\begin{eqnarray}
{\mathsf h}=\frac{2}{\sqrt {15}}{\mathsf H},~~~\overline{{\mathsf
h}}=\frac{2}{\sqrt {15}}\overline{{\mathsf H}},~~~ {\mathsf
h}^i=4\sqrt{\frac{ 2}{5}}{\mathsf H}^i,~~~{\mathsf
h}_i=4\sqrt{\frac{ 2}{5}}{\mathsf H}_i\nonumber\\
 {\mathsf
h}^{ijk}=\sqrt{\frac{ 2}{15}} \epsilon^{ijklm}{\mathsf
H}_{lm},~~~{\mathsf h}_{ijk}=\sqrt{\frac{ 2}{15}}
\epsilon_{ijklm}{\mathsf H}^{lm}\nonumber\\
 {\mathsf h}^i_{jklm}=
\sqrt{\frac{2}{15}}\epsilon_{jklmn}{\mathsf
H}^{ni}_{(S)},~~~{\mathsf h}_i^{jklm}=
\sqrt{\frac{2}{15}}\epsilon^{jklmn}{\mathsf
H}_{ni}^{(S)}\nonumber\\
 {\mathsf h}^i_{jk}=2\sqrt{\frac{2}{15}} {\mathsf
 H}^i_{jk},~~~{\mathsf h}_i^{jk}=2\sqrt{\frac{2}{15}} {\mathsf
 H}_i^{jk}\nonumber\\
  {\mathsf h}^{ijk}_{lm}=\frac{2}{\sqrt {15}} {\mathsf
  H}^{ijk}_{lm},~~~{\mathsf h}_{ijk}^{lm}=\frac{2}{\sqrt {15}} {\mathsf H}_{ijk}^{lm}
 \end{eqnarray}
The  kinetic energy for the $252$ plet field
$-\partial_{A}\Xi_{\mu\nu\lambda\rho\sigma}
\partial^{A}\Xi_{\mu\nu\lambda\rho\sigma}^{\dagger}$
in terms of the normalized fields is then given by
 \begin{eqnarray}
L_{kin}^{252-Higgs}=-\partial_{A} \overline{{\mathsf
H}}\partial^{A} \overline{{\mathsf H}}^{\dagger}-\partial_{A}
{\mathsf H}\partial^{A} {\mathsf H}^{\dagger}
 - \partial_{A} {\mathsf H}_{i}\partial^{A} {\mathsf
 H}^{\dagger}_i
 - \partial_{A} {\mathsf H}^{i}\partial^{A} {\mathsf
 H}^{i\dagger}\nonumber\\
- \frac{1}{2!}\partial_{A} {\mathsf H}^{ij}
\partial^{A} {\mathsf H}^{ij\dagger}
- \frac{1}{2!}\partial_{A} {\mathsf H}_{ij}
\partial^{A} {\mathsf H}_{ij}^{\dagger}
- \frac{1}{2!}\partial_{A} {\mathsf H}_{ij}^{(S)}
\partial^{A} {\mathsf H}^{(S)\dagger}_{ij}\nonumber\\
- \frac{1}{2!}\partial_{A} {\mathsf H}^{ij}_{(S)}
\partial^{A} {\mathsf H}^{ij\dagger}_{(S)}
-\frac{1}{2!}\partial_{A} {\mathsf H}^{jk}_i
\partial^{A} {\mathsf H}_{i}^{jk\dagger}
-\frac{1}{2!}\partial_{A} {\mathsf H}_{jk}^i
\partial^{A} {\mathsf H}_{jk}^{i\dagger}\nonumber\\
-\frac{1}{3!2!}\partial_{A} {\mathsf H}_{ijk}^{lm}
\partial^{A} {\mathsf H}^{lm\dagger}_{ijk}
-\frac{1}{3!2!}\partial_{A} {\mathsf H}^{ijk}_{lm}
\partial^{A} {\mathsf H}^{ijk\dagger}_{lm}
\end{eqnarray}
where $\overline{{\mathsf H}}({\mathsf H}), {\mathsf H}_{i},
{\mathsf H}^{i}, {\mathsf H}^{ij}, {\mathsf H}_{ij}, {\mathsf
H}_{ij}^{(S)}, {\mathsf H}^{ij}_{(S)}, {\mathsf H}^{jk}_i,
{\mathsf H}_{jk}^i, {\mathsf H}_{ijk}^{lm}, {\mathsf
H}^{ijk}_{lm}$ are the $1, \bar{5}, 5, 10, \overline{10},\\
\overline{15}, 15, 45, \overline{45}, \overline{50}, 50$ plet
representations of $SU(5)$.

Finally, Eq.(5.15) gives
\begin{eqnarray}
 {\mathsf
W}_{(++)}^{(\overline{126})}=
=i\sqrt{\frac{2}{15}}{f}^{^{(\overline{126})(+)}}_{\acute{a}\acute{b}}\left[
-\sqrt{2}\widehat{\bf M}_{\acute{a}}^{\bf T}\widehat{\bf
M}_{\acute{b}}{\mathsf H}+\widehat{\bf M}_{\acute{a}}^{\bf
T}\widehat{\bf M}_{\acute{b}}^{ij}{\mathsf H}_{ij}\right.\nonumber\\
\left.-\widehat{\bf M}_{\acute{a}i}^{\bf T}\widehat{\bf
M}_{\acute{b}j}{\mathsf H}_{(S)}^{ij}+\widehat{\bf
M}_{\acute{a}}^{ij\bf T}\widehat{\bf
M}_{\acute{b}k}{\mathsf H}_{ij}^{k}\right.\nonumber\\
\left.-\frac{1}{12\sqrt{2}}\epsilon_{ijklm}\widehat{\bf
M}_{\acute{a}}^{ij\bf
T}\widehat{\bf M}_{\acute{b}}^{rs}{\mathsf H}^{klm}_{rs}\right.\nonumber\\
\left.-\sqrt{3}\left(\widehat{\bf M}_{\acute{a}}^{\bf
T}\widehat{\bf M}_{\acute{b}m}
+\frac{1}{24}\epsilon_{ijklm}\widehat{\bf M}_{\acute{a}}^{ij\bf
T}\widehat{\bf M}_{\acute{b}}^{kl}\right){\mathsf H}^m\right]
\end{eqnarray}
As in the ${10}$ plet tensor case the couplings are symmetric
under the interchange of generation indices. Further, the
${16-16-\overline{126}}$ coupling has the $SU(5)$ structure
consisting of ${ 1_M-1_M-1_H}$, ${ 1_M-\bar 5_M-5_H}$, ${
1_M-10_M-\overline{10}_H}$, ${ 10_M-10_M-5_H}$, ${ \bar 5_M-\bar
5_M-15_H}$, ${ 10_M-\bar 5_M-\overline{45}_H}$ and
$10_M-10_M-50_M$.

\section{Large representations, textures and proton lifetime\cite{ns1}}
Proton decay is an important signal for grand unification and
detailed analyses for the proton lifetime exist in $SU(5)$
models\cite{nac,bajc}  and in $SO(10)$ models\cite{lucas,bpw}.
 In this section we discuss the possibility that couplings in the
 superpotential that involve large representations can drastically
 change the Higgs triplet textures and affect proton decay in
  a very significant manner.
These results are of significance in view of the recent data from
SuperKamiokande which has significantly improved the limit on the
proton decay mode $p\rightarrow \nu+K^+$. Thus the most recent
limit from SuperKamiokande gives $\tau/B(p\rightarrow \bar
\nu+K^+)> 1.9 \times 10^{33}$ yr\cite{totsuka}.
 At the same time
there is a new lattice gauge evaluation of the three quark matrix
elements $\alpha$ and $\beta$ of the nucleon wave
function\cite{aoki} (where $\alpha$ and $\beta$ are defined
by $\epsilon_{abc}<0|\epsilon_{\alpha\beta}
d^{\alpha}_{aR}u^{\beta}_{bR}u^{\gamma}_{cL}|p>$=$\alpha
u^{\gamma}_L$ and $\epsilon_{abc}<0|\epsilon_{\alpha\beta}
d^{\alpha}_{aL}u^{\beta}_{bL}u^{\gamma}_{cL}|p>$=$\beta
u^{\gamma}_L$).
 The more recent evaluation of Ref.\cite{aoki} gives
  $\alpha =-0.015(1) GeV^3$ and
$\beta =0.014(1) GeV^3$ which is a factor of about two and a half
times larger than the previous evaluations.
  The new experimental limit on the proton decay lifetime\cite{totsuka}
  combined with the new lattice gauge
  evaluations have begun to constrain
the SUSY GUT models prompting some reanalyses\cite{dmr,altarelli}.
In this context the enhancement of the proton lifetime by textures
is of interest. To make this idea more concrete we define textures
in the low energy theory in the quark lepton sector of the theory
just below the GUT scale as follows \beqn
W_Y= -M_HH_{1t}H_{2t}+ (l A^E e^c h_1 + q A^D d^c h_1 +h_2u^cA^Uq)\nonumber\\
+ (qB^El H_{1t}+ \epsilon_{abc}H_{1ta}d^c_bB^Du^c_c +
H_{2ta}u^c_aB^Ue^c +\epsilon_{abc}H_{2t}^au_bC^Ud_c) \eeqn where
$A^E$, $A^D$ and $A^U$  are the textures in the Higgs doublet
sector and $B^E, B^D$, $B^U$ and $C^U$ are the textures in the
Higgs triplet sector. A classification of the possible textures in
the Higgs doublet sector is given in Refs.\cite{gj,ross}. For our
purpose here we adopt the textures in the Higgs doublet sector in
the form \beq A^E= \left(\matrix{0 & f & 0 \cr f & -3e & 0\cr 0 &
0  & d}\right), A^D= \left(\matrix{0 & fe^{i\phi} & 0 \cr
fe^{-i\phi} & e & 0\cr 0 & 0  & d}\right), A^U= \left(\matrix{0 &
c & 0 \cr c & 0 & b\cr 0 & b  & a}\right)~~~~~~\eeq As is well
known\cite{gj} the appearance of -3 vs 1 in the 22 element of
$A^E$ vs $A^D$ is one of the important ingredients in achieving
the desired quark and lepton mass hierarchy and may provide an
insight into the nature of the fundamental coupling. Now the
textures in the Higgs triplet sector are generally different than
those in the Higgs doublet sector and they are sensitively
dependent on the nature of GUT and  Planck scale physics\cite{pn}.
The current experimental constraints on the proton lifetime leads
us to conjecture that the Higgs triplet sector contains additional
texture zeros over and above the texture zeros that appear in the
Higgs doublet sector and the coupling of the $\overline{126}$
tensor field plays an important role in this regard. In the
following we shall assume CP invariance and set the phases to zero
in Eq.(5.23). Since in this case the textures of Eq.(5.23) are
symmetric it is only the 10 plet and the $\overline{126}$ plet of
Higgs couplings that enter in the analysis and the 120 plet
couplings do not. To exhibit the above phenomenon more concretely
we consider on phenomenological grounds
 a superpotential in the Yukawa sector of the following type
\begin{eqnarray}
W_Y= f_{ij}^{(0)}(Y,M) \psi_i\psi_j \phi^{(0)}_{\bar{126}}
+f^{(d)}_{12}(Y, M)\psi_1\psi_2\phi^{(1)}_{10}
+f^{(u)}_{12}(Y, M)\psi_1\psi_2\phi^{(2)}_{10}\nonumber\\
+f^{(d)}_{22}(Y, M)\psi_2\psi_2\phi^{(1)}_{\bar{126}}
  + f^{(u)}_{23}(Y, M)\psi_2\psi_3\phi^{(2)}_{\bar{126}}
+f^{(d)}_{33}(Y, M)\psi_3\psi_3\phi^{(1)}_{10}\nonumber\\
+f^{(u)}_{33}(Y, M)\psi_3\psi_3\phi^{(2)}_{\bar{126}}~~~~~~
\end{eqnarray}
where $M$ is a superheavy scale and $f^{(d)}_{ij}(Y, M)$ and
$f^{(u)}_{ij}(Y, M)$ are functions of a set of scalar fields Y
which develop VEVs and the appropriate factors of $<Y^n\phi>/M^n$
generate the right sizes. The model of  Eq.(5.24) is of the
generic type discussed in refs.\cite{ hrr1,hrr2,cm}.
 We do not go into detail here regarding
the symmetry breaking mechanism, the doublet-triplet splitting and
the mass generation for the pseudo-goldstone bosons. All of these
topics have been dealt with at some length in the previous
literature\cite{doubtrip,hrr1,hrr2,babu,abmrs}. Further, while
models with large representations are not asymptotically free and
lead rapidly to non-perturbative physics above the unification
scale, the effective theories below the unification scale gotten
by integration over the heavy modes are nonetheless perfectly
normal and thus such theories are acceptable unified theories. For
our purpose  here we assume a pattern of VEV formation for the
neutral components of the Higgs so that $
<\phi^{(0)}_{\overline{126}}>$ develops a  VEV along the $SU(5)$
singlet direction (this corresponds to ${\mathsf H}$ in Eq.(5.21)
developing a VEV),
 $ <\phi^{(1)}_{10}>$ develops a VEV in the $\bar 5$ plet of
$SU(5)$ direction (this corresponds to ${\mathsf H}_j$ developing
a VEV in Eq.(5.6)),
 $ <\phi^{(2)}_{10}>$ develops a VEV in the 5 plet  direction
 (this corresponds to ${\mathsf H}^m$  developing a VEV in Eq.(5.6)),
 $<\phi^{(1)}_{\overline{126}}>$ develops a VEV  in the direction of
 $\overline{45}$ plet of Higgs (this corresponds to ${\mathsf H}^k_{ij}$ in Eq.(5.21) developing
 a VEV), and
  $<\phi^{(2)}_{\overline{126}}>$ develops a VEV  in the direction of
 5 plet of Higgs (this correspons to ${\mathsf H}^m$ in Eq.(5.21) developing
 a VEV).  It is the VEV of the 45 plet that leads to -3 and 1
 factors in $A^E$ vs $A^D$.
The superpotential of Eq.(5.24) with the above VEV alignments then
leads automatically to the textures in the Higgs doublet sector of
Eq.(5.23).
 One may now compute the textures in the Higgs triplet sector that
 result from  superpotential of Eq.(5.24). One finds
\beq B^E= \left(\matrix{0 & f & 0 \cr
 f & 0 & 0\cr
0 & 0  & d}\right), B^U= \left(\matrix{0 & c & 0 \cr c & 0 & 0\cr
0 & 0 & 0}\right) \eeq and $B^D=B^E$ and $C^U=B^U$. We note the
existence of the additional zeros in $B^E$ and $B^D$ relative to
$A^E$ and $A^D$ and in $B^U$ and $C^U$ relative to $A^U$. The
existence of the additional zeros in $B^E$, $B^D$,$B^U$ and $C^U$
increases the proton decay lifetime. We discuss  the origin of the
additional zeros.
 It is the coupling of the matter sector with
the $\overline{126}$ of Higgs which contributes a non-vanishing
element in the Higgs doublet sector produces a vanishing
contribution in the lepton and baryon number violating dimension
five operator or equivalently generates a corresponding zero in
the texture in the Higgs triplet sector. The reason for this is
rather straightforward. While one also needs a $126$ plet of Higgs
to cancel the D term generated by the VEV of the $\overline{126}$
of Higgs, the 126 plet of Higgs has no coupling with the ordinary
16 plet of matter. Since the only bilinear with the
$\overline{126}$ in the superpotential is of the form
$126\times\overline{126}$ (i.e., one cannot write a $(126)^2$ term
in the superpotential) one finds that
 no lepton and baryon number violating dimension five operators arise
as a consequence of integrating out the $126$ and $\overline{126}$
of Higgs which effectively corresponds to a texture zero in the
Higgs triplet sector. Of course, the extra zeros in the Higgs
triplet sector could also arise from accidental cancellations.
However, the group theoretic origin is more appealing.

The superpotential of Eq.(5.24) also generates  a Dirac neutrino
mass matrix.  However, a full analysis of neutrino masses requires
a model for the Majorana mass matrix $M_{Maj}$ to generate a
see-saw mechanism\cite{seesaw} so that $M_{\nu LL}=-M_{\nu
LR}^{T}M^{-1}_{Maj}M_{\nu LR}$ where the mass scale associated
with $ M_{Maj}$ is much larger than the mass scale that appears in
$M_{\nu LR}$. $ M_{Maj}$ depends on $f_{ij}^{(0)}$ in Eq.(5.24)
which are in general additional arbitrary parameters\cite{gn}.
[For a review see Ref.\cite{dorsner}]. The appearance of $M_{Maj}$
in the analysis of neutrino masses with additional arbitrary
parameters and a scale much larger than the mass scale that
appears in $M_{\nu LR}$ implies that there is not a rigid
relationship between proton decay and neutrino masses in $SO(10)$.
Nonetheless, it is interesting to investigate the correlation that
exists between these two important phenomena in specific models.
  For further application of these techniques in the context of
  proton decay see Ref.\cite{wie}

 \section{Appendix A}
 In this appendix we give a proof of -3 vs. 1 ratio in the
 textures $A^E$ vs. $A^D$.

 In the $\overline{126}$ couplings one has a $\overline{45}$ coupling with
 matter type
$$\widehat{\bf M}_{\acute{a}}^{ij\bf T}\widehat{\bf
M}_{\acute{b}k}{\mathsf H}_{ij}^{k}$$
where $i, j, k=1,.., 5$ and
$${\mathsf H}_{ij}^{k}=-{\mathsf H}_{ji}^{k};~~~~~\sum_k{\mathsf H}_{ik}^{k}=0$$
Let there be a VEV formation so that $<{\mathsf H}_{5j}^{k}>\ne
0$. Then anti-symmetry and traceless condition allows one to write
$$<{\mathsf H}_{5j}^{k}>= a\left(\frac{1}{4}\delta^k_j-\delta^k_4\delta^4_j\right)$$
The above implies that in the color sector $(\alpha, \beta=1, 2,
3)$ one has
$$<{\mathsf H}_{5\alpha}^{\beta}>= a\frac{1}{4}\delta^{\beta}_{\alpha}$$
and in the charged lepton sector one has
$$<{\mathsf H}_{54}^{4}>=-\frac{3}{4}a$$
This gives the -3 vs. 1 factors that appear in $A^E$ vs. $A^D$ in
Eq. (5.23)

 \section{Appendix B}
 In this appendix we give a proof of textures in the Higgs doublet
 sector.

 Consider the 10 plet and $\overline{126}$ plet couplings and
 assume that the 10 plet develops a VEV in the $SU(5)$ $\bar{5}$
 plet direction. This gives mass to the down quark and to the
 lepton. The hypercharge of the ${\mathsf H}_j$ Higgs is $-1$.
 From Eq. (5.6), we see that the mass growth for the down quark
 and the lepton are the same. Then one has
 $$M_d\sim\bar{f}^{^{(10)(+)}}_{\acute{a}\acute{b}}<10^->$$
$$M_e\sim\bar{f}^{^{(10)(+)}}_{\acute{a}\acute{b}}<10^->$$
From the analysis on the mass growth from the $\overline{126}$ of
Higgs one has
$$M_d\sim\bar{f}^{^{(\overline{126})(+)}}_{\acute{a}\acute{b}}<\overline{126}^->$$
$$M_e\sim-3\bar{f}^{^{(\overline{126})(+)}}_{\acute{a}\acute{b}}<\overline{126}^->$$
Combining the above results we have for the mass growth in the
down quark and lepton sector from the VEV growth of 10 plet and
$\overline{126}$
$$M_d\sim\bar{f}^{^{(10)(+)}}_{\acute{a}\acute{b}}<10^->+
\bar{f}^{^{(\overline{126})(+)}}_{\acute{a}\acute{b}}<\overline{126}^->$$
$$M_e\sim\bar{f}^{^{(10)(+)}}_{\acute{a}\acute{b}}<10^->
-3\bar{f}^{^{(\overline{126})(+)}}_{\acute{a}\acute{b}}<\overline{126}^->$$

Next let us look at the up quark and Dirac neutrino masses. Here
let us begin with a 10 plet of Higgs which develops a VEV in the 5
plet of $SU(5)$. The Higgs that develops a VEV has hypercharge
$+1$. Then from the 10 plet  couplings we find
$$M_u\sim\bar{f}^{^{(10)(+)}}_{\acute{a}\acute{b}}<10^+>$$
$$M_{\nu LR}\sim-\bar{f}^{^{(10)(+)}}_{\acute{a}\acute{b}}<10^+>$$
Next we look at the $\overline{126}$ coupling. Here let us assume
that $\overline{126}$ of Higgs develops a VEV along the direction
of 5 plet of Higgs. Then one has that the up quark and Dirac
neutrino masses are given by Then from the 10 plet  couplings we
find
$$M_u\sim\bar{f}^{^{(\overline{126})(+)}}_{\acute{a}\acute{b}}<\overline{126}^+>$$
$$M_{\nu LR}\sim 3\bar{f}^{^{(\overline{126})(+)}}_{\acute{a}\acute{b}}<\overline{126}^+>$$
Together, the $10^+$ and $<\overline{126}^+>$ VEV's produce the
following results
$$M_u\sim\bar{f}^{^{(10)(+)}}_{\acute{a}\acute{b}}<10^+>+
\bar{f}^{^{(\overline{126})(+)}}_{\acute{a}\acute{b}}<\overline{126}^+>$$
$$M_{\nu LR}\sim-\bar{f}^{^{(10)(+)}}_{\acute{a}\acute{b}}<10^+>+
3\bar{f}^{^{(\overline{126})(+)}}_{\acute{a}\acute{b}}<\overline{126}^+>$$
Now the overall sign in the $M_{\nu LR}$ case is opposite to that in Ref.\cite{cm}.
However, we can let $\nu^c \rightarrow
\nu^c$. In this case one finds the following four relations
$$M_d\sim\bar{f}^{^{(10)(+)}}_{\acute{a}\acute{b}}<10^->+
\bar{f}^{^{(\overline{126})(+)}}_{\acute{a}\acute{b}}<\overline{126}^->$$
$$M_e\sim\bar{f}^{^{(10)(+)}}_{\acute{a}\acute{b}}<10^->
-3\bar{f}^{^{(\overline{126})(+)}}_{\acute{a}\acute{b}}<\overline{126}^->$$

$$M_d\sim\bar{f}^{^{(10)(+)}}_{\acute{a}\acute{b}}<10^->+
\bar{f}^{^{(\overline{126})(+)}}_{\acute{a}\acute{b}}<\overline{126}^->$$
$$M_e\sim\bar{f}^{^{(10)(+)}}_{\acute{a}\acute{b}}<10^->
+3\bar{f}^{^{(\overline{126})(+)}}_{\acute{a}\acute{b}}<\overline{126}^->$$
These relations are the same as in Ref.\cite{cm}.

Next let us look at the superpotential given by Eq. (5.24) and
make the following identifications:
$$\phi_{10}^{(1)}= 10^-;~~~~~\phi_{10}^{(2)}= 10^+$$
$$\phi_{\overline{126}}^{(1)}= \overline{126}^-;~~~~~\phi_{\overline{126}}^{(2)}= \overline{126}^+$$
Then it follows that one has the textures in the Higgs doublet
sector of the type given by Eq.(5.23). This completes the proof.

\section{Appendix C}
In this appendix we give a proof of extra zeros in the Higgs
triplet sector.

We begin by considering Yukawa coupling in the superpotential so
that $126$ and $\overline{126}$ appear. Then the superpotential
reads
$$J_{\mu\nu\lambda\rho\sigma}\overline{\Phi}_{\mu\nu\lambda\rho\sigma}+
K_{\mu\nu\lambda\rho\sigma}\Phi_{\mu\nu\lambda\rho\sigma}$$ where
$J$ and $K$ depend on the matter fields. Now only the
$\overline{126}$ couples with the three generations of matter and
$126$ has no coupling with the three spinor generations. Thus
$$K_{\mu\nu\lambda\rho\sigma}=0$$

Next let us consider a mass term for the $126$ and
$\overline{126}$ of Higgs. We note that
$$126\times 126=(54+1050+2772+4125)_A+(945+8910)_S$$
One finds that there is no singlet in the product. So we cannot
form an $SO(10)$ invariant mass term out of the product of two
$126$'s. The same argument applies for the product of two
$\overline{126}$'s. However, consider the product of $126$ and
$\overline{126}$
$$126\times\overline{126}=1+45+210+770+5940+8910$$

This time we see that there is singlet and we can write a mass
term. Including the mass term the superpotential  with Yukawa
couplings is
$$W=M\Phi_{\mu\nu\lambda\rho\sigma}\overline{\Phi}_{\mu\nu\lambda\rho\sigma}
+J_{\mu\nu\lambda\rho\sigma}\overline{\Phi}{\mu\nu\lambda\rho\sigma}+
K_{\mu\nu\lambda\rho\sigma}\Phi{\mu\nu\lambda\rho\sigma}$$ Let us
now vary the superpotential with respect to
$\Phi_{\mu\nu\lambda\rho\sigma}$ and
$\overline{\Phi}_{\mu\nu\lambda\rho\sigma}$ and eliminate them
using field equations. Then one has the following superpotential
of dimension-5
$$W_5= -\frac{1}{M}J_{\mu\nu\lambda\rho\sigma}K_{\mu\nu\lambda\rho\sigma}$$
Next we use the fact that$K_{\mu\nu\lambda\rho\sigma}=0$. And thus
$$W_5=0$$
Normally, $W_5$ carries the lepton and baryon number violating
dimension five operators. Therefore, for the case of couplings
involving $\overline{126}$, the lepton and baryon violating
interactions vanish. This is directly due to the fact that
$\overline{126}$ plet has no couplings with matter. Alternatively,
one can say that the corresponding texture element in the Higgs
triplet sector is zero. This explains the zero in the 22 element
in $B^E$ and $B^D$ and the zeros in the 23+32 and 33 elements of
$B^U$ and $C^U$ (see Eq. (5.25)). These are the extra zeros over
and above the extra zeros in $A^E$, $A^D$ and $A^U$ of Eq. (5.23).

\chapter{Complete Cubic Couplings of Bilinear Spinors and 1, 45  and 210 plet
in  SO(10) Unification}\label{Ind_Detect}

Our focus in this chapter is the computation of cubic couplings in
the superpotential~\cite{ns2} of the form: $\overline{16}-16-1$,
$16-\overline{16}-1$, $\overline{16}-16-45$,
$16-\overline{16}-45$, $\overline{16}-16-210$ and
$16-\overline{16}-210$. The vector couplings\cite{ns2} of the
form: ${16}^{\dagger}- 16-1$,
$\overline{16}^{\dagger}-\overline{16}- 1$, ${16}^{\dagger}-
16-45$, $\overline{16}^{\dagger}-\overline{16}- 45$,
${16}^{\dagger}-16- 210$ and
$\overline{16}^{\dagger}-\overline{16}- 210$ are also analyzed.

\section{SO(10)-invariant cubic superpotential couplings involving semi-spinors ${\bf\overline{16}}$ and
$\bf{{16}}$ and tensors of dimensionality {\bf{1}}, {\bf{45}} and
{\bf{210}} }

The interactions of
  interest in the superpotential involving $16$ and $\overline{16}$ semispinors
 are of the form
\begin{equation}
{\mathsf W}^{(1)}_{(-+)}=h^{^{(1)}}_{\acute{a}\acute{b}}
<\widehat{\Psi}_{(-)\acute{a}}^*|B|\widehat{\Psi}_{(+)\acute{b}}>\Phi
\end{equation}
\begin{equation}
{\mathsf
W}^{(45)}_{(-+)}=\frac{1}{2!}h^{^{(45)}}_{\acute{a}\acute{b}}<\widehat{\Psi}_{(-)\acute{a}}^*|B
\Sigma_{\mu\nu}|\widehat{\Psi}_{(+)\acute{b}}>\Phi_{\mu\nu}
\end{equation}

\begin{equation}
{\mathsf W}^{(210)}_{(-+)}=\frac{1}{4!}
h^{^{(210)}}_{\acute{a}\acute{b}}<\widehat{\Psi}_{(-)\acute{a}}^*|B
\Gamma_{[\mu}\Gamma_{\nu}\Gamma_{\rho} \Gamma_{\lambda]}
|\widehat{\Psi}_{(+)\acute{b}}>\Phi_{\mu\nu\rho\lambda}
\end{equation}\\
\\
\textbf{(a)}~~\textsc{$\overline{16}\times 16\times 1$
\textnormal{and} $ 16\times\overline{16}\times 1$
couplings}\\
\\
We first present the result of the trivial ${\overline {16}\times
16\times 1}$ couplings. Eq.(6.1)
 at once gives
\begin{equation}
{\mathsf
W}^{(1)}_{(-+)}=ih^{^{(1)}}_{\acute{a}\acute{b}}\left(\widehat
{\bf N}_{\acute{a}}^{\bf{T}}\widehat {\bf
M}_{\acute{b}}-\frac{1}{2}\widehat {\bf
N}_{\acute{a}ij}^{\bf{T}}\widehat {\bf
M}_{\acute{b}}^{ij}+\widehat {\bf N}_{\acute{a}}^{i\bf{T}}\widehat
{\bf M}_{\acute{b}i}\right){\mathsf H}
\end{equation}
where ${\mathsf H}$ is an $SO(10)$ singlet. A  similar analysis
gives $ {\mathsf W}^{(1)}_{(+-)}$ and one has

\begin{eqnarray}
{\mathsf
W}^{(1)}_{(+-)}=h^{^{(1)}}_{\acute{a}\acute{b}}<\widehat{\Psi}_{(+)\acute{a}}^*|B|\widehat{\Psi}_{(-)\acute{ab}}>
\Phi
~~~~~~~~~~~~~~~~~~~~~~~\nonumber\\
=ih^{^{(1)}}_{\acute{a}\acute{b}}\left(-\widehat {\bf
M}_{\acute{a}}^{\bf{T}}\widehat {\bf
N}_{\acute{b}}+\frac{1}{2}\widehat {\bf
M}_{\acute{a}}^{ij\bf{T}}\widehat {\bf N}_{\acute{b}ij}-\widehat
{\bf M}_{\acute{a}i}^{\bf{T}}\widehat {\bf
N}_{\acute{b}}^i\right){\mathsf H}.
\end{eqnarray}\\
\\
\textbf{(b)}~~\textsc{$\overline{16}\times 16\times 45$
\textnormal{and} $ {16}\times\overline{16}\times 45$
couplings} \\
\\
 To compute the ${\overline {16}\times 16\times 45}$
couplings we expand the vertex $\Sigma_{\mu\nu}\Phi_{\mu\nu}$
 using Eq.(4.7)  where  $\Phi_{\mu\nu}$ is the 45 plet tensor field

\begin{equation}
\Sigma_{\mu\nu}\Phi_{\mu\nu}=\frac{1}{i}\left(b_ib_j
\Phi_{\overline c_i\overline c_j}+b_i^{\dagger}b_j^{\dagger}
\Phi_{c_ic_j}+2b_i^{\dagger}b_j\Phi_{c_i\overline c_j}-
\Phi_{c_n\overline c_n}\right).
\end{equation}
The reducible tensors that enter in the above expansion can be
decomposed into their irreducible parts as follows

 \begin{eqnarray}
\Phi_{c_n\overline c_n}={\mathsf h};~~~\Phi_{c_i\overline
c_j}={\mathsf h}_{j}^i+ \frac{1}{5}\delta_j^i{\mathsf h};~~~
\Phi_{c_ic_j}={\mathsf h}^{ij};~~~\Phi_{\overline c_i\overline
c_j}={\mathsf h}_{ij}
\end{eqnarray}
To normalize the $SU(5)$ Higgs fields contained in the tensor
$\Phi_{\mu\nu}$, we carry out a field redefinition
\begin{eqnarray}
{\mathsf h}=\sqrt {10}{\mathsf H};~~~{\mathsf h}_{ij}=\sqrt
2{\mathsf H}_{ij}; ~~~ {\mathsf h}^{ij}=\sqrt 2{\mathsf
H}^{ij};~~~{\mathsf h}_{j}^i=\sqrt 2 {\mathsf H}_{j}^i.
\end{eqnarray}
In terms of the normalized fields the kinetic energy of the
 45 plet of Higgs\\
$-\partial_A\Phi_{\mu\nu}\partial^A\Phi_{\mu\nu}^{\dagger}$ takes
the form
\begin{equation}
{\mathsf L}_{kin}^{45-Higgs}=-\partial^A{\mathsf
H}\partial_A{\mathsf H}^\dagger -\frac{1}{2!}
\partial^A{\mathsf H}_{ij}\partial_A{\mathsf H}_{ij}^{\dagger}
-\frac{1}{2!}
\partial^A{\mathsf H}^{ij}\partial_A{\mathsf H}^{ij\dagger}
-\partial_A{\mathsf H}_j^i
\partial^A{\mathsf H}_j^{i\dagger}.
\end{equation}
The terms in Eq.(6.9) are only exhibited for the purpose of
normalization and the remaining supersymmetric parts are not
exhibited as their normalizations are rigidly fixed relative to
the
 parts given above\cite{applied}.
Finally, straightforward evaluation of Eq.(6.2) using
Eqs.(6.6)-(6.8) gives
\begin{eqnarray}
{\mathsf W}^{(45)}_{(-+)}=\frac{1}{\sqrt
2}h^{^{(45)}}_{\acute{a}\acute{b}} \left[\sqrt
{5}\left(\frac{3}{5}\widehat {\bf N}_{\acute{a}}^{i\bf{T}}\widehat
{\bf M}_{\acute{b}i}+\frac{1}{10}\widehat {\bf
N}^{\bf{T}}_{\acute{a}ij}\widehat {\bf M}_{\acute{b}}^{ij}-
\widehat {\bf N}_{\acute{a}}^{\bf{T}}\widehat {\bf M}_{\acute{b}}\right){\mathsf H}\right.\nonumber\\
\left.+\left(-\widehat {\bf N}_{\acute{a}}^{\bf{T}}\widehat {\bf
M}_{\acute{b}}^{lm}+\frac{1}{2}\epsilon^{ijklm} \widehat
{\bf N}^{\bf{T}}_{\acute{a}ij}\widehat {\bf M}_{\acute{b}k}\right){\mathsf H}_{lm}\right.\nonumber\\
\left.+\left(-\widehat {\bf N}^{\bf{T}}_{\acute{a}lm}\widehat {\bf
M}_{\acute{b}}+\frac{1}{2}\epsilon_{ijklm}\widehat {\bf
N}_{\acute{a}}^{i\bf{T}}\widehat {\bf
M}_{\acute{b}}^{jk}\right){\mathsf
H}^{lm}\right.\nonumber\\
\left.+2\left(\widehat {\bf N}^{\bf{T}}_{\acute{a}ik}\widehat {\bf
M}_{\acute{b}}^{kj}-\widehat {\bf N}_{\acute{a}}^{j\bf{T}}\widehat
{\bf M}_{\acute{b}i}\right){\mathsf H}_j^i\right].
\end{eqnarray}
From Eq.(6.10), one finds that the $\overline {16}_N-16_M-45_H$
couplings  consist of the following $SU(5)$ invariant components:
 $5_N-\overline 5_M-1_H, \overline
{10}_N-10_M-1_H, 1_N-1_M-1_H, 1_N-10_M-\overline {10}_H, \overline
{10}_N-\overline 5_M-\overline {10}_H, \overline {10}_N-1_M-10_H,
5_N-10_M-10_H, \overline {10}_N-10_M-24_H$, and  $5_N-\overline
5_M-24_H$ couplings. One can carry out a similar analysis for
${\mathsf W}_{(+-)}^{(45)}$ and one finds

\begin{eqnarray}
{\mathsf
W}^{(45)}_{(+-)}=\frac{1}{2!}h^{^{(45)}}_{\acute{a}\acute{b}}<\widehat{\Psi}_{(+)\acute{a}}^*|B
\Sigma_{\mu\nu}|\widehat{\Psi}_{(-)\acute{b}}>\Phi_{\mu\nu}~~~~~~~~~~~~~~~~~
\nonumber\\
=\frac{1}{\sqrt 2}h^{^{(45)}}_{\acute{a}\acute{b}}\left[\sqrt
{5}\left(\frac{3}{5}\widehat {\bf M}_{\acute{a}i}^{\bf{T}}\widehat
{\bf N}_{\acute{b}}^i+\frac{1}{10}\widehat {\bf
M}^{ij\bf{T}}_{\acute{a}}\widehat {\bf N}_{\acute{b}ij}- \widehat
{\bf M}_{\acute{a}}^{\bf{T}}\widehat {\bf
N}_{\acute{b}}\right){\mathsf
H}\right.\nonumber\\
\left.+\left(-\widehat {\bf M}_{\acute{a}}^{lm\bf{T}}\widehat {\bf
N}_{\acute{b}}+\frac{1}{2}\epsilon^{ijklm} \widehat {\bf
M}^{\bf{T}}_{\acute{a}i}\widehat {\bf
N}_{\acute{b}jk}\right){\mathsf
H}_{lm}\right.\nonumber\\
\left.+\left(-\widehat {\bf M}^{\bf{T}}_{\acute{a}}\widehat {\bf
N}_{\acute{b}lm}+\frac{1}{2}\epsilon_{ijklm}\widehat {\bf
M}_{\acute{a}}^{ij\bf{T}}\widehat {\bf
N}_{\acute{b}}^{k}\right){\mathsf
H}^{lm}\right.\nonumber\\
\left.+2\left(\widehat {\bf M}^{jk\bf{T}}_{\acute{a}}\widehat {\bf
N}_{\acute{b}ki}-\widehat {\bf M}_{\acute{a}i}^{\bf{T}}\widehat
{\bf N}_{\acute{b}}^j\right){\mathsf H}_j^i\right].
\end{eqnarray}
\newpage
\noindent \textbf{(c)}~~\textsc{$\overline{16}\times 16\times 210$
\textnormal{and} ${16}\times \overline{16}\times 210$
couplings}\\
\\
 We now turn to the computation of the $\overline
{16}\times 16\times 210$ couplings. Using Eq.(4.7) we decompose
the vertex $\Gamma_{\mu}\Gamma_{\nu}\Gamma_{\rho}\Gamma_{\lambda}
\Phi_{\mu\nu\rho\lambda}$ so that

\begin{eqnarray}
\Gamma_{\mu}\Gamma_{\nu}\Gamma_{\rho}\Gamma_{\lambda}
\Phi_{\mu\nu\rho\lambda}=
4b_i^{\dagger}b_j^{\dagger}b_k^{\dagger}b_l\Phi_{c_ic_jc_k\overline
c_l} +4b_i^{\dagger}b_jb_kb_l\Phi_{c_i\overline c_j\overline
c_k\overline c_l} +b_i^{\dagger}b_j^{\dagger}b_k^{\dagger}
b_l^{\dagger}\Phi_{c_ic_jc_kc_l}\nonumber\\
+b_ib_jb_kb_l\Phi_{\overline c_i \overline c_j\overline
c_k\overline c_l}
-6b_i^{\dagger}b_j^{\dagger}\Phi_{c_ic_jc_m\overline c_m}+
6b_ib_j\Phi_{\overline c_i\overline c_j\overline c_mc_m}\nonumber\\
+3\Phi_{c_m\overline c_mc_n\overline c_n}\nonumber
-12b_i^{\dagger}b_j\Phi_{c_i\overline c_jc_m\overline c_m}
+6b_i^{\dagger}b_j^{\dagger}b_kb_l\Phi_{c_ic_j\overline
c_k\overline c_l}.
\nonumber\\
\end{eqnarray}
The tensors that appear above can be decomposed into their
irreducible parts as follows
\begin{eqnarray}
\Phi_{c_m\overline c_mc_n\overline c_n}={\mathsf
h};~~~\Phi_{\overline c_i\overline c_j \overline c_k\overline
c_l}=\frac{1}{24}\epsilon_{ijklm}{\mathsf h}^m;~~~
\Phi_{c_ic_jc_kc_l}=\frac{1}{24}\epsilon^{ijklm}{\mathsf h}_{m}\nonumber\\
\Phi_{c_ic_jc_m\overline c_m}={\mathsf h}^{ij};~~~\Phi_{\overline
c_i\overline c_j \overline c_mc_m}={\mathsf
h}_{ij};~~~\Phi_{c_i\overline
c_jc_m\overline c_m}={\mathsf h}_{j}^i+\frac{1}{5}\delta_j^i{\mathsf h}\nonumber\\
\Phi_{c_ic_j\overline c_k\overline c_l}={\mathsf
h}_{kl}^{ij}+\frac{1}{3} \left(\delta_l^i{\mathsf
h}_{k}^j-\delta_k^i{\mathsf h}_{l}^j+ \delta_k^j{\mathsf h}_{l}^i
-\delta_l^j{\mathsf h}_{k}^i\right)+\frac{1}{20}\left(\delta_l^i
\delta_k^j-\delta_k^i\delta_l^j\right){\mathsf h}\nonumber\\
\Phi_{c_ic_jc_k\overline c_l}={\mathsf h}_{l}^{ijk}+\frac{1}{3}
\left(\delta_l^k{\mathsf h}^{ij}-\delta_l^j{\mathsf h}^{ik}
+\delta_l^i{\mathsf h}^{jk}\right)\nonumber\\
\Phi_{\overline c_i\overline c_j\overline c_kc_l}={\mathsf
h}_{ijk}^l +\frac{1}{3}\left(\delta_k^l{\mathsf
h}_{ij}-\delta_j^l{\mathsf h}_{ik}+ \delta_i^l{\mathsf
h}_{jk}\right)
\end{eqnarray}
where ${\mathsf h}$, ${\mathsf h}^i$, ${\mathsf h}_{i}$, ${\mathsf
h}^{ij}$, ${\mathsf h}_{ij}$, ${\mathsf h}_{j}^i$, ${\mathsf
h}_{l}^{ijk}$; ${\mathsf h}_{jkl}^i$ and ${\mathsf h}_{kl}^{ij}$
are the 1-plet, 5-plet, $\overline 5$-plet, 10-plet, $\overline
{10}$-plet, 24-plet, 40-plet, $\overline {40}$-plet, and 75-plet
representations of $SU(5)$, respectively.  We carry out a field
redefinition such that

\begin{eqnarray}
{\mathsf h}=4\sqrt{\frac{5}{3}}{\mathsf H};~~~{\mathsf
h}^i=8\sqrt{6}{\mathsf H}^i;~~~{\mathsf h}_{i}
=8\sqrt{6}{\mathsf H}_{i}\nonumber\\
{\mathsf h}^{ij}=\sqrt{2}{\mathsf H}^{ij};~~~{\mathsf
h}_{ij}=\sqrt{2}{\mathsf H}_{ij};
~~~{\mathsf h}_{j}^i=\sqrt 2{\mathsf H}_{j}^i\nonumber\\
{\mathsf h}_{l}^{ijk}=\sqrt{\frac{2}{3}}{\mathsf
H}_{l}^{ijk};~~~{\mathsf h}_{jkl}^i =\sqrt{\frac{2}{3}}{\mathsf
H}_{jkl}^i;~~~{\mathsf h}_{kl}^{ij} =\sqrt{\frac{2}{3}}{\mathsf
H}_{kl}^{ij}.
\end{eqnarray}

\noindent Now the kinetic energy for the 210 dimensional Higgs
field is \\
$-\partial_A\Phi_{\mu\nu\rho\lambda}\partial^A
\Phi_{\mu\nu\rho\lambda}^{\dagger}$, which in terms of the
redefined fields takes the form
\begin{eqnarray}
{\mathsf L}_{kin}^{210-Higgs}=-\partial_A{\mathsf H}\partial^A
{\mathsf H}^{\dagger} -\partial_A{\mathsf H}^i\partial^A{\mathsf
H}^{i\dagger}-\partial_A{\mathsf H}_i\partial^A{\mathsf
H}_{i\dagger}\nonumber\\
-\frac{1}{2!}\partial_A{\mathsf H}^{ij}
\partial^A{\mathsf H}^{ij\dagger}
-\frac{1}{2!}\partial_A{\mathsf H}_{ij}
\partial^A{\mathsf H}_{ij}^{\dagger}
-\partial_A{\mathsf H}_j^i\partial^A{\mathsf
H}_j^{i\dagger}\nonumber\\
-\frac{1}{3!}\partial_A{\mathsf H}_l^{ijk}
\partial^A{\mathsf H}_l^{ijk\dagger}
-\frac{1}{3!}\partial_A{\mathsf H}^l_{ijk}
\partial^A{\mathsf H}^{l\dagger}_{ijk}
-\frac{1}{2!}\frac{1}{2!}
\partial_A{\mathsf H}_{kl}^{ij}\partial^A{\mathsf H}_{kl}^{ij\dagger}.
\end{eqnarray}
Evaluation of Eq.(6.3), using Eq.(6.12) and the normalization of
Eq.(6.14) gives,
\begin{eqnarray}
{\mathsf
W}^{(210)}_{(-+)}=i\sqrt{\frac{2}{3}}h_{\acute{a}\acute{b}}^{^{(210)}}
\left[\frac{1}{2}\sqrt {\frac{{5}}{2}}\left(\widehat {\bf
N}_{\acute{a}}^{\bf{T}}\widehat {\bf
M}_{\acute{b}}+{\frac{1}{10}}\widehat{\bf
N}_{\acute{a}ij}^{\bf{T}} \widehat {\bf
M}_{\acute{b}}^{ij}+\frac{1}{5}\widehat {\bf
N}_{\acute{a}}^{i\bf{T}}\widehat
{\bf M}_{\acute{b}i}\right){\mathsf H}\right.\nonumber\\
\left.+\frac{\sqrt 3}{4}\left(\widehat {\bf
N}^{\bf{T}}_{\acute{a}lm}\widehat {\bf M}_{\acute{b}} +
\frac{1}{6}\epsilon_{ijklm}\widehat{\bf
N}_{\acute{a}}^{i\bf{T}}\widehat {\bf M}_{\acute{b}}^{jk}
\right){\mathsf H}^{lm}\right.\nonumber\\
\left.-\frac{\sqrt{3}}{4}\left(\widehat {\bf
N}_{\acute{a}}^{\bf{T}}\widehat {\bf M}_{\acute{b}}^{lm}
+\frac{1}{6}\epsilon^{ijklm}\widehat{\bf
N}^{\bf{T}}_{\acute{a}ij}\widehat {\bf M}_{\acute{b}k}
\right){\mathsf H}_{lm}\right.\nonumber\\
\left.-\frac{\sqrt 3}{2}\left(\widehat {\bf
N}_{\acute{a}}^{j\bf{T}}\widehat {\bf M}_{\acute{b}i}
+\frac{1}{3}\widehat{\bf N}^{\bf{T}}_{\acute{a}ik}\widehat
{\bf M}_{\acute{b}}^{kj}\right){\mathsf H}_j^i\right.\nonumber\\
\left.+\frac{1}{6}\epsilon_{ijklm}\widehat {\bf
N}_{\acute{a}}^{i\bf{T}}\widehat {\bf M}_{\acute{b}}^{jn}{\mathsf
H}_n^{klm} +\frac{1}{6}\epsilon^{ijklm}\widehat
 {\bf N}^{\bf{T}}_{\acute{a}in}\widehat {\bf M}_{\acute{b}j}{\mathsf H}_{klm}^n\right.\nonumber\\
\left.+\frac{1}{4}\widehat {\bf N}^{\bf{T}}_{\acute{a}ij}\widehat
{\bf M}_{\acute{b}}^{kl}{\mathsf H}_{kl}^{ij} +\widehat {\bf
N}_{\acute{a}}^{\bf{T}}\widehat {\bf M}_{\acute{b}i}{\mathsf H}^i
+\widehat {\bf N}_{\acute{a}}^{i\bf{T}}\widehat {\bf
M}_{\acute{b}}{\mathsf H}_i\right].
\end{eqnarray}
We note that $\overline {16}_N-16_M-210_H$ couplings have the
$SU(5)$ invariant structure consisting of $1_N-1_M-1_H, \overline
{10}_N-10_M-1_H, 5_N-\overline 5_M-1_H, \overline
{10}_N-1_M-10_H,\\ 5_N-10_M-10_H, 1_N-10_M-\overline {10}_H,
\overline {10}_N-\overline 5_M-\overline {10}_H, 5_N-\overline
5_M-24_H, \overline {10}_N-10_M-24_H,5_N-10_M-40_H, \overline
{10}_N-\overline 5_M-\overline {40}_H, \overline {10}_N-10_M-75_H,
1_N-\overline 5_M-5_H, 5_N-1_M-\overline 5_H$.
 An analysis similar to that for Eq.(6.16) gives
${\mathsf W}^{(210)}_{(+-)}$
\begin{eqnarray}
{\mathsf
W}^{(210)}_{(+-)}=\frac{1}{4!}h^{^{(210)}}_{\acute{a}\acute{b}}<\widehat{\Psi}_{(+)\acute{a}}^*|B
\Gamma_{[\mu}\Gamma_{\nu}\Gamma_{\rho}
\Gamma_{\lambda]}|\widehat{\Psi}_{(-)\acute{b}}>\Phi_{\mu\nu\rho\lambda}~~~~~~~~~~~~~~~~~~~\nonumber\\
=i\sqrt{\frac{2}{3}}h_{\acute{a}\acute{b}}^{^{(210)}}
\left[-\frac{1}{2}\sqrt{\frac{{5}}{2}}\left( \widehat {\bf
M}_{\acute{a}}^{\bf{T}}\widehat {\bf
N}_{\acute{b}}+{\frac{1}{10}}\widehat{\bf
M}_{\acute{a}}^{ij\bf{T}} \widehat {\bf
N}_{\acute{b}ij}+\frac{1}{5}\widehat {\bf M}_{\acute{a}i}^{\bf{T}}
\widehat {\bf N}_{\acute{b}}^{i}\right){\mathsf H}\right.\nonumber\\
\left.-\frac{\sqrt{3}}{4}\left(\widehat {\bf
M}^{\bf{T}}_{\acute{a}}\widehat {\bf N}_{\acute{b}lm}
+\frac{1}{6}\epsilon_{ijklm}\widehat{\bf
M}_{\acute{a}}^{ij\bf{T}}\widehat {\bf N}_{\acute{b}}^{k}
\right){\mathsf H}^{lm}\right.\nonumber\\
\left.+\frac{\sqrt{3}}{4}\left(\widehat {\bf
M}_{\acute{a}}^{lm\bf{T}}\widehat {\bf N}_{\acute{b}}
+\frac{1}{6}\epsilon^{ijklm}\widehat{\bf
M}^{\bf{T}}_{\acute{a}i}\widehat {\bf N}_{\acute{b}jk}
\right){\mathsf H}_{lm}\right.\nonumber\\
\left.+\frac{\sqrt 3}{2}\left(\widehat {\bf
M}_{\acute{a}i}^{\bf{T}}\widehat {\bf N}_{\acute{b}}^j
+\frac{1}{3}\widehat{\bf M}^{jk\bf{T}}_{\acute{a}}\widehat
{\bf N}_{\acute{b}ki}\right){\mathsf H}_j^i\right.\nonumber\\
\left.+\frac{1}{12}\epsilon_{ijklm}\widehat {\bf
M}_{\acute{a}}^{ij\bf{T}}\widehat {\bf N}_{\acute{b}}^{n}{\mathsf
H}_n^{klm} -\frac{1}{12}\epsilon^{ijklm}\widehat
 {\bf M}^{\bf{T}}_{\acute{a}n}\widehat {\bf N}_{\acute{b}ij}{\mathsf
H}_{klm}^n\right.\nonumber\\
\left.-\frac{1}{4}\widehat {\bf M}^{kl\bf{T}}_{\acute{a}}\widehat
{A\bf N}_{\acute{b}ij}{\mathsf H}_{kl}^{ij} -\widehat {\bf
M}_{\acute{a}i}^{\bf{T}}\widehat {\bf N}_{\acute{b}}{\mathsf H}^i
-\widehat {\bf M}_{\acute{a}}^{\bf{T}}\widehat {\bf
N}_{\acute{b}}^i{\mathsf H}_i\right].
\end{eqnarray}
We note that the couplings of ${\mathsf W}^{(210)}_{(-+)}$ are in
general not the same as in ${\mathsf W}^{(210)}_{(+-)}$. Thus some
of the terms have signs which are opposite in the two sets.
Further, we note that there are in general two ways in which the
40 plet and the $\overline {40}$ plet can contract with the matter
fields. For the case of ${\mathsf W}^{(210)}_{(-+)}$ one of the 40
plet tensor index contracts with the tensor index of the 10 plet
of matter and similarly one  of the tensor index on the $\overline
{40}$ contracts with the tensor index in the $\overline{10}$  of
the $\overline{16}$ (see Eq.(6.16)). However, in the ${\mathsf
W}^{(210)}_{(+-)}$ couplings this is not the case. Here one of the
tensor index of 40 plet contracts with the tensor index in of the
5 plet of matter and similarly one  of the tensor index in
$\overline {40}$ contracts with the tensor index in the $\bar{5}$
plet of matter (see Eq.(6.17)).

\section{SO(10)-invariant cubic vector couplings involving semispinors
{\bf{16}} and  ${\bf{\overline{16}}}$ and tensors of
dimensionality {\bf{1}}, {\bf{45}} and {\bf{210}}}

For the construction of couplings of vector fields with
 $16$ and $\overline {16}$ plets,  it is natural to consider the couplings of the
 1 and 45 vector fields as abelian and Yang-Mills gauge interactions.
 However, one cannot do the same for the $\overline{16}-16-210$ and
 $16-\overline{16}-210$
 couplings. These interactions cannot be treated as gauge couplings
 as there are no corresponding Yang-Mills interactions for the 210 plet.
 For this reason we focus here first on the computation of the gauge
 couplings of the 1 and 45 plet of vector fields. The supersymmetric
 kinetic energy and gauge couplings of the chiral superfield
 $\widehat{\Phi}$ can be written in the usual superfield notation
\begin{equation}
 \int d^4\theta~tr(\widehat{\Phi}^{\dagger}e^{g\widehat{\mathsf {V}}}\widehat{\Phi})
\end{equation}
where $\widehat{\mathsf {V}}$ is the Lie valued vector superfield.
Similarly the supersymmetric Yang-Mills part of the Lagrangian can
be gotten from

\begin{eqnarray}
 \int d^2\theta~tr(\widehat{\cal
W}^{\tilde{\alpha}} \widehat{\cal W}_{\tilde{\alpha}}) + \int
d^2\bar{\theta}~tr(\widehat{\overline{\cal
W}}_{\dot{\tilde{\alpha}}}\widehat{\overline{\cal
W}}^{\dot{\tilde{\alpha}}})
\end{eqnarray}
where  $\widehat{\cal W}^{\tilde{\alpha}}$ is the field strength
 chiral spinor superfield. Since supersymmetry does not play any special
 role in the analysis of $SO(10)$ Clebsch-Gordon co-efficients,
 we will display in the analysis here only the parts of the Lagrangian
 relevant for our discussion. Thus the interactions of the
 $16$ with gauge vectors for the 1 and 45 plet cases are
 given by

\begin{equation}
{\mathsf
L}^{(1)}_{(++)}=g^{^{(1)}}_{\acute{a}\acute{b}}<\Psi_{(+)\acute{a}}|\gamma^0\gamma^A|\Psi_{(+)\acute{b}}>\Phi_A
\end{equation}
\begin{equation}
{\mathsf
L}^{(45)}_{(++)}=\frac{1}{i}\frac{1}{2!}g^{^{(45)}}_{\acute{a}\acute{b}}<\Psi_{(+)\acute{a}}|\gamma^0\gamma^A
\Sigma_{\mu\nu}|\Psi_{(+)\acute{b}}>\Phi_{A\mu\nu}
\end{equation}
where  $g$'s are the gauge coupling constants, and $\Phi_A$ and
$\Phi_{A\mu\nu}$ are gauge tensors of dimensionality 1 and 45,
respectively. Similarly one defines ${\mathsf L}^{(1)}_{(--)}$,
${\mathsf L}^{(45)}_{(--)}$ with
 $\Psi_{+}$ replaced by $\Psi_{-}$ in Eqs.(6.20) and (6.21).
\\

 \noindent \textbf{(a)}~~\textsc{${16}^{\dagger}\times 16\times 1$
\textnormal{and}$\overline{16}^{\dagger}\times\overline{16}\times
1$
couplings}\\

We first present the result of the trivial ${\overline {16}\times
16\times 1}$ couplings. Eqs.(6.20) and (4.7) at once give
\begin{equation}
{\mathsf
L}^{(1)}_{(++)}=g^{^{(1)}}_{\acute{a}\acute{b}}\left(\overline
{\bf M}_{\acute{a}}\gamma^A{\bf
M}_{\acute{b}}+\frac{1}{2}\overline {\bf
M}_{\acute{a}ij}\gamma^A{\bf M}_{\acute{b}}^{ij}+\overline {\bf
M}_{\acute{a}}^i\gamma^A{\bf M}_{\acute{b}i} \right){\mathsf G}_A.
\end{equation}
The barred matter fields are defined so that $~\overline {\bf
M}_{\acute{a}ij}={\bf M}_{\acute{a}ij}^{\dagger}\gamma ^0$ etc.

A  similar analysis gives ${\mathsf   L}^{(1)}_{(--)}$ and one has
\begin{eqnarray}
{\mathsf
L}^{(1)}_{(--)}=g^{^{(1)}}_{\acute{a}\acute{b}}<\Psi_{(-)\acute{a}}|\gamma^0\gamma^A|\Psi_{(-)\acute{b}}>\Phi_A
~~~~~~~~~~~~~~~~~~~~~~~~\nonumber\\
=g^{^{(1)}}_{\acute{a}\acute{b}}\left(\overline {\bf
N}_{\acute{a}}\gamma^A{\bf N}_{\acute{b}}+\frac{1}{2}\overline
{\bf N}^{ij}_{\acute{a}}\gamma^A{\bf N}_{\acute{b}ij}+\overline
{\bf N}_{\acute{a}i}\gamma^A{\bf N}_{\acute{b}}^i \right){\mathsf
G}_A.
\end{eqnarray}

\noindent \textbf{(b)}~~\textsc{${16}^{\dagger}\times 16\times 45$
\textnormal{and} $\overline{16}^{\dagger}\times\overline{16}\times
45$
couplings}\\

We next discuss the couplings of the 45 plet gauge tensor
$\Phi_{A\mu\nu}$ whose decomposition in terms of reducible $SU(5)$
tensors can be written similar to Eq.(6.6). This can be further
reduced into irreducible parts similar to Eq.(6.7) by
\begin{eqnarray}
\Phi_{Ac_n\overline c_n}={\mathsf g}_A;~~~\Phi_{Ac_i\overline
c_j}={\mathsf g}_{Aj}^i+ \frac{1}{5}\delta_j^i{\mathsf g}_A;~~~
\Phi_{Ac_ic_j}={\mathsf g}_A^{ij};~~~\Phi_{A\overline c_i\overline
c_j}={\mathsf g}_{Aij}
\end{eqnarray}
and normalized so that
\begin{eqnarray}
{\mathsf g}_A=2\sqrt 5 {\mathsf G}_A;~~~{\mathsf g}_{Aij}=\sqrt 2
{\mathsf G} _{Aij};~~~ {\mathsf g}_A^{ij}=\sqrt 2 {\mathsf
G}_A^{ij};~~~{\mathsf g}_{Aj}^i=\sqrt{2} {\mathsf G}_{Aj}^i.
\end{eqnarray}
The kinetic energy for the 45-plet is given by $-\frac{1}{4}{\cal
F}_{\mu\nu}^ {AB}{\cal F}_{AB\mu\nu}$,
 where ${\cal F}_{\mu\nu}^{AB}$ is the 45 of $SO(10)$ field strength tensor.
In terms of the redefined fields, 45-plet's kinetic energy takes
the form
\begin{equation}
{\mathsf L}_{kin}^{45-gauge}=-\frac{1}{2}{\cal G}_{AB}{\cal
G}^{AB\dagger}-\frac{1}{2!} \frac{1}{2}{\cal G}^{ABij}{\cal
G}_{AB}^{ij\dagger} -\frac{1}{4}{\cal G}_j^{ABi}{\cal G}_{ABi}^{j}
\end{equation}
where ${\cal F}_{\mu\nu}^{AB}$ is the 45 of $SO(10)$ field
strength tensor. As mentioned in the beginning of this section we
do not exhibit the gaugino and D terms needed for supersymmetry
since their normalization is fixed relative to terms exhibited in
Eq.(6.26). Using  Eqs.(6.21), (6.6) and the above normalizations
we find
\begin{eqnarray}
{\mathsf
L}^{(45)}_{(++)}=g^{^{(45)}}_{\acute{a}\acute{b}}\left[\sqrt
5\left(-\frac{3}{5}\overline {\bf M}_{\acute{a}}^i\gamma^A{\bf
M}_{\acute{b}i}+\frac{1}{10}\overline {\bf M}_{\acute{a}ij}
\gamma^A{\bf M}_{\acute{b}}^{ij}+
\overline{\bf M}_{\acute{a}}\gamma^A{\bf M}_{\acute{b}}\right){\mathsf G}_{A}\right.\nonumber\\
\left.+{\frac{1} {\sqrt 2}}\left(\overline {\bf
M}_{\acute{a}}\gamma^{A} {\bf M}_{\acute{b}}^{lm}
+{\frac{1}{2}}\epsilon^{ijklm}\overline {\bf
M}_{\acute{a}ij}\gamma^A{\bf M}_{\acute{b}k}\right)
{\mathsf G}_{Alm}\right.\nonumber\\
\left.-\frac{1}{\sqrt 2}\left(\overline {\bf
M}_{\acute{a}lm}\gamma^A{\bf
M}_{\acute{b}}+\frac{1}{2}\epsilon_{ijklm} \overline {\bf
M}_{\acute{a}}^i\gamma^A{\bf M}_{\acute{b}}^{jk}\right){\mathsf
G}_A^{lm}\right.\nonumber\\
\left.+\sqrt{2}\left(\overline {\bf M}_{\acute{a}ik}\gamma^A{\bf
M}_{\acute{b}}^{kj}+\overline {\bf M}_{\acute{a}}^j\gamma^A{\bf
M}_{\acute{b}i}\right){\mathsf G}_{Aj}^i\right].
\end{eqnarray}
A similar analysis gives
\begin{eqnarray}
{\mathsf L}^{(45)}_{(--)}=\frac{1}{i}
 \frac{1}{2!}g^{^{(45)}}_{\acute{a}\acute{b}}<\Psi_{(-)\acute{a}}|\gamma^0\gamma^A
\Sigma_{\mu\nu}|\Psi_{(-)\acute{b}}>\Phi_{A\mu\nu}~~~~~~~~~~~~~~~~~~~~~~~~~~\nonumber\\
=g^{^{(45)}}_{\acute{a}\acute{b}}\left[\sqrt
5\left(\frac{3}{5}\overline {\bf N}_{\acute{a}i}\gamma^A{\bf
N}_{\acute{b}}^i-\frac{1}{10}\overline {\bf N}_{\acute{a}}^{ij}
\gamma^A{\bf N}_{\acute{b}ij}-
\overline{\bf N}_{\acute{a}}\gamma^A{\bf N}_{\acute{b}}\right){\mathsf G}_{A}\right.\nonumber\\
\left.+{\frac{1} {\sqrt 2}}\left(\overline {\bf
N}_{\acute{a}}^{lm} \gamma^{A} {\bf N}_{\acute{b}}
+{\frac{1}{2}}\epsilon^{ijklm}\overline {\bf
N}_{\acute{a}i}\gamma^A{\bf N}_{\acute{b}jk}\right)
{\mathsf G}_{Alm}\right.\nonumber\\
\left.-\frac{1}{\sqrt 2}\left(\overline {\bf
N}_{\acute{a}}\gamma^A{\bf
N}_{\acute{b}lm}+\frac{1}{2}\epsilon_{ijklm} \overline {\bf
N}_{\acute{a}}^{ij}\gamma^A{\bf N}_{\acute{b}}^k\right){\mathsf
G}_A^{lm}\right.\nonumber\\
\left.-\sqrt{2}\left(\overline {\bf
N}_{\acute{a}}^{jk}\gamma^A{\bf N}_{\acute{b}ki}+\overline {\bf
N}_{\acute{a}i}\gamma^A{\bf N}_{\acute{b}}^j\right){\mathsf
G}_{Aj}^i\right].
\end{eqnarray}
\\
\noindent \textbf{(c)}~~\textsc{${16}^{\dagger}\times 16\times
210$ \textnormal{and}
$\overline{16}^{\dagger}\times\overline{16}\times 210$
couplings}\\

\noindent We discuss now the 210 vector multiplet. This vector
mutiplet is not a gauge multiplet with the usual Yang-Mills
interactions. This makes the multiplet rather pathological and it
cannot be treated in a normal fashion. Specifically Eq.(6.18) is
not valid for this case in any direct fashion. However, for the
sake of completeness, we present here the $SO(10)$ globally
invariant couplings corresponding to Eq.(6.21). Thus we
 have
\begin{equation}
{\mathsf
L}^{(210)}_{(++)}=\frac{1}{4!}g^{^{(210)}}_{\acute{a}\acute{b}}<\Psi_{(+)\acute{a}}
|\gamma^0\gamma^A\Gamma_{[\mu}\Gamma_{\nu}\Gamma_{\rho}
\Gamma_{\lambda]}|\Psi_{(+)\acute{b}}>\Phi_{A\mu\nu\rho\lambda}.
\end{equation}
To compute the couplings we carry out  expansions similar to
Eqs.(6.12) and (6.13) and to normalize the fields we carry out a
field redefinition
\begin{eqnarray}
{\mathsf g}_A=4\sqrt{\frac{10}{3}}{\mathsf G}_A;~~~{\mathsf
g}_A^i= 8\sqrt{6}{\mathsf G}_A^i;
~~~{\mathsf g}_{Ai}=8\sqrt{6}{\mathsf G}_{Ai}\nonumber\\
{\mathsf g}_A^{ij}=\sqrt{2}{\mathsf G}_A^{ij};~~~{\mathsf
g}_{Aij}= \sqrt{2}{\mathsf G}_{Aij}; ~~~{\mathsf g}_{Aj}^i
=\sqrt{2}{\mathsf G}_{Aj}^i\nonumber\\
{\mathsf g}_{Al}^{ijk}=\sqrt{\frac{2}{3}}{\mathsf G}_{Al}^{ijk};
~~~{\mathsf g}_{Ajkl}^i =\sqrt{\frac{2}{3}}{\mathsf
G}_{Ajkl}^i;~~~{\mathsf g}_{Akl}^{ij}=\frac{2} {\sqrt{3}}{\mathsf
G}_{Akl}^{ij}
\end{eqnarray}
so that the 210-plet's kinetic energy $-\frac{1}{4}{\cal
F}_{\mu\nu\rho\lambda}^{AB}{\cal F}_{AB\mu\nu\rho\lambda}$ takes
the form
\begin{eqnarray}
{\mathsf L}_{kin}^{210-gauge}=-\frac{1}{2}{\cal G}_{AB}{\cal
G}^{AB\dagger}-\frac{1}{2} {\cal G}_{AB}^i{\cal G}^{ABi\dagger}
-\frac{1}{2!}\frac{1}{2}{\cal G}_{AB}^{ij}{\cal G}^{ABij\dagger}\nonumber\\
-\frac{1}{4}{\cal G}_{ABj}^i{\cal G}_i^{ABj}
-\frac{1}{3!}\frac{1}{2}{\cal G}_{ABl}^{ijk} {\cal
G}_l^{ABijk\dagger} -\frac{1}{4}\frac{1}{2!}\frac{1}{2}{\cal
G}_{ABkl}^{ij}{\cal G}_{kl}^{ABij\dagger} .\end{eqnarray}

\noindent As discussed above, the 210 vector multiplet is not a
gauge multiplet and thus the quantity ${\cal G}_{AB}$ is just an
ordinary curl. Using Eqs.(6.29), (6.12) and the normalizations of
Eq.(6.30) one can compute ${\mathsf L}^{(210)}_{(++)}$. One finds
\begin{eqnarray}
{\mathsf
L}^{(210)}_{(++)}=\frac{1}{\sqrt{6}}g^{^{(210)}}_{\acute{a}\acute{b}}
\left[\sqrt{5}\left (\overline {\bf M}_{\acute{a}}\gamma^A{\bf
M}_{\acute{b}}-{\frac{1}{10}}\overline {\bf M}_{\acute{a}ij}
\gamma^A{\bf M}_{\acute{b}}^{ij}
+\frac{1}{5}\overline {\bf M}_{\acute{a}}^i\gamma^A{\bf M}_{\acute{b}i}\right){\mathsf G}_A\right.\nonumber\\
\left.+\frac{\sqrt{3}}{2}\left(-\overline{\bf
M}_{\acute{a}}\gamma^A{\bf M}_{\acute{b}}^{lm}
+\frac{1}{6}\epsilon^{ijklm}\overline{\bf M}_{{\acute{a}}ij}
\gamma^A{\bf M}_{\acute{b}k}\right){\mathsf G}_{Alm}\right.\nonumber\\
\left.+\frac{\sqrt{3}}{2}\left(-\overline{\bf
M}_{\acute{a}lm}\gamma^A{\bf M}_{\acute{b}}
+\frac{1}{6}\epsilon_{ijklm}\overline{\bf M}_{\acute{a}}^i
\gamma^A{\bf M}_{\acute{b}}^{jk}\right){\mathsf G}_A^{lm}\right.\nonumber\\
\left.+\sqrt{3}\left(-\overline {\bf M}_{\acute{a}}^j\gamma^A{\bf
M}_{\acute{b}i} +\frac{1}{3}\overline{\bf
M}_{\acute{a}ik}\gamma^A{\bf M}_{\acute{b}}^{kj}\right){\mathsf
G}_{Aj}^i
\right.\nonumber\\
\left.-\frac{1}{3}\epsilon^{ijklm}\overline {\bf M}_{\acute{a}in}
\gamma^A{\bf M}_{\acute{b}j}{\mathsf G}_{Aklm}^n
+\frac{1}{3}\epsilon_{ijklm}\overline {\bf M}_{\acute{a}}^i
\gamma^A{\bf M}_{\acute{b}}^{jn}{\mathsf G}_{An}^{klm}\right.\nonumber\\
\left.-\frac{1}{\sqrt{2}}\overline {\bf M}_{\acute{a}ij}
\gamma^A{\bf M}_{\acute{b}}^{kl}{\mathsf G}_{Akl}^{ij} +2\overline
{\bf M}_{\acute{a}}^i\gamma^A{\bf M}_{\acute{b}}G_{Ai} +2\overline
{\bf M}_{\acute{a}}\gamma^A{\bf M}_{\acute{b}i}{\mathsf
G}_A^i\right].
\end{eqnarray}
A similar analysis gives
\begin{eqnarray}
{\mathsf
L}^{(210)}_{(--)}=\frac{1}{4!}g^{^{(210)}}_{\acute{a}\acute{b}}<\Psi_{(-)\acute{a}}
|\gamma^0\gamma^A\Gamma_{[\mu}\Gamma_{\nu}\Gamma_{\rho}
\Gamma_{\lambda]}|\Psi_{(-)\acute{b}}>\Phi_{A\mu\nu\rho\lambda}~~~~~~~~~~~~~~~~~~~~
\nonumber\\
=\frac{1}{\sqrt{6}}g^{^{(210)}}_{\acute{a}\acute{b}}
\left[\sqrt{5}\left (\overline {\bf N}_{\acute{a}}\gamma^A{\bf
N}_{\acute{b}}-{\frac{1}{10}}\overline {\bf N}_{\acute{a}}^{ij}
\gamma^A{\bf N}_{\acute{b}ij} +\frac{1}{5}\overline {\bf
N}_{\acute{a}i}\gamma^A{\bf N}_{\acute{b}}^i\right){\mathsf
G}_A\right.\nonumber\\
\left.+\frac{\sqrt{3}}{2}\left(\overline{\bf
N}_{\acute{a}}^{lm}\gamma^A{\bf N}_{\acute{b}}
-\frac{1}{6}\epsilon^{ijklm}\overline{\bf N}_{\acute{a}i}
\gamma^A{\bf N}_{\acute{b}jk}\right){\mathsf G}_{Alm}\right.\nonumber\\
\left.+\frac{\sqrt{3}}{2}\left(\overline{\bf
N}_{\acute{a}}\gamma^A{\bf N}_{\acute{b}lm}
-\frac{1}{6}\epsilon_{ijklm}\overline{\bf N}_{\acute{a}}^{ij}
\gamma^A{\bf N}_{\acute{b}}^{k}\right){\mathsf G}_A^{lm}\right.\nonumber\\
\left.+\sqrt{3}\left(-\overline {\bf N}_{\acute{a}i}\gamma^A{\bf
N}_{\acute{b}}^j +\frac{1}{3}\overline{\bf
N}_{\acute{a}}^{jk}\gamma^A{\bf N}_{\acute{b}ki}\right){\mathsf
G}_{Aj}^i\right.\nonumber\\
\left.+\frac{1}{6}\epsilon^{ijklm}\overline {\bf N}_{\acute{a}n}
\gamma^A{\bf N}_{\acute{b}ij}{\mathsf G}_{Aklm}^n
+\frac{1}{6}\epsilon_{ijklm}\overline {\bf N}_{\acute{a}}^{ij}
\gamma^A{\bf N}_{\acute{b}}^{n}{\mathsf G}_{An}^{klm}\right.\nonumber\\
\left.-\frac{1}{\sqrt{2}}\overline {\bf N}_{\acute{a}}^{kl}
\gamma^A{\bf N}_{\acute{b}ij}{\mathsf G}_{Akl}^{ij} +2\overline
{\bf N}_{\acute{a}}\gamma^A{\bf N}_{\acute{b}}^iG_{Ai} +2\overline
{\bf N}_{\acute{a}i}\gamma^A{\bf N}_{\acute{b}}{\mathsf
G}_A^i\right]~~~~~~.
\end{eqnarray}
Supersymmetrizations of Eqs.(6.32) and (6.33) requires that we
deal with a massive vector multiplet and this topic will be dealt
in chapters 8 and 9\cite{ns3}.

\section{Appendix}
In this appendix we expand some of the $SO(10)$ interactions in
the familiar particle notation and exhibit the differences between
some of the $\overline{16}-16-45$ and the $\overline{16}-16-210$
couplings. We start by looking at the gauge interactions of the 24
plet of $SU(5)$ in $\overline{16}-16-45$ coupling. We can read
this off from the last term in Eq.(6.27). Disregarding the front
factor, this term is of the form
\begin{equation}
{\cal L}_{24/45}=g_{\acute{a}\acute{b}}^{^{(45)}} \left(\overline
{\bf M}_{aik}\gamma^A{\bf M}_b^{kj}+\overline {\bf
M}_{\acute{a}}^j\gamma^A{\bf M}_{bi}\right){\mathsf G}_{Aj}^i
\end {equation}
An expansion of Eq.(6.34)  using the SM particle states defined by
Eq.(2.273) gives
\begin{eqnarray}
{\cal L}_{24/45}=g_{\acute{a}\acute{b}}^{^{(45)}}
\sum_{x=1}^8\left[\overline U_{\acute{a}}\gamma^A{\bf V}_A^{x}
\frac{\lambda_x}{2}U_b+\overline D_{\acute{a}}\gamma^A{\bf
V}_A^{x}\frac{\lambda_x}{2}
D_b\right]\nonumber\\
+g_{\acute{a}\acute{b}}^{^{(45)}}\sum_{y=1}^3\left[\left(\matrix{{\overline
\nu}& {\overline E}^-}\right)_{aL}\gamma^A{\bf
W}_A^{y}\frac{\tau_y}{2} \left(\matrix{\nu\cr
E^-}\right)_{bL}\right.\nonumber\\
 \left.+\left(\matrix{{\overline U}&
{\overline D}}\right)_{aL}\gamma^A{\bf W}_A^{y}\frac{\tau_y}{2}
\left(\matrix{U\cr D}\right)_{bL}\right]\nonumber\\
+g_{\acute{a}\acute{b}}^{^{(45)}}
\sqrt{\frac{3}{5}}\left[-\frac{1}{2}\left({\overline
E}_{aL}^-\gamma^A {\bf B}_AE^-_{bL}+{\overline \nu}_{aL}\gamma^A
{\bf B}_A\nu_{bL}\right) \right.\nonumber\\
\left.+\frac{1}{6}\left({\overline U}_{aL}^-\gamma^A {\bf
B}_AU_{bL}+{\overline D}_{aL}\gamma^A {\bf B}_AD_{bL}\right)
+\frac{2}{3}{\overline U}_{aR}\gamma^A {\bf
B}_AU_{bR}\right.\nonumber\\
\left.-\frac{1}{3}{\overline D}_{aR}\gamma^A {\bf
B}_AD_{bR}-{\overline E}_{aR}^-\gamma^A
{\bf B}_AE^-_{bR}\right]\nonumber\\
+...~~~~~~~~~~~~~~~~~~~~~~~~~~~~~~~~~~~~~~~~~~~~~~~~~~~~~~~~~~~~
\end{eqnarray}
where ${\bf V}_A^{x}$ is an SU(3) octet of gluons, ${\bf W}_A^{y}$
is an SU(2) isovector of intermediate bosons, ${\bf B}_A$ is the
hypercharge boson, ${\tau_y}$ and   ${\lambda_x}$ are the usual
Pauli and Gell-Mann matrices, and the dots stand for the couplings
of the lepto-quark/diquark bosons to fermions. The above result,
of course, contains the SM interactions. Next, let us look at the
vector interaction of the 24 plet of $SU(5)$ in the
$\overline{16}-16-210$ coupling. This can be read off from
Eq.(6.32) and one has
\begin{eqnarray}
{\cal L}_{24/210}= g_{\acute{a}\acute{b}}^{^{(210)}}
\left(-\overline {\bf M}_{\acute{a}}^j\gamma^A{\bf M}_{bi}
+\frac{1}{3}\overline{\bf M}_{aik}\gamma^A{\bf
M}_b^{kj}\right){\mathsf
G}_{Aj}^i\nonumber\\
=\frac{1}{3}\frac{g_{\acute{a}\acute{b}}^{^{(210)}}}{g_{\acute{a}\acute{b}}^{^{(45)}}}{\cal
L}_{24/45}
-\frac{4}{3}\{g_{\acute{a}\acute{b}}^{^{(210)}}\sum_{x=1}^8
\overline D_{aR}\gamma^A{\bf V}_A^{x}\frac{\lambda_x}{2}
D_{bR}\nonumber\\
+g_{\acute{a}\acute{b}}^{^{(210)}}
\sqrt{\frac{3}{5}}\left[-\frac{1}{2}\left({\overline
E}_{aL}^-\gamma^A {\bf B}_AE^-_{bL}+{\overline \nu}_{aL}\gamma^A
{\bf B}_A\nu_{bL}\right) -\frac{1}{3}{\overline D}_{aR}\gamma^A
{\bf B}_AD_{bR}\right]\nonumber\\
+g_{\acute{a}\acute{b}}^{^{(210)}}
\sum_{y=1}^3\left(\matrix{{\overline \nu}& {\overline
E}^-}\right)_{aL}\gamma^A{\bf W}_A^{y}\frac{\tau_y}{2}
\left(\matrix{\nu\cr E^-}\right)_{bL}+...\}
\end{eqnarray}
Eq.(6.36) shows that the 24 plet of $SU(5)$ couplings in
$\overline {16}-16-210$, unlike the case of the 24 plet couplings
in $\overline {16}-16-45$, do not contain the same exact
interactions as in the Standard Model.

\chapter{{Quartic Couplings of the form} $[\overline {\bf{16}}~\bf{16}][\overline
{\bf{16}}~\bf{16}]$}

In this chapter we determine dimension five operators in the
superpotential  arising from the mediation of 1-, 45- and
210-dimensional representations\cite{ns2}.

\section {Quartic interactions of the type \\
$[\overline {\bf{16}}~\bf{16}][\overline {\bf{16}}~\bf{16}]$}

Here we give the technique for the elimination of heavy fields for
the case when the fields belong to a large tensor representation.
There are in fact three approaches one can use in affecting this
elimination. The first one is the direct approach where one
eliminates the heavy large Higgs representation in its $SO(10)$
form. While this is the most straightforward  approach the
disadvantage is that the analysis of dimension 4 operators cannot
be directly made use of and one has to carry out the entire
computation from scratch. An alternative possibility is that one
utilizes the result of computations of dimension 4 operators
already done to compute dimension five operators. In this case,
however, since all the heavy Higgs fields are in their $SU(5)$
irreducible representations the elimination of such fields would
involve cross cancellations which are quite delicate. Thus, for
example, in its $SU(5)$ decomposition $210=1+5+\bar 5+10+
\overline{10}+24+40+\overline{40}+75$ and elimination of these
involve cancellations between the 10 and the 40 plet
contributions, between the $\overline {10}$ and the $\overline
{40}$ plet contributions, and between the 1, 24 and 75 plet
contributions. Such cancellations make the analysis tedious once
again. It turns out that there is yet a third possibility which is
to derive the dimension 4 operators in $SU(5)$ decomposition
leaving the $SU(5)$ fields in their reducible form where possible,
i.e., to use Eq.(4.7) without further reduction of the tensor
fields in their irreducible components. Thus, for example, in this
case one would carry out the following $SU(5)$ decomposition of
the $SO(10)$ tensor, $210=5+\bar 5+50+\overline{50}+100$  where $
50, ~\overline{50},~100$ are reducible $SU(5)$ representations.
 After computing the dimension 4 operators in terms of these tensors
  one eliminates them. This procedure has the advantage of having
  the cancellations of procedure 2 already built in.
We give now more details of the three approaches.
\newpage

\noindent \textbf{(a)}~~\textsc{direct method}\\

We begin by discussion of the first approach\cite{ns2} where one
eliminates the heavy fields in the superpotential before one
carries out an $SU(5)$ decomposition.

In phenomenological analysis one generally needs more than one
Higgs representations. Hence to keep the analysis very general we
not only keep the generational indices but also allow for mixing
among Higgs representations. To that end, we assume several Higgs
representations of the same kind: $\Phi_{\cal X}, \Phi_{\mu\nu
{\cal Y}}, \Phi_{\mu\nu\rho\lambda {\cal Z}}$. Consider the
superpotential
\begin{equation}
{\mathsf W}^{^{(16 \times \overline {16})}}= {\mathsf
W}_{Higgs}^{^{(16 \times \overline {16})}} +{\mathsf
W}_{mass}^{^{(16 \times \overline {16})}}
\end{equation}
where
\begin{equation}
{\mathsf W}_{Higgs}^{^{(16 \times \overline {16})}} ={\mathsf
W}^{(1)'}_{(-+)}+{\mathsf W}^{(45)'}_{(-+)}+{\mathsf
W}^{(210)'}_{(-+)}
\end{equation}
with ${\mathsf W}^{(1)'}_{(-+)},~{\mathsf
W}^{(45)'}_{(-+)},~{\mathsf W}^{(210)'}_{(-+)}$ given by
\begin{eqnarray}
{\mathsf
W}^{(1)'}_{(-+)}=h^{^{(1)}}_{\acute{a}\acute{b}}<\widehat{\Psi}_{(-)\acute{a}}^*|B|\widehat{\Psi}_{(+)\acute{b}}>k_{_{{\cal
X}}}^{^{(1)}}\Phi_{\cal X}\nonumber\\
{\mathsf
W}^{(45)'}_{(-+)}=\frac{1}{2!}h^{^{(45)}}_{\acute{a}\acute{b}}<\widehat{\Psi}_{(-)\acute{a}}^*|B
\Sigma_{\mu\nu}|\widehat{\Psi}_{(+)\acute{b}}>k_{_{{\cal
Y}}}^{^{(45)}} \Phi_{\mu\nu {\cal Y}}\nonumber\\
{\mathsf W}^{(210)'}_{(-+)}=\frac{1}{4!}
h^{^{(210)}}_{\acute{a}\acute{b}}<\widehat{\Psi}_{(-)\acute{a}}^*|B
\Gamma_{[\mu}\Gamma_{\nu}\Gamma_{\rho} \Gamma_{\lambda]}
|\widehat{\Psi}_{(+)\acute{b}}>k_{_{{\cal Z}}}^{^{(210)}}
\Phi_{\mu\nu\rho\lambda {\cal Z}}
\end{eqnarray}
and
\begin{equation}
{\mathsf W}_{mass}^{^{(16 \times \overline {16})}}
=\frac{1}{2}\Phi_{\cal X}{\cal M}^{^{(1)}}_{{\cal X}{\cal
X}'}\Phi_{{\cal X}'} +\frac{1}{2}\Phi_{\mu\nu {\cal Y}} {\cal
M}^{^{(45)}}_{{\cal Y}{\cal Y}'} \Phi_{\mu\nu {\cal Y}'}
+\frac{1}{2}\Phi_{\mu\nu\rho\lambda {\cal Z}} {\cal M}^{^{(210)}}
_{{\cal Z}{\cal Z}'} \Phi_{\mu\nu\rho\lambda {\cal Z}'}.
\end{equation}

We next eliminate $\Phi_{\cal X}, \Phi_{\mu\nu {\cal Y}},
\Phi_{\mu\nu\rho\lambda {\cal Z}}$ as superheavy dimension-5
operators using the F-flatness conditions:
\begin{equation}
\frac{\partial {\mathsf W}^{^{(16 \times \overline
{16})}}}{\partial \Phi_{\cal X}}=0;~~~ \frac{\partial {\mathsf
W}^{^{(16 \times \overline {16})}}}{\partial \Phi_{\mu\nu {\cal
Y}}}=0;~~~ \frac{\partial {\mathsf W}^{^{(16 \times \overline
{16})}}}{\partial \Phi_{\mu\nu\rho\lambda {\cal Z}}}=0.
\end{equation}
The above leads to
\begin{equation}
{\mathsf W}^{(\overline {16} \times 16)}_{dim-5} ={\cal
I}_{1}+{\cal I}_{45}+{\cal I}_{210}.
\end{equation}
where
\begin{eqnarray}
{\cal
I}_{1}=2\lambda_{\acute{a}\acute{b},\acute{c}\acute{d}}^{^{(1)}}
<\widehat{\Psi}^*_{(-)\acute{a}}|B|\widehat{\Psi}_{(+)\acute{b}}><\widehat{\Psi}^*_{(-)\acute{c}}|B|
\widehat{\Psi}_{(+)\acute{d}}>
\end{eqnarray}
\begin{eqnarray}
{\cal
I}_{45}=-\frac{1}{2}\lambda_{\acute{a}\acute{b},\acute{c}\acute{d}}^{^{(45)}}
\left[<\widehat{\Psi}^*_{(-)\acute{a}}|B\Gamma_{\mu}\Gamma_{\nu}|\widehat{\Psi}_{(+)\acute{b}}>
<\widehat{\Psi}^*_{(-)\acute{c}}|B\Gamma_{\mu}\Gamma_{\nu}|\widehat{\Psi}_{(+)\acute{d}}>\right.\nonumber\\
\left.-10<\widehat{\Psi}^*_{(-)\acute{a}}|B|\widehat{\Psi}_{(+)\acute{b}}><\widehat{\Psi}^*_{(-)\acute{c}}|B|
\widehat{\Psi}_{(+)\acute{d}}>\right]
\end{eqnarray}
\begin{eqnarray}
{\cal
I}_{210}=\frac{1}{288}\lambda_{\acute{a}\acute{b},\acute{c}\acute{d}}^{^{(210)}}~~~~~~~~~~~~~~~~~~~~~~~~~~~~~~~~~~~~~~~~~~~~~~~~\nonumber\\
\times\left[<\widehat{\Psi}^*_{(-)\acute{a}}|B\Gamma_{\mu}\Gamma_{\nu}\Gamma_{\rho}
\Gamma_{\lambda}|\widehat{\Psi}_{(+)\acute{b}}>
<\widehat{\Psi}^*_{(-)\acute{c}}|B\Gamma_{\mu}\Gamma_{\nu}\Gamma_{\rho}
\Gamma_{\lambda}|
\widehat{\Psi}_{(+)\acute{d}}>\right.\nonumber\\
\left.-52<\widehat{\Psi}^*_{(-)\acute{a}}|B\Gamma_{\mu}\Gamma_{\nu}|\widehat{\Psi}_{(+)\acute{b}}>
<\widehat{\Psi}^*_{(-)\acute{c}}|B\Gamma_{\mu}\Gamma_{\nu}|\widehat{\Psi}_{(+)\acute{d}}>\right.
\nonumber\\
\left.+240<\widehat{\Psi}^*_{(-)\acute{a}}|B|\widehat{\Psi}_{(+)\acute{b}}><\widehat{\Psi}^*_{(-)\acute{c}}
|B|\widehat{\Psi}_{(+)\acute{d}}>\right]
\end{eqnarray}
Writing, ${\cal I}_{45}$ and ${\cal I}_{210}$ in terms of creation
and annihilation operators
\begin{eqnarray}
{\cal
I}_{45}=\lambda_{\acute{a}\acute{b},\acute{c}\acute{d}}^{^{(45)}}
\left[-4<\widehat{\Psi}^*_{(-)\acute{a}}|Bb_ib_j|\widehat{\Psi}_{(+)\acute{b}}>
<\widehat{\Psi}^*_{(-)\acute{c}}|Bb_i^{\dagger}b_j^{\dagger}|\widehat{\Psi}_{(+)\acute{d}}>
\right.\nonumber\\
\left.+4<\widehat{\Psi}^*_{(-)\acute{a}}|Bb_i^{\dagger}b_j|\widehat{\Psi}_{(+)\acute{b}}>
<\widehat{\Psi}^*_{(-)\acute{c}}|Bb_j^{\dagger}b_i|\widehat{\Psi}_{(+)\acute{d}}>
\right.\nonumber\\
\left.-4<\widehat{\Psi}^*_{(-)\acute{a}}|Bb_n^{\dagger}b_n|\widehat{\Psi}_{(+)\acute{b}}>
<\widehat{\Psi}^*_{(-)\acute{c}}|B|\widehat{\Psi}_{(+)\acute{d}}>\right.\nonumber\\
\left.+5<\widehat{\Psi}^*_{(-)\acute{a}}|B|\widehat{\Psi}_{(-+)\acute{b}}><\widehat{\Psi}^*_{(-)\acute{c}}|B|
\widehat{\Psi}_{(+)\acute{d}}>\right]
\end{eqnarray}

\begin{eqnarray}
{\cal
I}_{210}=-\frac{1}{18}\lambda_{\acute{a}\acute{b},cd}^{^{(210)}}
\left[8<\widehat{\Psi}^*_{(-)\acute{a}}|Bb_i^{\dagger}b_jb_kb_l|\widehat{\Psi}_{(+)\acute{b}}>
<\widehat{\Psi}^*_{(-)\acute{c}}|Bb_j^{\dagger}b_k^{\dagger}b_l^{\dagger}b_i
|\widehat{\Psi}_{(+)\acute{d}}>
\right.\nonumber\\
\left.-6<\widehat{\Psi}^*_{(-)\acute{a}}|Bb_i^{\dagger}b_j^{\dagger}b_kb_l
|\widehat{\Psi}_{(+)\acute{b}}>
<\widehat{\Psi}^*_{(-)\acute{c}}|Bb_k^{\dagger}b_l^{\dagger}b_ib_j
|\widehat{\Psi}_{(+)\acute{d}}>
\right.\nonumber\\
\left.-2<\widehat{\Psi}^*_{(-)\acute{a}}|Bb_ib_jb_kb_l
|\widehat{\Psi}_{(+)\acute{b}}>
<\widehat{\Psi}^*_{(-)\acute{c}}|Bb_i^{\dagger}b_j^{\dagger}b_k^{\dagger}
b_l^{\dagger} |\widehat{\Psi}_{(+)\acute{d}}>
\right.\nonumber\\
\left.+24<\widehat{\Psi}^*_{(-)\acute{a}}|Bb_i^{\dagger}b_j
|\widehat{\Psi}_{(+)\acute{b}}>
<\widehat{\Psi}^*_{(-)\acute{c}}|Bb_j^{\dagger}b_n^{\dagger}b_nb_i
|\widehat{\Psi}_{(+)\acute{d}}>
\right.\nonumber\\
\left.-12<\widehat{\Psi}^*_{(-)\acute{a}}|Bb_i^{\dagger}b_j^{\dagger}
|\widehat{\Psi}_{(+)\acute{b}}>
<\widehat{\Psi}^*_{(-)\acute{c}}|Bb_n^{\dagger}b_nb_ib_j
|\widehat{\Psi}_{(+)\acute{d}}>
\right.\nonumber\\
\left.-12<\widehat{\Psi}^*_{(-)\acute{a}}|Bb_ib_j
|\widehat{\Psi}_{(+)\acute{b}}>
<\widehat{\Psi}^*_{(-)\acute{c}}|Bb_i^{\dagger}b_j^{\dagger}b_n^{\dagger}b_n
|\widehat{\Psi}_{(+)\acute{d}}>
\right.\nonumber\\
\left.-6<\widehat{\Psi}^*_{(-)\acute{a}}|Bb_m^{\dagger}b_m
|\widehat{\Psi}_{(+)\acute{b}}>
<\widehat{\Psi}^*_{(-)\acute{c}}|Bb_n^{\dagger}b_n
|\widehat{\Psi}_{(+)\acute{d}}>
\right.\nonumber\\
\left.-6<\widehat{\Psi}^*_{(-)\acute{a}}|B
|\widehat{\Psi}_{(+)\acute{b}}>
<\widehat{\Psi}^*_{(-)\acute{c}}|Bb_m^{\dagger}b_n^{\dagger}b_nb_m
|\widehat{\Psi}_{(+)\acute{d}}>
\right.\nonumber\\
\left.+18<\widehat{\Psi}^*_{(-)\acute{a}}|Bb_ib_j
|\widehat{\Psi}_{(+)\acute{b}}>
<\widehat{\Psi}^*_{(-)\acute{c}}|Bb_i^{\dagger}b_j^{\dagger}
|\widehat{\Psi}_{(+)\acute{d}}>
\right.\nonumber\\
\left.-18<\widehat{\Psi}^*_{(-)\acute{a}}|Bb_i^{\dagger}b_j
|\widehat{\Psi}_{(+)\acute{b}}>
<\widehat{\Psi}^*_{(-)\acute{c}}|Bb_j^{\dagger}b_i
|\widehat{\Psi}_{(+)\acute{d}}>
\right.\nonumber\\
\left.+24<\widehat{\Psi}^*_{(-)\acute{a}}|B
|\widehat{\Psi}_{(+)\acute{b}}>
<\widehat{\Psi}^*_{(-)\acute{c}}|Bb_n^{\dagger}b_n
|\widehat{\Psi}_{(+)\acute{d}}>
\right.\nonumber\\
\left.-15<\widehat{\Psi}^*_{(-)\acute{a}}|B
|\widehat{\Psi}_{(+)\acute{b}}>
<\widehat{\Psi}^*_{(-)\acute{c}}|B |\widehat{\Psi}_{(+)\acute{d}}>\right]\nonumber\\
\end{eqnarray}

Finally, evaluating the matrix elements in terms of $SU(5)$
fields, we obtain
\begin{eqnarray}
{\cal I}_{1}
=\frac{1}{2}\lambda_{\acute{a}\acute{b},\acute{c}\acute{d}}^{^{(1)}}
\left[-\widehat{\bf N}_{\acute{a}ij}^{\bf{T}}\widehat {\bf
M}^{ij}_{\acute{b}}\widehat{\bf N}_{\acute{c}kl}^{\bf{T}}\widehat
{\bf M}^{kl}_{\acute{d}}+4\widehat{\bf
N}^{i\bf{T}}_{\acute{a}}\widehat {\bf M}_{\acute{b}i}\widehat{\bf
N}_{\acute{c}jk}^{\bf{T}}\widehat {\bf M}^{jk}_{\acute{d}}
-4\widehat{\bf N}^{i\bf{T}}_{\acute{a}}\widehat {\bf
M}_{\acute{b}i}\widehat{\bf N}^{j\bf{T}}_{\acute{c}}
\widehat {\bf M}_{\acute{d}j}\right.\nonumber\\
\left.+4\widehat{\bf N}^{\bf{T}}_{\acute{a}}\widehat {\bf
M}_{\acute{b}}\widehat{\bf N}^{\bf{T}}_{\acute{c}ij}\widehat {\bf
M}^{ij}_{\acute{d}} -8\widehat{\bf N}^{\bf{T}}_{\acute{a}}\widehat
{\bf M}_{\acute{b}}\widehat{\bf N}^{i\bf{T}}_{\acute{c}}\widehat
{\bf M}_{\acute{d}i} -4\widehat{\bf
N}_{\acute{a}}^{\bf{T}}\widehat {\bf M}_{\acute{b}}\widehat{\bf
N}_{\acute{c}}^{\bf{T}}\widehat {\bf
M}_{\acute{d}}\right]\nonumber\\
\end{eqnarray}
\begin{eqnarray}
{\cal I}_{45}=
\left(\lambda_{\acute{a}\acute{d},\acute{c}\acute{b}}^{^{(45)}}+\lambda_{\acute{a}\acute{b},\acute{c}\acute{d}}^{^{(45)}}
\right)\left(8\widehat{\bf N}^{i\bf{T}}_{\acute{a}}\widehat {\bf
M}_{\acute{b}j}\widehat{\bf N}_{\acute{c}ik}^{\bf{T}}\widehat {\bf
M}^{kj}_{\acute{d}}-\widehat{\bf N}^{i\bf{T}}_{\acute{a}}\widehat
{\bf M}_{\acute{b}i}\widehat{\bf
N}^{j\bf{T}}_{\acute{c}}\widehat {\bf M}_{\acute{d}j}\right)\nonumber\\
+\left(4\lambda_{\acute{a}\acute{d},\acute{c}\acute{b}}^{^{(45)}}+\lambda_{\acute{a}\acute{b},\acute{c}\acute{d}}^{^{(45)}}
\right)\left(\widehat{\bf N}_{\acute{a}}^{i\bf{T}}\widehat {\bf
M}_{\acute{b}i}\widehat{\bf N}_{\acute{c}jk}^{\bf{T}}\widehat {\bf
M}^{jk}_{\acute{d}}+ \widehat{\bf N}_{\acute{a}}^{\bf{T}}\widehat
{\bf M}_{\acute{b}}\widehat{\bf
N}_{\acute{c}ij}^{\bf{T}}\widehat {\bf M}^{ij}_{\acute{d}}\right)\nonumber\\
+\frac{1}{4}\lambda_{\acute{a}\acute{b},\acute{c}\acute{d}}^{^{(45)}}
\left[-8\epsilon^{ijklm}\widehat{\bf
N}_{\acute{a}ij}^{\bf{T}}\widehat {\bf M}_{\acute{b}k}\widehat{\bf
N}_{\acute{c}lm}^{\bf{T}}\widehat {\bf
M}_{\acute{d}}-8\epsilon_{ijklm}\widehat{\bf
N}_{\acute{a}}^{\bf{T}}\widehat {\bf
M}^{ij}_{\acute{b}}\widehat{\bf N}^{k\bf{T}}_{\acute{c}}\widehat {\bf M}^{lm}_{\acute{d}}\right.\nonumber\\
\left.-16\widehat{\bf N}_{\acute{a}ik}^{\bf{T}}\widehat {\bf
M}^{kj}_{\acute{b}}\widehat{\bf N}_{\acute{c}jl}^{\bf{T}}\widehat
{\bf M}^{li}_{\acute{d}} +3\widehat{\bf
N}_{\acute{a}ij}^{\bf{T}}\widehat {\bf
M}^{ij}_{\acute{b}}\widehat{\bf N}_{\acute{c}kl}^{\bf{T}}\widehat
{\bf M}^{kl}_{\acute{d}}+24\widehat{\bf
N}_{\acute{a}}^{\bf{T}}\widehat {\bf M}_{\acute{b}}\widehat{\bf
N}^{i\bf{T}}_{\acute{c}}\widehat {\bf M}_{\acute{d}i}\right.\nonumber\\
\left.-20\widehat{\bf N}_{\acute{a}}^{\bf{T}} \widehat {\bf
M}_{\acute{b}}\widehat{\bf N}^{\bf{T}}_{\acute{c}}\widehat {\bf
M}_{\acute{d}}\right]
\end{eqnarray}

\begin{eqnarray}
{\cal I}_{210}
=-\frac{1}{24}\left[4\left(6\lambda_{\acute{a}\acute{d},\acute{c}\acute{b}}^{^{(210)}}-
\lambda_{\acute{a}\acute{b},\acute{c}\acute{d}}^{^{(210)}}
\right)\left(\widehat{\bf N}^{i\bf{T}}_{\acute{a}}\widehat {\bf
M}_{\acute{b}i}\widehat{\bf N}^{j\bf{T}}_{\acute{c}}\widehat {\bf
M}_{\acute{d}j}-\widehat{\bf N}_{\acute{a}}^{i\bf{T}}\widehat {\bf
M}_{\acute{b}i}\widehat{\bf N}_{\acute{c}jk}^{\bf{T}}\widehat {\bf
M}^{jk}_{\acute{d}}\right.\right.\nonumber\\
\left.\left.-\widehat{\bf N}_{\acute{a}}^{\bf{T}}\widehat {\bf
M}_{\acute{b}}\widehat{\bf N}_{\acute{c}ij}^{\bf{T}}\widehat {\bf
M}^{ij}_{\acute{d}}\right)
+16\left(\lambda_{\acute{a}\acute{d},\acute{c}\acute{b}}^{^{(210)}}+\lambda_{\acute{a}\acute{b},\acute{c}\acute{d}}^{^{(210)}}
\right)\widehat{\bf N}^{i\bf{T}}_{\acute{a}}\widehat {\bf
M}_{\acute{b}j}\widehat{\bf N}_{\acute{c}ik}^{\bf{T}}\widehat {\bf
M}^{kj}_{\acute{d}}\right.\nonumber\\
\left.+\left(8\lambda_{\acute{a}\acute{d},\acute{c}\acute{b}}^{^{(210)}}+\lambda_{\acute{a}\acute{b},\acute{c}\acute{d}}^{^{(210)}}\right)
\left(\widehat{\bf N}_{\acute{a}ij}^{\bf{T}}\widehat {\bf
M}^{ij}_{\acute{b}}\widehat{\bf N}_{\acute{c}kl}^{\bf{T}}\widehat
{\bf M}^{kl}_{\acute{d}} +8 \widehat{\bf
N}_{\acute{a}}^{\bf{T}}\widehat {\bf M}_{\acute{b}}\widehat{\bf
N}^{i\bf{T}}_{\acute{c}}\widehat {\bf M}_{\acute{d}i}\right)\right.\nonumber\\
\left.+4\lambda_{\acute{a}\acute{b},\acute{c}\acute{d}}^{^{(210)}}
\left\{-\epsilon^{ijklm}\widehat{\bf
N}_{\acute{a}ij}^{\bf{T}}\widehat {\bf M}_{\acute{b}k}\widehat{\bf
N}_{\acute{c}lm}^{\bf{T}}\widehat {\bf M}_{\acute{d}}
-\epsilon_{ijklm}\widehat{\bf N}_{\acute{a}}^{\bf{T}}\widehat {\bf
M}^{ij}_{\acute{b}}\widehat{\bf
N}^{k\bf{T}}_{\acute{c}}\widehat {\bf M}^{lm}_{\acute{d}} \right.\right.\nonumber\\
\left.\left.-2\widehat{\bf N}_{\acute{a}ik}^{\bf{T}}\widehat {\bf
M}^{kj}_{\acute{b}}\widehat{\bf N}_{\acute{c}jl}^{\bf{T}}\widehat
{\bf M}^{li}_{\acute{d}} +5\widehat{\bf N}_{\acute{a}}^{\bf{T}}
\widehat {\bf M}_{\acute{b}}\widehat{\bf
N}^{\bf{T}}_{\acute{c}}\widehat {\bf
M}_{\acute{d}}\right\}\right]\nonumber\\
\end{eqnarray}

where
\begin{eqnarray}
\lambda_{\acute{a}\acute{b},\acute{c}\acute{d}}^{^{(1)}}=
h_{\acute{a}\acute{b}}^{^{(1)}}h_{\acute{c}\acute{d}}^{^{(1)}}k_{_{{\cal
X}}}^{^{(1)}}\left[\widetilde{{\cal M}}^{^{(1)}}\left\{{\cal
M}^{^{(1)}}\widetilde{{\cal M}}^{^{(1)}} -\bf{1}\right\}
\right]_{{\cal X}{\cal X}'}k_{_{{\cal X}'}}^{^{(1)}}\nonumber\\
\lambda_{\acute{a}\acute{b},\acute{c}\acute{d}}^{^{(45)}}=
h_{\acute{a}\acute{b}}^{^{(45)}}h_{\acute{c}\acute{d}}^{^{(45)}}k_{_{{\cal
Y}}}^{^{(45)}}\left[\widetilde{{\cal M}}^{^{(45)}}\left\{{\cal
M}^{^{(45)}}\widetilde{{\cal M}}^{^{(45)}} -\bf{1}\right\}
\right]_{{\cal Y}{\cal Y}'}k_{_{{\cal Y}'}}^{^{(45)}}\nonumber\\
\lambda_{\acute{a}\acute{b},\acute{c}\acute{d}}^{^{(210)}}=
h_{\acute{a}\acute{b}}^{^{(210)}}h_{\acute{c}\acute{d}}^{^{(210)}}k_{_{{\cal
Z}}}^{^{(210)}}\left[\widetilde{{\cal M}}^{^{(45)}}\left\{{\cal
M}^{^{(210)}}\widetilde{{\cal M}}^{^{(210)}} -\bf{1}\right\}
\right]_{{\cal Z}{\cal Z}'}k_{_{{\cal Z}'}}^{^{(210)}}\nonumber\\
\widetilde{{\cal M}}^{^{(.)}}=\left[{\cal M}^{^{(.)}}+\left({\cal
M}^{^{(.)}}\right)^{\bf {T}}\right]^{-1}
\end{eqnarray}
Although this is the most straight forward technique, one has to
 carry out the entire analysis ab initio and can be very labor intensive
 for the case of large tensor representations.

 After spontaneous breaking, the Higgs multiplet can develop vacuum
expectation values generating mass terms for some of the quark,
lepton and neutrino fields. An example of such operators can be
found in Ref.\cite{bpw}. A further discussion of model building is
given in Section 7.2.\\

\noindent \textbf{(b)}~~\textsc{indirect method I}\\

We discuss now the second approach\cite{ns2} where one decomposes
the large tensor representations in its irreducible $SU(5)$
components and utilizes the results of the cubic superpotential
already computed to derive  dimension five operators. For
illustration we consider the elimination of the 45 plet in the
$\overline {16}-16-45$ (see Eq.(6.10)) coupling and for simplicity
we consider only one generation of Higgs.
 We begin by displaying the 45 plet mass term in terms of its irreducible
$SU(5)$ components
\begin{eqnarray}
\frac{1}{2}{\cal M}^{^{(45)}}\Phi_{\mu\nu}\Phi_{\mu\nu}
=\frac{1}{2}{\cal M}^{^{(45)}} \left[{\mathsf H}^{ij}{\mathsf
H}_{ij} -{\mathsf H}^i_j{\mathsf H}^j_i -{\mathsf H}^2\right].
\end{eqnarray}
The superpotential is given by
\begin{eqnarray}
{\mathsf W}^{(45)}_{(-+)}= J^{(1/45)}{\mathsf H} +J^{(\overline
{10}/45)ij}{\mathsf H}_{ij} +J^{(10/45)}_{ij}{\mathsf H}^{ij}
+J^{(24/45)j}_i{\mathsf H}_j^i
\end{eqnarray}
where
\begin{eqnarray}
J^{(1/45)}
=\sqrt{\frac{5}{2}}h_{\acute{a}\acute{b}}^{^{(45)}}\left(\frac{3}{5}\widehat
{\bf N}_{\acute{a}}^{i\bf{T}}\widehat {\bf
M}_{bi}+\frac{1}{10}\widehat {\bf N}^{\bf{T}}_{aij}\widehat {\bf
M}_{\acute{b}}^{ij}-
\widehat {\bf N}_{\acute{a}}^{\bf{T}}\widehat {\bf M}_{\acute{b}}\right)\nonumber\\
J^{(\overline {10}/45)lm}
=\frac{h_{\acute{a}\acute{b}}^{^{(45)}}}{\sqrt 2}\left(-\widehat
{\bf N}_{\acute{a}}^{\bf{T}}\widehat {\bf
M}_{\acute{b}}^{lm}+\frac{1}{2}\epsilon^{ijklm} \widehat
{\bf N}^{\bf{T}}_{alm}\widehat {\bf M}_{bk}\right)\nonumber\\
J^{(10/45)}_{lm}
=\frac{h_{\acute{a}\acute{b}}^{^{(45)}}}{\sqrt2}\left(-\widehat
{\bf N}^{\bf{T}}_{alm}\widehat {\bf M}_{\acute{b}}+
\frac{1}{2}\epsilon_{ijklm}\widehat
{\bf N}_{\acute{a}}^{i\bf{T}}\widehat {\bf M}_{\acute{b}}^{jk}\right)\nonumber\\
J^{(24/45)j}_i=\sqrt {2}h_{\acute{a}\acute{b}}^{^{(45)}}
\left(\widehat {\bf N}^{\bf{T}}_{aik}\widehat {\bf
M}_{\acute{b}}^{kj}-\widehat {\bf N}_{\acute{a}}^{j\bf{T}}\widehat
{\bf M}_{bi}\right).
\end{eqnarray}
Eliminating the irreducible $SU(5)$ heavy Higgs fields through
F-flatness conditions taking care of the tracelessnes condition
for $H_i^j$ one gets
\begin{eqnarray}
{\mathsf I}_{45}=\frac{1}{{10\cal M}^{^{(45)}}}
[5J^{(1/45)}J^{(1/45)}
-20J^{(\overline {10}/45)ij}J^{(10/45)}_{ij}\nonumber\\
+5J^{(24/45)j}_iJ^{(24/45)i}_j -J^{(24/45)m}_mJ^{(24/45)n}_n].
\end{eqnarray}
${\mathsf I}_{45}$ computed above is the same as ${\cal I}_{45}$
given by Eq.(7.13) using the direct method with
$\frac{h_{\acute{a}\acute{b}}^{^{(45)}}h_{\acute{c}\acute{d}}^{^{(45)}}}{{\cal
M}^{^{(45)}}}$ replaced by
$-4\lambda_{\acute{a}\acute{b},cd}^{^{(45)}}$. As pointed out in
the beginning of this section one has cancellations in this
procedure between the contributions arising from elimination of
the 1 plet and the 24 plet. Such cancellations become more
abundant for the
210 plet case.\\

\noindent \textbf{(c)}~~\textsc{indirect method II}\\

It is more convenient to decompose the 210 plet into reducible
$SU(5)$ tensors\cite{ns2}.  We begin by exhibiting the mass term
for this case
\begin{eqnarray}
\frac{1}{2}{\cal
M}^{^{(210)}}\Phi_{\mu\nu\rho\lambda}\Phi_{\mu\nu\rho\lambda}
=\frac{1}{4}{\cal M}^{^{(210)}}\left[\frac{1}{4}{\mathsf
K}^{ijkl}{\mathsf K}_{ijkl} +{\mathsf K}^{jkl}_i{\mathsf
K}_{jkl}^i +\frac{3}{4}{\mathsf K}^{kl}_{ij}{\mathsf
K}_{kl}^{ij}\right]
\end{eqnarray}
where ${\mathsf K}^{ijkl}$, ${\mathsf K}_{ijkl}$, ${\mathsf
K}^{jkl}_i$, ${\mathsf K}_{jkl}^i$ and ${\mathsf K}_{kl}^{ij}$ are
the $5$ plet, $\bar 5$ plet, $50$ plet, $\overline {50}$ plet and
$100$ plet representations of $SU(5)$. As before we keep only one
generation of Higgs. The superpotential ${\mathsf
W}^{210)}_{(-+)}$ in this case may be written as
\begin{eqnarray}
{\mathsf W}^{210)}_{(-+)}= J^{(\overline {5}/210)}_{ijkl}{\mathsf
K}^{ijkl} +J^{(5/210)ijkl}{\mathsf K}_{ijkl}
+J^{(50/210)l}_{ijk}{\mathsf K}_l^{ijk}
+J^{(\overline {50}/210)ijk}_l{\mathsf K}^l_{ijk}\nonumber\\
+J^{(50/210)}_{ij}{\mathsf K}^{ijn}_n +J^{(\overline
{50}/210)ij}{\mathsf K}^n_{ijn} +J^{(100/210)ij}_{kl}{\mathsf
K}^{kl}_{ij}
+J^{(100/210)j}_{i}{\mathsf K}^{in}_{jn}\nonumber\\
+J^{(100/210)}{\mathsf K}^{mn}_{mn}~~~~~~
\end{eqnarray}
where
\begin{eqnarray}
J_{ijkl}^{(\bar 5/210)}
=\frac{h_{\acute{a}\acute{b}}^{^{(210)}}}{24}
<\widehat{\Psi}^*_{(-)\acute{a}}|Bb_i^{\dagger}b_j^{\dagger}b_k^{\dagger}
b_l^{\dagger}
|\widehat{\Psi}_{(+)\acute{b}}>\nonumber\\
J^{(5/210)ijkl}=
\frac{h_{\acute{a}\acute{b}}^{^{(210)}}}{24}<\widehat{\Psi}^*_{(-)\acute{a}}|Bb_ib_jb_kb_l
|\widehat{\Psi}_{(+)\acute{b}}>\nonumber\\
J^{(50/210)l}_{ijk}=\frac{h_{\acute{a}\acute{b}}^{^{(210)}}}{6}
<\widehat{\Psi}^*_{(-)\acute{a}}|Bb_i^{\dagger}b_j^{\dagger}
b_k^{\dagger}b_l |\widehat{\Psi}_{(+)\acute{b}}>
\nonumber\\
J^{(\overline
{50}/210)jkl}_i=-\frac{h_{\acute{a}\acute{b}}^{^{(210)}}}{6}
<\widehat{\Psi}^*_{(-)\acute{a}}|Bb_i^{\dagger}b_jb_kb_l|\widehat{\Psi}_{(+)\acute{b}}>
\nonumber\\
J^{(50/210)}_{ij}=-\frac{h_{\acute{a}\acute{b}}^{^{(210)}}}{4}
<\widehat{\Psi}^*_{(-)\acute{a}}|Bb_i^{\dagger}b_j^{\dagger}
|\widehat{\Psi}_{(+)\acute{b}}>\nonumber\\
J^{(\overline
{50}/210)ij}=\frac{h_{\acute{a}\acute{b}}^{^{(210)}}}{4}
<\widehat{\Psi}^*_{(-)\acute{a}}|Bb_ib_j
|\widehat{\Psi}_{(+)\acute{b}}>\nonumber\\
J^{(100/210)kl}_{ij}=\frac{h_{\acute{a}\acute{b}}^{^{(210)}}}{4}
\widehat{\Psi}^*_{(-)\acute{a}}|Bb_i^{\dagger}b_j^{\dagger}b_kb_l
|\widehat{\Psi}_{(+)\acute{b}}>\nonumber\\
J^{(100/210)j}_{i}= \frac{h_{\acute{a}\acute{b}}^{^{(210)}}}{2}
<\widehat{\Psi}^*_{(-)\acute{a}}|Bb_i^{\dagger}b_j
|\widehat{\Psi}_{(-+)\acute{b}}>\nonumber\\
J^{(100/210)}=-\frac{h_{\acute{a}\acute{b}}^{^{(210)}}}{8}
<\widehat{\Psi}^*_{(-)\acute{a}}|B
|\widehat{\Psi}_{(-+)\acute{b}}>.
\end{eqnarray}
Eliminating the reducible $SU(5)$ Higgs fields through the
F-flatness condition we get
\begin{eqnarray}
{\mathsf I}_{210}=-\frac{1}{{3\cal M}^{^{(210)}}}[4
J^{(100/210)ij}_{kl}J^{(100/210)kl}_{ij}+8
J^{(100/210)mi}_{mj}J^{(100/210)j}_{i}\nonumber\\
+8J^{(100/210)mn}_{mn}J^{(100/210)}
+3J^{(100/210)j}_{i}J^{(100/210)i}_{j}\nonumber\\
+J^{(100/210)m}_{m}J^{(100/210)n}_{n}
+16J^{(100/210)m}_{m}J^{(100/210)} \nonumber\\
+40J^{(100/210)}J^{(100/210)}
+48J^{(5/210)ijkl}J^{({\overline 5}/210)}_{ijkl}\nonumber\\
+12J^{(\overline{50}/210)ijk}_lJ^{(50/210)l}_{ijk}
+12J^{(\overline{50}/210)mij}_mJ^{(50/210)}_{ij}\nonumber\\
+12J^{(\overline{50}/210)ij}J^{(50/210)m}_{ijm}
+12J^{(\overline{50}/210)ij}J^{(50/210)}_{ij}].
\end{eqnarray}
One may now check that ${\mathsf I}_{210}$ derived  above
coincides with ${\cal I}_{210}$ given by Eq.(7.11) using the
direct method when we make the identification
$\frac{h_{\acute{a}\acute{b}}^{^{(210)}}h_{\acute{c}\acute{d}}^{^{(210)}}}{{\cal
M}^{^{(210)}}}$ with
$-4\lambda_{\acute{a}\acute{b},cd}^{^{(210)}}$.

\section{Possible role of large tensor
    representations in model building}
 Most of the model building in $SO(10)$\cite{ns2} has occured using small
 Higgs representations and large
 representations are generally avoided as they lead to
non-perturbative physics above the grand unified scale. However,
for the purposes of physics below the grand unified scale, the
existence of non-perturbativity above the unified scale is not a
central concern since the region above this scale in any case
cannot be fully understood without taking into account quantum
gravity effects. Thus there is no fundamental reason not to
consider
 model building which allows for couplings with
large tensor representations. Indeed large tensor representations
have some very interesting and desirable features. Thus, for
example, if the $\overline{126}$ develops a VEV in the direction
of $\overline{45}$ of $SU(5)$ one can get  the ratio 3:1 in the
"22" element of the lepton vs. the down quark sector in a natural
fashion  as desired in the Georgi-Jarlskog textures\cite{gj}. A
similar 3:1 ratio also appears in the 120 plet couplings. Because
of this feature the tensor representations $120$ and
$\overline{126}$ have already appeared in several analyses of
lepton and quark textures\cite{gn}. Further, it was pointed out in
Ref.\cite{ns1} that the tensor representation $\overline{126}$ may
also play a  role in suppressing proton decay arising from
dimension five operators in supersymmetric models. This is so
because couplings involving $\overline{126}$ plet of Higgs to 16
plet of matter do not give rise to dimension five operators. The
result derived here including the computation of cubic and quartic
couplings may find application also in the study of neutrino
masses and mixings. Thus, for example, one may consider
contributions to the neutrino mass (N) and to the up quark mass
(U) from the contraction
$[16_{\acute{a}}\overline{16}_H]_{45}[16_{\acute{b}}\overline{16}_H]_{45}$.
 From Eq.(7.13) we
find that a contribution to N arises from the fifth term in the
brackets of Eq.(7.13) while the contribution to U arises from the
second term in the brackets of Eq.(7.13). Now comparing the above
with Eq.(8) of Ref.\cite{ns1} for the 10 plet Higgs coupling which
gives a N:U ratio of 1:1 we find that the two couplings referred
to above in Eq.(7.13) give N:U=3:8 in agreement with
Ref.\cite{babu}. Regarding the 210 dimensional tensor, such a
mutiplet could play a role in the quark-lepton and neutrino mass
textures. The role of a 210 dimensional vector multiplet is less
clear.  One possible way it may surface in low energy physics is
as a condensate field. However, this topic needs further
exploration.
\section{Appendix}
In this appendix we prove that the
$[16_{\acute{a}}\overline{16}_H]_{45}[16_{\acute{b}}\overline{16}_H]_{45}$
interaction gives a ratio 3:8 for the Dirac neutrino mass, N vs
the up quark mass, U.

We begin with the well known result that the Dirac neutrino mass
and the up quark mass get equal contributions if the coupling is
via 10 plet of $SO(10)$. We will use this result in our proof.
Next consider Eq. (5.6):
$${\mathsf W}_{(++)}^{(10)}=i2\sqrt 2 {f}^{^{(10)(+)}}_{\acute{a}\acute{b}}[\widehat {\bf M}_{\acute{a}}^{ij\bf{T}}
  \widehat {\bf M}_{\acute{b}i}{\mathsf H}_{j}
  +\left(-\widehat {\bf M}_{\acute{a}}^{\bf{T}}\widehat {\bf M}_{\acute{b}m} + \frac{1}{8}\epsilon_{ijklm}
  \widehat {\bf M}_{\acute{a}}^{ij\bf{T}}\widehat {\bf M}_{\acute{b}}^{kl}\right)
  {\mathsf H}^{m}]$$

When ${\mathsf H}^{m}$ develops a VEV, we find the Dirac neutrino
mass from \\
$-\widehat {\bf M}_{\acute{a}}^{\bf{T}}\widehat {\bf
M}_{\acute{b}m}<{\mathsf H}^{m}>$ term and the up quark mass
from the last term \\
$\frac{1}{8}\epsilon_{ijklm}
  \widehat {\bf M}_{\acute{a}}^{ij\bf{T}}\widehat {\bf M}_{\acute{b}}^{kl}<{\mathsf
  H}^{m}>$.
  Let us write
  $$x\equiv -\widehat {\bf M}_{\acute{a}}^{\bf{T}}\widehat
{\bf M}_{\acute{b}m}{\mathsf H}^{m}$$
$$y\equiv \frac{1}{8}\epsilon_{ijklm}
  \widehat {\bf M}_{\acute{a}}^{ij\bf{T}}\widehat {\bf M}_{\acute{b}}^{kl}{\mathsf
  H}^{m}$$
  Then we see that
$$x+y\rightarrow N:U=1:1$$
Next we look at Eq. (7.13). Further we are looking for
combinations which involve $x$ and $y$. We see that the relevant
terms are the 2nd and the 5th terms in the brackets of Eq. (7.13)
i.e.,
$$-8\epsilon_{ijklm}\widehat{\bf
N}_{\acute{a}}^{\bf{T}}\widehat {\bf
M}^{ij}_{\acute{b}}\widehat{\bf N}^{k\bf{T}}_{\acute{c}}\widehat
{\bf M}^{lm}_{\acute{d}}+24\widehat{\bf
N}_{\acute{a}}^{\bf{T}}\widehat {\bf M}_{\acute{b}}\widehat{\bf
N}^{i\bf{T}}_{\acute{c}}\widehat {\bf M}_{\acute{d}i}\ $$

Using the $x$ and $y$ notation we can write the above as

$${\bf N}_{\acute{a}}^{\bf{T}}\left(64y+24x\right)$$

Thus when the singlet Higgs field in $16_H$ i.e., ${\bf
N}_{\acute{a}}^{\bf{T}}$ develops a VEV we have
$$24x+64y\rightarrow N:U=3:8$$

\chapter{SO(10) Supersymmetric Singlet and 45 plet vector couplings~}
In this chapter we present the complete supersymmetric
couplings\cite{ns3} containing the singlet and 45 of $SO(10)$.

\section{Singlet vector coupling }
We begin with the  supersymmetric vector couplings for the singlet
of $SO(10)$ in the Wess-Zumino gauge:
\begin{equation}
{\mathsf L}={\mathsf L}_{V}^{^{(1~K.E.)}} +{\mathsf
L}_{V+\Phi}^{^{(1~Interaction)}}+{\mathsf L}_{\mathsf W}^{(1)}
\end{equation}
where
\begin{equation}
{\mathsf L}_{V}^{^{(1~K.E.)}}=\frac{1}{64}\left[\widehat{\cal
W}^{\tilde{\alpha}} \widehat{\cal
W}_{\tilde{\alpha}}|_{\theta^2}+\widehat{\overline{\cal
W}}_{\dot{\tilde{\alpha}}}\widehat{\overline{\cal
W}}^{\dot{\tilde{\alpha}}}|_{\bar {\theta}^2}\right]
\end{equation}
\begin{equation}
\widehat{\cal W}^{\tilde{\alpha}} =\overline {\mathsf D}^2\mathsf
D_{\tilde{\alpha}}\widehat{\mathsf V}
\end{equation}
\begin{equation}
{\mathsf
D}_{\tilde{\alpha}}=\frac{\partial}{\partial\theta^{\tilde{\alpha}}}+
i\sigma_{\tilde{\alpha}{\dot{\tilde {\alpha}}}}^A\overline
{\theta}^{\dot{\tilde {\alpha}}}\partial_A;~~~~ \overline {\mathsf
D}_{\dot{\tilde{\alpha}}}=-\frac{\partial}{\partial\overline
{\theta}^{\dot{\tilde {\alpha}}}}-i\theta ^{\tilde{\alpha}}
\sigma_{\tilde{\alpha}\dot{\tilde{\alpha}}}^A\partial_A
\end{equation}
  and $\widehat{\mathsf V}$ is the vector superfield in the Wess-Zumino gauge and is given by
\begin{equation}
\widehat{\mathsf V}=-\theta\sigma^A\bar {\theta}{\cal V}_{A}
+i\theta^2\bar {\theta}\overline {\lambda} -i\bar
{\theta}^2\theta\lambda +\frac{1}{2}\theta^2\bar {\theta}^2D
\end{equation}
   Thus we have
\begin{eqnarray}
{\mathsf L}_{V}^{^{(1~K.E.)}}=-\frac{1}{4}{\cal V}_{AB}{\cal
V}^{AB}-\frac{i}{2}\overline {\Lambda}\gamma^A{\cal
D}_A\Lambda+{\mathsf L}_{(1)auxiliary}^{(1)}\\
{\cal V}^{AB}=\partial^A{\cal V}^B-\partial^B{\cal
V}^A\\
{\mathsf L}_{(1)auxiliary}^{(1)}=\frac{1}{2}D^2\\
\Lambda=\left(\matrix{\lambda_{\tilde{\alpha}}\cr
\overline{\lambda}^{\dot{\tilde\alpha}}}\right)
\end{eqnarray}
In the above the Greek letters with tilde's ($\tilde{\alpha}$,
$\dot{\tilde {\beta}}$, ...) are Weyl indices. Further,
\begin{eqnarray}
{\mathsf
L}_{V+\Phi}^{^{(1~Interaction)}}=h_{\acute{a}\acute{b}}^{^{(1+)}}<\widehat{\Phi}_{(+)\acute{a}}|
e^{{\mathsf g}^{^{(1)}}q^{^{(+)}}\widehat{{\mathsf V}}
}|\widehat{\Phi}_{(+)\acute{b}}>|_{\theta^2\bar
{\theta}^2}\nonumber\\
+h_{\acute{a}\acute{b}}^{^{(1-)}}<\widehat{\Phi}_{(-)\acute{a}}|e^{{\mathsf
g}^{^{(1)}}q^{^{(-)}}\widehat{{\mathsf V}}
}|\widehat{\Phi}_{(-)\acute{b}}>|_{\theta^2\bar {\theta}^2}
\end{eqnarray}
where $q^{^{(\pm)}}$ are the U(1) charges and
$\widehat{\Phi}_{(\pm)\acute{a}}$ are  $16$ and $\overline{16}$
chiral superfields and are given by
\begin{eqnarray}
\widehat{\Phi}_{(+)\acute{a}}=A_{\acute{a}}(x)+\sqrt{2}\theta\psi_{\acute{a}}+\theta^2F_{\acute{a}}(x)+i\theta\sigma^A\bar
{\theta}\partial_AA_{\acute{a}}(x)\nonumber\\
+\frac{i}{\sqrt{2}}\theta^2\bar {\theta}\overline
{\sigma}^A\partial_A\psi_{\acute{a}}(x)+\frac{1}{4}\theta^2\bar
{\theta}^2\partial_A\partial^AA_{\acute{a}}(x)\nonumber\\
\widehat{\Phi}_{(+
)\acute{a}}^{\dagger}=A_{\acute{a}}^{\dagger}(x)+\sqrt{2}\bar{\theta}\overline{\psi}
_{\acute{a}}+\bar{\theta}^2F_{\acute{a}}^{\dagger}(x)-i\theta\sigma^A\bar
{\theta}\partial_AA_{\acute{a}}^{\dagger}(x)\nonumber\\
+\frac{i}{\sqrt{2}}\bar{\theta}^2 {\theta}
{\sigma}^A\partial_A\overline{\psi}_{\acute{a}}(x)+\frac{1}{4}\theta^2\bar
{\theta}^2\partial_A\partial^AA_{\acute{a}}^{\dagger}(x)
\end{eqnarray}
Expanding $e^{{\mathsf g}^{^{(1)}}q^{^{(\pm)}}}$ we have
\begin{eqnarray}
{\mathsf
L}_{V+\Phi}^{^{(1~Interaction)}}=h_{\acute{a}\acute{b}}^{^{(1+)}}\left[<\widehat{\Phi}_{(+)\acute{a}}|
\widehat{\Phi}_{(+)\acute{b}}>+{\mathsf
g}^{^{(1)}}q^{^{(+)}}<\widehat{\Phi}_{(+)\acute{a}}|{\widehat{\mathsf
V}}|\widehat{\Phi}_{(+)\acute{b}}>\right.\nonumber\\
\left.+\frac{1}{2}{\mathsf g}^{^{(1)}2}q^{^{(+)}2}
<\widehat{\Phi}_{(+)\acute{a}}|{\widehat{\mathsf
V}}^2|\widehat{\Phi}_{(+)\acute{b}}>\right]|_{\theta^2\bar
{\theta}^2}
+h_{\acute{a}\acute{b}}^{^{(1-)}}\left[<\widehat{\Phi}_{(-)\acute{a}}|
\widehat{\Phi}_{(-)\acute{b}}>\right.\nonumber\\
\left.+{\mathsf
g}^{^{(1)}}q^{^{(-)}}<\widehat{\Phi}_{(-)\acute{a}}|{\widehat{\mathsf
V}}|\widehat{\Phi}_{(-)\acute{b}}> +\frac{1}{2}{\mathsf
g}^{^{(1)}2}q^{^{(-)}2}<\widehat{\Phi}_{(-)\acute{a}}|{\widehat{\mathsf
V}}^2|\widehat{\Phi}_{(-)\acute{b}}>\right]|_{\theta^2\bar
{\theta}^2}
\end{eqnarray}
where the quantities entering Eq.(8.12) are determined by
Eqs.(8.13) -(8.26) below
\begin{eqnarray}
h_{\acute{a}\acute{b}}^{^{(1+)}}<\widehat{\Phi}_{(+)\acute{a}}|
\widehat{\Phi}_{(+)\acute{b}}>|_{\theta^2\bar
{\theta}^2}=h_{\acute{a}\acute{b}}^{^{(1+)}}\left[-\partial_A{\bf
A}^{\dagger}_{(+)\acute{a}}\partial^A{\bf A}_{(+)\acute{b}}
-\partial_A{\bf A}^{i\dagger}_{(+)\acute{a}}\partial^A{\bf A}_{(+)\acute{b}i}\right.\nonumber\\
\left.-\partial_A{\bf A}^{\dagger}_{(+)\acute{a}ij}\partial^A{\bf
A}_{(+)\acute{b}}^{ij}-i\overline{\bf
{\Psi}}_{(+)\acute{a}L}\gamma^A\partial_A{\bf
{\Psi}}_{(+)\acute{b}L}\right.\nonumber\\
\left.-i\overline{\bf
{\Psi}}_{(+)\acute{a}L}^i\gamma^A\partial_A{\bf
{\Psi}}_{(+)\acute{b}iL}-i\overline{\bf
{\Psi}}_{(+)\acute{a}ijL}\gamma^A\partial_A{\bf
{\Psi}}_{(+)\acute{b}L}^{ij}\right] +{\mathsf
L}_{(2)auxiliary}^{(1)}~~~~~~\\
\nonumber\\
 {\mathsf
L}_{(2)auxiliary}^{(1)}\equiv
h_{\acute{a}\acute{b}}^{^{(1+)}}<F_{(+)\acute{a}}|F_{(+)\acute{b}}>~~~~~~
\end{eqnarray}
\begin{eqnarray}
h_{\acute{a}\acute{b}}^{^{(1-)}}<\widehat{\Phi}_{(-)\acute{a}}|
\widehat{\Phi}_{(-)\acute{b}}>|_{\theta^2\bar
{\theta}^2}=h_{\acute{a}\acute{b}}^{^{(1-)}}\left[-\partial_A{\bf
A}^{\dagger}_{(-)\acute{a}}\partial^A{\bf A}_{(-)\acute{b}}
-\partial_A{\bf A}^{\dagger}_{(-)\acute{a}i}\partial^A{\bf A}_{(-)\acute{b}}^i\right.\nonumber\\
\left.-\partial_A{\bf A}^{ij\dagger}_{(-)\acute{a}}\partial^A{\bf
A}_{(-)\acute{b}ij}-i\overline{\bf
{\Psi}}_{(-)\acute{a}L}\gamma^A\partial_A{\bf
{\Psi}}_{(-)\acute{b}L}\right.\nonumber\\
\left.-i\overline{\bf
{\Psi}}_{(-)\acute{a}iL}\gamma^A\partial_A{\bf
{\Psi}}_{(-)\acute{b}L}^i-i\overline{\bf
{\Psi}}_{(-)\acute{a}L}^{ij}\gamma^A\partial_A{\bf
{\Psi}}_{(-)\acute{b}ijL}\right] +{\mathsf L}_{(3)auxiliary}^{(1)}~~~~~~\\
\nonumber\\
 {\mathsf
L}_{(3)auxiliary}^{(1)}\equiv
h_{\acute{a}\acute{b}}^{^{(1-)}}<F_{(-)\acute{a}}|F_{(-)\acute{b}}>~~~~~~
\end{eqnarray}

\begin{equation}
{\bf {\Psi}}_{(\pm)\acute{a}}=\left(\matrix{{\bf
\psi}_{(\pm)\acute{a}\tilde{\alpha}}\cr
\overline{{\bf\psi}}^{\dot{\tilde\alpha}}_{(\pm)\acute{a}}}\right)
\end{equation}
\begin{equation}
{\bf{\Psi}}_{R,L}=\frac{1\pm \gamma_5}{2}\bf{\Psi}
\end{equation}
\begin{equation}
\overline{\bf {\Psi}}_{(+)\acute{a}}={\bf
{\Psi}}_{(+)\acute{a}}^{\dagger}\gamma^0
\end{equation}
\begin{eqnarray}
h_{\acute{a}\acute{b}}^{^{(1+)}}{\mathsf
g}^{^{(1)}}q^{^{(+)}}<\widehat{\Phi}_{(+)\acute{a}}|{\widehat{\mathsf
V}}|\widehat{\Phi}_{(+)\acute{b}}>|_{\theta^2\bar {\theta}^2}~~~~~\nonumber\\
=h_{\acute{a}\acute{b}}^{^{(1+)}}{\mathsf
g}^{^{(1)}}q^{^{(+)}}\left\{\left[\frac{ 1}{2}\left(i{\bf
A}^{\dagger}_{(+)\acute{a}}\stackrel{\leftrightarrow}{\partial}_A
{\bf A}_{(+)\acute{b}}-\overline{\bf
{\Psi}}_{(+)\acute{a}L}\gamma_A{\bf
{\Psi}}_{(+)\acute{b}L}\right)\right.\right.\nonumber\\
\left.\left.+\frac{1}{4}\left(i{\bf
A}^{\dagger}_{(+)\acute{a}ij}\stackrel{\leftrightarrow}{\partial}_A{\bf
A}_{(+)\acute{b}}^{ij}-\overline{\bf
{\Psi}}_{(+)\acute{a}ijL}\gamma_A{\bf
{\Psi}}_{(+)\acute{b}L}^{ij}\right)\right.\right.\nonumber\\
\left.\left.+\frac{1}{2}\left(i{\bf
A}^{i\dagger}_{(+)\acute{a}}\stackrel{\leftrightarrow}{\partial}_A{\bf
A}_{(+)\acute{b}i}-\overline{\bf
{\Psi}}_{(+)\acute{a}L}^i\gamma_A{\bf
{\Psi}}_{(+)\acute{b}iL}\right)\right]{\cal V}^{A}\right.\nonumber\\
\left.+\frac{i}{\sqrt {2}}\left[{\bf
A}^{\dagger}_{(+)\acute{a}}\overline{\bf
{\Psi}}_{(+)\acute{b}R}+\frac{1}{2}{\bf
A}^{\dagger}_{(+)\acute{a}ij}\overline{\bf
{\Psi}}_{(+)\acute{b}R}^{ij}+{\bf
A}^{i\dagger}_{(+)\acute{a}}\overline{\bf
{\Psi}}_{(+)\acute{b}iR}\right]{\Lambda}_{L}\right.\nonumber\\
\left.-\frac{i}{\sqrt{2}}\left[\overline{\bf
{\Psi}}_{(+)\acute{a}L}{\bf
A}_{(+)\acute{b}}+\frac{1}{12}\overline{\bf
{\Psi}}_{(+)\acute{a}ijL}{\bf A}^{ij}_{(+)\acute{b}}+\overline{\bf
{\Psi}}_{(+)\acute{a}L}^i{\bf
A}_{(+)\acute{b}i}\right]{\Lambda}_{R}\right\}\nonumber\\
+{\mathsf L}_{(4)auxiliary}^{(1)}\\
\nonumber\\
 {\mathsf
L}_{(4)auxiliary}^{(1)}\equiv
\frac{h_{\acute{a}\acute{b}}^{^{(1+)}}{\mathsf
g}^{^{(1)}}q^{^{(+)}}}{2}<A_{(+)\acute{a}}|A_{(+)\acute{b}}>D
\end{eqnarray}
where
\begin{equation}
{\bf
A}^{i\dagger}_{(+)\acute{a}}\stackrel{\leftrightarrow}{\partial}_A{\bf
A}_{(+)\acute{b}i}\stackrel{def}={\bf
A}^{i\dagger}_{(+)\acute{a}}{\partial}_A{\bf
A}_{(+)\acute{b}i}-\left({\partial}_A{\bf
A}^{i\dagger}_{(+)\acute{a}}\right) {\bf A}_{(+)\acute{b}i}
\end{equation}
\begin{eqnarray}
h_{\acute{a}\acute{b}}^{^{(1-)}}{\mathsf
g}^{^{(1)}}q^{^{(-)}}<\widehat{\Phi}_{(-)\acute{a}}|{\widehat{\mathsf
V}}|\widehat{\Phi}_{(-)\acute{b}}>|_{\theta^2\bar
{\theta}^2}~~~~~\nonumber\\
 =h_{\acute{a}\acute{b}}^{^{(1-)}}{\mathsf
g}^{^{(1)}}q^{^{(-)}}\left\{\left[\frac{1}{2}\left(i{\bf
A}^{\dagger}_{(-)\acute{a}}\stackrel{\leftrightarrow}{\partial}_A
{\bf A}_{(-)\acute{b}}-\overline{\bf
{\Psi}}_{(-)\acute{a}L}\gamma_A{\bf
{\Psi}}_{(-)\acute{b}L}\right)\right.\right.\nonumber\\
\left.\left.+\frac{1}{4}\left({\bf
A}^{ij\dagger}_{(-)\acute{a}}\stackrel{\leftrightarrow}{\partial}_A{\bf
A}_{(-)\acute{b}ij}+i\overline{\bf
{\Psi}}_{(-)\acute{a}L}^{ij}\gamma_A{\bf
{\Psi}}_{(-)\acute{b}ijL}\right)\right.\right.\nonumber\\
\left.\left.+\frac{1}{2}\left({\bf
A}^{\dagger}_{(-)\acute{a}i}\stackrel{\leftrightarrow}{\partial}_A{\bf
A}_{(-)\acute{b}}^{i}+i\overline{\bf
{\Psi}}_{(-)\acute{a}iL}\gamma_A{\bf
{\Psi}}_{(-)\acute{b}L}^i\right)\right]{\cal V}^{A}\right.\nonumber\\
\left.+\frac{i}{\sqrt{2}}\left[{\bf
A}^{\dagger}_{(-)\acute{a}}\overline{\bf
{\Psi}}_{(-)\acute{b}R}+\frac{1}{2}{\bf
A}^{ij\dagger}_{(-)\acute{a}}\overline{\bf
{\Psi}}_{(-)\acute{b}ijR}+{\bf
A}^{\dagger}_{(-)\acute{a}i}\overline{\bf
{\Psi}}_{(-)\acute{b}R}^i\right]{\Lambda}_{L}\right.\nonumber\\
\left.-\frac{i}{\sqrt{2}}\left[\overline{\bf
{\Psi}}_{(-)\acute{a}L}{\bf
A}_{(-)\acute{b}}+\frac{1}{2}\overline{\bf
{\Psi}}_{(-)\acute{a}L}^{ij}{\bf A}_{(-)\acute{b}ij}+\overline{\bf
{\Psi}}_{(-)\acute{a}iL}{\bf
A}_{(-)\acute{b}}^i\right]{\Lambda}_{R}\right\}\nonumber\\
+{\mathsf L}_{(5)auxiliary}^{(1)}\\
\nonumber\\
 {\mathsf
L}_{(5)auxiliary}^{(1)}\equiv
\frac{h_{\acute{a}\acute{b}}^{^{(1-)}}{\mathsf
g}^{^{(1)}}q^{^{(-)}}}{2}<A_{(-)\acute{a}}|A_{(-)\acute{b}}>D
\end{eqnarray}
\begin{eqnarray}
\frac{1}{2} h_{\acute{a}\acute{b}}^{^{(1+)}}{\mathsf
g}^{^{(1)}2}q^{^{(+)}2}<\widehat{\Phi}_{(+)\acute{a}}|{\widehat{\mathsf
V}}^2|\widehat{\Phi}_{(+)\acute{b}}>|_{\theta^2\bar
{\theta}^2}~~~~~\nonumber\\
=\frac{h_{\acute{a}\acute{b}}^{^{(1+)}}{\mathsf
g}^{^{(1)}2}q^{^{(+)}2}}{4} \left[{\bf
A}^{\dagger}_{(+)\acute{a}}{\bf A}_{(+)\acute{b}} +\frac{1}{2}{\bf
A}^{\dagger}_{(+)\acute{a}ij}{\bf A}_{(+)\acute{b}}^{ij}+{\bf
A}^{i\dagger}_{(+)\acute{a}}{\bf A}_{(+)\acute{b}i}\right]{\cal
V}_A{\cal V}^{A}
\end{eqnarray}
\begin{eqnarray}
\frac{1}{2} h_{\acute{a}\acute{b}}^{^{(1-)}}{\mathsf
g}^{^{(1)}2}q^{^{(-)}2}<\widehat{\Phi}_{(-)\acute{a}}|{\widehat{\mathsf
V}}^2|\widehat{\Phi}_{(-)\acute{b}}>|_{\theta^2\bar
{\theta}^2}~~~~~~\nonumber\\
= \frac{h_{\acute{a}\acute{b}}^{^{(1-)}}{\mathsf
g}^{^{(1)}2}q^{^{(-)}2}}{4} \left[{\bf
A}^{\dagger}_{(-)\acute{a}}{\bf A}_{(-)\acute{b}} +\frac{1}{2}{\bf
A}^{ij\dagger}_{(-)\acute{a}}{\bf A}_{(-)\acute{b}ij}+{\bf
A}^{\dagger}_{(-)\acute{a}i}{\bf A}_{(-)\acute{b}}^i\right]{\cal
V}_A{\cal V}^{A}
\end{eqnarray}
Finally, ${\mathsf L}_{\mathsf W}^{(1)}$ appearing in Eq.(8.1) is
given by
\begin{eqnarray}
{\mathsf L}_{\mathsf
W}^{(1)}=\mu_{\acute{a}\acute{b}}<\widehat{\Phi}_{(-)\acute{a}}^*|B|\widehat{\Phi}_{(+)\acute{b}}>|_{\theta^2}+
h.c.
\end{eqnarray}
Evaluation of Eq.(8.27) gives
\begin{eqnarray}
{\mathsf L}_{\mathsf W}^{(1)}=-i\mu_{\acute{a}\acute{b}}\left(
\overline{\bf {\Psi}}_{(-)\acute{a}R}{\bf
{\Psi}}_{(+)\acute{b}L}+\overline{\bf
{\Psi}}_{(-)\acute{a}R}^i{\bf
{\Psi}}_{(+)\acute{b}iL}-\frac{1}{2}\overline{\bf
{\Psi}}_{(-)\acute{a}ijR}{\bf
{\Psi}}_{(+)\acute{b}L}^{ij}\right)\nonumber\\
+i\mu_{\acute{a}\acute{b}}^{*}\left( \overline{\bf
{\Psi}}_{(-)\acute{a}L}{\bf {\Psi}}_{(+)\acute{b}R}+\overline{\bf
{\Psi}}_{(-)\acute{a}iL}{\bf
{\Psi}}_{(+)\acute{b}R}^i-\frac{1}{2}\overline{\bf
{\Psi}}_{(-)\acute{a}L}^{ij}{\bf
{\Psi}}_{(+)\acute{b}ijR}\right)\nonumber\\
+{\mathsf L}_{(6)auxiliary}^{(1)}\\
\nonumber\\
 {\mathsf
L}_{(6)auxiliary}^{(1)}\equiv i\mu_{\acute{a}\acute{b}}\left[{\bf
F}_{(-)\acute{a}}{\bf A}_{(+)\acute{b}}+{\bf
A}_{(-)\acute{a}}^{\bf{T}}{\bf F}_{(+)\acute{b}}-\frac{1}{2}{\bf
F}_{(-)\acute{a}ij}{\bf A}_{(+)\acute{b}}^{ij}\right.\nonumber\\
\left.-\frac{1}{2}{\bf A}_{(-)\acute{a}ij}^{\bf{T}}{\bf
F}_{(+)\acute{b}}^{ij} + {\bf F}_{(-)\acute{a}}^i{\bf
A}_{(+)\acute{b}i}+{\bf A}_{(-)\acute{a}}^{i\bf{T}}{\bf
F}_{(+)\acute{b}i}\right] +h.c.
\end{eqnarray}
Elimination of the auxiliary fields ${\bf F}_{(\pm)}$ through
their field equations gives
\begin{eqnarray}
{\mathsf L}_{(2)auxiliary}^{(1)}+{\mathsf
L}_{(3)auxiliary}^{(1)}+{\mathsf
L}_{(6)auxiliary}^{(1)}~~~~~~~~~~~\nonumber\\
=-\left(\mu^*
[h^{^{(1-)}}]^{\bf{-1}}[h^{^{(1-)}}]^{\bf{T}}[h^{^{(1-)}}]^{\bf{-1}}\mu\right)_{\acute{a}\acute{b}}\nonumber\\
\times \left[{\bf A}^{\dagger}_{(+)\acute{a}}{\bf
A}_{(+)\acute{b}}+\frac{1}{4}{\bf
A}^{\dagger}_{(+)\acute{a}ij}{\bf A}_{(+)\acute{b}}^{ij}+{\bf
A}^{i\dagger}_{(+)\acute{a}}{\bf A}_{(+)\acute{b}i}\right]\nonumber\\
-\left(\mu[h^{^{(1+)}}]^{\bf{-1T}}h^{^{(1+)}}[h^{^{(1+)}}]^{\bf{-1T}}\mu^*\right)_{\acute{a}\acute{b}}\nonumber\\
\times\left[ {\bf A}^{\bf{T}}_{(-)\acute{a}}{\bf
A}_{(-)\acute{b}}^*+\frac{1}{4}{\bf
A}^{\bf{T}}_{(-)\acute{a}ij}{\bf A}_{(-)\acute{b}}^{ij*}+{\bf
A}^{i\bf{T}}_{(-)\acute{a}}{\bf A}_{(-)\acute{b}i}^*\right]~~~~~~
\end{eqnarray}

Similarly, after eliminating the field $D$ we get
\begin{eqnarray}
{\mathsf L}_{(1)auxiliary}^{(1)}+{\mathsf
L}_{(4)auxiliary}^{(1)}+{\mathsf
L}_{(5)auxiliary}^{(1)}~~~~~~~~~~~~~\nonumber\\
=-\frac{1}{8} {\mathsf
g}^{^{(1)}2}h_{\acute{a}\acute{b}}^{^{(1+)}}h_{\acute{c}\acute{d}}^{^{(1+)}}q^{^{(+)}2}<A_{(+)\acute{a}}|A_{(+)\acute{b}}><A_{(+)\acute{c}}|
A_{(+)\acute{d}}>\nonumber\\
-\frac{1}{8} {\mathsf
g}^{^{(1)}2}h_{\acute{a}\acute{b}}^{^{(1-)}}h_{\acute{c}\acute{d}}^{^{(1-)}}q^{^{(-)}2}<A_{(-)\acute{a}}|A_{(-)\acute{b}}><A_{(-)\acute{c}}|A_{(-)\acute{d}}>\nonumber\\
-\frac{1}{4} {\mathsf
g}^{^{(45)}2}h_{\acute{a}\acute{b}}^{^{(1+)}}
h_{\acute{c}\acute{d}}^{^{(1-)}}q^{^{(+)}}q^{^{(-)}}
<A_{(+)\acute{a}}|A_{(+)\acute{b}}><A_{(-)\acute{c}}|A_{(-)\acute{d}}>
\end{eqnarray}

\begin{eqnarray}
-\frac{1}{8} {\mathsf
g}^{^{(1)}2}h_{\acute{a}\acute{b}}^{^{(1+)}}h_{\acute{c}\acute{d}}^{^{(1+)}}q^{^{(+)}2}<A_{(+)\acute{a}}|A_{(+)\acute{b}}><A_{(+)\acute{c}}|
A_{(+)\acute{d}}>\nonumber\\
=-\frac{1}{8} {\mathsf
g}^{^{(1)}2}h_{\acute{a}\acute{b}}^{^{(1+)}}h_{\acute{c}\acute{d}}^{^{(1+)}}q^{^{(+)}2}\left[{\bf
A}^{\dagger}_{(+)\acute{a}}{\bf A}_{(+)\acute{b}}{\bf
A}^{\dagger}_{(+)\acute{c}}{\bf
A}_{(+)\acute{d}}\right.\nonumber\\
\left.+\frac{1}{4}{\bf A}^{\dagger}_{(+)\acute{a}ij}{\bf
A}_{(+)\acute{b}}^{ij}{\bf A}^{\dagger}_{(+)\acute{c}kl}{\bf
A}_{(+)\acute{d}}^{kl}
 +{\bf A}^{i\dagger}_{(+)\acute{a}}{\bf
A}_{(+)\acute{b}i}{\bf A}^{j\dagger}_{(+)\acute{c}}{\bf
A}_{(+)\acute{d}j}\right.\nonumber\\
 \left.+{\bf
A}^{\dagger}_{(+)\acute{a}}{\bf A}_{(+)\acute{b}}{\bf
A}^{\dagger}_{(+)\acute{c}ij}{\bf A}_{(+)\acute{d}}^{ij} +2{\bf
A}^{\dagger}_{(+)\acute{a}}{\bf A}_{(+)\acute{b}}{\bf
A}^{i\dagger}_{(+)\acute{c}}{\bf
A}_{(+)\acute{d}i}\right.\nonumber\\
\left.+{\bf A}^{i\dagger}_{(+)\acute{a}}{\bf
A}_{(+)\acute{b}i}{\bf A}^{\dagger}_{(+)\acute{c}jk}{\bf
A}_{(+)\acute{d}}^{jk}\right]
\end{eqnarray}

\begin{eqnarray}
-\frac{1}{8} {\mathsf
g}^{^{(1)}2}h_{\acute{a}\acute{b}}^{^{(1-)}}h_{\acute{c}\acute{d}}^{^{(1-)}}q^{^{(-)}2}<A_{(-)\acute{a}}|A_{(-)\acute{b}}><A_{(-)\acute{c}}|
A_{(-)\acute{d}}>\nonumber\\
=-\frac{1}{8} {\mathsf
g}^{^{(1)}2}h_{\acute{a}\acute{b}}^{^{(1-)}}h_{\acute{c}\acute{d}}^{^{(1-)}}q^{^{(-)}2}\left[{\bf
A}^{\dagger}_{(-)\acute{a}}{\bf A}_{(-)\acute{b}}{\bf
A}^{\dagger}_{(-)\acute{c}}{\bf
A}_{(-)\acute{d}}\right.\nonumber\\
\left.+\frac{1}{4}{\bf A}^{ij\dagger}_{(-)\acute{a}}{\bf
A}_{(-)\acute{b}ij}{\bf A}^{kl\dagger}_{(-)\acute{c}}{\bf
A}_{(-)\acute{d}kl} +{\bf A}^{\dagger}_{(-)\acute{a}i}{\bf
A}_{(-)\acute{b}}^i{\bf A}^{\dagger}_{(-)\acute{c}j}{\bf
A}_{(-)\acute{d}}^j\right.\nonumber\\
\left.+{\bf A}^{\dagger}_{(-)\acute{a}}{\bf A}_{(-)\acute{b}}{\bf
A}^{ij\dagger}_{(-)\acute{c}}{\bf A}_{(-)\acute{d}ij} +2{\bf
A}^{\dagger}_{(-)\acute{a}}{\bf A}_{(-)\acute{b}}{\bf
A}^{\dagger}_{(-)\acute{c}i}{\bf
A}_{(-)\acute{d}}^i\right.\nonumber\\
\left.+{\bf A}^{\dagger}_{(-)\acute{a}i}{\bf
A}_{(-)\acute{b}}^i{\bf A}^{jk\dagger}_{(-)\acute{c}}{\bf
A}_{(-)\acute{d}jk}\right]
\end{eqnarray}

\begin{eqnarray}
-\frac{1}{4} {\mathsf
g}^{^{(1)}2}h_{\acute{a}\acute{b}}^{^{(1+)}}h_{\acute{c}\acute{d}}^{^{(1-)}}q^{^{(+)}}q^{^{(-)}}<A_{(+)\acute{a}}|A_{(+)\acute{b}}><A_{(-)\acute{c}}|
A_{(-)\acute{d}}>\nonumber\\
=-\frac{1}{4} {\mathsf
g}^{^{(1)}2}h_{\acute{a}\acute{b}}^{^{(1-)}}h_{\acute{c}\acute{d}}^{^{(1-)}}q^{^{(+)}}q^{^{(-)}}\left[{\bf
A}^{\dagger}_{(+)\acute{a}}{\bf A}_{(+)\acute{b}}{\bf
A}^{\dagger}_{(-)\acute{c}}{\bf
A}_{(-)\acute{d}}\right.\nonumber\\
\left.+\frac{1}{2}{\bf A}^{\dagger}_{(+)\acute{a}}{\bf
A}_{(+)\acute{b}}{\bf A}^{ij\dagger}_{(-)\acute{c}}{\bf
A}_{(-)\acute{d}ij} +{\bf A}^{\dagger}_{(+)\acute{a}}{\bf
A}_{(+)\acute{b}}{\bf A}^{\dagger}_{(-)\acute{c}i}{\bf
A}_{(-)\acute{d}}^i\right.\nonumber\\
\left.+\frac{1}{2}{\bf A}^{\dagger}_{(+)\acute{a}ij}{\bf
A}_{(+)\acute{b}}^{ij}{\bf A}^{\dagger}_{(-)\acute{c}}{\bf
A}_{(-)\acute{d}}
 +\frac{1}{4}{\bf
A}^{\dagger}_{(+)\acute{a}ij}{\bf A}_{(+)\acute{b}}^{ij}{\bf
A}^{kl\dagger}_{(-)\acute{c}}{\bf
A}_{(-)\acute{d}kl}\right.\nonumber\\
\left.+\frac{1}{2}{\bf A}^{\dagger}_{(+)\acute{a}ij}{\bf
A}_{(+)\acute{b}}^{ij}{\bf A}^{\dagger}_{(-)\acute{c}k}{\bf
A}_{(-)\acute{d}}^k +{\bf A}^{i\dagger}_{(+)\acute{a}}{\bf
A}_{(+)\acute{b}i}{\bf A}^{\dagger}_{(-)\acute{c}}{\bf
A}_{(-)\acute{d}}\right.\nonumber\\
\left.+\frac{1}{2}{\bf A}^{i\dagger}_{(+)\acute{a}}{\bf
A}_{(+)\acute{b}i}{\bf A}^{kl\dagger}_{(-)\acute{c}}{\bf
A}_{(-)\acute{d}kl} +{\bf A}^{i\dagger}_{(+)\acute{a}}{\bf
A}_{(+)\acute{b}i}{\bf A}^{\dagger}_{(-)\acute{c}j}{\bf
A}_{(-)\acute{d}}^j\right]
\end{eqnarray}

\section{{\bf{45}} plet vector coupling}
In this section we give the complete supersymmetric vector
couplings for the 45-dimensional tensor of $SO(10)$ in the
Wess-Zumino gauge:
\begin{equation}
{\mathsf L}^{45}={\mathsf L}_{V}^{^{(45~K.E.)}} +{\mathsf
L}_{V+\Phi}^{^{(45~Interaction)}}+{\mathsf L}_{\mathsf W}^{(45)}
\end{equation}
where
\begin{eqnarray}
{\mathsf L}_{V}^{^{(45~K.E.)}}=\frac{1}{64}\left[\widehat{\cal
W}^{\tilde{\alpha}} \widehat{\cal
W}_{\tilde{\alpha}}|_{\theta^2}+\widehat{\overline{\cal
W}}_{\dot{\tilde{\alpha}}}\widehat{\overline{\cal
W}}^{\dot{\tilde{\alpha}}}|_{\bar {\theta}^2}\right]\\
 \widehat{\cal
W}_{\tilde{\alpha}}=\frac{1}{{\mathsf g}^{^{(45)}}}\overline
{\mathsf D}^2e^{-{\mathsf g}^{^{(45)}}\widehat{\mathsf
V}_{\mu\nu}M_{\mu\nu}}{\mathsf D}_{\tilde{\alpha}}e^{{\mathsf
g}^{^{(45)}}\widehat{\mathsf V}_{\rho\lambda}M_{\rho\lambda}}
\end{eqnarray}
where $M_{\mu\nu}$ are the 45 generators in the vector
(10-dimensional) representation that satisfy the following Lie
algebra
\begin{equation}
[M_{\alpha\beta},M_{\gamma\rho}]=-i\left(\delta_{\beta\gamma}M_{\alpha\rho}+\delta_{\alpha\rho}M_{\beta\gamma}
-\delta_{\alpha\gamma}M_{\beta\rho}-\delta_{\beta\rho}M_{\alpha\gamma}\right)
\end{equation}
and take on the form
\begin{equation}
\left(M_{\mu\nu}\right)_{\alpha\beta}=-i\left(\delta_{\mu\alpha}\delta_{\nu\beta}-\delta_{\mu\beta}\delta_{\nu\alpha}
\right)
\end{equation}
and $\widehat{\mathsf V}_{\mu\nu}$ is in the Wess-Zumino gauge.
 Finally we have
\begin{eqnarray}
{\mathsf L}_{V}^{^{(45~K.E.)}}=-\frac{1}{4}{\cal
V}_{AB\mu\nu}{\cal V}^{AB}_{\mu\nu}-\frac{i}{2}\overline
{\Lambda}_{\mu\nu}\gamma^A{\cal D}_A\Lambda_{\mu\nu}+{\mathsf
L}_{(1)auxiliary}^{(45)}\\
{\cal V}_{\mu\nu}^{AB}=\partial^A{\cal
V}^B_{\mu\nu}-\partial^B{\cal V}^A_{\mu\nu}+{\mathsf
g}^{^{(45)}}\left({\cal V}^A_{\mu\alpha}{\cal
V}^B_{\alpha\nu}-{\cal V}^B_{\mu\alpha}{\cal
V}^A_{\alpha\nu}\right)\\
{\cal D}^A=\partial^A+\frac{i{\mathsf
g}^{^{(45)}}}{2}M_{\mu\nu}{\cal V}_{\mu\nu}^A\\
{\cal D}^A\Lambda_{\mu\nu}=\partial^A\Lambda_{\mu\nu}+{\mathsf
g}^{^{(45)}}\left({\cal V}^A_{\mu\alpha}\Lambda_{\alpha\nu}-{\cal
V}^A_{\nu\alpha}\Lambda_{\nu\alpha}\right)\\
{\mathsf L}_{(1)auxiliary}^{(45)}=\frac{1}{2}D_{\mu\nu}D_{\mu\nu}\\
\Lambda_{\mu\nu}=\left(\matrix{\lambda_{\tilde{\alpha}\mu\nu}\cr
\overline{\lambda}_{\mu\nu}^{\dot{\tilde\alpha}}}\right)
\end{eqnarray}
To normalize the $SU(5)$ fields appearing in $-\frac{1}{4}{\cal
V}_{AB\mu\nu}{\cal V}^{AB}_{\mu\nu}$, we carry out a field
redefinition
\begin{eqnarray}
{\cal V}_A=2\sqrt{5}{\cal V}^{'}_A;~~~~~{\cal
V}^{j}_{\acute{a}i}=\sqrt{2}{\cal V}^{'j}_{\acute{a}i};~~~~~ {\cal
V}^{ij}_A=\sqrt{2}{\cal V}^{'ij}_A;~~~~
 {\cal V}_{\acute{a}ij}=\sqrt{2}{\cal V}^{'}_{\acute{a}ij}\nonumber\\
{\Lambda}=\sqrt{10}{\Lambda}^{'};~~~~~{\Lambda}^{j}_{i}=
\sqrt{2}{\Lambda}^{'j}_{i};~~~~~
{\Lambda}^{ij}=\sqrt{2}{\Lambda}^{'ij};~~~~
 {\Lambda}_{ij}=\sqrt{2}{\Lambda}^{'}_{ij}
\end{eqnarray}
so that
\begin{eqnarray}
-\frac{1}{4}{\cal V}^{AB} _{\mu\nu}{\cal V}
_{AB\mu\nu}=-\frac{1}{2}{\cal V}'_{AB}{\cal
V}^{'AB\dagger}-\frac{1}{2!}\frac{1}{2}{\cal V}^{'ij}_{AB}{\cal
V}^{'ABij\dagger} -\frac{1}{4}{\cal V}^{'i}_{ABj}{\cal V}^{'ABj}_i\nonumber\\
-\frac{i}{2}\overline {\Lambda}_{\mu\nu}\gamma^A{\cal
D}_A\Lambda_{\mu\nu}=-\frac{i}{2}\frac{1}{2!}\overline
{\Lambda}_{ij}^{'}\gamma^A{\cal
D}_A\Lambda_{ij}^{'}-\frac{i}{2}\frac{1}{2!}\overline
{\Lambda}^{'ij}\gamma^A{\cal
D}_A\Lambda^{'ij}\nonumber\\
-\frac{i}{2}\overline {\Lambda}^{'i}_j\gamma^A{\cal
D}_A\Lambda^{'i}_j-\frac{i}{2}\overline {\Lambda}^{'}\gamma^A{\cal
D}_A\Lambda^{'}
\end{eqnarray}
Next we look at the second term in Eq.(8.35)
\begin{eqnarray}
{\mathsf
L}_{V+\Phi}^{^{(45~Interaction)}}=h_{\acute{a}\acute{b}}^{^{(45+)}}<\widehat{\Phi}_{(+)\acute{a}}|
e^{\frac{1}{2!}{\mathsf g}^{^{(45)}}\widehat{{\mathsf
V}}_{\mu\nu}\Sigma_{\mu\nu}
}|\widehat{\Phi}_{(+)\acute{b}}>|_{\theta^2\bar
{\theta}^2}\nonumber\\
+h_{\acute{a}\acute{b}}^{^{(45-)}}<\widehat{\Phi}_{(-)\acute{a}}|e^{\frac{1}{2!}{\mathsf
g}^{^{(45)}}\widehat{{\mathsf V}}_{\mu\nu}\Sigma_{\mu\nu}
}|\widehat{\Phi}_{(-)\acute{b}}>|_{\theta^2\bar {\theta}^2}
\end{eqnarray}
$\Sigma_{\mu\nu}$ being the 45 generators in the spinorial
representation. We find
\begin{eqnarray}
{\mathsf
L}_{V+\Phi}^{^{(45~Interaction)}}=h_{\acute{a}\acute{b}}^{^{(45+)}}\left.[<\widehat{\Phi}_{(+)\acute{a}}|
\widehat{\Phi}_{(+)\acute{b}}>+\frac{1}{2}{\mathsf
g}^{^{(45)}}<\widehat{\Phi}_{(+)\acute{a}}|{\widehat{\mathsf
V}}_{\mu\nu}\Sigma_{\mu\nu}|\widehat{\Phi}_{(+)\acute{b}}>\right.\nonumber\\
\left.+\frac{1}{8}{\mathsf
g}^{^{(45)}2}<\widehat{\Phi}_{(+)\acute{a}}|{\widehat{\mathsf
V}}_{\mu\nu}\Sigma_{\mu\nu}{\widehat{\mathsf
V}}_{\rho\lambda}\Sigma_{\rho\lambda}|\widehat{\Phi}_{(+)\acute{b}}>\right]|_{\theta^2\bar
{\theta}^2}\nonumber\\
+h_{\acute{a}\acute{b}}^{^{(45-)}}\left[<\widehat{\Phi}_{(-)\acute{a}}|
\widehat{\Phi}_{(-)\acute{b}}> +\frac{1}{2}{\mathsf
g}^{^{(45)}}<\widehat{\Phi}_{(-)\acute{a}}|{\widehat{\mathsf
V}}_{\mu\nu}\Sigma_{\mu\nu}|\widehat{\Phi}_{(-)\acute{b}}>\right.\nonumber\\
\left.+\frac{1}{8}{\mathsf
g}^{^{(45)}2}<\widehat{\Phi}_{(-)\acute{a}}|{\widehat{\mathsf
V}}_{\mu\nu}\Sigma_{\mu\nu}{\widehat{\mathsf
V}}_{\rho\lambda}\Sigma_{\rho\lambda}|\widehat{\Phi}_{(-)\acute{b}}>\right]|_{\theta^2\bar
{\theta}^2}\nonumber\\
\end{eqnarray}
where
\begin{eqnarray}
h_{\acute{a}\acute{b}}^{^{(45+)}}<\widehat{\Phi}_{(+)\acute{a}}|
\widehat{\Phi}_{(+)\acute{b}}>|_{\theta^2\bar
{\theta}^2}=h_{\acute{a}\acute{b}}^{^{(45+)}}\left[-\partial_A{\bf
A}^{\dagger}_{(+)\acute{a}}\partial^A{\bf
A}_{(+)\acute{b}}\right.\nonumber\\
\left.-\partial_A{\bf A}^{i\dagger}_{(+)\acute{a}}\partial^A{\bf
A}_{(+)\acute{b}i}
 -\partial_A{\bf
A}^{\dagger}_{(+)\acute{a}ij}\partial^A{\bf
A}_{(+)\acute{b}}^{ij}\right.\nonumber\\
\left.-i\overline{\bf
{\Psi}}_{(+)\acute{a}L}\gamma^A\partial_A{\bf
{\Psi}}_{(+)\acute{b}L} -i\overline{\bf
{\Psi}}_{(+)\acute{a}L}^i\gamma^A\partial_A{\bf
{\Psi}}_{(+)\acute{b}iL}\right.\nonumber\\
\left.-i\overline{\bf
{\Psi}}_{(+)\acute{a}ijL}\gamma^A\partial_A{\bf
{\Psi}}_{(+)\acute{b}L}^{ij}\right] +{\mathsf
L}_{(2)auxiliary}^{(45)}\\
\nonumber\\
 {\mathsf
L}_{(2)auxiliary}^{(45)}=h_{\acute{a}\acute{b}}^{^{(45+)}}<F_{(+)\acute{a}}|F_{(+)\acute{b}}>~~~~~~
\end{eqnarray}
\begin{eqnarray}
h_{\acute{a}\acute{b}}^{^{(45-)}}<\widehat{\Phi}_{(-)\acute{a}}|
\widehat{\Phi}_{(-)\acute{b}}>|_{\theta^2\bar
{\theta}^2}=h_{\acute{a}\acute{b}}^{^{(45-)}}\left[-\partial_A{\bf
A}^{\dagger}_{(-)\acute{a}}\partial^A{\bf
A}_{(-)\acute{b}}\right.\nonumber\\
\left.-\partial_A{\bf A}^{\dagger}_{(-)\acute{a}i}\partial^A{\bf
A}_{(-)\acute{b}}^i -\partial_A{\bf
A}^{ij\dagger}_{(-)\acute{a}}\partial^A{\bf
A}_{(-)\acute{b}ij}\right.\nonumber\\
\left.-i\overline{\bf
{\Psi}}_{(-)\acute{a}L}\gamma^A\partial_A{\bf
{\Psi}}_{(-)\acute{b}L} -i\overline{\bf
{\Psi}}_{(-)\acute{a}iL}\gamma^A\partial_A{\bf
{\Psi}}_{(-)\acute{b}L}^i\right.\nonumber\\
\left.-i\overline{\bf
{\Psi}}_{(-)\acute{a}L}^{ij}\gamma^A\partial_A{\bf
{\Psi}}_{(-)\acute{b}ijL}\right]
+{\mathsf L}_{(3)auxiliary}^{(45)}\\
\nonumber\\
 {\mathsf
L}_{(3)auxiliary}^{(45)}=h_{\acute{a}\acute{b}}^{^{(45-)}}<F_{(-)\acute{a}}|F_{(-)\acute{b}}>~~~~~~
\end{eqnarray}
\begin{eqnarray}
\frac{1}{2}h_{\acute{a}\acute{b}}^{^{(45+)}}{\mathsf
g}^{^{(45)}}<\widehat{\Phi}_{(+)\acute{a}}|{\widehat{\mathsf
V}}_{\mu\nu}\Sigma_{\mu\nu}|\widehat{\Phi}_{(+)\acute{b}}>|_{\theta^2\bar
{\theta}^2}~~~~~\nonumber\\
=h_{\acute{a}\acute{b}}^{^{(45+)}}{\mathsf
g}^{^{(45)}}\left\{\left[-\frac{\sqrt 5}{2}\left({\bf
A}^{\dagger}_{(+)\acute{a}}\stackrel{\leftrightarrow}{\partial}_A
{\bf A}_{(+)\acute{b}}+i\overline{\bf
{\Psi}}_{(+)\acute{a}L}\gamma_A{\bf
{\Psi}}_{(+)\acute{b}L}\right)\right.\right.\nonumber\\
\left.\left.-\frac{1}{4\sqrt 5}\left({\bf
A}^{\dagger}_{(+)\acute{a}ij}\stackrel{\leftrightarrow}{\partial}_A{\bf
A}_{(+)\acute{b}}^{ij}+i\overline{\bf
{\Psi}}_{(+)\acute{a}ijL}\gamma_A{\bf
{\Psi}}_{(+)\acute{b}L}^{ij}\right)\right.\right.\nonumber\\
\left.\left.+\frac{3}{2\sqrt 5}\left({\bf
A}^{i\dagger}_{(+)\acute{a}}\stackrel{\leftrightarrow}{\partial}_A{\bf
A}_{(+)\acute{b}i}+i\overline{\bf
{\Psi}}_{(+)\acute{a}L}^i\gamma_A{\bf
{\Psi}}_{(+)\acute{b}iL}\right)\right]{\cal V}^{'A}\right.\nonumber\\
\left.+\left[\frac{1}{2\sqrt 2}\left({\bf
A}^{\dagger}_{(+)\acute{a}lm}\stackrel{\leftrightarrow}{\partial}_A{\bf
A}_{(+)\acute{b}}+i\overline{\bf
{\Psi}}_{(+)\acute{a}lmL}\gamma_A{\bf
{\Psi}}_{(+)\acute{b}L}\right)\right.\right.\nonumber\\
\left.\left.+\frac{1}{4\sqrt 2}\epsilon_{ijklm}\left({\bf
A}^{i\dagger}_{(+)\acute{a}}\stackrel{\leftrightarrow}{\partial}_A{\bf
A}_{(+)\acute{b}}^{jk}+i\overline{\bf
{\Psi}}_{(+)\acute{a}L}^i\gamma_A{\bf
{\Psi}}_{(+)\acute{b}L}^{jk}\right)\right]{\cal V}^{'Alm}\right.\nonumber\\
\left.+\left[-\frac{1}{2\sqrt 2}\left({\bf
A}^{\dagger}_{(+)\acute{a}}\stackrel{\leftrightarrow}{\partial}_A{\bf
A}_{(+)\acute{b}}^{lm}+i\overline{\bf
{\Psi}}_{(+)\acute{a}L}\gamma_A{\bf
{\Psi}}_{(+)\acute{b}L}^{lm}\right)\right.\right.\nonumber\\
\left.\left.-\frac{1}{4\sqrt 2}\epsilon^{ijklm}\left({\bf
A}^{\dagger}_{(+)\acute{a}ij}\stackrel{\leftrightarrow}{\partial}_A{\bf
A}_{(+)\acute{b}k}+i\overline{\bf
{\Psi}}_{(+)\acute{a}ijL}\gamma_A{\bf
{\Psi}}_{(+)\acute{b}kL}\right)\right]{\cal V}^{'A}_{lm}\right.\nonumber\\
 \left.+\left[-\frac{1}{\sqrt
2}\left({\bf
A}^{\dagger}_{(+)\acute{a}ik}\stackrel{\leftrightarrow}{\partial}_A{\bf
A}_{(+)\acute{b}}^{kj}+i\overline{\bf
{\Psi}}_{(+)\acute{a}ikL}\gamma_A{\bf
{\Psi}}_{(+)\acute{b}L}^{kj}\right)\right.\right.\nonumber\\
\left.\left.-\frac{1}{\sqrt 2}\left({\bf
A}^{j\dagger}_{(+)\acute{a}}\stackrel{\leftrightarrow}{\partial}_A{\bf
A}_{(+)\acute{b}i}+i\overline{\bf
{\Psi}}_{(+)\acute{a}L}^j\gamma_A{\bf
{\Psi}}_{(+)\acute{b}iL}\right)\right]{\cal V}^{'iA}_j\right.\nonumber\\
\left.+\frac{\sqrt 5}{2}\left[-{\bf
A}^{\dagger}_{(+)\acute{a}}\overline{\bf
{\Psi}}_{(+)\acute{b}R}-\frac{1}{10}{\bf
A}^{\dagger}_{(+)\acute{a}ij}\overline{\bf
{\Psi}}_{(+)\acute{b}R}^{ij}+\frac{3}{5}{\bf
A}^{i\dagger}_{(+)\acute{a}}\overline{\bf
{\Psi}}_{(+)\acute{b}iR}\right]{\Lambda}^{'}_{L}\right.\nonumber\\
\left.+\frac{1}{2}\left[{\bf
A}^{\dagger}_{(+)\acute{a}lm}\overline{\bf
{\Psi}}_{(+)\acute{b}R}+\frac{1}{2}\epsilon_{ijklm}{\bf
A}^{i\dagger}_{(+)\acute{a}}\overline{\bf
{\Psi}}_{(+)\acute{b}R}^{jk}\right]{\Lambda}^{'lm}_{L}\right.\nonumber\\
\left.-\frac{1}{2}\left[{\bf
A}^{\dagger}_{(+)\acute{a}}\overline{\bf
{\Psi}}_{(+)\acute{b}R}^{lm}+\frac{1}{2}\epsilon^{ijklm}{\bf
A}^{\dagger}_{(+)\acute{a}ij}\overline{\bf
{\Psi}}_{(+)\acute{b}kR}\right]{\Lambda}^{'}_{lmL}\right.\nonumber\\
\left.-\left[{\bf A}^{\dagger}_{(+)\acute{a}ik}\overline{\bf
{\Psi}}_{(+)\acute{b}R}^{kj}+{\bf
A}^{j\dagger}_{(+)\acute{a}}\overline{\bf
{\Psi}}_{(+)\acute{b}iR}\right]{\Lambda}^{'i}_{jL}\right.\nonumber\\
\left.-\frac{\sqrt 5}{2}\left[-\overline{\bf
{\Psi}}_{(+)\acute{a}L}{\bf
A}_{(+)\acute{b}}-\frac{1}{10}\overline{\bf
{\Psi}}_{(+)\acute{a}ijL}{\bf
A}^{ij}_{(+)\acute{b}}+\frac{3}{5}\overline{\bf
{\Psi}}_{(+)\acute{a}L}^i{\bf
A}_{(+)\acute{b}i}\right]{\Lambda}^{'}_{R}\right.\nonumber\\
\left.-\frac{1}{2}\left[\overline{\bf
{\Psi}}_{(+)\acute{a}lmL}{\bf
A}_{(+)\acute{b}}+\frac{1}{2}\epsilon_{ijklm}\overline{\bf
{\Psi}}_{(+)\acute{a}L}^i{\bf A}^{jk}_{(+)\acute{b}}\right]{\Lambda}^{'lm}_{R}\right.\nonumber\\
\left.+\frac{1}{2}\left[\overline{\bf {\Psi}}_{(+)\acute{a}L}{\bf
A}_{(+)\acute{b}}^{lm}+\frac{1}{2}\epsilon^{ijklm}\overline{\bf
{\Psi}}_{(+)\acute{a}ijL}{\bf A}_{(+)\acute{b}k}\right]{\Lambda}^{'}_{lmR}\right.\nonumber\\
\left.+\left[\overline{\bf {\Psi}}_{(+)\acute{a}ikL}{\bf
A}_{(+)\acute{b}}^{kj}+\overline{\bf {\Psi}}_{(+)\acute{a}L}^j{\bf
A}_{(+)\acute{b}i}\right]{\Lambda}^{'i}_{jR}\right\}
+{\mathsf L}_{(4)auxiliary}^{(45)}\\
\nonumber\\
 {\mathsf
L}_{(4)auxiliary}^{(45)}=\frac{h_{\acute{a}\acute{b}}^{^{(45+)}}{\mathsf
g}^{^{(45)}}}{4}<A_{(+)\acute{a}}|\Sigma_{\mu\nu}|A_{(+)\acute{b}}>D_{\mu\nu}~~~~~~
\end{eqnarray}

Similarly for $\overline{16}_{-}16_{-}-$ couplings we have
\begin{eqnarray}
\frac{1}{2}h_{\acute{a}\acute{b}}^{^{(45-)}}{\mathsf
g}^{^{(45)}}<\widehat{\Phi}_{(-)\acute{a}}|{\widehat{\mathsf
V}}_{\mu\nu}\Sigma_{\mu\nu}|\widehat{\Phi}_{(-)\acute{b}}>|_{\theta^2\bar
{\theta}^2}~~~~~\nonumber\\
 =h_{\acute{a}\acute{b}}^{^{(45-)}}{\mathsf
g}^{^{(45)}}\left\{\left[\frac{\sqrt 5}{2}\left({\bf
A}^{\dagger}_{(-)\acute{a}}\stackrel{\leftrightarrow}{\partial}_A
{\bf A}_{(-)\acute{b}}+i\overline{\bf
{\Psi}}_{(-)\acute{a}L}\gamma_A{\bf
{\Psi}}_{(-)\acute{b}L}\right)\right.\right.\nonumber\\
\left.\left.+\frac{1}{4\sqrt 5}\left({\bf
A}^{ij\dagger}_{(-)\acute{a}}\stackrel{\leftrightarrow}{\partial}_A{\bf
A}_{(-)\acute{b}ij}+i\overline{\bf
{\Psi}}_{(-)\acute{a}L}^{ij}\gamma_A{\bf
{\Psi}}_{(-)\acute{b}ijL}\right)\right.\right.\nonumber\\
\left.\left.-\frac{3}{2\sqrt 5}\left({\bf
A}^{\dagger}_{(-)\acute{a}i}\stackrel{\leftrightarrow}{\partial}_A{\bf
A}_{(-)\acute{b}}^{i}+i\overline{\bf
{\Psi}}_{(-)\acute{a}iL}\gamma_A{\bf
{\Psi}}_{(-)\acute{b}L}^i\right)\right]{\cal V}^{'A}\right.\nonumber\\
\left.+\left[\frac{1}{2\sqrt 2}\left({\bf
A}^{\dagger}_{(-)\acute{a}}\stackrel{\leftrightarrow}{\partial}_A{\bf
A}_{(-)\acute{b}lm}+i\overline{\bf
{\Psi}}_{(-)\acute{a}L}\gamma_A{\bf
{\Psi}}_{(-)\acute{b}lmL}\right)\right.\right.\nonumber\\
\left.\left.+\frac{1}{4\sqrt 2}\epsilon_{ijklm}\left({\bf
A}^{ij\dagger}_{(-)\acute{a}}\stackrel{\leftrightarrow}{\partial}_A{\bf
A}_{(-)\acute{b}}^{k}+i\overline{\bf
{\Psi}}_{(-)\acute{a}L}^{ij}\gamma_A{\bf
{\Psi}}_{(-)\acute{b}L}^{k}\right)\right]{\cal V}^{'Alm}\right.\nonumber\\
\left.+\left[-\frac{1}{2\sqrt 2}\left({\bf
A}^{lm\dagger}_{(-)\acute{a}}\stackrel{\leftrightarrow}{\partial}_A{\bf
A}_{(-)\acute{b}}+i\overline{\bf
{\Psi}}_{(-)\acute{a}L}^{lm}\gamma_A{\bf
{\Psi}}_{(-)\acute{b}L}\right)\right.\right.\nonumber\\
\left.\left.-\frac{1}{4\sqrt 2}\epsilon^{ijklm}\left({\bf
A}^{\dagger}_{(-)\acute{a}i}\stackrel{\leftrightarrow}{\partial}_A{\bf
A}_{(-)\acute{b}jk}+i\overline{\bf
{\Psi}}_{(-)\acute{a}iL}\gamma_A{\bf
{\Psi}}_{(-)\acute{b}jkL}\right)\right]{\cal V}^{'A}_{lm}\right.\nonumber\\
 \left.+\left[\frac{1}{\sqrt
2}\left({\bf
A}^{jk\dagger}_{(-)\acute{a}}\stackrel{\leftrightarrow}{\partial}_A{\bf
A}_{(-)\acute{b}ki}+i\overline{\bf
{\Psi}}_{(-)\acute{a}L}^{jk}\gamma_A{\bf
{\Psi}}_{(-)\acute{b}kiL}\right)\right.\right.\nonumber\\
\left.\left.+\frac{1}{\sqrt 2}\left({\bf
A}^{\dagger}_{(-)\acute{a}i}\stackrel{\leftrightarrow}{\partial}_A{\bf
A}_{(-)\acute{b}}^{j}+i\overline{\bf
{\Psi}}_{(-)\acute{a}iL}\gamma_A{\bf
{\Psi}}_{(-)\acute{b}L}^j\right)\right]{\cal V}^{'iA}_j\right.\nonumber\\
\left.-\frac{\sqrt 5}{2}\left[-{\bf
A}^{\dagger}_{(-)\acute{a}}\overline{\bf
{\Psi}}_{(-)\acute{b}R}-\frac{1}{10}{\bf
A}^{ij\dagger}_{(-)\acute{a}}\overline{\bf
{\Psi}}_{(-)\acute{b}ijR}+\frac{3}{5}{\bf
A}^{\dagger}_{(-)\acute{a}i}\overline{\bf
{\Psi}}_{(-)\acute{b}R}^i\right]{\Lambda}^{'}_{L}\right.\nonumber\\
\left.+\frac{1}{2}\left[{\bf
A}^{\dagger}_{(-)\acute{a}}\overline{\bf
{\Psi}}_{(-)\acute{b}lmR}+\frac{1}{2}\epsilon_{ijklm}{\bf
A}^{ij\dagger}_{(-)\acute{a}}\overline{\bf
{\Psi}}_{(-)\acute{b}R}^{k}\right]{\Lambda}^{'lm}_{L}\right.\nonumber\\
\left.-\frac{1}{2}\left[{\bf
A}^{lm\dagger}_{(-)\acute{a}}\overline{\bf
{\Psi}}_{(-)\acute{b}R}+\frac{1}{2}\epsilon^{ijklm}{\bf
A}^{\dagger}_{(-)\acute{a}i}\overline{\bf
{\Psi}}_{(-)\acute{b}jkR}\right]{\Lambda}^{'}_{lmL}\right.\nonumber\\
\left.+\left[{\bf A}^{jk\dagger}_{(-)\acute{a}}\overline{\bf
{\Psi}}_{(-)\acute{b}kiR}+{\bf
A}^{\dagger}_{(-)\acute{a}i}\overline{\bf
{\Psi}}_{(-)\acute{b}R}^j\right]{\Lambda}^{'i}_{jL}\right.\nonumber\\
\left.+\frac{\sqrt 5}{2}\left[-\overline{\bf
{\Psi}}_{(-)\acute{a}L}{\bf
A}_{(-)\acute{b}}-\frac{1}{10}\overline{\bf
{\Psi}}_{(-)\acute{a}L}^{ij}{\bf
A}_{(-)\acute{b}ij}+\frac{3}{5}\overline{\bf
{\Psi}}_{(-)\acute{a}iL}{\bf
A}_{(-)\acute{b}}^i\right]{\Lambda}^{'}_{R}\right.\nonumber\\
\left.-\frac{1}{2}\left[\overline{\bf {\Psi}}_{(-)\acute{a}L}{\bf
A}_{(-)\acute{b}lm}+\frac{1}{2}\epsilon_{ijklm}\overline{\bf
{\Psi}}_{(-)\acute{a}L}^{ij}{\bf A}^{k}_{(-)\acute{b}}\right]{\Lambda}^{'lm}_{R}\right.\nonumber\\
\left.+\frac{1}{2}\left[\overline{\bf
{\Psi}}_{(-)\acute{a}L}^{lm}{\bf
A}_{(-)\acute{b}}+\frac{1}{2}\epsilon^{ijklm}\overline{\bf
{\Psi}}_{(-)\acute{a}iL}{\bf A}_{(-)\acute{b}jk}\right]{\Lambda}^{'}_{lmR}\right.\nonumber\\
\left.-\left[\overline{\bf {\Psi}}_{(-)\acute{a}L}^{jk}{\bf
A}_{(-)\acute{b}ki}+\overline{\bf {\Psi}}_{(-)\acute{a}iL}{\bf
A}_{(-)\acute{b}}^j\right]{\Lambda}^{'i}_{jR}\right\}
+{\mathsf L}_{(5)auxiliary}^{(45)}\\
\nonumber\\
 {\mathsf
L}_{(5)auxiliary}^{(45)}=\frac{h_{\acute{a}\acute{b}}^{^{(45-)}}{\mathsf
g}^{^{(45)}}}{4}<A_{(-)\acute{a}}|\Sigma_{\mu\nu}|A_{(-)\acute{b}}>D_{\mu\nu}~~~~~~
\end{eqnarray}

Next, we evaluate couplings to  $\overline{16}_+16_+$ of  matter
 which are quadratic in the vector
multiplet  fields.   We have
\begin{eqnarray}
\frac{1}{8} h_{\acute{a}\acute{b}}^{^{(45+)}}{\mathsf
g}^{^{(45)}2}<\widehat{\Phi}_{(+)\acute{a}}|{\widehat{\mathsf
V}}_{\mu\nu}\Sigma_{\mu\nu}{\widehat{\mathsf
V}}_{\rho\lambda}\Sigma_{\rho\lambda}|\widehat{\Phi}_{(+)\acute{b}}>|_{\theta^2\bar
{\theta}^2}\nonumber\\
= h_{\acute{a}\acute{b}}^{^{(45+)}}{\mathsf g}^{^{(45)}2}
\left\{-\frac{5}{4}\left[{\bf A}^{\dagger}_{(+)\acute{a}}{\bf
A}_{(+)\acute{b}} +\frac{1}{50}{\bf
A}^{\dagger}_{(+)\acute{a}ij}{\bf
A}_{(+)\acute{b}}^{ij}\right.\right.\nonumber\\
\left.\left.+\frac{9}{25}{\bf
A}^{i\dagger}_{(+)\acute{a}}{\bf A}_{(+)\acute{b}i}\right]{\cal V}^{'}_A{\cal V}^{'A}\right.\nonumber\\
\left.+\frac{1}{2}\left[-{\bf A}^{m\dagger}_{(+)\acute{a}}{\bf
A}_{(+)\acute{b}m}\delta_i^l\delta_j^k+\left({\bf
A}^{\dagger}_{(+)\acute{a}im}{\bf A}_{(+)\acute{b}}^{ml}-{\bf
A}^{l\dagger}_{(+)\acute{a}}{\bf
A}_{(+)\acute{b}i}\right)\delta_j^k\right.\right.\nonumber\\
\left.\left.+{\bf A}^{\dagger}_{(+)\acute{a}ij}{\bf
A}_{(+)\acute{b}}^{kl}\right]{\cal V}^{'i}_{\acute{a}k}{\cal
V}^{'Aj}_{l}\right.\nonumber\\
\left.+\frac{1}{4}\left[\left(2{\bf
A}^{\dagger}_{(+)\acute{a}}{\bf
A}_{(+)\acute{b}}^{km}+\epsilon^{ijpkm}{\bf
A}^{\dagger}_{(+)\acute{a}ij}{\bf
A}_{(+)\acute{b}p}\right)\delta_n^l\right.\right.\nonumber\\
\left.\left.-2\epsilon^{ijklm}{\bf
A}^{\dagger}_{(+)\acute{a}in}{\bf A}_{(+)\acute{b}j}\right]{\cal
V}^{'}_{\acute{a}kl}{\cal
V}^{'An}_{m}\right.\nonumber\\
\left.+\frac{1}{4}\left[\left(-2{\bf
A}^{\dagger}_{(+)\acute{a}km}{\bf
A}_{(+)\acute{b}}-\epsilon_{ijpkm}{\bf
A}^{i\dagger}_{(+)\acute{a}}{\bf
A}_{(+)\acute{b}}^{jp}\right)\delta_l^n\right.\right.\nonumber\\
\left.\left.+2\epsilon_{ijklm}{\bf
A}^{i\dagger}_{(+)\acute{a}}{\bf
A}_{(+)\acute{b}}^{nj}\right]{\cal V}^{'kl}_{\acute{a}}{\cal
V}^{'Am}_{n}\right.\nonumber\\
\left.+\frac{1}{4\sqrt {10}}\left[-6{\bf
A}^{\dagger}_{(+)\acute{a}}{\bf
A}_{(+)\acute{b}}^{lm}+\epsilon^{ijklm}{\bf
A}^{\dagger}_{(+)\acute{a}ij}{\bf A}_{(+)\acute{b}k}\right]{\cal
V}^{'}_{\acute{a}lm}{\cal
V}^{'A}\right.\nonumber\\
\left.+\frac{1}{4\sqrt {10}}\left[6{\bf
A}^{\dagger}_{(+)\acute{a}lm}{\bf
A}_{(+)\acute{b}}-\epsilon_{ijklm}{\bf
A}^{i\dagger}_{(+)\acute{a}}{\bf
A}_{(+)\acute{b}}^{jk}\right]{\cal V}^{'lm}_{\acute{a}}{\cal
V}^{'A}\right.\nonumber\\
\left.+\frac{1}{\sqrt{10}}\left[3{\bf
A}^{j\dagger}_{(+)\acute{a}}{\bf A}_{(+)\acute{b}i}+5{\bf
A}^{\dagger}_{(+)\acute{a}ik}{\bf
A}_{(+)\acute{b}}^{kj}\right]{\cal V}^{'i}_{\acute{a}j}{\cal
V}^{'A}\right.\nonumber\\
\left.-\frac{1}{8}\left[\epsilon^{ijklm}{\bf
A}^{\dagger}_{(+)\acute{a}}{\bf A}_{(+)\acute{b}i}\right]{\cal
V}^{'}_{\acute{a}jk}{\cal V}^{'A}_{lm}
-\frac{1}{8}\left[\epsilon_{ijklm}{\bf
A}^{i\dagger}_{(+)\acute{a}}{\bf
A}_{(+)\acute{b}}\right]{\cal V}^{'jk}_{\acute{a}}{\cal V}^{'Alm}\right.\nonumber\\
\left.+\frac{1}{8}\left[\left( 2{\bf A}^{\dagger}_{(+)\acute{a}}
{\bf A}_{(+)\acute{b}}+{\bf A}^{\dagger}_{(+)\acute{a}mn} {\bf
A}_{(+)\acute{b}}^{mn}\right)\delta_i^k\delta_j^l +2{\bf
A}^{\dagger}_{(+)\acute{a}ij} {\bf A}_{(+)\acute{b}}^{kl}\right.\right.\nonumber\\
\left.\left.+2\left({\bf A}^{\dagger}_{(+)\acute{a}im} {\bf
A}_{(+)\acute{b}}^{ml}-{\bf A}^{l\dagger}_{(+)\acute{a}} {\bf
A}_{(+)\acute{b}i}\right)\delta_j^k\right]{\cal
V}^{'ij}_{\acute{a}}{\cal V}^{'A}_{kl}\right\}\nonumber\\
\end{eqnarray}
Similarly couplings to  $\overline{16}_-16_-$ of  matter
 which are quadratic in the vector
multiplet  fields are given by
\begin{eqnarray}
\frac{1}{8} h_{\acute{a}\acute{b}}^{^{(45-)}}{\mathsf
g}^{^{(45)}2}<\widehat{\Phi}_{(-)\acute{a}}|{\widehat{\mathsf
V}}_{\mu\nu}\Sigma_{\mu\nu}{\widehat{\mathsf
V}}_{\rho\lambda}\Sigma_{\rho\lambda}|\widehat{\Phi}_{(-)\acute{b}}>|_{\theta^2\bar
{\theta}^2}\nonumber\\
= h_{\acute{a}\acute{b}}^{^{(45-)}}{\mathsf g}^{^{(45)}2}
 \left\{-\frac{5}{4}\left[{\bf
A}^{\dagger}_{(-)\acute{a}}{\bf A}_{(-)\acute{b}}
+\frac{1}{50}{\bf A}^{ij\dagger}_{(-)\acute{a}}{\bf
A}_{(-)\acute{b}ij}\right.\right.\nonumber\\
\left.\left.+\frac{9}{25}{\bf
A}^{\dagger}_{(-)\acute{a}i}{\bf A}_{(-)\acute{b}}^i\right]{\cal V}^{'}_A{\cal V}^{'A}\right.\nonumber\\
\left.+\frac{1}{2}\left[\left({\bf
A}^{im\dagger}_{(-)\acute{a}}{\bf A}_{(-)\acute{b}ml}-{\bf
A}^{\dagger}_{(-)\acute{a}l}{\bf
A}_{(-)\acute{b}}^i\right)\delta_k^j-{\bf
A}^{ij\dagger}_{(-)\acute{a}}{\bf A}_{(-)\acute{b}kl}\right]{\cal
V}^{'k}_{\acute{a}i}{\cal
V}^{'Al}_{j}\right.\nonumber\\
\left.+\frac{1}{4}\left[\left(2{\bf
A}^{km\dagger}_{(-)\acute{a}}{\bf A}_{(-)\acute{b}}
+\epsilon^{ijpkm}{\bf A}^{\dagger}_{(-)\acute{a}i}{\bf
A}_{(-)\acute{b}jp}\right)\delta_n^l\right.\right.\nonumber\\
\left.\left.+\epsilon^{ijklm}{\bf A}^{\dagger}_{(-)\acute{a}n}{\bf
A}_{(-)\acute{b}ij}\right]{\cal V}^{'}_{\acute{a}kl}{\cal
V}^{'An}_{m}\right.\nonumber\\
\left.-\frac{1}{4}\left[\left(2{\bf
A}^{\dagger}_{(-)\acute{a}}{\bf A}_{(-)\acute{b}km}
+\epsilon_{ijpkm}{\bf A}^{ij\dagger}_{(-)\acute{a}}{\bf
A}_{(-)\acute{b}}^{p}\right)\delta_l^n\right.\right.\nonumber\\
\left.\left.+\epsilon_{ijklm}{\bf
A}^{ij\dagger}_{(-)\acute{a}}{\bf
A}_{(-)\acute{b}}^{n}\right]{\cal V}^{'kl}_{\acute{a}}{\cal
V}^{'Am}_{n}\right.\nonumber\\
\left.+\frac{1}{4\sqrt {10}}\left[6{\bf
A}^{lm\dagger}_{(-)\acute{a}}{\bf
A}_{(-)\acute{b}}-\epsilon^{ijklm}{\bf
A}^{\dagger}_{(-)\acute{a}i}{\bf A}_{(-)\acute{b}jk}\right]{\cal
V}^{'}_{\acute{a}lm}{\cal
V}^{'A}\right.\nonumber\\
\left.+\frac{1}{4\sqrt {10}}\left[-6{\bf
A}^{\dagger}_{(-)\acute{a}}{\bf
A}_{(-)\acute{b}lm}+\epsilon_{ijklm}{\bf
A}^{ij\dagger}_{(-)\acute{a}}{\bf
A}_{(-)\acute{b}}^{k}\right]{\cal V}^{'lm}_{\acute{a}}{\cal
V}^{'A}\right.\nonumber\\
\left.+\frac{1}{\sqrt{10}}\left[3{\bf
A}^{\dagger}_{(-)\acute{a}i}{\bf A}_{(-)\acute{b}^j}+{\bf
A}^{jk\dagger}_{(-)\acute{a}}{\bf A}_{(-)\acute{b}ki}\right]{\cal
V}^{'i}_{\acute{a}j}{\cal
V}^{'A}\right.\nonumber\\
\left.-\frac{1}{8}\left[\epsilon^{ijklm}{\bf
A}^{\dagger}_{(-)\acute{a}i}{\bf A}_{(-)\acute{b}}\right]{\cal
V}^{'}_{\acute{a}jk}{\cal V}^{'A}_{lm}
-\frac{1}{8}\left[\epsilon_{ijklm}{\bf
A}^{\dagger}_{(-)\acute{a}}{\bf
A}_{(-)\acute{b}}^i\right]{\cal V}^{'jk}_{\acute{a}}{\cal V}^{'Alm}\right.\nonumber\\
\left.+\frac{1}{4}\left[\left({\bf A}^{\dagger}_{(-)\acute{a}}
{\bf A}_{(-)\acute{b}}+{\bf A}^{mn\dagger}_{(-)\acute{a}}{\bf
A}_{(-)\acute{b}mn}+{\bf A}^{\dagger}_{(-)\acute{a}m} {\bf
A}_{(-)\acute{b}}^m\right)\delta_k^i\delta_l^j+{\bf
A}^{ij\dagger}_{(-)\acute{a}} {\bf A}_{(-)\acute{b}kl}\right.\right.\nonumber\\
\left.\left.+\left(-3{\bf A}^{im\dagger}_{(-)\acute{a}} {\bf
A}_{(-)\acute{b}ml}+{\bf A}^{\dagger}_{(-)\acute{a}l} {\bf
A}_{(-)\acute{b}}^i\right)\delta_k^j\right] {\cal
V}^{'}_{\acute{a}ij}{\cal V}^{'Akl}\right\}\nonumber\\
\end{eqnarray}

\begin{eqnarray}
{\mathsf L}_{\mathsf W}^{(45)}={\mathsf
W}(\widehat{\Phi}_{(+)},\widehat{\Phi}_{(-)})|_{\theta^2}+ h.c.
\end{eqnarray}
where
\begin{eqnarray}
{\mathsf W}(\widehat{\Phi}_{(+)},\widehat{\Phi}_{(-)})=
\mu_{\acute{a}\acute{b}}<\widehat{\Phi}_{(-)\acute{a}}^*|B|\widehat{\Phi}_{(+)\acute{b}}>\\
{\mathsf W}({\bf A}_{(+)},{\bf A}_{(-)})
=i\mu_{\acute{a}\acute{b}}\left({\bf
A}_{(-)\acute{a}}^{\bf{T}}{\bf A}_{(+)\acute{b}}-\frac{1}{2}{\bf
A}_{(-)\acute{a}ij}^{\bf{T}}{\bf A}_{(+)\acute{b}}^{ij}+{\bf
A}_{(-)\acute{a}}^{i\bf{T}}{\bf A}_{(+)\acute{b}i}\right)
\end{eqnarray}
and where $\mu_{\acute{a}\acute{b}}$ is taken to be a symmetric
tensor. Thus we have
\begin{eqnarray}
{\mathsf L}_{\mathsf W}^{(45)}=-i\mu_{\acute{a}\acute{b}}\left(
\overline{\bf {\Psi}}_{(-)\acute{a}R}{\bf
{\Psi}}_{(+)\acute{b}L}+\overline{\bf
{\Psi}}_{(-)\acute{a}R}^i{\bf
{\Psi}}_{(+)\acute{b}iL}-\frac{1}{2}\overline{\bf
{\Psi}}_{(-)\acute{a}ijR}{\bf
{\Psi}}_{(+)\acute{b}L}^{ij}\right)\nonumber\\
+i\mu_{\acute{a}\acute{b}}^{*}\left( \overline{\bf
{\Psi}}_{(-)\acute{a}L}{\bf {\Psi}}_{(+)\acute{b}R}+\overline{\bf
{\Psi}}_{(-)\acute{a}iL}{\bf
{\Psi}}_{(+)\acute{b}R}^i-\frac{1}{2}\overline{\bf
{\Psi}}_{(-)\acute{a}L}^{ij}{\bf
{\Psi}}_{(+)\acute{b}ijR}\right)\nonumber\\
+{\mathsf L}_{(6)auxiliary}^{(45)}\\
\nonumber\\
 {\mathsf
L}_{(6)auxiliary}^{(45)}=i\mu_{\acute{a}\acute{b}}\left[{\bf
F}_{(-)\acute{a}}{\bf A}_{(+)\acute{b}}+{\bf
A}_{(-)\acute{a}}^{\bf{T}}{\bf F}_{(+)\acute{b}}-\frac{1}{2}{\bf
F}_{(-)\acute{a}ij}{\bf A}_{(+)\acute{b}}^{ij}\right.\nonumber\\
\left.-\frac{1}{2}{\bf A}_{(-)\acute{a}ij}^{\bf{T}}{\bf
F}_{(+)\acute{b}}^{ij} + {\bf F}_{(-)\acute{a}}^i{\bf
A}_{(+)\acute{b}i}+{\bf A}_{(-)\acute{a}}^{i\bf{T}}{\bf
F}_{(+)\acute{b}i}\right] +h.c.
\end{eqnarray}
Eliminating the fields, ${\bf F}_{(\pm)}$ through their field
equations we get
\begin{eqnarray}
{\mathsf L}_{(2)auxiliary}^{(45)}+{\mathsf
L}_{(3)auxiliary}^{(45)}+{\mathsf
L}_{(6)auxiliary}^{(45)}\nonumber\\
=-\left(\mu^*
[h^{^{(45-)}}]^{\bf{-1}}[h^{^{(45-)}}]^{\bf{T}}[h^{^{(45-)}}]^{\bf{-1}}\mu\right)_{\acute{a}\acute{b}}\nonumber\\
\times\left[{\bf A}^{\dagger}_{(+)\acute{a}}{\bf
A}_{(+)\acute{b}}+\frac{1}{4}{\bf
A}^{\dagger}_{(+)\acute{a}ij}{\bf A}_{(+)\acute{b}}^{ij}+{\bf
A}^{i\dagger}_{(+)\acute{a}}{\bf A}_{(+)\acute{b}i}\right]\nonumber\\
-\left(\mu[h^{^{(45+)}}]^{\bf{-1T}}h^{^{(45+)}}[h^{^{(45+)}}]^{\bf{-1T}}\mu^*\right)_{\acute{a}\acute{b}}\nonumber\\
\times\left[ {\bf A}^{\bf{T}}_{(-)\acute{a}}{\bf
A}_{(-)\acute{b}}^*+\frac{1}{4}{\bf
A}^{\bf{T}}_{(-)\acute{a}ij}{\bf A}_{(-)\acute{b}}^{ij*}+{\bf
A}^{i\bf{T}}_{(-)\acute{a}}{\bf A}_{(-)\acute{b}i}^*\right]~~~~~~
\end{eqnarray}

Similarly, eliminating the auxiliary field $D_{\mu\nu}$ we get
\begin{eqnarray}
{\mathsf L}_{(1)auxiliary}^{(45)}+{\mathsf
L}_{(4)auxiliary}^{(45)}+{\mathsf
L}_{(5)auxiliary}^{(45)}\nonumber\\
=-\frac{1}{32} {\mathsf
g}^{^{(45)}2}h_{\acute{a}\acute{b}}^{^{(45+)}}h_{\acute{c}\acute{d}}^{^{(45+)}}<A_{(+)\acute{a}}|\Sigma_{\mu\nu}|A_{(+)\acute{b}}><A_{(+)\acute{c}}|\Sigma_{\mu\nu}|A_{(+)\acute{d}}>
\nonumber\\
-\frac{1}{32} {\mathsf
g}^{^{(45)}2}h_{\acute{a}\acute{b}}^{^{(45-)}}h_{\acute{c}\acute{d}}^{^{(45-)}}<A_{(-)\acute{a}}|\Sigma_{\mu\nu}|A_{(-)\acute{b}}><A_{(-)\acute{c}}|\Sigma_{\mu\nu}|A_{(-)\acute{d}}>\nonumber\\
-\frac{1}{16} {\mathsf
g}^{^{(45)}2}h_{\acute{a}\acute{b}}^{^{(45+)}}
h_{\acute{c}\acute{d}}^{^{(45-)}}
<A_{(+)\acute{a}}|\Sigma_{\mu\nu}|A_{(+)\acute{b}}><A_{(-)\acute{c}}|\Sigma_{\mu\nu}|A_{(-)\acute{d}}>~~~~~~
\end{eqnarray}
The terms above when expanded in terms of $SU(5)$ fields give
\begin{eqnarray}
-\frac{1}{32} {\mathsf
g}^{^{(45)}2}h_{\acute{a}\acute{b}}^{^{(45+)}}h_{\acute{c}\acute{d}}^{^{(45+)}}<A_{(+)\acute{a}}|\Sigma_{\mu\nu}|A_{(+)\acute{b}}><A_{(+)\acute{c}}|\Sigma_{\mu\nu}|A_{(+)\acute{d}}>
\nonumber\\
={\mathsf
g}^{^{(45)}2}\left\{-\frac{1}{16}\left(\eta_{\acute{a}\acute{b},\acute{c}\acute{d}}^{^{(45++)}}+4\eta_{\acute{a}\acute{d},\acute{c}\acute{b}}^{^{(45++)}}\right)
\left({\bf A}^{\dagger}_{(+)\acute{a}}{\bf A}_{(+)\acute{b}}{\bf
A}^{\dagger}_{(+)\acute{c}ij}{\bf
A}_{(+)\acute{d}}^{ij}\right.\right.\nonumber\\
\left.\left.+{\bf A}^{i\dagger}_{(+)\acute{a}}{\bf
A}_{(+)\acute{b}i}{\bf A}^{j\dagger}_{(+)\acute{c}}{\bf
A}_{(+)\acute{d}j} +{\bf A}^{i\dagger}_{(+)\acute{a}}{\bf
A}_{(+)\acute{b}i}{\bf
A}^{\dagger}_{(+)\acute{c}jk}{\bf A}_{(+)\acute{d}}^{jk}\right)\right.\nonumber\\
\left.-\frac{1}{2}\left(\eta_{\acute{a}\acute{b},\acute{c}\acute{d}}^{^{(45++)}}+\eta_{\acute{a}\acute{d},\acute{c}\acute{b}}^{^{(45++)}}\right){\bf
A}^{i\dagger}_{(+)\acute{a}}{\bf A}_{(+)\acute{b}j}{\bf
A}^{\dagger}_{(+)\acute{c}ik}{\bf
A}_{(+)\acute{d}}^{kj}\right.\nonumber\\
\left.+\eta_{\acute{a}\acute{b},\acute{c}\acute{d}}^{^{(45++)}}\left[-\frac{1}{8}\epsilon_{ijklm}{\bf
A}^{\dagger}_{(+)\acute{a}}{\bf A}_{(+)\acute{b}}^{ij}{\bf
A}^{k\dagger}_{(+)\acute{c}}{\bf
A}_{(+)\acute{d}}^{lm}\right.\right.\nonumber\\
\left.\left.-\frac{1}{8}\epsilon^{ijklm}{\bf
A}^{\dagger}_{(+)\acute{a}ij}{\bf A}_{(+)\acute{b}k}{\bf
A}^{\dagger}_{(+)\acute{c}lm}{\bf
A}_{(+)\acute{d}}\right.\right.\nonumber\\
\left.\left.-\frac{1}{4}{\bf A}^{\dagger}_{(+)\acute{a}ik}{\bf
A}_{(+)\acute{b}}^{kj}{\bf A}^{\dagger}_{(+)\acute{c}jl}{\bf
A}_{(+)\acute{d}}^{li}+\frac{3}{64}{\bf
A}^{\dagger}_{(+)\acute{a}ij}{\bf A}_{(+)\acute{b}}^{ij}{\bf
A}^{\dagger}_{(+)\acute{c}kl}{\bf A}_{(+)\acute{d}}^{kl}\right.\right.\nonumber\\
\left.\left.+\frac{3}{8}{\bf A}^{\dagger}_{(+)\acute{a}}{\bf
A}_{(+)\acute{b}}{\bf A}^{i\dagger}_{(+)\acute{c}}{\bf
A}_{(+)\acute{d}i}-\frac{5}{16}{\bf
A}^{\dagger}_{(+)\acute{a}}{\bf A}_{(+)\acute{b}}{\bf
A}^{\dagger}_{(+)\acute{c}}{\bf
A}_{(+)\acute{d}}\right]\right\}~~~~~~
\end{eqnarray}
\newpage
\begin{eqnarray}
-\frac{1}{32} {\mathsf
g}^{^{(45)}2}h_{\acute{a}\acute{b}}^{^{(45-)}}h_{\acute{c}\acute{d}}^{^{(45-)}}<A_{(-)\acute{a}}|\Sigma_{\mu\nu}|A_{(-)\acute{b}}><A_{(-)\acute{c}}|\Sigma_{\mu\nu}|A_{(-)\acute{d}}>
\nonumber\\
={\mathsf
g}^{^{(45)}2}\left\{-\frac{1}{16}\left(\eta_{\acute{a}\acute{b},\acute{c}\acute{d}}^{^{(45--)}}+4\eta_{\acute{a}\acute{d},\acute{c}\acute{b}}^{^{(45--)}}\right)
\left({\bf A}^{\dagger}_{(-)\acute{a}i}{\bf
A}_{(-)\acute{b}}^{i}{\bf A}^{jk\dagger}_{(-)\acute{c}}{\bf
A}_{(-)\acute{d}jk}
\right.\right.\nonumber\\
\left.\left.+{\bf A}^{\dagger}_{(-)\acute{a}i}{\bf
A}_{(-)\acute{b}}^{i}{\bf
A}^{\dagger}_{(-)\acute{c}j}{\bf A}_{(-)\acute{d}}^{j}\right)\right.\nonumber\\
\left.-\frac{1}{16}\left(-11\eta_{\acute{a}\acute{b},\acute{c}\acute{d}}^{^{(45--)}}+4\eta_{\acute{a}\acute{d},\acute{c}\acute{b}}^{^{(45--)}}\right)
{\bf A}^{\dagger}_{(-)\acute{a}}{\bf A}_{(-)\acute{b}}{\bf
A}^{ij\dagger}_{(-)\acute{c}}{\bf A}_{(-)\acute{d}ij}\right.\nonumber\\
\left.-\frac{1}{2}\left(\eta_{\acute{a}\acute{b},\acute{c}\acute{d}}^{^{(45--)}}+\eta_{\acute{a}\acute{d},\acute{c}\acute{b}}^{^{(45--)}}\right){\bf
A}^{\dagger}_{(-)\acute{a}i}{\bf A}_{(-)\acute{b}}^{j}{\bf
A}^{ik\dagger}_{(-)\acute{c}}{\bf
A}_{(-)\acute{d}kj}\right.\nonumber\\
\left.+\eta_{\acute{a}\acute{b},\acute{c}\acute{d}}^{^{(45--)}}\left[-\frac{1}{8}\epsilon_{ijklm}{\bf
A}^{ij\dagger}_{(-)\acute{a}}{\bf A}_{(-)\acute{b}}{\bf
A}^{kl\dagger}_{(-)\acute{c}}{\bf
A}_{(-)\acute{d}}^{m}\right.\right.\nonumber\\
\left.\left.-\frac{1}{8}\epsilon^{ijklm}{\bf
A}^{\dagger}_{(-)\acute{a}i}{\bf A}_{(-)\acute{b}jk}{\bf
A}^{\dagger}_{(-)\acute{c}}{\bf
A}_{(-)\acute{d}lm}\right.\right.\nonumber\\
\left.\left.-\frac{1}{4}{\bf A}^{ik\dagger}_{(-)\acute{a}}{\bf
A}_{(-)\acute{b}kj}{\bf A}^{jl\dagger}_{(-)\acute{c}}{\bf
A}_{(-)\acute{d}li}+\frac{3}{64}{\bf
A}^{ij\dagger}_{(-)\acute{a}}{\bf A}_{(-)\acute{b}ij}{\bf
A}^{kl\dagger}_{(-)\acute{c}}{\bf A}_{(-)\acute{d}kl}\right.\right.\nonumber\\
\left.\left.+\frac{7}{8}{\bf A}^{\dagger}_{(-)\acute{a}}{\bf
A}_{(-)\acute{b}}{\bf A}^{\dagger}_{(-)\acute{c}i}{\bf
A}_{(-)\acute{d}}^i+\frac{15}{16}{\bf
A}^{\dagger}_{(-)\acute{a}}{\bf A}_{(-)\acute{b}}{\bf
A}^{\dagger}_{(-)\acute{c}}{\bf
A}_{(-)\acute{d}}\right]\right\}~~~~~~
\end{eqnarray}

\begin{eqnarray}
-\frac{1}{16} {\mathsf
g}^{^{(45)}2}h_{\acute{a}\acute{b}}^{^{(45+)}}h_{\acute{c}\acute{d}}^{^{(45-)}}<A_{(+)\acute{a}}|\Sigma_{\mu\nu}|A_{(+)\acute{b}}><A_{(-)\acute{c}}|\Sigma_{\mu\nu}|A_{(-)\acute{d}}>
\nonumber\\
={\mathsf
g}^{^{(45)}2}\eta_{\acute{a}\acute{b},\acute{c}\acute{d}}^{^{(45+-)}}\left\{\frac{5}{8}{\bf
A}^{\dagger}_{(+)\acute{a}}{\bf A}_{(+)\acute{b}}{\bf
A}^{\dagger}_{(-)\acute{c}}{\bf A}_{(-)\acute{d}}-\frac{3}{32}{\bf
A}^{\dagger}_{(+)\acute{a}ij}{\bf A}_{(+)\acute{b}}^{ij}{\bf
A}^{kl\dagger}_{(-)\acute{c}}{\bf
A}_{(-)\acute{d}kl}\right.\nonumber\\
\left.+\frac{1}{8}{\bf A}^{i\dagger}_{(+)\acute{a}}{\bf
A}_{(+)\acute{b}i}{\bf A}^{\dagger}_{(-)\acute{c}j}{\bf
A}_{(-)\acute{d}}^j+\frac{1}{2}{\bf
A}^{j\dagger}_{(+)\acute{a}}{\bf A}_{(+)\acute{b}i}{\bf
A}^{\dagger}_{(-)\acute{c}j}{\bf
A}_{(-)\acute{d}}^i\right.\nonumber\\
\left.+\frac{1}{2}{\bf A}^{\dagger}_{(+)\acute{a}ik}{\bf
A}_{(+)\acute{b}}^{kj}{\bf A}^{il\dagger}_{(-)\acute{c}}{\bf
A}_{(-)\acute{d}lj}-\frac{3}{8}{\bf
A}^{\dagger}_{(+)\acute{a}}{\bf A}_{(+)\acute{b}}{\bf
A}^{\dagger}_{(-)\acute{c}i}{\bf
A}_{(-)\acute{d}}^i\right.\nonumber\\
\left.+\frac{13}{8}{\bf A}^{i\dagger}_{(+)\acute{a}}{\bf
A}_{(+)\acute{b}i}{\bf A}^{\dagger}_{(-)\acute{c}}{\bf
A}_{(-)\acute{d}}+\frac{1}{16}{\bf A}^{\dagger}_{(+)\acute{a}}{\bf
A}_{(+)\acute{b}}{\bf A}^{ij\dagger}_{(-)\acute{c}}{\bf
A}_{(-)\acute{d}ij}\right.\nonumber\\
\left.+\frac{9}{16}{\bf A}^{\dagger}_{(+)\acute{a}ij}{\bf
A}_{(+)\acute{b}}^{ij}{\bf A}^{\dagger}_{(-)\acute{c}}{\bf
A}_{(-)\acute{d}}-\frac{15}{16}{\bf
A}^{i\dagger}_{(+)\acute{a}}{\bf A}_{(+)\acute{b}i}{\bf
A}^{jk\dagger}_{(-)\acute{c}}{\bf
A}_{(-)\acute{d}jk}\right.\nonumber\\
\left.-\frac{1}{16}{\bf A}^{\dagger}_{(+)\acute{a}ij}{\bf
A}_{(+)\acute{b}}^{ij}{\bf A}^{\dagger}_{(-)kc}{\bf
A}_{(-)\acute{d}}^k+\frac{1}{2}{\bf
A}^{j\dagger}_{(+)\acute{a}}{\bf A}_{(+)\acute{b}i}{\bf
A}^{ik\dagger}_{(-)\acute{c}}{\bf
A}_{(-)\acute{d}kj}\right.\nonumber\\
\left.+\frac{1}{2}{\bf A}^{\dagger}_{(+)\acute{a}ik}{\bf
A}_{(+)\acute{b}}^{kj}{\bf A}^{\dagger}_{(-)jc}{\bf
A}_{(-)\acute{d}}^i-\frac{1}{4}{\bf
A}^{\dagger}_{(+)\acute{a}ij}{\bf A}_{(+)\acute{b}}{\bf
A}^{ij\dagger}_{(-)\acute{c}}{\bf
A}_{(-)\acute{d}}\right.\nonumber\\
\left.-\frac{1}{4}{\bf A}^{\dagger}_{(+)\acute{a}}{\bf
A}_{(+)\acute{b}}^{ij}{\bf A}^{\dagger}_{(-)\acute{c}}{\bf
A}_{(-)\acute{d}ij}-\frac{1}{4}{\bf
A}^{i\dagger}_{(+)\acute{a}}{\bf A}_{(+)\acute{b}}^{jk}{\bf
A}^{\dagger}_{(-)\acute{c}i}{\bf
A}_{(-)\acute{d}jk}\right.\nonumber\\
\left.-\frac{1}{4}{\bf A}^{\dagger}_{(+)\acute{a}ij}{\bf
A}_{(+)\acute{b}k}{\bf A}^{ij\dagger}_{(-)\acute{c}}{\bf
A}_{(-)\acute{d}}^k-\frac{1}{2}{\bf
A}^{i\dagger}_{(+)\acute{a}}{\bf A}_{(+)\acute{b}}^{jk}{\bf
A}^{\dagger}_{(-)\acute{c}j}{\bf
A}_{(-)\acute{d}ki}\right.\nonumber\\
\left.-\frac{1}{2}{\bf A}^{\dagger}_{(+)\acute{a}ij}{\bf
A}_{(+)\acute{b}k}{\bf A}^{ki\dagger}_{(-)\acute{c}}{\bf
A}_{(-)\acute{d}}^j-\frac{1}{8}\epsilon^{ijklm}{\bf
A}^{\dagger}_{(+)\acute{a}ij}{\bf A}_{(+)\acute{b}}{\bf
A}^{\dagger}_{(-)\acute{c}k}{\bf
A}_{(-)\acute{d}lm}\right.\nonumber\\
\left.-\frac{1}{8}\epsilon_{ijklm}{\bf
A}^{\dagger}_{(+)\acute{a}}{\bf A}_{(+)\acute{b}}^{ij}{\bf
A}^{kl\dagger}_{(-)\acute{c}}{\bf
A}_{(-)\acute{d}}^m-\frac{1}{8}\epsilon_{ijklm}{\bf
A}^{i\dagger}_{(+)\acute{a}}{\bf A}_{(+)\acute{b}}^{jk}{\bf
A}^{lm\dagger}_{(-)\acute{c}}{\bf
A}_{(-)\acute{d}}\right.\nonumber\\
\left.-\frac{1}{8}\epsilon^{ijklm}{\bf
A}^{\dagger}_{(+)\acute{a}ij}{\bf A}_{(+)\acute{b}k}{\bf
A}^{\dagger}_{(-)\acute{c}}{\bf A}_{(-)\acute{d}lm}\right\}~~~~~~
\end{eqnarray}
where $\eta$'s are defined by
\begin{eqnarray}
\eta_{\acute{a}\acute{b},\acute{c}\acute{d}}^{^{(45++)}}=h_{\acute{a}\acute{b}}^{^{(45+)}}h_{\acute{c}\acute{d}}^{^{(45+)}};~~~\eta_{\acute{a}\acute{b},\acute{c}\acute{d}}^{^{(45--)}}=h_{\acute{a}\acute{b}}^{^{(45-)}}h_{\acute{c}\acute{d}}^{^{(45-)}}
;~~~ \eta_{\acute{a}\acute{b},\acute{c}\acute{d}}^{^{(45+-)}}=
h_{\acute{a}\acute{b}}^{^{(45+)}}h_{\acute{c}\acute{d}}^{^{(45-)}}~~~~~~
\end{eqnarray}

\chapter{Coupling the Supersymmetric 210 Vector Multiplet to Matter
in SO(10)}\label{Dir_Detect}

\section{Coupling of {$\bf{210}$} vector multiplet to $\bf{16_{\pm}}$
plet of matter} In usual formulations of particle interactions the
vectors belong to either singlets or the adjoint representations
of the gauge group of the theory under consideration. To couple
the vectors to
 matter one forms a Lie valued quantity ${\widehat {\mathsf V}}^aT_a$
where $T_a$ are the generators of the gauge group satisfying the
algebra $[T_a,T_b]=if_{abc} T_c$. Then one couples the Lie valued
quantity to the matter fields in the form
$\Phi^{\dagger}e^{g{{\widehat {\mathsf V}}}}\Phi$ which can be
shown to be a gauge invariant combination. For the representations
$1$ and $45$ one can carry out this construction straightforwardly
(see chapter 8). However, this construction does not work for the
$210$ vector multiplet as one cannot write a gauge invariant
Yang-Mills theory for  it.
 Further, for the same reason one cannot write a
gauge invariant coupling of the $210$ vector  with matter. To
construct the 210 vector couplings, the technique we adopt is to
carry out a direct expansion in powers of the vector superfield.
Thus we have\cite{ns3}.

 \begin{eqnarray}
{\mathsf L}_{V+\Phi}^{^{(210~
Interaction)}}=h_{\acute{a}\acute{b}}^{^{(210+)}}\left[<\widehat{\Phi}_{(+)\acute{a}}|
\widehat{\Phi}_{(+)\acute{b}}>\right.\nonumber\\
\left.+\frac{{\mathsf
g}^{^{(210)}}}{4!}<\widehat{\Phi}_{(+)\acute{a}}|\widehat{\mathsf
V}_{\mu\nu\rho\lambda}\Gamma_{[\mu}\Gamma_{\nu}\Gamma_{\rho}
\Gamma_{\lambda]}|\widehat{\Phi}_{(+)\acute{b}}>\right.\nonumber\\
\left.+\frac{1}{2}\left(\frac{{\mathsf
g}^{^{(210)}}}{4!}\right)^2<\widehat{\Phi}_{(+)\acute{a}}|\widehat{\mathsf
V}_{\mu\nu\rho\lambda}\Gamma_{[\mu}\Gamma_{\nu}\Gamma_{\rho}
\Gamma_{\lambda]}\widehat{\mathsf
V}_{\alpha\beta\gamma\delta}\Gamma_{[\alpha}\Gamma_{\beta}\Gamma_{\gamma}
\Gamma_{\delta]}|\widehat{\Phi}_{(+)\acute{b}}>\right]|_{\theta^2\bar
{\theta}^2}\nonumber\\
 +h_{\acute{a}\acute{b}}^{^{(210-)}}\left[<\widehat{\Phi}_{(-)\acute{a}}|
\widehat{\Phi}_{(-)\acute{b}}>+\frac{{\mathsf
g}^{^{(210)}}}{4!}<\widehat{\Phi}_{(-)\acute{a}}|\widehat{\mathsf
V}_{\mu\nu\rho\lambda}\Gamma_{[\mu}\Gamma_{\nu}\Gamma_{\rho}
\Gamma_{\lambda]}|\widehat{\Phi}_{(-)\acute{b}}>
\right.\nonumber\\
\left.+\frac{1}{2}\left(\frac{{\mathsf
g}^{^{(210)}}}{4!}\right)^2<\widehat{\Phi}_{(-)\acute{a}}|\widehat{\mathsf
V}_{\mu\nu\rho\lambda}\Gamma_{[\mu}\Gamma_{\nu}\Gamma_{\rho}
\Gamma_{\lambda]}\widehat{\mathsf
V}_{\alpha\beta\gamma\delta}\Gamma_{[\alpha}\Gamma_{\beta}\Gamma_{\gamma}
\Gamma_{\delta]}|\widehat{\Phi}_{(-)\acute{b}}>\right]|_{\theta^2\bar
{\theta}^2} \nonumber\\
+...~~~~~~~~\nonumber\\
\end{eqnarray}
 $\widehat{\mathsf V}_{\mu\nu\rho\lambda}$ ($\mu$,
$\nu$, $\lambda$, $\rho$=1,2,..,10) is the vector superfield in
the Wess-Zumino gauge. In the following we give a full exhibition
of the couplings to only linear order in the vector superfield in
terms of its $SU(5)\times U(1)$ decompostion but a similar
analysis can be done for couplings involving higher powers of the
superfield.
 Thus using the analysis of Ref.\cite{ns1,ns2} we find that
 the ${16}^{\dagger}~16$ couplings can be decomposed as follows

\begin{eqnarray}
h_{\acute{a}\acute{b}}^{^{(210+)}}<\widehat{\Phi}_{(+)\acute{a}}|\widehat{\Phi}_{(+)\acute{b}}>|_{\theta^2\bar
{\theta}^2}=h_{\acute{a}\acute{b}}^{^{(210+)}}\left[-\partial_A{\bf
A}^{\dagger}_{(+)\acute{a}}\partial^A{\bf
A}_{(+)\acute{b}}\right.\nonumber\\
\left.-\partial_A{\bf A}^{i\dagger}_{(+)\acute{a}}\partial^A{\bf
A}_{(+)\acute{b}i} -\partial_A{\bf
A}^{\dagger}_{(+)\acute{a}ij}\partial^A{\bf
A}_{(+)\acute{b}}^{ij}\right.\nonumber\\
\left.-i\overline{\bf
{\Psi}}_{(+)\acute{a}L}\gamma^A\partial_A{\bf
{\Psi}}_{(+)\acute{b}L} -i\overline{\bf
{\Psi}}_{(+)\acute{a}L}^i\gamma^A\partial_A{\bf
{\Psi}}_{(+)\acute{b}iL}\right.\nonumber\\
\left.-i\overline{\bf
{\Psi}}_{(+)\acute{a}ijL}\gamma^A\partial_A{\bf
{\Psi}}_{(+)\acute{b}L}^{ij}\right]
+{\mathsf L}^{(210)}_{(1)auxiliary}\\
\nonumber\\
 {\mathsf
L}^{(210)}_{(1)auxiliary}\equiv
h_{\acute{a}\acute{b}}^{^{(210+)}}<F_{(+)\acute{a}}|F_{(+)\acute{b}}>~~~~~~
\end{eqnarray}

\begin{eqnarray}
h_{\acute{a}\acute{b}}^{^{(210-)}}<\widehat{\Phi}_{(-)\acute{a}}|
\widehat{\Phi}_{(-)\acute{b}}>|_{\theta^2\bar
{\theta}^2}=h_{\acute{a}\acute{b}}^{^{(210-)}}\left[-\partial_A{\bf
A}^{\dagger}_{(-)\acute{a}}\partial^A{\bf
A}_{(-)\acute{b}}\right.\nonumber\\
\left.-\partial_A{\bf A}^{\dagger}_{(-)\acute{a}i}\partial^A{\bf
A}_{(-)\acute{b}}^i -\partial_A{\bf
A}^{ij\dagger}_{(-)\acute{a}}\partial^A{\bf
A}_{(-)\acute{b}ij}\right.\nonumber\\
\left.-i\overline{\bf
{\Psi}}_{(-)\acute{a}L}\gamma^A\partial_A{\bf
{\Psi}}_{(-)\acute{b}L} -i\overline{\bf
{\Psi}}_{(-)\acute{a}iL}\gamma^A\partial_A{\bf
{\Psi}}_{(-)\acute{b}L}^i\right.\nonumber\\
\left.-i\overline{\bf
{\Psi}}_{(-)\acute{a}L}^{ij}\gamma^A\partial_A{\bf
{\Psi}}_{(-)\acute{b}ijL}\right]
+{\mathsf L}^{(210)}_{(2)auxiliary}\\
\nonumber\\
 {\mathsf
L}^{(210)}_{(2)auxiliary} \equiv
h_{\acute{a}\acute{b}}^{^{(210-)}}<F_{(-)\acute{a}}|F_{(-)\acute{b}}>~~~~~~
\end{eqnarray}

\begin{equation}
{\bf {\Psi}}_{(\pm)\acute{a}}=\left(\matrix{{\bf
\psi}_{(\pm)\acute{a}\tilde{\alpha}}\cr
\overline{{\bf\psi}}^{\dot{\tilde\alpha}}_{(\pm)\acute{a}}}\right)
\end{equation}

 Similarly we find
\begin{eqnarray}
\frac{h_{\acute{a}\acute{b}}^{^{(210+)}}{\mathsf
g}^{^{(210)}}}{4!}<\widehat{\Phi}_{(+)\acute{a}}|\widehat{\mathsf
V}_{\mu\nu\rho\lambda}\Gamma_{[\mu}\Gamma_{\nu}\Gamma_{\rho}
\Gamma_{\lambda]}|\widehat{\Phi}_{(+)\acute{b}}>|_{\theta^2\bar
{\theta}^2}~~~~~\nonumber\\
=h_{\acute{a}\acute{b}}^{^{(210+)}}{\mathsf
g}^{^{(210)}}\left\{\left[\frac{1}{2}\sqrt{\frac{5}{6}}\left(i{\bf
A}^{\dagger}_{(+)\acute{a}}\stackrel{\leftrightarrow}{\partial}_A
{\bf A}_{(+)\acute{b}}-\overline{\bf
{\Psi}}_{(+)\acute{a}L}\gamma_A{\bf
{\Psi}}_{(+)\acute{b}L}\right)\right.\right.\nonumber\\
\left.\left.-\frac{1}{4\sqrt {30}}\left(i{\bf
A}^{\dagger}_{(+)\acute{a}ij}\stackrel{\leftrightarrow}{\partial}_A{\bf
A}_{(+)\acute{b}}^{ij}-\overline{\bf
{\Psi}}_{(+)\acute{a}ijL}\gamma_A{\bf
{\Psi}}_{(+)\acute{b}L}^{ij}\right)\right.\right.\nonumber\\
\left.\left.+\frac{1}{2\sqrt {30}}\left(i{\bf
A}^{i\dagger}_{(+)\acute{a}}\stackrel{\leftrightarrow}{\partial}_A{\bf
A}_{(+)\acute{b}i}-\overline{\bf
{\Psi}}_{(+)\acute{a}L}^i\gamma_A{\bf
{\Psi}}_{(+)\acute{b}iL}\right)\right]{\cal V}^{'A}\right.\nonumber\\
\left.+\frac{1}{\sqrt {6}}\left(i{\bf
A}^{i\dagger}_{(+)\acute{a}}\stackrel{\leftrightarrow}{\partial}_A{\bf
A}_{(+)\acute{b}}-\overline{\bf
{\Psi}}_{(+)\acute{a}L}^i\gamma_A{\bf
{\Psi}}_{(+)\acute{b}L}\right){\cal V}^{'A}_i\right.\nonumber\\
\left.+\frac{1}{\sqrt {6}}\left(i{\bf
A}^{\dagger}_{(+)\acute{a}}\stackrel{\leftrightarrow}{\partial}_A{\bf
A}_{(+)\acute{b}i}-\overline{\bf
{\Psi}}_{(+)\acute{a}L}\gamma_A{\bf
{\Psi}}_{(+)\acute{b}iL}\right){\cal V}^{'Ai}\right.\nonumber\\
\left.+\left[-\frac{1}{4\sqrt 2}\left(i{\bf
A}^{\dagger}_{(+)\acute{a}lm}\stackrel{\leftrightarrow}{\partial}_A{\bf
A}_{(+)\acute{b}}-\overline{\bf
{\Psi}}_{(+)\acute{a}lmL}\gamma_A{\bf
{\Psi}}_{(+)\acute{b}L}\right)\right.\right.\nonumber\\
\left.\left.+\frac{1}{24\sqrt 2}\epsilon_{ijklm}\left(i{\bf
A}^{i\dagger}_{(+)\acute{a}}\stackrel{\leftrightarrow}{\partial}_A{\bf
A}_{(+)\acute{b}}^{jk}-\overline{\bf
{\Psi}}_{(+)\acute{a}L}^i\gamma_A{\bf
{\Psi}}_{(+)\acute{b}L}^{jk}\right)\right]{\cal V}^{'Alm}\right.\nonumber\\
\left.+\left[-\frac{1}{4\sqrt 2}\left(i{\bf
A}^{\dagger}_{(+)\acute{a}}\stackrel{\leftrightarrow}{\partial}_A{\bf
A}_{(+)\acute{b}}^{lm}-\overline{\bf
{\Psi}}_{(+)\acute{a}L}\gamma_A{\bf
{\Psi}}_{(+)\acute{b}L}^{lm}\right)\right.\right.\nonumber\\
\left.\left.+\frac{1}{24\sqrt 2}\epsilon^{ijklm}\left(i{\bf
A}^{\dagger}_{(+)\acute{a}ij}\stackrel{\leftrightarrow}{\partial}_A{\bf
A}_{(+)\acute{b}k}-\overline{\bf
{\Psi}}_{(+)\acute{a}ijL}\gamma_A{\bf
{\Psi}}_{(+)\acute{b}kL}\right)\right]{\cal V}^{'A}_{lm}\right.\nonumber\\
 \left.+\left[\frac{1}{6\sqrt
2}\left(i{\bf
A}^{\dagger}_{(+)\acute{a}ik}\stackrel{\leftrightarrow}{\partial}_A{\bf
A}_{(+)\acute{b}}^{kj}-\overline{\bf
{\Psi}}_{(+)\acute{a}ikL}\gamma_A{\bf
{\Psi}}_{(+)\acute{b}L}^{kj}\right)\right.\right.\nonumber\\
\left.\left.-\frac{1}{2\sqrt 2}\left({\bf
A}^{j\dagger}_{(+)\acute{a}}\stackrel{\leftrightarrow}{\partial}_A{\bf
A}_{(+)\acute{b}i}+i\overline{\bf
{\Psi}}_{(+)\acute{a}L}^j\gamma_A{\bf
{\Psi}}_{(+)\acute{b}iL}\right)\right]{\cal V}^{'iA}_j\right.\nonumber\\
\left.+\frac{1}{6\sqrt {6}}\epsilon_{ijklm}\left(i{\bf
A}^{i\dagger}_{(+)\acute{a}}\stackrel{\leftrightarrow}{\partial}_A{\bf
A}_{(+)\acute{b}}^{jn}-\overline{\bf
{\Psi}}_{(+)\acute{a}L}^i\gamma_A{\bf
{\Psi}}_{(+)\acute{b}L}^{jn}\right){\cal V}^{'Aklm}_n\right.\nonumber\\
\left.+-\frac{1}{6\sqrt {6}}\epsilon^{ijklm}\left(i{\bf
A}^{\dagger}_{(+)\acute{a}in}\stackrel{\leftrightarrow}{\partial}_A{\bf
A}_{(+)\acute{b}j}-\overline{\bf
{\Psi}}_{(+)\acute{a}inL}\gamma_A{\bf
{\Psi}}_{(+)\acute{b}jL}\right){\cal V}^{'An}_{klm}\right.\nonumber\\
\left.+-\frac{1}{4\sqrt {6}}\left(i{\bf
A}^{\dagger}_{(+)\acute{a}ij}\stackrel{\leftrightarrow}{\partial}_A{\bf
A}_{(+)\acute{b}}^{kl}-\overline{\bf
{\Psi}}_{(+)\acute{a}ijL}\gamma_A{\bf
{\Psi}}_{(+)\acute{b}L}^{kl}\right){\cal V}^{'Aij}_{kl}\right.\nonumber\\
\left.-\frac{i}{2}\sqrt{\frac{5}{6}}\left[-{\bf
A}^{\dagger}_{(+)\acute{a}}\overline{\bf
{\Psi}}_{(+)\acute{b}R}+\frac{1}{10}{\bf
A}^{\dagger}_{(+)\acute{a}ij}\overline{\bf
{\Psi}}_{(+)\acute{b}R}^{ij}-\frac{1}{5}{\bf
A}^{i\dagger}_{(+)\acute{a}}\overline{\bf
{\Psi}}_{(+)\acute{b}iR}\right]{\Lambda}^{'}_{L}\right.\nonumber\\
\left.+\frac{i}{\sqrt {3}}\left[ {\bf
A}^{i\dagger}_{(+)\acute{a}}\overline{\bf
{\Psi}}_{(+)\acute{b}R}\right]{\Lambda}^{'}_{iL}+\frac{i}{\sqrt
{3}}\left[ {\bf A}^{\dagger}_{(+)\acute{a}}\overline{\bf
{\Psi}}_{(+)\acute{b}iR}\right]{\Lambda}^{'i}_{L}\right.\nonumber\\
\left.-\frac{i}{4}\left[{\bf
A}^{\dagger}_{(+)\acute{a}lm}\overline{\bf
{\Psi}}_{(+)\acute{b}R}-\frac{1}{6}\epsilon_{ijklm}{\bf
A}^{i\dagger}_{(+)\acute{a}}\overline{\bf
{\Psi}}_{(+)\acute{b}R}^{jk}\right]{\Lambda}^{'lm}_{L}\right.\nonumber\\
\left.-\frac{i}{4}\left[{\bf
A}^{\dagger}_{(+)\acute{a}}\overline{\bf
{\Psi}}_{(+)\acute{b}R}^{lm}-\frac{1}{6}\epsilon^{ijklm}{\bf
A}^{\dagger}_{(+)\acute{a}ij}\overline{\bf
{\Psi}}_{(+)\acute{b}kR}\right]{\Lambda}^{'}_{lmL}\right.\nonumber\\
\left.-\frac{i}{2}\left[-\frac{1}{3}{\bf
A}^{\dagger}_{(+)\acute{a}ik}\overline{\bf
{\Psi}}_{(+)\acute{b}R}^{kj}+{\bf
A}^{j\dagger}_{(+)\acute{a}}\overline{\bf
{\Psi}}_{(+)\acute{b}iR}\right]{\Lambda}^{'i}_{jL}\right.\nonumber\\
\left.+\frac{i}{6\sqrt {3}}\epsilon_{ijklm}\left[ {\bf
A}^{i\dagger}_{(+)\acute{a}}\overline{\bf
{\Psi}}_{(+)\acute{b}R}^{jn}\right]{\Lambda}^{'klm}_{nL}-\frac{i}{6\sqrt
{3}}\epsilon^{ijklm}\left[ {\bf
A}^{\dagger}_{(+)\acute{a}in}\overline{\bf
{\Psi}}_{(+)\acute{b}jR}\right]{\Lambda}^{'n}_{klmL}\right.\nonumber\\
\left.-\frac{i}{4\sqrt {3}}\left[ {\bf
A}^{\dagger}_{(+)\acute{a}ij}\overline{\bf
{\Psi}}_{(+)\acute{b}R}^{kl}\right]{\Lambda}^{'ij}_{klL}\right.\nonumber\\
 \left.-\frac{i}{2}\sqrt{\frac{
5}{6}}\left[-\overline{\bf {\Psi}}_{(+)\acute{a}L}{\bf
A}_{(+)\acute{b}}+\frac{1}{10}\overline{\bf
{\Psi}}_{(+)\acute{a}ijL}{\bf
A}^{ij}_{(+)\acute{b}}-\frac{1}{5}\overline{\bf
{\Psi}}_{(+)\acute{a}L}^i{\bf
A}_{(+)\acute{b}i}\right]{\Lambda}^{'}_{R}\right.\nonumber\\
\left.-\frac{i}{\sqrt {3}}\left[ \overline{\bf
{\Psi}}_{(+)\acute{a}L}^{i}{\bf
A}_{(+)\acute{b}}\right]{\Lambda}^{'}_{iR}-\frac{i}{\sqrt
{3}}\left[ {\overline{\bf
{\Psi}}_{(+)\acute{a}L}\bf A}_{(+)\acute{b}i}\right]{\Lambda}^{'i}_{R}\right.\nonumber\\
\left.+\frac{i}{4}\left[\overline{\bf
{\Psi}}_{(+)\acute{a}lmL}{\bf
A}_{(+)\acute{b}}-\frac{1}{6}\epsilon_{ijklm}\overline{\bf
{\Psi}}_{(+)\acute{a}L}^i{\bf A}^{jk}_{(+)\acute{b}}\right]{\Lambda}^{'lm}_{R}\right.\nonumber\\
\left.+\frac{i}{4}\left[\overline{\bf {\Psi}}_{(+)\acute{a}L}{\bf
A}_{(+)\acute{b}}^{lm}-\frac{1}{6}\epsilon^{ijklm}\overline{\bf
{\Psi}}_{(+)\acute{a}ijL}{\bf A}_{(+)\acute{b}k}\right]{\Lambda}^{'}_{lmR}\right.\nonumber\\
\left.+\frac{i}{2}\left[-\frac{1}{3}\overline{\bf
{\Psi}}_{(+)\acute{a}ikL}{\bf A}_{(+)\acute{b}}^{kj}+\overline{\bf
{\Psi}}_{(+)\acute{a}L}^j{\bf
A}_{(+)\acute{b}i}\right]{\Lambda}^{'i}_{jR}\right.\nonumber\\
\left.-\frac{i}{6\sqrt {3}}\epsilon_{ijklm}\left[ \overline{\bf
{\Psi}}_{(+)\acute{a}L}^{i}{\bf
A}^{jn}_{(+)\acute{b}}\right]{\Lambda}^{'klm}_{nR}+\frac{i}{6\sqrt
{3}}\epsilon^{ijklm}\left[ {\overline{\bf
{\Psi}}_{(+)\acute{a}inL}\bf A}_{(+)\acute{b}j}\right]{\Lambda}^{'n}_{klmR}\right.\nonumber\\
\left.+\frac{i}{4\sqrt {3}}\left[ \overline{\bf
{\Psi}}_{(+)\acute{a}ijL}{\bf
A}^{kl}_{(+)\acute{b}}\right]{\Lambda}^{'ij}_{klR}\right\}\nonumber\\
+{\mathsf L}^{(210)}_{(3)auxiliary}~~~~~~\\
\nonumber\\
 {\mathsf
L}^{(210)}_{(3)auxiliary}=\frac{h_{\acute{a}\acute{b}}^{^{(210+)}}{\mathsf
g}^{^{(210)}}}{4!2}<A_{(+)\acute{a}}|\Gamma_{[\mu}\Gamma_{\nu}\Gamma_{\rho}
\Gamma_{\lambda]}|A_{(+)\acute{b}}>D_{\mu\nu\rho\lambda}~~~~~~
\end{eqnarray}
where
\begin{eqnarray}
{\bf
A}^{i\dagger}_{(+)\acute{a}}\stackrel{\leftrightarrow}{\partial}_A{\bf
A}_{(+)\acute{b}i}\stackrel{def}={\bf
A}^{i\dagger}_{(+)\acute{a}}{\partial}_A{\bf
A}_{(+)\acute{b}i}-\left({\partial}_A{\bf
A}^{i\dagger}_{(+)\acute{a}}\right) {\bf
A}_{(+)\acute{b}i}\\
\Lambda_{R,L}=\frac{1\pm \gamma_5}{2}\Lambda\\
\Lambda_{jkl}^i=\left(\matrix{\lambda_{\tilde{\alpha}jkl}^i\cr
\overline{\lambda}^{\dot{\tilde\alpha}i}_{jkl}}\right)
\end{eqnarray}
and so on. Similarly for the $\overline{16}_-16_-$ couplings we
find
\begin{eqnarray}
\frac{h_{\acute{a}\acute{b}}^{^{(210-)}}{\mathsf
g}^{^{(210)}}}{4!}<\widehat{\Phi}_{(-)\acute{a}}|\widehat{\mathsf
V}_{\mu\nu\rho\lambda}\Gamma_{[\mu}\Gamma_{\nu}\Gamma_{\rho}
\Gamma_{\lambda]}|\widehat{\Phi}_{(-)\acute{b}}>|_{\theta^2\bar
{\theta}^2}~~~~~\nonumber\\
 =h_{\acute{a}\acute{b}}^{^{(210-)}}{\mathsf
g}^{^{(210)}}\left\{\left[\frac{1}{2}\sqrt{\frac{5}{6}}\left(i{\bf
A}^{\dagger}_{(-)\acute{a}}\stackrel{\leftrightarrow}{\partial}_A
{\bf A}_{(-)\acute{b}}-\overline{\bf
{\Psi}}_{(-)\acute{a}L}\gamma_A{\bf
{\Psi}}_{(-)\acute{b}L}\right)\right.\right.\nonumber\\
\left.\left.-\frac{1}{4\sqrt {30}}\left(i{\bf
A}^{ij\dagger}_{(-)\acute{a}}\stackrel{\leftrightarrow}{\partial}_A{\bf
A}_{(-)\acute{b}ij}-\overline{\bf
{\Psi}}_{(-)\acute{a}L}^{ij}\gamma_A{\bf
{\Psi}}_{(-)\acute{b}ijL}\right)\right.\right.\nonumber\\
\left.\left.+\frac{1}{2\sqrt {30}}\left(i{\bf
A}^{\dagger}_{(-)\acute{a}i}\stackrel{\leftrightarrow}{\partial}_A{\bf
A}_{(-)\acute{b}}^{i}+i\overline{\bf
{\Psi}}_{(-)\acute{a}iL}\gamma_A{\bf
{\Psi}}_{(-)\acute{b}L}^i\right)\right]{\cal V}^{'A}\right.\nonumber\\
\left.+\frac{1}{\sqrt {6}}\left(i{\bf
A}^{\dagger}_{(-)\acute{a}}\stackrel{\leftrightarrow}{\partial}_A{\bf
A}_{(-)\acute{b}}^{i}-\overline{\bf
{\Psi}}_{(-)\acute{a}L}\gamma_A{\bf
{\Psi}}_{(-)\acute{b}L}^i\right){\cal V}^{'A}_i\right.\nonumber\\
\left.+\frac{1}{\sqrt {6}}\left(i{\bf
A}^{\dagger}_{(-)\acute{a}i}\stackrel{\leftrightarrow}{\partial}_A{\bf
A}_{(-)\acute{b}}-\overline{\bf
{\Psi}}_{(-)\acute{a}iL}\gamma_A{\bf
{\Psi}}_{(-)\acute{b}L}\right){\cal V}^{'Ai}\right.\nonumber\\
\left.+\left[\frac{1}{4\sqrt 2}\left({\bf
A}^{\dagger}_{(-)\acute{a}}\stackrel{\leftrightarrow}{\partial}_A{\bf
A}_{(-)\acute{b}lm}+i\overline{\bf
{\Psi}}_{(-)\acute{a}L}\gamma_A{\bf
{\Psi}}_{(-)\acute{b}lmL}\right)\right.\right.\nonumber\\
\left.\left.-\frac{1}{24\sqrt 2}\epsilon_{ijklm}\left(i{\bf
A}^{ij\dagger}_{(-)\acute{a}}\stackrel{\leftrightarrow}{\partial}_A{\bf
A}_{(-)\acute{b}}^{k}-\overline{\bf
{\Psi}}_{(-)\acute{a}L}^{ij}\gamma_A{\bf
{\Psi}}_{(-)\acute{b}L}^{k}\right)\right]{\cal V}^{'Alm}\right.\nonumber\\
\left.+\left[\frac{1}{4\sqrt 2}\left(i{\bf
A}^{lm\dagger}_{(-)\acute{a}}\stackrel{\leftrightarrow}{\partial}_A{\bf
A}_{(-)\acute{b}}-\overline{\bf
{\Psi}}_{(-)\acute{a}L}^{lm}\gamma_A{\bf
{\Psi}}_{(-)\acute{b}L}\right)\right.\right.\nonumber\\
\left.\left.-\frac{1}{24\sqrt 2}\epsilon^{ijklm}\left(i{\bf
A}^{\dagger}_{(-)\acute{a}i}\stackrel{\leftrightarrow}{\partial}_A{\bf
A}_{(-)\acute{b}jk}-\overline{\bf
{\Psi}}_{(-)\acute{a}iL}\gamma_A{\bf
{\Psi}}_{(-)\acute{b}jkL}\right)\right]{\cal V}^{'A}_{lm}\right.\nonumber\\
\left.+\left[\frac{1}{6\sqrt 2}\left(i{\bf
A}^{jk\dagger}_{(-)\acute{a}}\stackrel{\leftrightarrow}{\partial}_A{\bf
A}_{(-)\acute{b}ki}-\overline{\bf
{\Psi}}_{(-)\acute{a}L}^{jk}\gamma_A{\bf
{\Psi}}_{(-)\acute{b}kiL}\right)\right.\right.\nonumber\\
\left.\left.-\frac{1}{2\sqrt 2}\left(i{\bf
A}^{\dagger}_{(-)\acute{a}i}\stackrel{\leftrightarrow}{\partial}_A{\bf
A}_{(-)\acute{b}}^{j}-\overline{\bf
{\Psi}}_{(-)\acute{a}iL}\gamma_A{\bf
{\Psi}}_{(-)\acute{b}L}^j\right)\right]{\cal V}^{'iA}_j\right.\nonumber\\
\left.+\frac{1}{12\sqrt 6}\epsilon_{ijklm}\left(i{\bf
A}^{ij\dagger}_{(-)\acute{a}}\stackrel{\leftrightarrow}{\partial}_A{\bf
A}_{(-)\acute{b}}^n-\overline{\bf
{\Psi}}_{(-)\acute{a}L}^{ij}\gamma_A{\bf
{\Psi}}_{(-)\acute{b}L}^n\right){\cal V}^{'Aklm}_{n}\right.\nonumber\\
\left.+\frac{1}{12\sqrt 6}\epsilon^{ijklm}\left(i{\bf
A}^{\dagger}_{(-)\acute{a}n}\stackrel{\leftrightarrow}{\partial}_A{\bf
A}_{(-)\acute{b}ij}-\overline{\bf
{\Psi}}_{(-)\acute{a}nL}\gamma_A{\bf
+{\Psi}}_{(-)\acute{b}ijL}\right){\cal V}^{'An}_{klm}\right.\nonumber\\
\left.+-\frac{1}{4\sqrt 6}\epsilon_{ijklm}\left(i{\bf
A}^{kl\dagger}_{(-)\acute{a}}\stackrel{\leftrightarrow}{\partial}_A{\bf
A}_{(-)\acute{b}ij}-\overline{\bf
{\Psi}}_{(-)\acute{a}L}^{kl}\gamma_A{\bf
{\Psi}}_{(-)\acute{b}ijL}\right){\cal V}^{'Aij}_{kl}\right.\nonumber\\
\left.-\frac{i}{2}\sqrt{\frac{5}{6}}\left[-{\bf
A}^{\dagger}_{(-)\acute{a}}\overline{\bf
{\Psi}}_{(-)\acute{b}R}+\frac{1}{10}{\bf
A}^{ij\dagger}_{(-)\acute{a}}\overline{\bf
{\Psi}}_{(-)\acute{b}ijR}-\frac{1}{5}{\bf
A}^{\dagger}_{(-)\acute{a}i}\overline{\bf
{\Psi}}_{(-)\acute{b}R}^i\right]{\Lambda}^{'}_{L}\right.\nonumber\\
\left.+\frac{i}{\sqrt 3}\left[{\bf
A}^{\dagger}_{(-)\acute{a}}\overline{\bf
{\Psi}}_{(-)\acute{b}R}^i\right]{\Lambda}^{'}_{iL}+\frac{i}{\sqrt
3}\left[{\bf A}^{\dagger}_{(-)\acute{a}i}\overline{\bf
{\Psi}}_{(-)\acute{b}R}\right]{\Lambda}^{'i}_{L}\right.\nonumber\\
\left.+\frac{i}{4}\left[{\bf
A}^{\dagger}_{(-)\acute{a}}\overline{\bf
{\Psi}}_{(-)\acute{b}lmR}-\frac{1}{6}\epsilon_{ijklm}{\bf
A}^{ij\dagger}_{(-)\acute{a}}\overline{\bf
{\Psi}}_{(-)\acute{b}R}^{k}\right]{\Lambda}^{'lm}_{L}\right.\nonumber\\
\left.+\frac{i}{4}\left[{\bf
A}^{lm\dagger}_{(-)\acute{a}}\overline{\bf
{\Psi}}_{(-)\acute{b}R}-\frac{1}{6}\epsilon^{ijklm}{\bf
A}^{\dagger}_{(-)\acute{a}i}\overline{\bf
{\Psi}}_{(-)\acute{b}jkR}\right]{\Lambda}^{'}_{lmL}\right.\nonumber\\
\left.-\frac{i}{2}\left[-\frac{1}{3}{\bf
A}^{jk\dagger}_{(-)\acute{a}}\overline{\bf
{\Psi}}_{(-)\acute{b}kiR}+{\bf
A}^{\dagger}_{(-)\acute{a}i}\overline{\bf
{\Psi}}_{(-)\acute{b}R}^j\right]{\Lambda}^{'i}_{jL}\right.\nonumber\\
\left.+\frac{i}{6\sqrt 3}\epsilon_{ijklm}\left[{\bf
A}^{ij\dagger}_{(-)\acute{a}}\overline{\bf
{\Psi}}_{(-)\acute{b}R}^{n}\right]{\Lambda}^{'klm}_{nL}
+\frac{i}{6\sqrt 3}\epsilon^{ijklm}\left[{\bf
A}^{\dagger}_{(-)\acute{a}n}\overline{\bf
{\Psi}}_{(-)\acute{b}ijR}\right]{\Lambda}^{'n}_{klmL}\right.\nonumber\\
\left.-\frac{i}{4\sqrt 3}\left[{\bf
A}^{kl\dagger}_{(-)\acute{a}}\overline{\bf
{\Psi}}_{(-)\acute{b}ijR}\right]{\Lambda}^{'ij}_{klL}\right.\nonumber\\
\left.+\frac{i}{2}\sqrt{\frac{5}{6}}\left[-\overline{\bf
{\Psi}}_{(-)\acute{a}L}{\bf
A}_{(-)\acute{b}}+\frac{1}{10}\overline{\bf
{\Psi}}_{(-)\acute{a}L}^{ij}{\bf
A}_{(-)\acute{b}ij}-\frac{1}{5}\overline{\bf
{\Psi}}_{(-)\acute{a}iL}{\bf
A}_{(-)\acute{b}}^i\right]{\Lambda}^{'}_{R}\right.\nonumber\\
\left.-\frac{i}{\sqrt 3}\left[\overline{\bf
{\Psi}}_{(-)\acute{a}L}{\bf
A}_{(-)\acute{b}}^i\right]{\Lambda}^{'}_{iR} -\frac{i}{\sqrt
3}\left[\overline{\bf {\Psi}}_{(-)\acute{a}iL}{\bf
A}_{(-)\acute{b}}\right]{\Lambda}^{'i}_{R}\right.\nonumber\\
\left.-\frac{i}{4}\left[\overline{\bf {\Psi}}_{(-)\acute{a}L}{\bf
A}_{(-)\acute{b}lm}-\frac{1}{6}\epsilon_{ijklm}\overline{\bf
{\Psi}}_{(-)\acute{a}L}^{ij}{\bf A}^{k}_{(-)\acute{b}}\right]{\Lambda}^{'lm}_{R}\right.\nonumber\\
\left.-\frac{i}{4}\left[\overline{\bf
{\Psi}}_{(-)\acute{a}L}^{lm}{\bf
A}_{(-)\acute{b}}-\frac{1}{6}\epsilon^{ijklm}\overline{\bf
{\Psi}}_{(-)\acute{a}iL}{\bf A}_{(-)\acute{b}jk}\right]{\Lambda}^{'}_{lmR}\right.\nonumber\\
\left.+\frac{i}{2}\left[-\frac{1}{3}\overline{\bf
{\Psi}}_{(-)\acute{a}L}^{jk}{\bf A}_{(-)\acute{b}ki}+\overline{\bf
{\Psi}}_{(-)\acute{a}iL}{\bf A}_{(-)\acute{b}}^j\right]{\Lambda}^{'i}_{jR}\right.\nonumber\\
\left.-\frac{i}{6\sqrt{3}}\epsilon_{ijklm}\left[\overline{\bf
{\Psi}}_{(-)\acute{a}L}^{ij}{\bf
A}^{n}_{(-)\acute{b}}\right]{\Lambda}^{'klm}_{nR} -\frac{i}{6\sqrt
{3}}\epsilon^{ijklm}\left[\overline{\bf
{\Psi}}_{(-)\acute{a}nL}{\bf A}_{(-)\acute{b}ij}\right]{\Lambda}^{'n}_{klmR}\right.\nonumber\\
\left.+\frac{i}{4\sqrt
{3}}\left[\overline{\bf{\Psi}}_{(-)\acute{a}L}^{kl}{\bf
A}_{(-)\acute{b}ij}\right]{\Lambda}^{'ij}_{klR}\right\}\nonumber\\
+{\mathsf
L}^{(210)}_{(4)auxiliary}~~~~~~\\
\nonumber\\ {\mathsf
L}^{(210)}_{(4)auxiliary}=\frac{h_{\acute{a}\acute{b}}^{^{(210-)}}{\mathsf
g}^{^{(210)}}}{4!2
}<A_{(-)\acute{a}}|\Gamma_{[\mu}\Gamma_{\nu}\Gamma_{\rho}
\Gamma_{\lambda]}|A_{(-)\acute{b}}>D_{\mu\nu\rho\lambda}~~~~~~
\end{eqnarray}

Further, the kinetic energy for the vector multiplet is given by
\begin{eqnarray}
 {\mathsf
L}_{V}^{^{(210~K.E.)}}=\frac{1}{64}\left[\widehat{\cal
W}^{\tilde{\alpha}}_{\mu\nu\rho\lambda} \widehat{\cal
W}_{\tilde{\alpha}\mu\nu\rho\lambda}|_{\theta^2}+\widehat{\overline{\cal
W}}_{\dot{\tilde{\alpha}}\mu\nu\rho\lambda}\widehat{\overline{\cal
W}}^{\dot{\tilde{\alpha}}}_{\mu\nu\rho\lambda}|_{\bar
{\theta}^2}\right]\\
\widehat{\cal W}^{\tilde{\alpha}}_{\mu\nu\rho\lambda} =\overline
{\mathsf D}^2\mathsf D_{\tilde{\alpha}}\widehat{\mathsf
V}_{\mu\nu\rho\lambda},~~~ \widehat{\overline{\cal
W}}_{\dot{\tilde{\alpha}}\mu\nu\rho\lambda}={\mathsf D}^2\overline
{\mathsf D}_{\dot{\tilde{\alpha}}}\widehat{\mathsf
V}_{\mu\nu\rho\lambda}
\end{eqnarray}
Explicit evaluation of Eq.(9.15) gives
\begin{eqnarray}
{\mathsf L}_{V}^{^{(210~K.E.)}}=-\frac{1}{4}{\cal
V}_{AB\mu\nu\rho\sigma}{\cal
V}^{AB}_{\mu\nu\rho\sigma}-\frac{i}{2}\overline
{\Lambda}_{\mu\nu\rho\sigma}\gamma^A\partial_A\Lambda_{\mu\nu\rho\sigma}+{\mathsf L}^{(210)}_{(5)auxiliary}\\
{\cal V}^{AB}_{\mu\nu\rho\sigma}=\partial^A{\cal
V}^B_{\mu\nu\rho\sigma}-\partial^B{\cal
V}^A_{\mu\nu\rho\sigma}\\
{\mathsf L}^{(210)}_{(5)auxiliary}=\frac{1}{2}D_{\mu\nu\rho\sigma}D_{\mu\nu\rho\sigma}\\
\Lambda_{\mu\nu\rho\sigma}=\left(\matrix{\lambda_{\tilde{\alpha}_{\mu\nu\rho\sigma}}\cr
\overline{\lambda}^{\dot{\tilde\alpha}}_{\mu\nu\rho\sigma}}\right)
\end{eqnarray}
Finally, the superpotential of the theory is taken to be
\begin{equation}
{\mathsf L}^{(210)}_{\mathsf W}={\mathsf
W}(\widehat{\Phi}_{(+)},\widehat{\Phi}_{(-)})|_{\theta^2}+ h.c.
\end{equation}
\begin{equation}
 {\mathsf
W}(\widehat{\Phi}_{(+)},\widehat{\Phi}_{(-)})=
\mu_{\acute{a}\acute{b}}<\widehat{\Phi}_{(-)\acute{a}}^*|B|\widehat{\Phi}_{(+)\acute{b}}>
\end{equation}
\begin{equation}
 {\mathsf W}({\bf A}_{(+)},{\bf A}_{(-)})
=i\mu_{\acute{a}\acute{b}}\left({\bf
A}_{(-)\acute{a}}^{\bf{T}}{\bf A}_{(+)\acute{b}}-\frac{1}{2}{\bf
A}_{(-)\acute{a}ij}^{\bf{T}}{\bf A}_{(+)\acute{b}}^{ij}+{\bf
A}_{(-)\acute{a}}^{i\bf{T}}{\bf A}_{(+)\acute{b}i}\right)
\end{equation}
where $\mu_{\acute{a}\acute{b}}$ is taken to be a symmetric
tensor. Thus we have
\begin{eqnarray}
{\mathsf L}_{\mathsf W}=-i\mu_{\acute{a}\acute{b}}\left(
\overline{\bf {\Psi}}_{(-)\acute{a}R}{\bf
{\Psi}}_{(+)\acute{b}L}+\overline{\bf
{\Psi}}_{(-)\acute{a}R}^i{\bf
{\Psi}}_{(+)\acute{b}iL}-\frac{1}{2}\overline{\bf
{\Psi}}_{(-)\acute{a}ijR}{\bf
{\Psi}}_{(+)\acute{b}L}^{ij}\right)\nonumber\\
+i\mu_{\acute{a}\acute{b}}^{*}\left( \overline{\bf
{\Psi}}_{(-)\acute{a}L}{\bf {\Psi}}_{(+)\acute{b}R}+\overline{\bf
{\Psi}}_{(-)\acute{a}iL}{\bf
{\Psi}}_{(+)\acute{b}R}^i-\frac{1}{2}\overline{\bf
{\Psi}}_{(-)\acute{a}L}^{ij}{\bf
{\Psi}}_{(+)\acute{b}ijR}\right)\nonumber\\
+{\mathsf L}^{(210)}_{(6)auxiliary}~~~~~~\\
\nonumber\\
 {\mathsf
L}^{(210)}_{(6)auxiliary}=i\mu_{\acute{a}\acute{b}}\left[{\bf
F}_{(-)\acute{a}}{\bf A}_{(+)\acute{b}}+{\bf
A}_{(-)\acute{a}}^{\bf{T}}{\bf F}_{(+)\acute{b}}-\frac{1}{2}{\bf
F}_{(-)\acute{a}ij}{\bf A}_{(+)\acute{b}}^{ij}\right.\nonumber\\
\left.-\frac{1}{2}{\bf A}_{(-)\acute{a}ij}^{\bf{T}}{\bf
F}_{(+)\acute{b}}^{ij} + {\bf F}_{(-)\acute{a}}^i{\bf
A}_{(+)\acute{b}i}+{\bf A}_{(-)\acute{a}}^{i\bf{T}}{\bf
F}_{(+)\acute{b}i}\right] +h.c.~~~~~~
\end{eqnarray}

We now eliminate the auxiliary fields, $D_{\mu\nu\rho\sigma}$ and
${\bf F}_{(\pm)}$. We find
\begin{eqnarray}
{\mathsf L}^{(210)}_{(3)auxiliary}+{\mathsf
L}^{(210)}_{(4)auxiliary}+{\mathsf
L}^{(210)}_{(5)auxiliary}\nonumber\\
=-\frac{1}{4608} {\mathsf
g}^{^{(210)}2}h_{\acute{a}\acute{b}}^{^{(210+)}}h_{\acute{c}\acute{d}}^{^{(210+)}}<A_{(+)\acute{a}}|\Gamma_{[\mu}\Gamma_{\nu}\Gamma_{\rho}
\Gamma_{\lambda]}|A_{(+)\acute{b}}>\nonumber\\
\times<A_{(+)\acute{c}}|\Gamma_{[\mu}\Gamma_{\nu}\Gamma_{\rho}
\Gamma_{\lambda]}|A_{(+)\acute{d}}>
\nonumber\\
-\frac{1}{4608} {\mathsf
g}^{^{(210)}2}h_{\acute{a}\acute{b}}^{^{(210-)}}h_{\acute{c}\acute{d}}^{^{(210-)}}
<A_{(-)\acute{a}}|\Gamma_{[\mu}\Gamma_{\nu}\Gamma_{\rho}
\Gamma_{\lambda]}|A_{(-)\acute{b}}>\nonumber\\
\times<A_{(-)\acute{c}}|\Gamma_{[\mu}\Gamma_{\nu}\Gamma_{\rho}
\Gamma_{\lambda]}|A_{(-)\acute{d}}>\nonumber\\
-\frac{1}{2304} {\mathsf
g}^{^{(210)}2}h_{\acute{a}\acute{b}}^{^{(210+)}}h_{\acute{c}\acute{d}}^{^{(210-)}}
<A_{(+)\acute{a}}|\Gamma_{[\mu}\Gamma_{\nu}\Gamma_{\rho}
\Gamma_{\lambda]}|A_{(+)\acute{b}}>\nonumber\\
\times<A_{(-)\acute{c}}|\Gamma_{[\mu}\Gamma_{\nu}\Gamma_{\rho}
\Gamma_{\lambda]}|A_{(-)\acute{d}}>
\end{eqnarray}

$SU(5)$ expansion of these expressions gives
\begin{eqnarray}
-\frac{1}{4608} {\mathsf
g}^{^{(210)}2}h_{\acute{a}\acute{b}}^{^{(210+)}}h_{\acute{c}\acute{d}}^{^{(210+)}}<A_{(+)\acute{a}}|\Gamma_{[\mu}\Gamma_{\nu}\Gamma_{\rho}
\Gamma_{\lambda]}|A_{(+)\acute{b}}>\nonumber\\
\times<A_{(+)\acute{c}}|\Gamma_{[\mu}\Gamma_{\nu}\Gamma_{\rho}
\Gamma_{\lambda]}|A_{(+)\acute{d}}>\nonumber\\
={\mathsf
g}^{^{(210)}2}\left\{-\frac{1}{6144}\left(\eta_{\acute{a}\acute{b},\acute{c}\acute{d}}^{^{(210++)}}+8\eta_{\acute{a}\acute{d},\acute{c}\acute{b}}^{^{(210++)}}\right){\bf
A}^{\dagger}_{(+)\acute{a}ij}{\bf A}_{(+)\acute{b}}^{ij}{\bf
A}^{\dagger}_{(+)\acute{c}kl}{\bf A}_{(+)\acute{d}}^{kl}\right.\nonumber\\
\left.-\frac{1}{768}\left(\eta_{\acute{a}\acute{b},\acute{c}\acute{d}}^{^{(210++)}}+8\eta_{\acute{a}\acute{d},\acute{c}\acute{b}}^{^{(210++)}}\right){\bf
A}^{\dagger}_{(+)\acute{a}}{\bf A}_{(+)\acute{b}}{\bf
A}^{i\dagger}_{(+)\acute{c}}{\bf A}_{(+)\acute{d}i}
\right.\nonumber\\
\left.+\frac{1}{512}\left(11\eta_{\acute{a}\acute{b},\acute{c}\acute{d}}^{^{(210++)}}-2\eta_{\acute{a}\acute{d},\acute{c}\acute{b}}^{^{(210++)}}\right){\bf
A}^{\dagger}_{(+)\acute{a}ij}{\bf A}_{(+)\acute{b}}^{ij}{\bf
A}^{k\dagger}_{(+)\acute{c}}{\bf A}_{(+)\acute{d}k}\right.\nonumber\\
\left.+\frac{1}{192}\left(\eta_{\acute{a}\acute{b},\acute{c}\acute{d}}^{^{(210++)}}+2\eta_{\acute{a}\acute{d},\acute{c}\acute{b}}^{^{(210++)}}\right){\bf
A}^{\dagger}_{(+)\acute{a}ij}{\bf A}_{(+)\acute{b}}^{jk}{\bf
A}^{i\dagger}_{(+)\acute{c}}{\bf
A}_{(+)\acute{d}k}\right.\nonumber\\
\left.+\frac{1}{1536}\left(25\eta_{\acute{a}\acute{b},\acute{c}\acute{d}}^{^{(210++)}}+18\eta_{\acute{a}\acute{d},\acute{c}\acute{b}}^{^{(210++)}}\right)
{\bf A}^{i\dagger}_{(+)\acute{a}}{\bf A}_{(+)\acute{b}i}{\bf
A}^{j\dagger}_{(+)\acute{c}}{\bf A}_{(+)\acute{d}j}\right.\nonumber\\
\left.+\frac{1}{1536}\left(\eta_{\acute{a}\acute{b},\acute{c}\acute{d}}^{^{(210++)}}-6\eta_{\acute{a}\acute{d},\acute{c}\acute{b}}^{^{(210++)}}\right){\bf
A}^{\dagger}_{(+)\acute{a}}{\bf A}_{(+)\acute{b}}{\bf
A}^{\dagger}_{(+)\acute{c}ij}{\bf A}_{(+)\acute{d}}^{ij}
\right.\nonumber\\
\left.+\eta_{\acute{a}\acute{b},\acute{c}\acute{d}}^{^{(210++)}}\left[\frac{1}{1536}\epsilon_{ijklm}{\bf
A}^{\dagger}_{(+)\acute{a}}{\bf A}_{(+)\acute{b}}^{ij}{\bf
A}^{k\dagger}_{(+)\acute{c}}{\bf
A}_{(+)\acute{d}}^{lm}\right.\right.\nonumber\\
\left.\left.+\frac{1}{1536}\epsilon^{ijklm}{\bf
A}^{\dagger}_{(+)\acute{a}ij}{\bf A}_{(+)\acute{b}k}{\bf
A}^{\dagger}_{(+)\acute{c}lm}{\bf A}_{(+)\acute{d}}\right.\right.\nonumber\\
\left.\left.+\frac{1}{768}{\bf A}^{\dagger}_{(+)\acute{a}ij}{\bf
A}_{(+)\acute{b}}^{jk}{\bf A}^{\dagger}_{(+)\acute{c}kl}{\bf
A}_{(+)\acute{d}}^{li}\right.\right.\nonumber\\
\left.\left.-\frac{5}{1536}{\bf A}^{\dagger}_{(+)\acute{a}}{\bf
A}_{(+)\acute{b}}{\bf A}^{\dagger}_{(+)\acute{c}}{\bf
A}_{(+)\acute{d}}\right]\right\}~~~~~~
\end{eqnarray}

\begin{eqnarray}
-\frac{1}{4608} {\mathsf
g}^{^{(210)}2}h_{\acute{a}\acute{b}}^{^{(210-)}}h_{\acute{c}\acute{d}}^{^{(210-)}}<A_{(-)\acute{a}}|\Gamma_{[\mu}\Gamma_{\nu}\Gamma_{\rho}
\Gamma_{\lambda]}|A_{(-)\acute{b}}>\nonumber\\
\times<A_{(-)\acute{c}}|\Gamma_{[\mu}\Gamma_{\nu}\Gamma_{\rho}
\Gamma_{\lambda]}|A_{(-)\acute{d}}>\nonumber\\
={\mathsf
g}^{^{(210)}2}\left\{-\frac{1}{6144}\left(\eta_{\acute{a}\acute{b},\acute{c}\acute{d}}^{^{(210--)}}+16\eta_{\acute{a}\acute{d},\acute{c}\acute{b}}^{^{(210--)}}\right){\bf
A}^{ij\dagger}_{(-)\acute{a}}{\bf A}_{(-)\acute{b}ij}{\bf
A}^{kl\dagger}_{(-)\acute{c}}{\bf A}_{(-)\acute{d}}^{kl}\right.\nonumber\\
\left.-\frac{1}{768}\left(11\eta_{\acute{a}\acute{b},\acute{c}\acute{d}}^{^{(210--)}}+8\eta_{\acute{a}\acute{d},\acute{c}\acute{b}}^{^{(210--)}}\right){\bf
A}^{\dagger}_{(-)\acute{a}}{\bf A}_{(-)\acute{b}}{\bf
A}^{\dagger}_{(-)\acute{c}i}{\bf A}_{(-)\acute{d}}^i
\right.\nonumber\\
\left.-\frac{1}{1536}\left(5\eta_{\acute{a}\acute{b},\acute{c}\acute{d}}^{^{(210--)}}+6\eta_{\acute{a}\acute{d},\acute{c}\acute{b}}^{^{(210--)}}\right){\bf
A}^{ij\dagger}_{(-)\acute{a}}{\bf A}_{(-)\acute{b}ij}{\bf
A}^{\dagger}_{(-)\acute{c}k}{\bf A}_{(-)\acute{d}}^k\right.\nonumber\\
\left.+\frac{1}{384}\left(\eta_{\acute{a}\acute{b},\acute{c}\acute{d}}^{^{(210--)}}+\eta_{\acute{a}\acute{d},\acute{c}\acute{b}}^{^{(210--)}}\right){\bf
A}^{ij\dagger}_{(-)\acute{a}}{\bf A}_{(-)\acute{b}jk}{\bf
A}^{\dagger}_{(-)\acute{c}i}{\bf
A}_{(-)\acute{d}}^k\right.\nonumber\\
\left.+\frac{1}{1536}\left(\eta_{\acute{a}\acute{b},\acute{c}\acute{d}}^{^{(210--)}}-6\eta_{\acute{a}\acute{d},\acute{c}\acute{b}}^{^{(210--)}}\right)
{\bf A}^{\dagger}_{(-)\acute{a}i}{\bf A}_{(-)\acute{b}}^i{\bf
A}^{\dagger}_{(-)\acute{c}j}{\bf A}_{(-)\acute{d}}^j\right.\nonumber\\
\left.-\frac{1}{1536}\left(29\eta_{\acute{a}\acute{b},\acute{c}\acute{d}}^{^{(210-)}}+18\eta_{\acute{a}\acute{d},\acute{c}\acute{b}}^{^{(210--)}}\right){\bf
A}^{\dagger}_{(-)\acute{a}}{\bf A}_{(-)\acute{b}}{\bf
A}^{ij\dagger}_{(-)\acute{c}}{\bf A}_{(-)\acute{d}ij}
\right.\nonumber\\
\left.+\eta_{\acute{a}\acute{b},\acute{c}\acute{d}}^{^{(210--)}}\left[\frac{5}{1536}\epsilon_{ijklm}{\bf
A}^{ij\dagger}_{(-)\acute{a}}{\bf A}_{(-)\acute{b}}{\bf
A}^{kl\dagger}_{(-)\acute{c}}{\bf
A}_{(-)\acute{d}}^{m}\right.\right.\nonumber\\
\left.\left.+\frac{5}{1536}\epsilon^{ijklm}{\bf
A}^{\dagger}_{(-)\acute{a}i}{\bf A}_{(-)\acute{b}jk}{\bf
A}^{\dagger}_{(-)\acute{c}}{\bf
A}_{(-)\acute{d}lm}\right.\right.\nonumber\\
\left.\left.-\frac{1}{256}{\bf A}^{ij\dagger}_{(-)\acute{a}}{\bf
A}_{(-)\acute{b}jk}{\bf A}^{kl\dagger}_{(-)\acute{c}}{\bf
A}_{(-)\acute{d}li} \right.\right.\nonumber\\
\left.\left.+\frac{25}{1536}{\bf A}^{\dagger}_{(-)\acute{a}}{\bf
A}_{(-)\acute{b}}{\bf A}^{\dagger}_{(-)\acute{c}}{\bf
A}_{(-)\acute{d}}\right]\right\}
\end{eqnarray}

\begin{eqnarray}
-\frac{1}{2304} {\mathsf
g}^{^{(210)}2}h_{\acute{a}\acute{b}}^{^{(210+)}}h_{\acute{c}\acute{d}}^{^{(210-)}}<A_{(+)\acute{a}}|\Gamma_{[\mu}\Gamma_{\nu}\Gamma_{\rho}
\Gamma_{\lambda]}|A_{(+)\acute{b}}> \nonumber\\
\times<A_{(-)\acute{c}}|\Gamma_{[\mu}\Gamma_{\nu}\Gamma_{\rho}
\Gamma_{\lambda]}|A_{(-)\acute{d}}>\nonumber\\
=\frac{{\mathsf
g}^{^{(210)}2}\eta_{\acute{a}\acute{b},\acute{c}\acute{d}}^{^{(210+-)}}}{192}
\left\{ -10{\bf A}^{\dagger}_{(+)\acute{a}}{\bf
A}_{(+)\acute{b}}{\bf A}^{\dagger}_{(-)\acute{c}}{\bf
A}_{(-)\acute{d}}+2{\bf A}^{\dagger}_{(+)\acute{a}}{\bf
A}_{(+)\acute{b}}{\bf A}^{ij\dagger}_{(-)\acute{c}}{\bf A}_{(-)\acute{d}ij}\right.\nonumber\\
\left.-4{\bf A}^{\dagger}_{(+)\acute{a}}{\bf A}_{(+)\acute{b}}{\bf
A}^{\dagger}_{(-)\acute{c}i}{\bf A}_{(-)\acute{d}}^i-32{\bf
A}^{\dagger}_{(+)\acute{a}}{\bf
A}_{(+)\acute{b}i}{\bf A}^{\dagger}_{(-)\acute{c}}{\bf A}_{(-)\acute{d}}^i\right.\nonumber\\
\left.-4{\bf A}^{\dagger}_{(+)\acute{a}}{\bf
A}_{(+)\acute{b}}^{ij}{\bf A}^{\dagger}_{(-)\acute{c}}{\bf
A}_{(-)\acute{d}ij}-32{\bf A}^{i\dagger}_{(+)\acute{a}}{\bf
A}_{(+)\acute{b}}{\bf A}^{\dagger}_{(-)\acute{c}i}{\bf A}_{(-)\acute{d}}\right.\nonumber\\
\left.+12{\bf A}^{i\dagger}_{(+)\acute{a}}{\bf
A}_{(+)\acute{b}}^{jk}{\bf A}^{\dagger}_{(-)\acute{c}i}{\bf
A}_{(-)\acute{d}jk}-8{\bf A}^{i\dagger}_{(+)\acute{a}}{\bf
A}_{(+)\acute{b}}^{jk}{\bf A}^{\dagger}_{(-)\acute{c}j}{\bf A}_{(-)\acute{d}ki}\right.\nonumber\\
\left.+18{\bf A}^{i\dagger}_{(+)\acute{a}}{\bf
A}_{(+)\acute{b}i}{\bf A}^{jk\dagger}_{(-)\acute{c}}{\bf
A}_{(-)\acute{d}jk}+4{\bf A}^{i\dagger}_{(+)\acute{a}}{\bf
A}_{(+)\acute{b}i}{\bf A}^{\dagger}_{(-)\acute{c}j}{\bf A}_{(-)\acute{d}}^j\right.\nonumber\\
\left.-56{\bf A}^{i\dagger}_{(+)\acute{a}}{\bf
A}_{(+)\acute{b}j}{\bf A}^{jk\dagger}_{(-)\acute{c}}{\bf
A}_{(-)\acute{d}ki}-24{\bf A}^{i\dagger}_{(+)\acute{a}}{\bf
A}_{(+)\acute{b}j}{\bf A}^{\dagger}_{(-)\acute{c}i}{\bf A}_{(-)\acute{d}}^j\right.\nonumber\\
\left.-164{\bf A}^{i\dagger}_{(+)\acute{a}}{\bf
A}_{(+)\acute{b}i}{\bf A}^{\dagger}_{(-)\acute{c}}{\bf
A}_{(-)\acute{d}}-4{\bf A}^{\dagger}_{(+)\acute{a}ij}{\bf
A}_{(+)\acute{b}}{\bf A}^{ij\dagger}_{(-)\acute{c}}{\bf A}_{(-)\acute{d}}\right.\nonumber\\
\left.-8{\bf A}^{\dagger}_{(+)\acute{a}ij}{\bf
A}_{(+)\acute{b}k}{\bf A}^{jk\dagger}_{(-)\acute{c}}{\bf
A}_{(-)\acute{d}}^i+8{\bf A}^{\dagger}_{(+)\acute{a}ij}{\bf
A}_{(+)\acute{b}k}{\bf A}^{ij\dagger}_{(-)\acute{c}}{\bf A}_{(-)\acute{d}}^k\right.\nonumber\\
\left.-{\bf A}^{\dagger}_{(+)\acute{a}ij}{\bf
A}_{(+)\acute{b}}^{ij}{\bf A}^{kl\dagger}_{(-)\acute{c}}{\bf
A}_{(-)\acute{d}kl}+8{\bf A}^{\dagger}_{(+)\acute{a}ij}{\bf
A}_{(+)\acute{b}}^{jk}{\bf A}^{il\dagger}_{(-)\acute{c}}{\bf A}_{(-)\acute{d}lk}\right.\nonumber\\
\left.+8{\bf A}^{\dagger}_{(+)\acute{a}ij}{\bf
A}_{(+)\acute{b}}^{jk}{\bf A}^{\dagger}_{(-)\acute{c}k}{\bf
A}_{(-)\acute{d}}^i-38{\bf A}^{\dagger}_{(+)\acute{a}ij}{\bf
A}_{(+)\acute{b}}^{ij}{\bf A}^{\dagger}_{(-)\acute{c}}{\bf A}_{(-)\acute{d}}\right.\nonumber\\
\left.+2{\bf A}^{\dagger}_{(+)\acute{a}ij}{\bf
A}_{(+)\acute{b}}^{ij}{\bf A}^{\dagger}_{(-)\acute{c}k}{\bf
A}_{(-)\acute{d}}^k-8{\bf A}^{\dagger}_{(+)\acute{a}ij}{\bf
A}_{(+)\acute{b}}^{kl}{\bf A}^{ij\dagger}_{(-)\acute{c}}{\bf A}_{(-)\acute{d}kl}\right.\nonumber\\
\left.+14\epsilon^{ijklm}{\bf A}^{\dagger}_{(+)\acute{a}ij}{\bf
A}_{(+)\acute{b}k}{\bf A}^{\dagger}_{(-)\acute{c}}{\bf
A}_{(-)\acute{d}lm}\right.\nonumber\\
\left.+14\epsilon_{ijklm}{\bf A}^{i\dagger}_{(+)\acute{a}}{\bf
A}_{(+)\acute{b}}^{jk}{\bf
A}^{lm\dagger}_{(-)\acute{c}}{\bf A}_{(-)\acute{d}}\right.\nonumber\\
\left.-2\epsilon^{ijklm}{\bf A}^{\dagger}_{(+)\acute{a}ij}{\bf
A}_{(+)\acute{b}}{\bf A}^{\dagger}_{(-)\acute{c}k}{\bf
A}_{(-)\acute{d}lm}\right.\nonumber\\
\left.-2\epsilon_{ijklm}{\bf A}^{\dagger}_{(+)\acute{a}}{\bf
A}_{(+)\acute{b}}^{ij}{\bf A}^{kl\dagger}_{(-)\acute{c}}{\bf
A}_{(-)\acute{d}}^m \right\}\nonumber\\
\end{eqnarray}
where $\eta$'s are defined by
\begin{eqnarray}
\eta_{\acute{a}\acute{b},\acute{c}\acute{d}}^{^{(210++)}}=h_{\acute{a}\acute{b}}^{^{(210+)}}h_{\acute{c}\acute{d}}^{^{(210+)}};~~~\eta_{\acute{a}\acute{b},\acute{c}\acute{d}}^{^{(210--)}}=h_{\acute{a}\acute{b}}^{^{(210-)}}
h_{\acute{c}\acute{d}}^{^{(210-)}} ;~~
\eta_{\acute{a}\acute{b},\acute{c}\acute{d}}^{^{(210+-)}}=
h_{\acute{a}\acute{b}}^{^{(210+)}}h_{\acute{c}\acute{d}}^{^{(210-)}}\nonumber\\
\end{eqnarray}
We also find
\begin{eqnarray}
{\mathsf L}^{(210)}_{(1)auxiliary}+{\mathsf
L}^{(210)}_{(2)auxiliary}+{\mathsf
L}^{(210)}_{(6)auxiliary}\nonumber\\
=-\left(\mu^*
[h^{^{(210-)}}]^{\bf{-1}}[h^{^{(210-)}}]^{\bf{T}}[h^{^{(210-)}}]^{\bf{-1}}\mu\right)_{\acute{a}\acute{b}}\nonumber\\
\times\left[{\bf A}^{\dagger}_{(+)\acute{a}}{\bf
A}_{(+)\acute{b}}+\frac{1}{4}{\bf
A}^{\dagger}_{(+)\acute{a}ij}{\bf A}_{(+)\acute{b}}^{ij}+{\bf
A}^{i\dagger}_{(+)\acute{a}}{\bf A}_{(+)\acute{b}i}\right]\nonumber\\
-\left(\mu[h^{^{(210+)}}]^{\bf{-1T}}h^{^{(210+)}}[h^{^{(210+)}}]^{\bf{-1T}}\mu^*\right)_{\acute{a}\acute{b}}\nonumber\\
\times \left[ {\bf A}^{\bf{T}}_{(-)\acute{a}}{\bf
A}_{(-)\acute{b}}^*+\frac{1}{4}{\bf
A}^{\bf{T}}_{(-)\acute{a}ij}{\bf A}_{(-)\acute{b}}^{ij*}+{\bf
A}^{i\bf{T}}_{(-)\acute{a}}{\bf A}_{(-)\acute{b}i}^*\right]~~~~~~
\end{eqnarray}

\section{A more general analysis of  ${\bf{16^{\dagger}-{16}-210}}$
vector couplings}

In this section we consider the couplings of the unconstrained
$210$ vector multiplet with matter, i.e., in the analysis we use
the full vector multiplet rather than the truncated one under the
constraint of the Wess-Zumino gauge.  An illustration of this
procedure is given in Appendix C for the $U(1)$ case. The
Lagrangian that governs the interactions of the 210 multiplet
consists of the kinetic energy term for the 210 plet, self
interactions, and interactions of the 210 plet with $16$ and
$\overline{16}$ of matter. For generality we also include
 a mass term for the 210 vector multiplet. As in Appendix C we will not
 impose the Wess-Zumino gauge but keep the full multiplet. Thus we take
 the Lagrangian governing the 210 vector multiplet to be

\begin{eqnarray}
{\mathsf L}^{(210)}={\mathsf L}_{V}^{'^{(210~K.E.)}}+{\mathsf
L}_{V}^{'^{(210~Mass)}}~~~~~~~~~~~~~~~~~~~~~~\nonumber\\
+ {\mathsf
L}_V^{'^{(210~Self-Interaction)}} +{\mathsf
L}_{V+\Phi}^{'^{(210~Interaction)}}
+{\mathsf L}^{'^{(210~Self-Interaction)}}_{\Phi}~~~~~~\\
\nonumber\\
 {\mathsf
L}_{V}^{'^{(210~K.E.)}}=\frac{1}{64}\left[\widehat{\cal
W}^{\tilde{\alpha}}_{\mu\nu\rho\lambda} \widehat{\cal
W}_{\tilde{\alpha}\mu\nu\rho\lambda}|_{\theta^2}+\widehat{\overline{\cal
W}}_{\dot{\tilde{\alpha}}\mu\nu\rho\lambda}\widehat{\overline{\cal
W}}^{\dot{\tilde{\alpha}}}_{\mu\nu\rho\lambda}|_{\bar
{\theta}^2}\right]~~~~~~\\
\nonumber\\
 {\mathsf L}_{V}^{'^{(210~Mass)}}=m^2\widehat{\mathsf
V}_{\mu\nu\rho\lambda}\widehat{\mathsf
V}_{\mu\nu\rho\lambda}|_{\theta^2\bar {\theta}^2}~~~~~~\\
\nonumber\\
 {\mathsf L}_V^{'^{(210~Self-Interaction)}}={\alpha}_1
\widehat{\mathsf V}_{\mu\nu\rho\lambda}\widehat{\mathsf
V}_{\rho\lambda\alpha\beta} \widehat{\mathsf
V}_{\alpha\beta\mu\nu} |_{\theta^2\bar{\theta}^2}\nonumber\\
+{\alpha}_2\widehat{\mathsf V}_{\mu\nu\rho\lambda}
\widehat{\mathsf V}_{\rho\lambda\alpha\beta} \widehat{\mathsf
V}_{\alpha\beta\eta\tau}\widehat{\mathsf V}_{\eta\tau\mu\nu}
|_{\theta^2\bar{\theta}^2}~~~~~~\\
\nonumber\\
 {\mathsf
L}_{V+\Phi}^{'^{(210~Interaction)}}=\frac{h_{\acute{a}\acute{b}}^{^{(210)}}}{4!}\widehat{\Phi}_{\acute{a}}^{\dagger}
\widehat{\mathsf
V}_{\mu\nu\rho\lambda}\Gamma_{[\mu}\Gamma_{\nu}\Gamma_{\rho}
\Gamma_{\lambda]}\widehat{\Phi}_{\acute{b}}|_{\theta^2\bar {\theta}^2}~~~~~~\\
\nonumber\\
 \widehat{\mathsf
L}_{\Phi}^{'^{(210~Self-Interaction)}}=\widehat{\Phi}^{\dagger}_{\acute{a}}\widehat{\Phi}_{\acute{a}}|_{\theta^2\bar
{\theta}^2}~~~~~~
\end{eqnarray}
\begin{eqnarray}
\widehat{\mathsf V}=C(x)+i\theta\chi (x)-i\bar {\theta}\overline
{\chi}(x) +\frac
{i}{2}\theta^2\left[M(x)+iN(x)\right]\nonumber\\
-\frac{i}{2}\bar
{\theta}^2
\left[M(x)-iN(x)\right] -\theta\sigma^A\bar {\theta}{\cal V}_A(x)\nonumber\\
+i\theta^2\bar {\theta}\left[\overline
{\lambda}(x)+\frac{i}{2}\overline{\sigma}^A\partial_A\chi(x)\right]
-i\bar
{\theta}^2\theta\left[\lambda(x)+\frac{i}{2}\sigma^A\partial_A
\overline{\chi}(x)\right] \nonumber\\
+\frac{1}{2}\theta^2\bar
{\theta}^2\left[D(x)+\frac{1}{2}\partial_A\partial^AC(x)\right]~~~~~~
\end{eqnarray}

 One could, of course, add more interactions, for
example, in \\
${\mathsf L}_V^{'^{(210~Self-Interaction)}}$ such as
$\mathsf{V}^5$ etc. which are allowed once one gives up the
Wess-Zumino gauge. Similarly in ${\mathsf
L}_{V+\Phi}^{'^{(210~Interaction)}}$ one may add additional terms
as well. However, the line of construction remains unchanged and
the inclusion of additional terms only brings in more complexity.
Thus to  keep the analysis simple we omit such terms. Evaluating
Eq.(9.31) we get
\begin{eqnarray}
{\mathsf L}^{(210)}= -\frac{1}{4}{\cal V}_{AB\mathtt{XY}}{\cal
V}^{AB}_{\mathtt{XY}}-\frac{1}{2}m^2{\cal
V}_{\acute{a}\mathtt{XY}}{\cal
V}^{A}_{\mathtt{XY}}-\frac{1}{2}\partial^AB_{\mathtt{XY}}\partial_AB_{\mathtt{XY}}\nonumber\\
-i\overline
{\Lambda}_{\mathtt{XY}}\gamma^A\partial_A\Lambda_{\mathtt{XY}}-m\overline
{\Lambda}_{\mathtt{XY}}\Lambda_{\mathtt{XY}}\nonumber\\
-\partial^AA^{\dagger}_{\acute{a}}\partial_AA_{\acute{a}}-i\overline
{\Psi}_{\acute{a}L}\gamma^A\partial_A\Psi_{\acute{a}L}\nonumber\\
 -\frac{h_{\acute{a}\acute{b}}^{^{(210)}}}{24m}B_{\mathtt{XY}}(\partial^AA^{\dagger}_{\acute{a}})\widetilde{\Gamma}_{\mathtt{XY}}\partial_AA_{\acute{b}}
-\frac{h_{\acute{a}\acute{b}}^{^{(210)}}}{96m}\partial^A\left(A_{\acute{a}}^{\dagger}\widetilde{\Gamma}_{\mathtt{XY}}A_{\acute{b}}\right)
\partial_AB_{\mathtt{XY}}\nonumber\\
+\frac{ih_{\acute{a}\acute{b}}^{^{(210)}}}{48}\left[A^{\dagger}_{\acute{a}}\widetilde{\Gamma}_{\mathtt{XY}}
\partial^AA_{\acute{b}}-(\partial^A
A_{\acute{a}}^{\dagger})\widetilde{\Gamma}_{\mathtt{XY}}A_{\acute{b}}\right]
{\cal V}_{\acute{a}\mathtt{XY}}\nonumber\\
+\frac{h_{\acute{a}\acute{b}}^{^{(210)}}}{48m\sqrt{2}}\left[\left(\overline
{\Psi}_{\acute{a}L}\gamma^A\widetilde{\Gamma}_{\mathtt{XY}}\Lambda_{L\mathtt{XY}}\right)\partial_AA_{\acute{b}}+\partial_AA_{\acute{a}}^{\dagger}
\left(\overline
{\Lambda}_{L\mathtt{XY}}\gamma^A\widetilde{\Gamma}_{\mathtt{XY}}\Psi_{\acute{b}L}\right)\right]\nonumber\\
+\frac{h_{\acute{a}\acute{b}}^{^{(210)}}}{24m\sqrt{2}}\left[\left(\overline
{\Psi}_{\acute{a}L}\gamma^A\widetilde{\Gamma}_{\mathtt{XY}}\partial_A\Lambda_{L\mathtt{XY}}\right)A_{\acute{b}}-A_{\acute{a}}^{\dagger}\left(\overline
{\Lambda}_{L\mathtt{XY}}\gamma^A\widetilde{\Gamma}_{\mathtt{XY}}\partial_A\Psi_{\acute{b}L}\right)\right]\nonumber\\
-\frac{ih_{\acute{a}\acute{b}}^{^{(210)}}}{24\sqrt{2}}\left[\left(\overline
{\Psi}_{\acute{a}L}\widetilde{\Gamma}_{\mathtt{XY}}\Lambda_{R\mathtt{XY}}\right)A_{\acute{b}}-A_{\acute{a}}^{\dagger}\left(\overline
{\Lambda}_{R\mathtt{XY}}\widetilde{\Gamma}_{\mathtt{XY}}\Psi_{\acute{b}L}\right)\right]\nonumber\\
-\frac{h_{\acute{a}\acute{b}}^{^{(210)}}}{48}\overline
{\Psi}_{\acute{a}L}\gamma^A\widetilde{\Gamma}_{\mathtt{XY}}\Psi_{\acute{b}L}{\cal
V}_{\acute{a}\mathtt{XY}}-\frac{ih_{\acute{a}\acute{b}}^{^{(210)}}}{24m}\overline
{\Psi}_{\acute{a}L}\gamma^A\widetilde{\Gamma}_{\mathtt{XY}}\partial_A(\Psi_{\acute{b}L})B_{\mathtt{XY}}\nonumber\\
-\frac{1}{m^3}\left(\frac{3\alpha_1}{4}B_{\mathtt{WY}}+\frac{\alpha_2}{m}B_{\mathtt{WX}}B_{\mathtt{XY}}\right)
\partial_AB_{\mathtt{YZ}}\partial^AB_{\mathtt{ZW}}\nonumber\\
-\frac{3}{m}\left(\frac{\alpha_1}{2}B_{\mathtt{WY}}+\frac{\alpha_2}{m}B_{\mathtt{WX}}B_{\mathtt{XY}}\right)
{\cal V}_{
A\mathtt{YZ}}{\cal V}^A_{\mathtt{ZW}}\nonumber\\
+\frac{3}{m^3}\left(\alpha_1B_{\mathtt{WY}}+\frac{2\alpha_2}{m}B_{\mathtt{WX}}B_{\mathtt{XY}}\right)
\left(i\overline
{\Lambda}_{L\mathtt{YZ}}\gamma^A\partial_A\Lambda_{L\mathtt{ZW}}
-m\overline{\Lambda}_{\mathtt{YZ}}\Lambda_{\mathtt{ZW}}\right)\nonumber\\
+\frac{3}{m^2}\left(\frac{\alpha_1}{2}\delta_{\mathtt{WX}}+\frac{2\alpha_2}{m}B_{\mathtt{WX}}\right)
\left(\overline{\Lambda}_{L\mathtt{XY}}\gamma^A\Lambda_{L\mathtt{YZ}}\right){\cal
V}_{\acute{a}\mathtt{ZW}}\nonumber\\
 + \frac{3\alpha_2}{2m^4}
\left(\overline{\Lambda}^c_{R\mathtt{WX}}\Lambda_{L\mathtt{XY}}\right)
\left(\overline{\Lambda}_{L\mathtt{YZ}}\Lambda_{R\mathtt{ZW}}^c\right)
+{\mathsf L}_{auxiliary}^{'~210}~~~~~~
\end{eqnarray}
Where we have defined for brevity
\begin{eqnarray}
\widetilde{\Gamma}_{\mathtt{XY}}=\Gamma_{[\mu}\Gamma_{\nu}\Gamma_{\rho}
\Gamma_{\lambda]};~~~B_{\mathtt{XY}}=B_{\mu\nu\rho\sigma};~~~{\cal
V}_{\acute{a}\mathtt{XY}}{\cal V}^A_{\mathtt{YZ}}={\cal
V}_{\acute{a}\mu\nu\rho\sigma}{\cal V}^A_{\rho\sigma\lambda\tau}
\end{eqnarray}
and so on. Further
\begin{eqnarray}
\Lambda=\left(\matrix{m\chi_{\tilde{\alpha}}\cr
\overline{\lambda}^{\dot{\tilde\alpha}}}\right),~~~
\Psi_{\acute{a}}=\left(\matrix{\psi_{\acute{a}\tilde{\alpha}}\cr
\overline{\psi}^{\dot{\tilde\alpha}}_{\acute{a}}}\right),~~~ B=mC,\nonumber\\
 \Lambda^c={\cal C}\overline{\Lambda}^T,~~~ {\cal
C}=\left(\matrix{i\sigma^2&0\cr 0&i\overline{\sigma}^2}\right),~~~
 \overline{\Lambda}=\Lambda^{\dagger}\gamma^0
\end{eqnarray}
We will expand some of the terms appearing in Eq.(9.39) in
appendix B.

 The Lagrangian containing the auxiliary fields
is given by
\begin{eqnarray}
{\mathsf
L}_{auxilliary}^{'~210}=\left(mB_{\mathtt{ZW}}+\frac{3\alpha_1}{2m^2}B_{\mathtt{WX}}B_{\mathtt{XZ}}
+\frac{2\alpha_2}{m^3}B_{\mathtt{WX}}B_{\mathtt{XY}}B_{\mathtt{YZ}}+\frac{h_{\acute{a}\acute{b}}^{^{(210)}}}{48}A^{\dagger}_{\acute{a}}
\widetilde{\Gamma}_{\mathtt{ZW}}A_{\acute{b}}
\right)D_{\mathtt{ZW}}\nonumber\\
+\frac{1}{2}D_{\mathtt{XY}}D_{\mathtt{XY}}
+\left(\frac{1}{2}m^2\delta_{\mathtt{YW}}+\frac{3\alpha_1}{2m}B_{\mathtt{YW}}+\frac{3\alpha_2}
{m^2}B_{\mathtt{WX}}B_{\mathtt{XY}}\right)
\left(M_{\mathtt{YZ}}M_{\mathtt{ZW}}+N_{\mathtt{YZ}}N_{\mathtt{ZW}}\right)\nonumber\\
+i\left[\frac{h_{\acute{a}\acute{b}}^{^{(210)}}}{48}F^{\dagger}_{\acute{a}}\widetilde{\Gamma}_{\mathtt{ZW}}A_{\acute{b}}+\frac{3}{m^2}\left(
\frac{\alpha_1}{4}\delta_{\mathtt{WX}}+
\frac{\alpha_2}{m}B_{\mathtt{WX}}\right)\left(\overline{\Lambda}_{L\mathtt{XY}}\Lambda_{R\mathtt{YZ}}^c\right)
\right]\left(M_{\mathtt{ZW}}+iN_{\mathtt{ZW}}\right)\nonumber\\
-i\left[\frac{h_{\acute{a}\acute{b}}^{^{(210)}}}{48}A^{\dagger}_{\acute{a}}\widetilde{\Gamma}_{\mathtt{ZW}}F_{\acute{b}}+\frac{3}{m^2}\left(
\frac{\alpha_1}{4}\delta_{\mathtt{WX}}
+\frac{\alpha_2}{m}B_{\mathtt{WX}}\right)\left(\overline{\Lambda}^c_{R\mathtt{XY}}\Lambda_{L\mathtt{YZ}}\right)
\right]\left(M_{\mathtt{ZW}}-iN_{\mathtt{ZW}}\right)\nonumber\\
 +\frac{ih_{\acute{a}\acute{b}}^{^{(210)}}}{24m\sqrt 2}\left(\overline{\Lambda}_{L\mathtt{XY}}\Psi_{\acute{a}R}\widetilde{\Gamma}_{\mathtt{XY}}
 \right)F_{\acute{b}}
 -\frac{ih_{\acute{b}a}^{^{(210)}}}{24m\sqrt 2}
\left(\widetilde{\Gamma}_{\mathtt{XY}}\overline{\Psi}_{\acute{a}R}\Lambda_{L\mathtt{XY}}\right)F_{\acute{b}}^{\dagger}\nonumber\\
+\frac{h_{\acute{a}\acute{b}}^{^{(210)}}}{24m}B_{\mathtt{XY}}
F^{\dagger}_{\acute{a}}\widetilde{\Gamma}_{\mathtt{XY}}F_{\acute{b}}+F^{\dagger}_{\acute{a}}F_{\acute{a}}~~~~~~~
\end{eqnarray}

Finally, eliminating the auxiliary fields we obtain
\begin{eqnarray}
{\mathsf L}_{auxilliary}^{'~210}=
-\frac{1}{2}m^2B_{\mathtt{XY}}B_{\mathtt{YX}}-\frac{3\alpha_1}{2m}B_{\mathtt{XY}}B_{\mathtt{YZ}}B_{\mathtt{ZX}}\nonumber\\
-\left(\frac{9{\alpha_1}^2}{8m^4}+\frac{2\alpha_2}{m^2}\right)B_{\mathtt{WX}}
B_{\mathtt{XY}}B_{\mathtt{YZ}}B_{\mathtt{ZW}}
-\frac{3\alpha_1\alpha_2}{m^5}B_{\mathtt{VW}}B_{\mathtt{WX}}B_{\mathtt{XY}}B_{\mathtt{YZ}}B_{\mathtt{ZV}}\nonumber\\
-\frac{2{\alpha_2}^2}{m^6}B_{\mathtt{UV}}
B_{\mathtt{VW}}B_{\mathtt{WX}}B_{\mathtt{XY}}B_{\mathtt{YZ}}B_{\mathtt{ZU}}
-\frac{h_{\acute{a}\acute{b}}^{^{(210)}}h_{\acute{c}\acute{d}}^{^{(210)}}}{4608}
\left(A^{\dagger}_{\acute{a}}\widetilde{\Gamma}_{\mathtt{XY}}A_{\acute{b}}\right)
\left(A^{\dagger}_{\acute{c}} \widetilde{\Gamma}_{\mathtt{XY}}
A_{\acute{d}}\right)\nonumber\\
-\frac{mh_{\acute{a}\acute{b}}^{^{(210)}}}{48}
\left(A^{\dagger}_{\acute{a}}\widetilde{\Gamma}_{\mathtt{XY}}A_{\acute{b}}\right)B_{\mathtt{XY}}
-\frac{h_{\acute{a}\acute{b}}^{^{(210)}}\alpha_1}{32m^2}
\left(A^{\dagger}_{\acute{a}}\widetilde{\Gamma}_{\mathtt{XY}}A_{\acute{b}}\right)B_{\mathtt{XZ}}
B_{\mathtt{ZY}}\nonumber\\
-\frac{h_{\acute{a}\acute{b}}^{^{(210)}}\alpha_2}{24m^3}
\left(A^{\dagger}_{\acute{a}}\widetilde{\Gamma}_{\mathtt{XY}}A_{\acute{b}}\right)
B_{\mathtt{XW}}B_{\mathtt{WZ}}B_{\mathtt{ZY}}-\frac{ih_{\acute{a}\acute{b}}^{^{(210)}}}{24m\sqrt
2}({\mathbf Q}^{-1})_{\acute{a}\acute{c}}{\mathbf R}_{\acute{c}}
\widetilde{\Gamma}_{\mathtt{XY}}\overline{\Psi}_{\acute{b}R}\Lambda_{L\mathtt{XY}}\nonumber\\
-\frac{1}{2}m^2\left[{\mathbf K}_{\mathtt{WX}}({\mathbf
P}^{-1})_{\mathtt{WXYZ}}{\mathbf K}_{\mathtt{YZ}} +{\mathbf
J}_{\mathtt{WX}}({\mathbf P}^{-1})_{\mathtt{WXYZ}}{\mathbf
J}_{\mathtt{YZ}}\right]\nonumber\\
+\frac{h_{\acute{a}\acute{b}}^{^{(210)}}}{8m^4}A^{\dagger}_{\acute{a}}\widetilde{\Gamma}_{\mathtt{UV}}({\mathbf
S}^{-1})_{\acute{b}\acute{c}}{\mathbf T}_{\acute{c}}({\mathbf
P}^{-1})_{\mathtt{UVXY}}\left(
\frac{\alpha_1}{4}\delta_{\mathtt{YW}}+
\frac{\alpha_2}{m}B_{\mathtt{YW}}\right)\overline{\Lambda}_{L\mathtt{WZ}}\Lambda_{R\mathtt{ZX}}^c
\end{eqnarray}
where
\begin{eqnarray}
{\mathbf
P}_{\mathtt{UVXY}}=\delta_{\mathtt{UX}}\delta_{\mathtt{VY}}+\frac{3\alpha_1}{2m^3}\left(\delta_{\mathtt{UY}}
B_{\mathtt{VX}}+B_{\mathtt{UX}}\delta_{\mathtt{VY}}\right)\nonumber\\
+\frac{3\alpha_2}{m^4}\left(\delta_{\mathtt{UY}}
B_{\mathtt{VW}}B_{\mathtt{WX}}+B_{\mathtt{UW}}B_{\mathtt{WY}}
\delta_{\mathtt{XV}}\right)\\
{\mathbf
Q}_{\acute{b}\acute{c}}=\delta_{\acute{b}\acute{c}}+\frac{h_{\acute{b}\acute{c}}^{^{(210)}}}{24m}\widetilde{\Gamma}_{\mathtt{XY}}B_{\mathtt{XY}}
-\frac{h_{\acute{a}\acute{c}}^{^{(210)}}h_{\acute{b}\acute{d}}^{^{(210)}}}{1152m^2}A^{\dagger}_{\acute{a}}\widetilde{\Gamma}_{\mathtt{UV}}({\mathbf
P}^{-1})_{\mathtt{UVXY}}\widetilde{\Gamma}_{\mathtt{XY}}A_{\acute{d}}~~~~~~\\
{\mathbf
S}_{\acute{b}\acute{c}}=\delta_{\acute{b}\acute{c}}+\frac{h_{\acute{c}\acute{b}}^{^{(210)}}}{24m}\widetilde{\Gamma}_{\mathtt{XY}}B_{\mathtt{XY}}
-\frac{h_{\acute{c}\acute{d}}^{^{(210)}}h_{\acute{a}\acute{b}}^{^{(210)}}}{1152m^2}\widetilde{\Gamma}_{\mathtt{UV}}A_{\acute{d}}({\mathbf
P}^{-1})_{\mathtt{UVXY}}A^{\dagger}_{\acute{a}}\widetilde{\Gamma}_{\mathtt{XY}}~~~~~~\\
{\mathbf
R}_{\acute{c}}=\frac{h_{\acute{a}\acute{c}}^{^{(210)}}}{8m^4}
A^{\dagger}_{\acute{a}}\widetilde{\Gamma}_{\mathtt{UV}} ({\mathbf
P}^{-1})_{\mathtt{UVXY}}\left(
\frac{\alpha_1}{4}\delta_{\mathtt{YW}}+
\frac{\alpha_2}{m}B_{\mathtt{YW}}\right)\overline{\Lambda}_{L\mathtt{WZ}}\Lambda_{R\mathtt{ZX}}^c\nonumber\\
-\frac{ih_{\acute{a}\acute{c}}^{^{(210)}}}{24m\sqrt
2}\overline{\Lambda}_{L\mathtt{XY}}\Psi_{\acute{a}R}\widetilde{\Gamma}_{\mathtt{XY}}~~~~~~\\
{\mathbf
T}_{\acute{c}}=+\frac{h_{\acute{c}\acute{a}}^{^{(210)}}}{8m^4}
\widetilde{\Gamma}_{\mathtt{UV}}A_{\acute{a}} ({\mathbf
P}^{-1})_{\mathtt{UVXY}}\left(
\frac{\alpha_1}{4}\delta_{\mathtt{YW}}+
\frac{\alpha_2}{m}B_{\mathtt{YW}}\right)\overline{\Lambda}_{R\mathtt{WZ}}^c\Lambda_{L\mathtt{ZX}}\nonumber\\
\frac{ih_{\acute{c}\acute{a}}^{^{(210)}}}{24m\sqrt
2}\widetilde{\Gamma}_{\mathtt{XY}}\overline{\Psi}_{\acute{a}R}\Lambda_{L\mathtt{XY}}~~~~~~\\
{\mathbf
K}_{\mathtt{XY}}=-\frac{ih_{\acute{a}\acute{b}}^{^{(210)}}}{48m^2}\left[({\mathbf
Q}^{-1})_{\acute{a}\acute{c}}{\mathbf
R}_{\acute{c}}\widetilde{\Gamma}_{\mathtt{XY}}A_{\acute{b}}-A^{\dagger}_{\acute{a}}\widetilde{\Gamma}_{\mathtt{XY}}({\mathbf
S}^{-1})_{\acute{b}\acute{c}}{\mathbf T}_{\acute{c}} \right]\nonumber\\
-\frac{3i}{m^4}\left( \frac{\alpha_1}{4}\delta_{\mathtt{XU}}+
\frac{\alpha_2}{m}B_{\mathtt{XU}}\right)\left(\overline{\Lambda}_{L\mathtt{UV}}\Lambda_{R\mathtt{VY}}^c
-\overline{\Lambda}_{R\mathtt{UV}}^c\Lambda_{L\mathtt{VY}}\right)~~~~~~\\
{\mathbf
J}_{\mathtt{XY}}=\frac{h_{\acute{a}\acute{b}}^{^{(210)}}}{48m^2}\left[({\mathbf
Q}^{-1})_{\acute{a}\acute{c}}{\mathbf
R}_{\acute{c}}\widetilde{\Gamma}_{\mathtt{XY}}A_{\acute{b}}+A^{\dagger}_{\acute{a}}\widetilde{\Gamma}_{\mathtt{XY}}({\mathbf
S}^{-1})_{\acute{b}\acute{c}}{\mathbf T}_{\acute{c}} \right]\nonumber\\
+\frac{3}{m^4}\left( \frac{\alpha_1}{4}\delta_{\mathtt{XU}}+
\frac{\alpha_2}{m}B_{\mathtt{XU}}\right)\left(\overline{\Lambda}_{L\mathtt{UV}}\Lambda_{R\mathtt{VY}}^c
+\overline{\Lambda}_{R\mathtt{UV}}^c\Lambda_{L\mathtt{VY}}\right)~~~~~~
\end{eqnarray}

\section{Appendix A}
In this appendix we normalize the irreducible $SU(5)$ tensors
contained in a $210$ vector ${\cal V}^{A} _{\mu\nu\rho\sigma}$, a
$210$ scalar $B_{\mu\nu\rho\sigma}$, and a $210$ spinor
$\Lambda_{\mu\nu\rho\sigma}$. Latin letters ($i, j, k, ...$) are
used to denote the $SU(5)$ indices.
 The normalized $SU(5)$ gauge tensors
appearing in ${\cal V}^A _{\mu\nu\rho\sigma}$ are
\begin{eqnarray}
{\cal V}_A=4\sqrt{\frac{10}{3}}{\cal V}^{'}_A;~~~~{\cal
V}_A^i=8\sqrt{6}{\cal V}^{'i}_A;~~~~{\cal
V}_{\acute{a}i}=8\sqrt{6}{\cal
V}^{'}_{\acute{a}i}\nonumber\\
{\cal V}^{ij}_A=\sqrt{2}{\cal V}^{'ij}_A;~~~~
 {\cal
V}_{\acute{a}ij}=\sqrt{2}{\cal V}^{'}_{\acute{a}ij};~~~~{\cal
V}^{j}_{\acute{a}i}=\sqrt{2}{\cal V}^{'j}_{\acute{a}i}\nonumber\\
{\cal V}^{ijk}_{\acute{a}l}=\sqrt{\frac{2}{3}}{\cal
V}^{'ijk}_{\acute{a}l};~~~~{\cal
V}^{i}_{\acute{a}jkl}=\sqrt{\frac{2}{3}}{\cal
V}^{'i}_{\acute{a}jkl};~~~~{\cal
V}^{ij}_{\acute{a}kl}=\sqrt{\frac{2}{3}}{\cal
V}^{'ij}_{\acute{a}kl}
\end{eqnarray}
so that
\begin{eqnarray}
-\frac{1}{4}{\cal V}^{AB} _{\mu\nu\rho\sigma}{\cal V}
_{AB\mu\nu\rho\sigma}=-\frac{1}{2}{\cal V}'_{AB}{\cal
V}^{'AB\dagger}-\frac{1}{2}{\cal V}^{'i}_{AB}{\cal
V}^{'ABi\dagger}-\frac{1}{2!}\frac{1}{2}{\cal V}^{'ij}_{AB}{\cal
V}^{'ABij\dagger}\nonumber\\
-\frac{1}{4}{\cal V}^{'i}_{ABj}{\cal
V}^{'ABj}_i-\frac{1}{3!}\frac{1}{2}{\cal V}^{'ijk}_{ABl}{\cal
V}^{'ABijk\dagger}_l-\frac{1}{2!}\frac{1}{2!}\frac{1}{4}{\cal
V}^{'ij}_{ABkl}{\cal V}^{'ABkl}_{ij}
\end{eqnarray}
The normalized $SU(5)$ fields appearing in $B _{\mu\nu\rho\sigma}$
are
\begin{eqnarray}
B=4\sqrt{\frac{10}{3}}B^{'};~~~~B^i=8\sqrt{6}B^{'i};~~~~B_{i}=8\sqrt{6}B^{'}_{i}\nonumber\\
B^{ij}=\sqrt{2}B^{'ij};~~~~ B_{ij}=\sqrt{2}B^{'}_{ij};~~~~B^{j}_{i}=\sqrt{2}B^{'j}_{i}\nonumber\\
B^{ijk}_{l}=\sqrt{\frac{2}{3}}B^{'ijk}_{l};~~~~B_{jkl}=\sqrt{\frac{2}{3}}B^{'i}_{jkl};~~~~B^{ij}_{kl}=\sqrt{\frac{2}{3}}
B^{'ij}_{kl}
\end{eqnarray}
so that
\begin{eqnarray}
-\frac{1}{2}\partial^AB_{\mu\nu\rho\sigma}
\partial_AB_{\mu\nu\rho\sigma}=-\partial^AB^{'}\partial_AB^{'\dagger}-\partial^AB^{'i}\partial_AB
^{'i\dagger}-\frac{1}{2!}\partial^AB^{'ij}\partial_AB^{'ij\dagger}\nonumber\\
-\frac{1}{2}\partial^AB^{'i}_{j}\partial_AB^{'j}_i-\frac{1}{3!}\partial^AB^{'ijk}_{l}\partial_AB^{'ijk\dagger}_l
-\frac{1}{2!}\frac{1}{2!}\frac{1}{2}\partial^AB^{'ij}_{kl}\partial_AB^{'kl}_{ij}\nonumber\\
\end{eqnarray}
The normalized $SU(5)$ fields appearing in $\Lambda
_{\mu\nu\rho\sigma}$ are
\begin{eqnarray}
\Lambda=4\sqrt{\frac{5}{3}}\Lambda^{'};~~~~\Lambda_i=8\sqrt{6}\Lambda^{'i};~~~~\Lambda_{i}=8\sqrt{6}\Lambda^{'}_{i}\nonumber\\
\Lambda^{ij}=\sqrt{2}\Lambda^{'ij};~~~~\Lambda_{ij}=\sqrt{2}\Lambda^{'}_{ij};~~~~\Lambda^{j}_{i}=\sqrt{2}\Lambda^{'j}_{i}\nonumber\\
\Lambda^{ijk}_{l}=\sqrt{\frac{2}{3}}\Lambda^{'ijk}_{l};~~~~\Lambda_{jkl}=\sqrt{\frac{2}{3}}\Lambda^{'i}_{jkl}
;~~~~\Lambda^{ij}_{kl}=\sqrt{\frac{2}{3}} \Lambda^{'ij}_{kl}
\end{eqnarray}
so that
\begin{eqnarray}
-i\overline{\Lambda}_{\mu\nu\rho\sigma}\gamma^A
\partial_A\Lambda_{\mu\nu\rho\sigma}=-i\overline{\Lambda}^{'}\gamma^A\partial_A\Lambda^{'}
-i\overline{\Lambda}^{'}_{i}\gamma^A\partial_A\Lambda{'}_{i}-i\overline{\Lambda}^{'i}\gamma^A\partial_A\Lambda^{'i}\nonumber\\
-\frac{1}{2!}i\overline{\Lambda}^{'}_{ij}\gamma^A\partial_A\Lambda^{'}_{ij}
-\frac{1}{2!}i\overline{\Lambda}^{'ij}\gamma^A\partial_A\Lambda^{'ij}
-i\overline{\Lambda}^{'i}_j\gamma^A\partial_A\Lambda^{'i}_j\nonumber\\
-\frac{1}{3!}i\overline{\Lambda}^{'ijk}_l\gamma^A\partial_A\Lambda^{'ijk}_l
-\frac{1}{3!}i\overline{\Lambda}^{'l}_{ijk}\gamma^A\partial_A\Lambda^{'l}_{ijk}
-\frac{1}{2!}\frac{1}{2!}i\overline{\Lambda}^{'ij}_{kl}\gamma^A\partial_A\Lambda^{'ij}_{kl}
\end{eqnarray}

\section{Appendix B}
In this appendix we exhibit, for the benefit of the reader, a few
$SU(5)$ expansions of the terms appearing in our final Lagrangian
Eq.(9.38). We begin by noting that any of the chiral fields $S$
($\equiv A, \Psi, F$) can be expanded in terms of its $SU(5)$
components as
\begin{equation}
|S_{\acute{a}}>=|0>{\bf
S}_{\acute{a}}+\frac{1}{2}b_i^{\dagger}b_j^{\dagger}|0>{\bf
S}_{\acute{a}}^{ij}+\frac{1}{24}\epsilon^{ijklm}b_j^{\dagger}b_k^{\dagger}b_l^{\dagger}b_m^{\dagger}|0>{\bf
S}_{\acute{a}i}
\end{equation}
Together with the normalizations of appendix A and the basic
theorem given in Ref.\cite{ns1}, we can expand terms such as
\begin{eqnarray}
\frac{h_{\acute{a}\acute{b}}^{^{(210)}}}{48m\sqrt
2}\left(\overline
{\Psi}_{\acute{a}L}\gamma^A\widetilde{\Gamma}_{\mathtt{XY}}\Lambda_{L\mathtt{XY}}\right)\partial_AA_{\acute{b}}=
\frac{h_{\acute{a}\acute{b}}^{^{(210)}}}{48m\sqrt
2}\left(\overline
{\Psi}_{\acute{a}L}\gamma^A\Gamma_{[\mu}\Gamma_{\nu}\Gamma_{\rho}
\Gamma_{\lambda]}\partial_AA_{\acute{b}}\right)\Lambda_{L\mu\nu\rho\sigma}\nonumber\\
=\frac{h_{\acute{a}\acute{b}}^{^{(210)}}}{m}\left[\frac{1}{8\sqrt{30}}\left(10\overline
{\bf{\Psi}}_{\acute{a}L}\gamma^A\partial_A{\bf
A}_{\acute{b}}-\overline
{\bf{\Psi}}_{\acute{a}ijL}\gamma^A\partial_A{\bf
A}_{\acute{b}}^{ij}+2\overline
{\bf{\Psi}}_{\acute{a}L}^i\gamma^A\partial_A{\bf A}_{\acute{b}i}\right)\Lambda^{'}_{L}\right.\nonumber\\
\left.+\frac{1}{2\sqrt 3}\left(\overline
{\bf{\Psi}}_{\acute{a}L}^i\gamma^A\partial_A{\bf
A}_{\acute{b}}\right)\Lambda^{'}_{iL}+\frac{1}{2\sqrt
3}\left(\overline {\bf{\Psi}}_{\acute{a}L}\gamma^A\partial_A{\bf
A}_{\acute{b}i}\right)\Lambda^{'i}_{L}\right.\nonumber\\
\left.+\frac{1}{48}\left(-6\overline
{\bf{\Psi}}_{\acute{a}lmL}^i\gamma^A\partial_A{\bf
A}_{\acute{b}}+\epsilon_{ijklm}\overline
{\bf{\Psi}}_{\acute{a}L}^i\gamma^A\partial_A{\bf
A}_{\acute{b}}^{jk}\right)\Lambda^{'lm}_{L}\right.\nonumber\\
 \left.+\frac{1}{48}\left(-6\overline
{\bf{\Psi}}_{\acute{a}L}\gamma^A\partial_A{\bf
A}_{\acute{b}}^{lm}+\epsilon^{ijklm}\overline
{\bf{\Psi}}_{\acute{a}ijL}\gamma^A\partial_A{\bf
A}_{\acute{b}k}\right)\Lambda^{'}_{lmL}\right.\nonumber\\
 \left.+\frac{1}{12}\left(-3\overline
{\bf{\Psi}}_{\acute{a}L}^j\gamma^A\partial_A{\bf
A}_{\acute{b}}^{i}+\overline
{\bf{\Psi}}_{\acute{a}ikL}\gamma^A\partial_A{\bf
A}_{\acute{b}}^{kj}\right)\Lambda^{'i}_{jL}\right.\nonumber\\
\left.+\frac{1}{12\sqrt 3}\left(\epsilon_{ijklm}\overline
{\bf{\Psi}}_{\acute{a}L}^i\gamma^A\partial_A{\bf
A}_{\acute{b}}^{jn}\right)\Lambda^{'klm}_{nL}-\frac{1}{12\sqrt 3
}\left(\epsilon^{ijklm}\overline
{\bf{\Psi}}_{\acute{a}inL}\gamma^A\partial_A{\bf
A}_{\acute{b}j}\right)\Lambda^{'n}_{klmL}\right.\nonumber\\
\left.-\frac{1}{8\sqrt 3}\left(\overline
{\bf{\Psi}}_{\acute{a}ijL}\gamma^A\partial_A{\bf
A}_{\acute{b}}^{kl}\right)\Lambda^{'ij}_{klL}\right]~~~~~~~
\end{eqnarray}
\begin{eqnarray}
B_{\mathtt{WY}}{\cal V}_{\mathtt{YZ}}^A{\cal
V}_{\acute{a}\mathtt{ZW}}=B_{\mu\nu\rho\sigma}{\cal
V}_{\rho\sigma\lambda\tau}^A{\cal
V}_{\acute{a}\lambda\tau\mu\nu}\nonumber
=B_{\mu\nu\rho\sigma}{\mathit V}_{\mu\nu\rho\sigma}\nonumber\\
=\frac{1}{16}\left\{B_{c_ic_jc_kc_l}{\mathit V}_{{\bar c}_i{\bar
c}_j{\bar c}_k{\bar c}_l}+B_{{\bar c}_i{\bar c}_j{\bar c}_k{\bar
c}_l}{\mathit V}_{c_ic_jc_kc_l}+4B_{c_ic_jc_k{\bar c}_l}{\mathit
V}_{{\bar c}_i{\bar c}_j{\bar c}_k c_l}\right.\nonumber\\
\left.+4B_{{\bar c}_i{\bar c}_j{\bar c}_k c_l}{\mathit
V}_{c_ic_jc_k{\bar c}_l}+6B_{c_ic_j{\bar c}_k{\bar c}_l}{\mathit
V}_{{\bar c}_i{\bar c}_jc_k c_l}\right\}
\end{eqnarray}
 where${\mathit V}_{\mu\nu\rho\sigma}={\cal
V}_{\rho\sigma\lambda\tau}^A{\cal
V}_{\acute{a}\lambda\tau\mu\nu}$. Thus we find
\begin{eqnarray}
B_{c_ic_jc_kc_l}{\mathit V}_{{\bar c}_i{\bar c}_j{\bar c}_k{\bar
c}_l}=\frac{1}{2}B_{c_ic_jc_kc_l}\left({\cal V}_{\acute{a}{\bar
c}_i{\bar c}_jc_mc_n}{\cal V}_{{\bar c}_m{\bar c}_n{\bar c}_k{\bar
c}_l}^A +{\cal V}_{\acute{a}{\bar c}_i{\bar c}_j{\bar
c}_mc_n}{\cal V}_{c_m{\bar
c}_n{\bar c}_k{\bar c}_l}^A \right)\nonumber\\
=\frac{1}{144}B_i{\cal V}_A^j{\cal V}_j^{Ai}-\frac{1}{480}B_i{\cal
V}_A^i{\cal V}^{A}-\frac{1}{48}\epsilon^{ijklm}B_i{\cal
V}_{\acute{a}jkn}^p{\cal V}_{plm}^{An}\nonumber\\
+\frac{1}{36}\epsilon^{ijklm}B_i{\cal V}_{\acute{a}jkl}^n{\cal
V}_{nm}^{A} -\frac{1}{144}\epsilon^{ijklm}B_i{\cal
V}_{\acute{a}jk}{\cal V}_{lm}^{A}~~~~~~~~
\end{eqnarray}
\begin{eqnarray}
B_{{\bar c}_i{\bar c}_j{\bar c}_k{\bar c}_l}{\mathit
V}_{c_ic_jc_kc_l}=\frac{1}{2}B_{{\bar c}_i{\bar c}_j{\bar
c}_k{\bar c}_l}\left({\cal V}_{\acute{a} c_i c_jc_mc_n}{\cal
V}_{{\bar c}_m{\bar c}_n c_k c_l}^A +{\cal
V}_{\acute{a}\acute{c}_i c_j{\bar c}_mc_n}{\cal V}_{c_m{\bar
c}_nc_kc_l}^A \right)\nonumber\\
=\frac{1}{144}B^i{\cal V}_{\acute{a}j}{\cal
V}_i^{Aj}-\frac{1}{480}B^i{\cal V}_{\acute{a}i}{\cal
V}^{A}-\frac{1}{48}\epsilon_{ijklm}B^i{\cal
V}^{jkn}_{\acute{a}p}{\cal V}^{Aplm}_{n}\nonumber\\
+\frac{1}{36}\epsilon_{ijklm}B^i{\cal V}^{jkl}_{\acute{a}n}{\cal
V}^{Anm} -\frac{1}{144}\epsilon_{ijklm}B^i{\cal V}^{ij}_A{\cal
V}^{Akl}~~~~~~~~
\end{eqnarray}
\begin{eqnarray}
4B_{c_ic_jc_k{\bar c}_l}{\mathit V}_{{\bar c}_i{\bar c}_j{\bar
c}_k c_l}=B_{c_ic_jc_k{\bar c}_l}\left({\cal V}_{\acute{a} {\bar
c}_i{\bar c}_jc_mc_n}{\cal V}_{{\bar c}_m{\bar c}_n {\bar c}_k
c_l}^A +{\cal V}_{\acute{a}{\bar c}_i{\bar c}_j{\bar c}_m{\bar
c_n}}{\cal V}_{c_m
c_n{\bar c}_kc_l}^A\right.\nonumber\\
\left.+2{\cal V}_{\acute{a}{\bar c}_i{\bar c}_j{\bar c}_m
c_n}{\cal V}_{c_m
{\bar c}_n{\bar c}_kc_l}^A\right)\nonumber\\
=B^{ijk}_l{\cal V}_{\acute{a}ij}^{mn}{\cal
V}_{mnk}^{Al}+2B^{ijk}_l{\cal V}_{\acute{a}kn}^{lm}{\cal
V}_{mij}^{An}-\frac{2}{3}B^{ijk}_l{\cal
V}_{\acute{a}jk}^{lm}{\cal V}_{mi}^{A}\nonumber\\
-\frac{2}{3}B^{ijk}_l{\cal V}_{\acute{a}jkm}^{l}{\cal
V}_{i}^{Am}+\frac{2}{3}B^{ijk}_l{\cal V}_{\acute{a}ijk}^{m}{\cal
V}_{m}^{Al}+\frac{2}{9}B^{ijk}_l{\cal V}_{\acute{a}jk}{\cal
V}_{i}^{Al}\nonumber\\
-\frac{2}{3}B^{ij}{\cal V}_{\acute{a}jlm}^{k}{\cal
V}_{ki}^{Alm}-\frac{2}{3}B^{ij}{\cal V}_{\acute{a}ijl}^{k}{\cal
V}_{k}^{Al}+\frac{7}{9}B^{ij}{\cal V}_{\acute{a}ij}^{kl}{\cal
V}_{kl}^{A}\nonumber\\
+\frac{28}{27}B^{ij}{\cal V}_{\acute{a}jk}{\cal
V}_{i}^{Ak}-\frac{1}{10}B^{ij}{\cal V}_{\acute{a}ij}{\cal
V}^{A}-\frac{1}{24}\epsilon_{ijklm}B^{ijn}_p{\cal
V}_{\acute{a}}^{k}{\cal
V}_{n}^{Almp}\nonumber\\
+\frac{1}{36}\epsilon_{ijklm}B^{ijk}_n{\cal
V}_{\acute{a}}^{l}{\cal
V}^{Amn}-\frac{1}{36}\epsilon_{ijklm}B^{in}{\cal
V}_{\acute{a}}^{j}{\cal
V}^{Aklm}_n+\frac{1}{36}\epsilon_{ijklm}B^{ij}{\cal
V}_{\acute{a}}^{k}{\cal V}^{Alm}\nonumber\\
\end{eqnarray}
\begin{eqnarray}
4B_{{\bar c}_i{\bar c}_j{\bar c}_k c_l}{\mathit V}_{c_ic_jc_k{\bar
c}_l}=B_{{\bar c}_i{\bar c}_j{\bar c}_k c_l}\left({\cal
V}_{\acute{a} c_i c_jc_mc_n}{\cal V}_{{\bar c}_m{\bar c}_n c_k
{\bar c}_l}^A +{\cal V}_{\acute{a} c_ic_j{\bar c}_m{\bar
c_n}}{\cal V}_{c_m
c_nc_k{\bar c}_l}^A\right.\nonumber\\
\left.+2{\cal V}_{\acute{a}\acute{c}_ic_j{\bar c}_m c_n}{\cal
V}_{c_m
{\bar c}_n c_k{\bar c}_l}^A\right)\nonumber\\
=B_{ijk}^l{\cal V}^{Aij}_{mn}{\cal
V}^{mnk}_{\acute{a}l}+2B_{ijk}^l{\cal V}^{Akn}_{lm}{\cal
V}^{mij}_{\acute{a}n}-\frac{2}{3}B_{ijk}^l{\cal
V}^{Ajk}_{lm}{\cal V}^{mi}_{\acute{a}}\nonumber\\
-\frac{2}{3}B_{ijk}^l{\cal V}^{Ajkm}_{l}{\cal
V}^{i}_{\acute{a}m}+\frac{2}{3}B_{ijk}^l{\cal V}^{Aijk}_{m}{\cal
V}^{m}_{\acute{a}l}+\frac{2}{9}B_{ijk}^l{\cal V}^{Ajk}{\cal
V}^{i}_{\acute{a}l}\nonumber\\
-\frac{2}{3}B_{ij}{\cal V}^{Ajlm}_{k}{\cal
V}^{ki}_{\acute{a}lm}-\frac{2}{3}B_{ij}{\cal V}^{Aijl}_{k}{\cal
V}^{k}_{\acute{a}l}+\frac{7}{9}B_{ij}{\cal V}^{Aij}_{kl}{\cal
V}^{kl}_{\acute{a}}\nonumber\\
+\frac{28}{27}B_{ij}{\cal V}^{Ajk}{\cal
V}^{i}_{\acute{a}k}-\frac{1}{10}B_{ij}{\cal V}^{Aij}{\cal
V}_{\acute{a}}-\frac{1}{24}\epsilon^{ijklm}B_{ijn}^p{\cal
V}^{A}_{k}{\cal
V}^{n}_{\acute{a}lmp}\nonumber\\
+\frac{1}{36}\epsilon^{ijklm}B_{ijk}^n{\cal V}^{A}_{l}{\cal
V}_{\acute{a}mn}-\frac{1}{36}\epsilon^{ijklm}B_{in}{\cal
V}^{A}_{j}{\cal
V}_{\acute{a}klm}^n+\frac{1}{36}\epsilon^{ijklm}B_{ij}{\cal
V}^{A}_{k}{\cal V}_{\acute{a}lm}\nonumber\\
\end{eqnarray}
\begin{eqnarray}
6B_{c_ic_j{\bar c}_k{\bar c}_l}{\mathit V}_{{\bar c}_i{\bar
c}_jc_k c_l}=B_{c_ic_j{\bar c}_k{\bar c}_l}\left(\frac{3}{2}{\cal
V}_{\acute{a} {\bar c}_i {\bar c}_jc_mc_n}{\cal V}_{{\bar
c}_m{\bar c}_n c_k c_l}^A +\frac{3}{2}{\cal V}_{\acute{a} {\bar
c}_i{\bar c}_j{\bar c}_m{\bar c_n}}{\cal V}_{c_m
c_nc_k c_l}^A\right.\nonumber\\
\left.+3{\cal V}_{\acute{a}{\bar c}_i{\bar c}_j{\bar c}_m
c_n}{\cal V}_{c_m
{\bar c}_n c_kc_l}^A\right)\nonumber\\
=\frac{3}{2}B_{kl}^{ij}{\cal V}^{Amn}_{ij}{\cal
V}^{kl}_{\acute{a}mn}+2B_{kl}^{ij}{\cal V}^{Aln}_{ij}{\cal
V}^{k}_{\acute{a}n}-\frac{3}{20}B_{kl}^{ij}{\cal
V}^{Akl}_{ij}{\cal
V}_{\acute{a}}\nonumber\\
+2B_{kl}^{ij}{\cal V}^{Akl}_{jm}{\cal
V}^{m}_{\acute{a}i}+\frac{4}{3}B_{kl}^{ij}{\cal V}^{Al}_{j}{\cal
V}^{k}_{\acute{a}i}+2B_{j}^{i}{\cal V}^{Akl}_{im}{\cal
V}_{\acute{a}kl}^{mj}\nonumber\\
+\frac{4}{3}B_{j}^{i}{\cal V}^{Ajk}_{il}{\cal
V}^{l}_{\acute{a}k}-\frac{4}{9}B_{j}^{i}{\cal V}^{Aj}_{k}{\cal
V}^{k}_{\acute{a}i}-\frac{2}{15}B_{j}^{i}{\cal V}^{Aj}_{i}{\cal
V}_{\acute{a}}\nonumber\\
-\frac{3}{20}B{\cal V}^{Aij}_{kl}{\cal
V}^{kl}_{\acute{a}ij}-\frac{2}{15}B{\cal V}^{Ai}_{j}{\cal
V}^{j}_{\acute{a}i}-\frac{3}{200}B{\cal V}^{A}{\cal
V}_{\acute{a}}\nonumber\\
+\frac{1}{48}B_{j}^{i}{\cal V}^{Aj}{\cal
V}_{\acute{a}i}-\frac{1}{480}B{\cal V}^{Ai}{\cal
V}_{\acute{a}i}+3B^{ij}_{kl}{\cal V}^{An}_{ijm}{\cal
V}_{\acute{a}n}^{mkl}\nonumber\\
+2B_{kl}^{ij}{\cal V}^{Al}_{ijm}{\cal
V}_{\acute{a}}^{mk}+2B^{ij}_{kl}{\cal V}^{Aklm}_j{\cal
V}_{\acute{a}mi}+\frac{1}{3}B^{ij}_{kl}{\cal V}^{A}_{ij}{\cal
V}_{\acute{a}}^{kl}\nonumber\\
-4B_{m}^{i}{\cal V}^{Aj}_{ikl}{\cal
V}_{\acute{a}j}^{klm}-\frac{4}{3}B^{i}_{j}{\cal V}^{Aj}_{ikl}{\cal
V}_{\acute{a}}^{kl}-\frac{4}{3}B^{i}_{j}{\cal V}^{Ajkl}_{i}{\cal
V}_{\acute{a}kl}\nonumber\\
+2B_{k}^{i}{\cal V}^{A}_{ij}{\cal
V}_{\acute{a}}^{jk}-\frac{3}{10}B{\cal V}^{Ai}_{jkl}{\cal
V}_{\acute{a}i}^{jkl}-\frac{1}{6}B{\cal V}^{A}_{ij}{\cal
V}_{\acute{a}}^{ij}~~~~~~~~
\end{eqnarray}

\section{Appendix C}
In this appendix we discuss the coupling of the U(1) vector with
matter without imposition of the constraint of the Wess-Zumino
gauge. We consider the following Lagrangian  which couples the
vector multiplet with a scalar multiplet $\widehat{\Phi}$.
\begin{eqnarray}
{\mathsf L}^{(U(1))}={\mathsf L}_{V}^{^{(U(1)~K.E.)}}+{\mathsf
L}_{V}^{^{(U(1)~Mass)}}+ {\mathsf
L}_V^{^{(U(1)~Self-Interaction)}}\nonumber\\
+{\mathsf L}_{V+\Phi}^{^{(U(1)~Interaction)}}+{\mathsf L}_{\Phi}^{(U(1))}~~~~~~\\
\nonumber\\
 {\mathsf
L}_{V}^{^{(U(1)~K.E.)}}=\frac{1}{4}\left[\widehat{\cal
W}^{\tilde{\alpha}} \widehat{\cal
W}_{\tilde{\alpha}}|_{\theta^2}+\widehat{\overline{\cal
W}}_{\dot{\tilde{\alpha}}}\widehat{\overline{\cal
W}}^{\dot{\tilde{\alpha}}}|_{\bar {\theta}^2}\right]~~~~~~\\
{\mathsf L}_{V}^{^{(U(1)~Mass)}}=m^2\widehat{\mathsf
V}^2|_{\theta^2\bar
{\theta}^2}~~~~~~\\
{\mathsf
L}_V^{^{(U(1)~Self-Interaction)}}={\alpha}_1\widehat{\mathsf
V}^3|_{\theta^2\bar{\theta}^2}+{\alpha}_2\widehat{\mathsf
V}^4|_{\theta^2\bar{\theta}^2}~~~~~~\\
{\mathsf
L}_{V+\Phi}^{^{(U(1)~Interaction)}}=h\widehat{\Phi}^{\dagger}_{\acute{a}}\widehat{\mathsf
V}\widehat{\Phi}_{\acute{a}}|_{\theta^2\bar {\theta}^2}~~~~~~\\
{\mathsf
L}_{\Phi}^{(U(1))}=\widehat{\Phi}^{\dagger}_{\acute{a}}\widehat{\Phi}_{\acute{a}}|_{\theta^2\bar
{\theta}^2}+\left[{\mathsf
W}(\widehat{\Phi})|_{\theta^2}+h.c.\right]~~~~~~
\end{eqnarray}
where

\begin{eqnarray}
\widehat{\cal W}_{\tilde{\alpha}}=-\frac{1}{4}\overline {\mathsf
D}^2\mathsf D_{\tilde{\alpha}}\widehat{\mathsf V};~~~~
\widehat{\overline{\cal
W}}_{\dot{\tilde{\alpha}}}=-\frac{1}{4}{\mathsf D}^2\overline
{\mathsf D}_{\dot{\tilde{\alpha}}}\widehat{\mathsf V}
\end{eqnarray}

Finally, the superpotential ${\mathsf W}(\widehat{\Phi})$ of the
theory is
\begin{equation}
{\mathsf W}(\widehat{\Phi})={\cal
F}_{\acute{a}}\widehat{\Phi}_{\acute{a}}+\frac{1}{2} {\cal
M}_{\acute{a}\acute{b}}\widehat{\Phi}_{\acute{a}}\widehat{\Phi}_{\acute{b}}+\frac{1}{3}
{\cal
G}_{\acute{a}\acute{b}\acute{c}}\widehat{\Phi}_{\acute{a}}\widehat{\Phi}_{\acute{b}}\widehat{\Phi}_{\acute{c}}
\end{equation}
The couplings ${\cal M}_{\acute{a}\acute{b}}$ and ${\cal
G}_{\acute{a}\acute{b}\acute{c}}$ are taken to be completely
symmetric tensors.
 Expansion in component form gives
\begin{equation}
{\mathsf L}_{V}^{^{(U(1)~K.E.)}}=-\frac{1}{4}{\cal V}_{AB}{\cal
V}^{AB}+\frac{1}{2}D^2-i\lambda\sigma^A\partial_A\overline
{\lambda}
\end{equation}
\begin{eqnarray}
{\mathsf
L}_{V}^{^{(U(1)~Mass)}}=m^2CD+\frac{1}{2}m^2\left(M^2+N^2\right)-m^2\left(\lambda\chi+
\overline{\lambda}\overline{\chi}\right)-\frac{1}{2}m^2\partial^AC\partial_AC
\nonumber\\
-im^2\chi\sigma^A\partial_A\overline{\chi}-\frac{1}{2}m^2{\cal
V}_{\acute{a}}{\cal V}^A~~~~~~
\end{eqnarray}
\begin{eqnarray}
{\mathsf L}_{V+\Phi}^{^{(U(1)~Interaction)}}=-\frac{h}{2\sqrt
2}\left(\chi\sigma^A\overline{\psi}_{\acute{a}}\right)\partial_AA_{\acute{a}}+\frac{h}{\sqrt
2}
\left(\chi\sigma^A\partial\overline{\psi}_{\acute{a}}\right)A_{\acute{a}}\nonumber\\
-\frac{ih}{\sqrt
2}\left(\overline{\lambda}\overline{\psi}_{\acute{a}}\right)A_{\acute{a}}
-\frac{h}{2\sqrt
2}\left(\psi_{\acute{a}}\sigma^A\overline{\chi}\right)\partial_AA^{\dagger}_{\acute{a}}\nonumber\\
-\frac{h}{\sqrt
2}
\left(\psi_{\acute{a}}\sigma^A\partial\overline{\chi}\right)A^{\dagger}_{\acute{a}}+\frac{ih}{\sqrt
2}\left(\lambda\psi_{\acute{a}}\right)A^{\dagger}_{\acute{a}}\nonumber\\
+\frac{ih}{2}{\cal
V}_A\left[\left(\partial^AA_{\acute{a}}\right)A^{\dagger}_{\acute{a}}-
\left(\partial^AA^{\dagger}_{\acute{a}}\right)A_{\acute{a}}\right]+\frac{h}{2}{\cal
V}_A
\left(\psi_{\acute{a}}\sigma^A\overline{\psi}_{\acute{a}}\right)\nonumber\\
-hC\left(\partial_AA^{\dagger}_{\acute{a}}\right)\left(\partial_AA_{\acute{a}}\right)
-ihC\left(\psi_{\acute{a}}\sigma^A\partial_A\overline{\psi}_{\acute{a}}\right)-
\frac{h}{4}\partial^A\left(A_{\acute{a}}A^{\dagger}_{\acute{a}}\right)\partial_AC\nonumber\\
+hCF_{\acute{a}}F^{\dagger}_{\acute{a}}-\frac{ih}{\sqrt
2}\left(\chi\psi_{\acute{a}}\right)F^{\dagger}_{\acute{a}}
+\frac{ih}{\sqrt
2}\left(\overline{\chi}\overline{\psi}_{\acute{a}}\right)F_{\acute{a}}
+\frac{h}{2}DA_{\acute{a}}A^{\dagger}_{\acute{a}}\nonumber\\
+\frac{ih}{2}\left(M+iN\right)A_{\acute{a}}F^{\dagger}_{\acute{a}}-\frac{ih}{2}\left(M-iN\right)A^{\dagger}_{\acute{a}}
F_{\acute{a}}~~~~~~
\end{eqnarray}
\begin{eqnarray}
{\mathsf
L}_V^{^{(U(1)~Self-Interaction)}}=-3\left(\frac{\alpha_1}{2}C+\alpha_2C^2\right)
\left[{\cal V}_A{\cal V}^A+2\left(\lambda\chi+
\overline{\lambda}\overline{\chi}\right)\right.\nonumber\\
\left.+2i\chi\sigma^A\partial_A\overline{\chi}
-\left(M^2+N^2\right)\right]
+3\left(\frac{\alpha_1}{4}+\alpha_2C\right)\left[i\left(M+iN\right)
\left(\overline{\chi}\overline{\chi}\right)\right.\nonumber\\
\left.-i\left(M-iN\right)\left(\chi\chi\right)
-2\left(\chi\sigma^A\overline{\chi}\right) {\cal V}_A\right] +
\frac{3\alpha_2}{2}
\left(\chi\chi\right)\left(\overline{\chi}\overline{\chi}\right)\nonumber\\
+\left(\frac{3\alpha_1}{4}C+\alpha_2C^2\right)\left[2CD-\partial_AC\partial^AC
\right]
\end{eqnarray}
\begin{eqnarray}
{\mathsf
L}_{\Phi}^{(U(1))}=-\partial_AA^{\dagger}_{\acute{a}}\partial^AA_{\acute{a}}-i\overline{\psi}_{\acute{a}}
\overline{\sigma}^A\partial_A\psi_{\acute{a}}+F^{\dagger}_{\acute{a}}F_{\acute{a}}\nonumber\\
-\left(\frac{1}{2}{\cal M}_{\acute{a}\acute{b}}+{\cal
G}_{\acute{a}\acute{b}\acute{c}}A_{\acute{c}}\right)\psi_{\acute{a}}\psi_{\acute{b}}
-\left(\frac{1}{2}{\cal M}_{\acute{a}\acute{b}}^*+{\cal
G}_{\acute{a}\acute{b}\acute{c}}^*A_{\acute{c}}^{\dagger}
\right)\overline{\psi}_{\acute{a}}\overline{\psi}_{\acute{b}}\nonumber\\
+\left( {\cal F}_{\acute{a}}+{\cal
M}_{\acute{a}\acute{b}}A_{\acute{b}}+{\cal
G}_{\acute{a}\acute{b}\acute{c}}A_{\acute{c}}A_{\acute{c}}\right)F_{\acute{a}}
+\left({\cal F}_{\acute{a}}^*+{\cal
M}_{\acute{a}\acute{b}}^*A_{\acute{b}}^{\dagger}+{\cal
G}_{\acute{a}\acute{b}\acute{c}}^*A_{\acute{c}}^{\dagger}A_{\acute{c}}^{\dagger}
\right)F_{\acute{a}}^{\dagger}~~~~~~
\end{eqnarray}
Evaluation of Eq.(9.64) using Eqs.(9.65) - (9.76) gives in the
four-component notation
\begin{eqnarray}
{\mathsf L}^{(U(1))}= -\frac{1}{4}{\cal V}_{AB}{\cal
V}^{AB}-\frac{1}{2}m^2{\cal V}_{\acute{a}}{\cal
V}^{A}-\frac{1}{2}\partial^AB\partial_AB -i\overline
{\Lambda}\gamma^A\partial_A\Lambda\nonumber\\
-m\overline
{\Lambda}\Lambda-\partial^AA^{\dagger}_{\acute{a}}\partial_AA_{\acute{a}}
 -\frac{h}{m}B\partial^AA^{\dagger}_{\acute{a}}\partial_AA_{\acute{a}}
\nonumber\\
-\frac{h}{4m}\partial^A\left(A_{\acute{a}}A^{\dagger}_{\acute{a}}\right)
\partial_AB+\frac{ih}{2}\left(A^{\dagger}_{\acute{a}}\partial^AA_{\acute{a}}-A_{\acute{a}}\partial^A
A_{\acute{a}}^{\dagger}\right)
{\cal V}_A\nonumber\\
+\frac{h}{2m\sqrt{2}}\left[\left(\overline
{\Psi}_{\acute{a}L}\gamma^A\Lambda_{L}\right)\partial_AA_{\acute{a}}+\left(\overline
{\Lambda}_{L}\gamma^A\Psi_{\acute{a}L}\right)\partial_AA_{\acute{a}}^{\dagger}\right]\nonumber\\
+\frac{h}{m\sqrt{2}}\left[\left(\overline
{\Psi}_{\acute{a}L}\gamma^A\partial_A\Lambda_{L}\right)A_{\acute{a}}-\left(\overline
{\Lambda}_{L}\gamma^A\partial_A\Psi_{\acute{a}L}\right)A_{\acute{a}}^{\dagger}\right]\nonumber\\
-\frac{ih}{\sqrt{2}}\left[\left(\overline
{\Psi}_{\acute{a}L}\Lambda_{R}\right)A_{\acute{a}}-\left(\overline
{\Lambda}_{R}\Psi_{\acute{a}L}\right)A_{\acute{a}}^{\dagger}\right]\nonumber\\
-\left[\left(\frac{1}{2}{\cal M}_{\acute{a}\acute{b}}+{\cal
G}_{\acute{a}\acute{b}\acute{c}}A_{\acute{d}}\right)\overline{\Psi}_{\acute{a}R}\Psi_{\acute{b}L}
+\left(\frac{1}{2}{\cal M}_{\acute{a}\acute{b}}^*+{\cal
G}_{\acute{a}\acute{b}\acute{c}}^*A_{\acute{d}}^{\dagger}
\right)\overline{\Psi}_{\acute{a}L}\Psi_{\acute{b}R}\right]\nonumber\\
-i\left(1+\frac{h}{m}B\right)\overline {\Psi}_{\acute{a}L}\gamma^A
{\cal D}_A\Psi_{\acute{a}L}\nonumber\\
-\frac{1}{m^3}\left(\frac{3\alpha_1}{4}B+\frac{\alpha_2}{m}B^2\right)
\partial_AB\partial^AB\nonumber\\
-\frac{3}{m}\left(\frac{\alpha_1}{2}B+\frac{\alpha_2}{m}B^2\right)
{\cal V}_A{\cal V}^A\nonumber\\
+\frac{3}{m^3}\left(\alpha_1B+\frac{2\alpha_2}{m}B^2\right)
\left(i\overline {\Lambda}_L\gamma^A\partial_A\Lambda_L
-m\overline{\Lambda}\Lambda\right)\nonumber\\
+\frac{3}{m^2}\left(\frac{\alpha_1}{2}+\frac{2\alpha_2}{m}B\right)
\left(\overline{\Lambda}_L\gamma^A\Lambda_L\right){\cal
V}_A\nonumber\\
+ \frac{3\alpha_2}{2m^4}
\left(\overline{\Lambda}^c_R\Lambda_L\right)
\left(\overline{\Lambda}_L\Lambda_R^c\right) +{\mathsf
L}_{auxiliary}^{(U(1))}
\end{eqnarray}
where we have defined
\begin{eqnarray}
\Lambda=\left(\matrix{m\chi_{\tilde{\alpha}}\cr
\overline{\lambda}^{\dot{\tilde\alpha}}}\right),~~~
\Psi_{\acute{a}}=\left(\matrix{\psi_{\acute{a}\tilde{\alpha}}\cr
\overline{\psi}^{\dot{\tilde\alpha}}_{\acute{a}}}\right),~~~
B=mC,\nonumber\\
 \Lambda^c={\cal C}\overline{\Lambda}^T,~~~ {\cal
C}=\left(\matrix{i\sigma^2&0\cr
0&i\overline{\sigma}^2}\right),\nonumber\\
 \overline{\Lambda}=\Lambda^{\dagger}\gamma^0,~~~\Lambda_{R,L}=\frac{1\pm
 \gamma_5}{2}\Lambda,\nonumber\\
 {\cal D}_A=\partial_A-\frac{ig}{2}\left(1+\frac{g}{m}B\right)^{-1}{\cal
V}_A~~~~~~
\end{eqnarray}
and
\begin{eqnarray}
{\mathsf
L}_{auxiliary}^{(U(1))}=\left(mB+\frac{3\alpha_1}{2m^2}B^2
+\frac{2\alpha_2}{m^3}B^3+\frac{h}{2}A^{\dagger}_{\acute{a}}A_{\acute{a}}\right)D
+\frac{1}{2}D^2\nonumber\\
+\left(\frac{1}{2}m^2+\frac{3\alpha_1}{2m}B+\frac{3\alpha_2}{m^2}B^2\right)
\left(M^2+N^2\right)\nonumber\\
+i\left[\frac{h}{2}A_{\acute{a}}F^{\dagger}_{\acute{a}}+\frac{3}{m^2}\left(\frac{\alpha_1}{4}+
\frac{\alpha_2}{m}B\right)\left(\overline{\Lambda}_L\Lambda_R^c\right)
\right]\left(M+iN\right)\nonumber\\
-i\left[\frac{h}{2}A^{\dagger}_{\acute{a}}F_{\acute{a}}+\frac{3}{m^2}\left(\frac{\alpha_1}{4}+
\frac{\alpha_2}{m}B\right)\left(\overline{\Lambda}^c_R\Lambda_L\right)
\right]\left(M-iN\right)\nonumber\\
+\left[{\cal F}_{\acute{a}}+{\cal
M}_{\acute{a}\acute{b}}A_{\acute{b}}+{\cal
G}_{\acute{a}\acute{b}\acute{c}}A_{\acute{b}}A_{\acute{c}}
 +\frac{ih}{m\sqrt 2}\left(\overline{\Lambda}_{L}\Psi_{\acute{a}R}\right)\right]F_{\acute{a}}
\nonumber\\
+\left[{\cal F}_{\acute{a}}^*+{\cal
M}_{\acute{a}\acute{b}}^*A_{\acute{b}}^{\dagger}+{\cal
G}_{\acute{a}\acute{b}\acute{c}}^*A_{\acute{c}}^{\dagger}A_{\acute{d}}^{\dagger}-\frac{ih}{m\sqrt
2}
\left(\overline{\Psi}_{\acute{a}R}\Lambda_L\right)\right]F_{\acute{a}}^{\dagger}\nonumber\\
+\left(\frac{h}{m}B+1\right)F^{\dagger}_{\acute{a}}F_{\acute{a}}
\end{eqnarray}

We next eliminate the auxiliary fields $M$, $N$, and $D$ through
their field equations  to get
\begin{eqnarray}
{\mathsf
L}_{auxiliary}^{(U(1))}=-\frac{1}{2}m^2B^2-\frac{3\alpha_1}{2m}B^3
-\left(\frac{9{\alpha_1}^2}{8m^4}+\frac{2\alpha_2}{m^2}\right)B^4\nonumber\\
-\frac{3\alpha_1\alpha_2}{m^5}B^5-\frac{2{\alpha_2}^2}{m^6}B^6
-\frac{h^2}{8}\left(A^{\dagger}_{\acute{a}}A_{\acute{a}}\right)\left(A^{\dagger}_{\acute{b}}A_{\acute{b}}\right)\nonumber\\
-\frac{mh}{2}B\left(A^{\dagger}_{\acute{a}}A_{\acute{a}}\right)
-\frac{3h\alpha_1}{4m^2}B^2\left(A^{\dagger}_{\acute{a}}A_{\acute{a}}\right)
-\frac{h\alpha_2}{m^3}B^3\left(A^{\dagger}_{\acute{a}}A_{\acute{a}}\right)\nonumber\\
+ \left[{\cal F}_{\acute{a}}^*+{\cal
M}_{\acute{a}\acute{b}}^*A_{\acute{b}}^{\dagger}+{\cal
G}_{\acute{a}\acute{b}\acute{c}}^*A_{\acute{b}}^{\dagger}A_{\acute{c}}^{\dagger}-\frac{ih}{m\sqrt
2} \left(\overline{\Psi}_{\acute{a}R}\Lambda_L\right)
-\frac{hf_1(B)}{f_2(B)}
\left(\overline{\Lambda}^c_R\Lambda_L\right)\right]F_{\acute{a}}^{\dagger}\nonumber\\
-\frac{2f_1^2(B)}{f_2(B)}
\left(\overline{\Lambda}^c_R\Lambda_L\right)
\left(\overline{\Lambda}_L\Lambda_R^c\right)~~~~~~
\end{eqnarray}
where
\begin{eqnarray}
f_1(B)=\frac{3}{m^2}\left(\frac{\alpha_1}{4}+\frac{\alpha_2}{m}B\right),~~
f_2(B)=m^2+\frac{3\alpha_1}{m}B+\frac{6\alpha_2}{m^2}B^2
\end{eqnarray}
and the auxiliary field $F$ satisfies the field equation
\begin{eqnarray}
F^{\dagger}_{\acute{b}}\left[\delta_{\acute{a}\acute{b}}\left(1+\frac{h}{m}B\right)-\frac{h^2}{2f_2(B)}
A^{\dagger}_{\acute{a}}A_{\acute{b}}\right]= -\frac{ih}{m\sqrt
2}\overline{\Lambda}_{L}\Psi_{\acute{a}R}\nonumber\\
+\frac{hf_1(B)}{f_2(B)}\left(\overline{\Lambda}_L\Lambda_R^c\right)
A^{\dagger}_{\acute{a}} -{\cal F}_{\acute{a}}-{\cal
M}_{\acute{a}\acute{b}}A_{\acute{b}}-{\cal
G}_{\acute{a}\acute{b}\acute{c}}A_{\acute{b}}A_{\acute{c}}
\end{eqnarray}

Inverting this last equation we obtain
\begin{eqnarray}
F^{\dagger}_{\acute{a}}=
\left(1+\frac{h}{m}B\right)^{-1}\left[\delta_{\acute{a}\acute{b}}+
\frac{h^2A_{\acute{a}}^{\dagger}A_{\acute{b}}}{2f_2^2(B)\left(1+\frac{h}{m}B\right)-h^2A_{\acute{c}}^{\dagger}A_{\acute{c}}}
\right]\nonumber\\
\times \left[-\frac{ih}{m\sqrt
2}\overline{\Lambda}_{L}\Psi_{\acute{b}R}+
\frac{hf_1(B)}{f_2(B)}\left(\overline{\Lambda}_L\Lambda_R^c\right)
A^{\dagger}_{\acute{b}} -{\cal F}_{\acute{b}}-{\cal
M}_{\acute{b}\acute{d}}A_{\acute{d}}-{\cal
G}_{\acute{b}\acute{d}e}A_{\acute{d}}A_e \right]~~~~~~
\end{eqnarray}

For the case when self-interactions of the vector multiplet are
absent (i.e., $\alpha_1=\alpha_2=0$), we get
\begin{eqnarray}
{\mathsf L}^{(U(1))}= -\frac{1}{4}{\cal V}_{AB}{\cal
V}^{AB}-\frac{1}{2}m^2{\cal V}_{\acute{a}}{\cal
V}^{A}-\frac{1}{2}\partial^AB\partial_AB -i\overline
{\Lambda}\gamma^A\partial_A\Lambda-m\overline
{\Lambda}\Lambda\nonumber\\
-\partial^AA^{\dagger}_{\acute{a}}\partial_AA_{\acute{a}}
 -\frac{h}{m}B\partial^AA^{\dagger}_{\acute{a}}\partial_AA_{\acute{a}}
\nonumber\\
-\frac{h}{4m}\partial^A\left(A_{\acute{a}}A^{\dagger}_{\acute{a}}\right)
\partial_AB+\frac{ih}{2}\left(A^{\dagger}_{\acute{a}}\partial^AA_{\acute{a}}-A_{\acute{a}}\partial^A
A_{\acute{a}}^{\dagger}\right)
{\cal V}_A\nonumber\\
+\frac{h}{2m\sqrt{2}}\left[\left(\overline
{\Psi}_{\acute{a}L}\gamma^A\Lambda_{L}\right)\partial_AA_{\acute{a}}+\left(\overline
{\Lambda}_{L}\gamma^A\Psi_{\acute{a}L}\right)\partial_AA_{\acute{a}}^{\dagger}\right]\nonumber\\
+\frac{h}{m\sqrt{2}}\left[\left(\overline
{\Psi}_{\acute{a}L}\gamma^A\partial_A\Lambda_{L}\right)A_{\acute{a}}-\left(\overline
{\Lambda}_{L}\gamma^A\partial_A\Psi_{\acute{a}L}\right)A_{\acute{a}}^{\dagger}\right]\nonumber\\
-\frac{ih}{\sqrt{2}}\left[\left(\overline
{\Psi}_{\acute{a}L}\Lambda_{R}\right)A_{\acute{a}}-\left(\overline
{\Lambda}_{R}\Psi_{\acute{a}L}\right)A_{\acute{a}}^{\dagger}\right]\nonumber\\
-\left[\left(\frac{1}{2}{\cal M}_{\acute{a}\acute{b}}+{\cal
G}_{\acute{a}\acute{b}\acute{c}}A_{\acute{c}}\right)\overline{\Psi}_{\acute{a}R}\Psi_{\acute{b}L}
+\left(\frac{1}{2}{\cal M}_{\acute{a}\acute{b}}^*+{\cal
G}_{\acute{a}\acute{b}\acute{c}}^*A_{\acute{c}}^{\dagger}
\right)\overline{\Psi}_{\acute{a}L}\Psi_{\acute{b}R}\right]\nonumber\\
-i\left(1+\frac{h}{m}B\right)\overline {\Psi}_{\acute{a}L}\gamma^A
{\cal D}_A\Psi_{\acute{a}L}\nonumber\\
 -\frac{1}{2}m^2B^2-\frac{h^2}{8}\left(A^{\dagger}_{\acute{a}}A_{\acute{a}}\right)
\left(A^{\dagger}_{\acute{b}}A_{\acute{b}}\right)-\frac{h}{2}mB\left(A^{\dagger}_{\acute{a}}A_{\acute{a}}\right)
\nonumber\\
-\left[\frac{\delta_{\acute{a}\acute{b}}}{1+\frac{h}{m}B}+
\frac{h^2A_{\acute{a}}^{\dagger}A_{\acute{b}}}{2m^2-h^2\left(1+\frac{h}{m}B\right)A_{\acute{c}}^{\dagger}A_{\acute{c}}}
\right]\nonumber\\
\times \left[{\cal F}_{\acute{a}}+{\cal
M}_{\acute{a}d}A_{\acute{d}}+{\cal
G}_{\acute{a}de}A_{\acute{d}}A_e +\frac{ih}{m\sqrt
2}\overline{\Lambda}_{L}\Psi_{\acute{a}R}
\right]\nonumber\\
\times \left[{\cal F}_{\acute{b}}^*+{\cal
M}_{\acute{b}f}^*A_f^{\dagger}+{\cal
G}_{\acute{b}fg}^*A_f^{\dagger}A_g^{\dagger}-\frac{ih}{m\sqrt 2}
\left(\overline{\Psi}_{\acute{b}R}\Lambda_L\right)\right]~~
\end{eqnarray}
As is evident the $U(1)$ invariant effective Lagrangian above is
highly nonlinear with infinite order nonlinearities.

\chapter{Conclusions}

\vspace{-0.15in}

In conclusion, we have developed a technique which allows the
explicit computation of the $SO(2N)$ invariant couplings in terms
of $SU(N)$ invariant couplings. The technique is specially useful
in the analysis of interactions involving large tensor
representations. We have illustrated the technique by carrying out
a complete analysis of the $SO(10)$ invariant superpotential at
the trilinear level involving interactions of matter with Higgs
which consists of the $ 16_{(+)\acute{a}}\times
16_{(+)\acute{b}}\times 10_s$, $16_{(+)\acute{a}}\times
16_{(+)\acute{b}}\times 120_{as}$, $16_{+)\acute{a}}\times
16_{(+)\acute{b}}\times \overline{126}_s$, ${16}_{(\pm)\acute{a}}
 \times{16}_{(\mp)\acute{b}}\times 1$, ${16}_{(\pm)\acute{a}}
 \times{16}_{(\mp)\acute{b}}\times 45$, and ${16}_{(\pm)\acute{a}}
 \times{16}_{(\mp)\acute{b}}\times 210$ couplings in their $SU(5)$
 decomposed form, where $16_{+}\equiv 16$ and $16_{-}\equiv\overline{16}$.
 It would be very straightforward now to expand all the $SU(5)$
invariants in terms of $SU(3)_C\times SU(2)_L\times U(1)_Y$
invariants using the particle assignments. We note that the
decomposition of $SO(10)$ into multiplets of  $SU(5)$ is merely a
convenient device for expanding the $SO(10)$ interaction in a
compact form and does not necessarily imply a preference for the
symmetry breaking pattern. Indeed one can compute the SO(10)
interactions using the technique used in this thesis and then use
any symmetry breaking scheme one wishes to get to the low energy
theory. An analysis of vector couplings in the Lagrangian
${16}_{(\pm)\acute{a}}^{\dagger}\times 16_{(\pm)\acute{b}}\times
1$, ${16}_{(\pm)\acute{a}}^{\dagger}\times
16_{(\pm)\acute{b}}\times 45$ and
${16}_{(\pm)\acute{a}}^{\dagger}\times 16_{(\pm)\acute{b}}\times
210$ was also given. The technique discussed in this thesis is
easily extendible to models with $SO(2N+1)$ invariance.

Further, we have computed  quartic interactions in
 the superpotential and Lagrangian of the type
$[{{16}}_{\acute{a}(-)}\times{16}_{\acute{b}(+)}]_1[{16}_{(-)\acute{c}}\times{16}_{(+)\acute{d}}]_1$,
$[{{16}}_{\acute{a}(-)}\times{16}_{\acute{b}(+)}]_{45}[{16}_{(-)\acute{c}}\times{16}_{(+)\acute{d}}]_{45}$,
$[{{16}}_{\acute{a}(-)}\times{16}_{\acute{b}(+)}]_{210}[{16}_{(-)\acute{c}}\times{16}_{(+)\acute{d}}]_{210}$,
by eliminating the SO(10) tensor directly . We also exhibited a
technique which is much simpler and involves elimination of heavy
fields in cubic couplings.

Further, in this thesis we computed the supersymmetric vector
multiplet couplings containing the 1 and 45. We have also given a
computation of the couplings of the 210 dimensional SO(10) vector
mutliplet. Specifically, we have computed the vector couplings
$\overline{16}_{\pm}-16_{\pm}-210$ in terms of its $SU(5)\times
U(1)$ decomposition. We approached this coupling from two view
points. First, we use the conventional approach of using the
Wess-Zumino gauge. However, since the $210$ couplings are not
expected to be  gauge invariant and hence such interactions are
not expected to be renormalizable, we  also consider a nonlinear
sigma model type couplings of $210$ with matter. Such couplings
arise when we consider the  full $210$ multiplet without using the
Wess-Zumino gauge.  Here elimination of the auxiliary fields leads
to interactions of the vector multiplet with the chiral fields
with nonlinearities of infinite order as in a nonlinear sigma
model.
 Although couplings of the type discussed do not thus far
 appear in theories of fundamental interactions, 210 vector multiplet
 may arise as a condensate in effective theories.

Very recently the techniques discussed here have been used in the
formulation of a new class of $SO(10)$ theories where one can
achieve spontaneous breaking of an $SO(10)$ gauge group down to
$SU(3)_C\times SU(2)_L \times U(1)_Y$ in a single step\cite{ns5}.
The analysis utilizes a 144 plet of Higgs which is a spinor vector
$|\Psi_{\mu}>$ with a constraint, that is
$\Gamma_{\mu}|\Psi_{\mu}>=0$. In model of Ref.\cite{ns5}, one
needs to fine tune the Higgs doublet mass to be light which
requires a string type landscape type scenario for its
implementation\cite{landscape1,landscape2}. The techniques
developed here are found very useful in such analyses.

\end{document}